\pdfoutput=1

\documentclass[11pt,twoside,a4paper,cmspaper,final,collab]{cms-tdr}
\def\svnVersion{296813}\def\cmsCernNo{2013-037}\def\cmsCernDate{\today}\def\cmsMessage{Published in Physical Review D as \href{http://dx.doi.org/10.1103/PhysRevD.92.012004}{\doi{10.1103/PhysRevD.92.012004.}}}

\begin{document}\cmsNoteHeader{HIG-14-018}

\hyphenation{had-ron-i-za-tion}
\hyphenation{cal-or-i-me-ter}
\hyphenation{de-vices}
\RCS$Revision: 296812 $
\RCS$HeadURL: svn+ssh://alverson@svn.cern.ch/reps/tdr2/papers/HIG-14-018/trunk/HIG-14-018.tex $
\RCS$Id: HIG-14-018.tex 296812 2015-07-16 15:19:27Z alverson $
\newlength\cmsFigWidth
\ifthenelse{\boolean{cms@external}}{\setlength\cmsFigWidth{0.35\textwidth}}{\setlength\cmsFigWidth{0.5\textwidth}}
\ifthenelse{\boolean{cms@external}}{\providecommand{\cmsLeft}{top\xspace}}{\providecommand{\cmsLeft}{left\xspace}}
\ifthenelse{\boolean{cms@external}}{\providecommand{\cmsRight}{bottom\xspace}}{\providecommand{\cmsRight}{right\xspace}}
\ifthenelse{\boolean{cms@external}}{\providecommand{\CL}{C.L.\xspace}}{\providecommand{\CL}{CL\xspace}}
\ifthenelse{\boolean{cms@external}}{\providecommand{\CLend}{C.L.\xspace}}{\providecommand{\CLend}{CL.\xspace}}
\newcommand{\CLs}{\ensuremath{CL_\mathrm{s}}\xspace}

\newcommand\sss{} 
\newcommand{\F}{\ensuremath{\cmsSymbolFace{F}}}
\newcommand{\V}{\ensuremath{\cmsSymbolFace{V}}\xspace}
\newcommand{\VVV}{\cmsSymbolFace{VVV}\xspace}
\newcommand{\Wjets}{\ensuremath{\PW\text{+jets}}\xspace}
\newcommand{\X}{\ensuremath{\cmsSymbolFace{X}}\xspace}
\newcommand{\delphill}{\ensuremath{\Delta\phi_{\ell\ell}}}
\newcommand{\delphillmet}{\ensuremath{\Delta\phi(\ell\ell,\VEtmiss)}}
\newcommand{\dyll}{\ensuremath{\cPZ/\gamma^*\to \ell^+\ell^-}\xspace}
\newcommand{\dymm}{\ensuremath{\cPZ/\gamma^*\to\Pgmp\Pgmm}}
\newcommand{\dytt}{\ensuremath{\cPZ/\gamma^* \to\tau^+\tau^-}}
\newcommand{\jp}{\ensuremath{J^{P}}\xspace}
\newcommand{\mll}{\ensuremath{m_{\ell\ell}}\xspace}
\providecommand{\mt}{\ensuremath{m_\mathrm{T}}\xspace}
\newcommand{\psvectorKD}{\ensuremath{\mathcal{D}_{1^-}}\xspace}
\newcommand{\ptl}{\ensuremath{p_\perp^{\ell}}\xspace}
\newcommand{\ptll}{\ensuremath{\pt^{\ell\ell}}\xspace}
\newcommand{\qq}{\ensuremath{\Pq\Pq}\xspace}
\providecommand{\qqbar}{\ensuremath{\Pq\Paq}\xspace}
\newcommand{\superKD}{\ensuremath{\mathcal{D}_\text{bkg}} }
\newcommand{\tw}{\ensuremath{\cPqt\PW}\xspace}
\newcommand{\usedLumiA}{5.1\fbinv}
\newcommand{\usedLumiB}{19.7\fbinv}
\newcommand{\vectorKD}{\ensuremath{\mathcal{D}_{1^+}} }
\newcommand{\wgamma}{\ensuremath{\PW\gamma}\xspace}

\cmsNoteHeader{HIG-14-018}
\title{Constraints on the spin-parity and anomalous \texorpdfstring{$\PH\V\V$ couplings of the Higgs boson in proton collisions at 7 and 8\TeV}{HVV couplings of the Higgs boson in proton collisions at 7 and 8 TeV}}
\date{\today}

\abstract{
The study of the spin-parity and tensor structure of the interactions of the recently discovered Higgs boson
is performed using the $\PH\to \cPZ\cPZ,\cPZ\gamma^*,\gamma^*\gamma^*\to 4\ell$,
$\PH\to \PW\PW\to\ell\Pgn\ell\Pgn$, and $\PH\to\gamma\gamma$ decay modes.
The full dataset recorded by the CMS experiment during the LHC Run 1 is used, corresponding to an integrated luminosity of
up to 5.1\fbinv at a center-of-mass energy of 7\TeV and up to 19.7\fbinv at 8\TeV.
A wide range of spin-two models is excluded at a 99\% confidence level or higher,
or at a 99.87\% confidence level for the minimal gravity-like couplings,
regardless of whether assumptions are made on the production mechanism.
Any mixed-parity spin-one state is excluded in the $\cPZ\cPZ$ and $\PW\PW$ modes at a greater than 99.999\% confidence level.
Under the hypothesis that the resonance is a spin-zero boson, the tensor structure of the interactions of the Higgs boson
with two vector bosons $\cPZ\cPZ$, $\cPZ\gamma$, $\gamma\gamma$, and $\PW\PW$ is investigated and limits on eleven anomalous
contributions are set. Tighter constraints on anomalous $\PH\V\V$ interactions are obtained by combining the $\PH\cPZ\cPZ$ and $\PH\PW\PW$
measurements. All observations are consistent with the expectations for the standard model Higgs boson
with the quantum numbers $J^{PC}=0^{++}$.
}

\hypersetup{%
pdfauthor={CMS Collaboration},%
pdftitle={Constraints on the spin-parity and anomalous HVV couplings of the Higgs boson in proton collisions at 7 and 8 TeV},%
pdfsubject={CMS},%
pdfkeywords={Higgs, exotic spin, anomalous coupling}}

\maketitle

\ifthenelse{\boolean{cms@external}}{}{
\tableofcontents
\clearpage
}
\section{Introduction} \label{sec:Introduction}
The observation of a new boson~\cite{Aad:2012tfa,Chatrchyan:2012ufa,Chatrchyan:2013lba}
with a mass around $125\GeV$ and properties
consistent with the standard model (SM) Higgs boson~\cite{StandardModel67_1,
Englert:1964et,Higgs:1964ia,Higgs:1964pj,Guralnik:1964eu,StandardModel67_2,StandardModel67_3}
was reported by the ATLAS and CMS Collaborations in 2012. The discovery was followed by a comprehensive
set of measurements~\cite{Chatrchyan:2012jja,Chatrchyan:2013mxa,CMS-HIG-14-002,Chatrchyan:2013iaa,Khachatryan:2014ira,
Chatrchyan:2013zna,Chatrchyan:2014nva,Chatrchyan:2013yea,Chatrchyan:2013vaa,Chatrchyan:2014tja,Khachatryan:2014jba,
Aad:2013xqa,Aad:2013wqa,Aad:2014aba,Aad:2014fia,Aad:2014iia,Aad:2014eha}
of its properties to determine if the new boson follows the SM predictions or if there are
indications for physics beyond the SM (BSM).

The CMS experiment analyzed the full dataset collected during the CERN LHC Run 1 and measured
the properties of the Higgs-like boson, $\PH$, using its decay modes to two electroweak gauge bosons
$\PH\to\Z\Z\to4\ell$~\cite{Chatrchyan:2012jja, Chatrchyan:2013mxa,CMS-HIG-14-002},
$\PH\to\PW\PW\to\ell\nu\ell\nu$~\cite{Chatrchyan:2013iaa},
and $\PH\to\gamma\gamma$~\cite{Khachatryan:2014ira},
where $\ell$ denotes $\Pe^\pm$ or $\Pgm^\pm$, and $\PW\PW$ denotes $\PWp\PWm$.
The results showed that the spin-parity properties of the new boson are consistent with
the expectations for the scalar SM Higgs boson. In particular, the hypotheses of a pseudoscalar,
vector, and pseudovector boson were excluded at a 99.95\% confidence level (\CL) or higher,
and several spin-two boson hypotheses were excluded at a 98\% \CL or higher.
The investigated spin-two models included two bosons with graviton-like interactions
and two bosons with higher-dimension operators and opposite parity.
The spin-zero results included the first constraint of the $f_{a3}$ parameter, which probes the tensor
structure of the $\PH\Z\Z$ interactions and is defined as the fractional pseudoscalar cross section,
with $f_{a3}=1$ corresponding to the pure pseudoscalar hypothesis.
The ATLAS experiment has also excluded at a 98\% \CL or higher the hypotheses of
a pseudoscalar, vector, pseudovector, and graviton-inspired spin-two boson with minimal couplings
and several assumptions on the boson production mechanisms~\cite{Aad:2013xqa}.

In this paper, an extended study of the spin-parity properties of the Higgs boson and
of the tensor structure of its interactions with electroweak gauge bosons is presented using the
$\PH\to\Z\Z$, $\Z\gamma^*$, $\gamma^*\gamma^*\to4\ell$,
where the interference between the three intermediate states is included,
and $\PH\to\PW\PW\to\ell\nu\ell\nu$~decay modes at the CMS experiment.
The study focuses on testing for the presence of anomalous effects in $\PH\Z\Z$ and $\PH\PW\PW$
interactions under spin-zero, -one, and -two hypotheses.
The  $\PH\Z \gamma$ and $\PH\gamma\gamma$ interactions are probed for the first time using the $4\ell$ final state.
Constraints are set on eleven anomalous coupling contributions to the $\PH\V\V$ interactions,
where $\V$ is a gauge vector boson, under the spin-zero assumption of the Higgs boson,
extending the original measurement of the $f_{a3}$ parameter~\cite{Chatrchyan:2012jja, Chatrchyan:2013mxa}.
The exotic-spin study is extended to the analysis of mixed spin-one states, beyond the pure parity states studied
earlier~\cite{Chatrchyan:2013mxa, Chatrchyan:2013iaa}, and ten spin-two hypotheses of the boson under the assumption
of production either via gluon fusion or quark-antiquark annihilation, or without such an assumption.
This corresponds to thirty spin-two models,
beyond the six production and decay models studied earlier~\cite{Chatrchyan:2012jja, Chatrchyan:2013mxa, Chatrchyan:2013iaa}.
The $\PH\to\gamma\gamma$ decay channel is also studied in the context of exotic spin-two
scenarios, and the results presented in Ref.~\cite{Khachatryan:2014ira} are combined with those
obtained in the $\PH\to\Z\Z$ and $\PH\to\PW\PW$ channels~\cite{Chatrchyan:2013mxa,Chatrchyan:2013iaa}.

The experimental approaches used here are similar to those used by CMS to study the spin-parity and
other properties of the new
resonance~\cite{Chatrchyan:2012jja, Chatrchyan:2013mxa,CMS-HIG-14-002,Chatrchyan:2013iaa,Khachatryan:2014ira},
and use the techniques developed for such
measurements~\cite{Gao:2010qx, Bolognesi:2012mm, Chen:2012jy, Anderson:2013afp, Chen:2013waa, Chen:2014pia}.
The analysis is based on theoretical and phenomenological studies that describe
the couplings of a Higgs-like boson to two gauge bosons. They provide techniques and ideas for measuring the spin
and $CP$ properties of a particle interacting with vector bosons~\cite{Dell'Aquila:1985vb,
Nelson:1986ki,Soni:1993jc,Chang:1993jy,Barger:1993wt,Arens:1994wd,Han:2000mi, Plehn:2001nj,Choi:2002jk,Buszello:2002uu,
Accomando:2006ga,Godbole:2007cn,Antipin:2008hj,Keung:2008ve,Hagiwara:2009wt,Gao:2010qx,DeRujula:2010ys,Christensen:2010pf,
Gainer:2011xz,Bolognesi:2012mm,Ellis:2012xd,Chen:2012jy,Gainer:2013rxa,Artoisenet:2013puc,Anderson:2013afp,
Chen:2013waa,Chen:2013ejz,Chen:2014pia,Chen:2014gka,Gainer:2014hha}.
Historically, such techniques have been applied to the analysis of meson decays to four-body final
states~\cite{Samios:1962zza,Cabibbo:1965zz,Kramer:1991xw,Abouzaid:2008cd,Aubert:2008zza}.

The paper is organized as follows. First, the phenomenology of spin-parity and anomalous $\PH\V\V$ interactions
is described in Section~\ref{sec:Pheno}. The experimental apparatus, simulation,
and reconstruction techniques are discussed in Section~\ref{sec:CMS}.
The analysis techniques are introduced in Section~\ref{sec:AnalysisStrategyIntro}.
The exclusion of exotic spin-one and spin-two scenarios is shown in Section~\ref{sec:ResultsExotic}.
Finally, for the spin-zero scenario, comprehensive studies of the tensor structure of $\PH\V\V$ interactions
are presented in Section~\ref{sec:ResultsSpinZero}. The results are summarized in Section~\ref{sec:Conclusion}.

\section{Phenomenology of spin-parity and anomalous \texorpdfstring{$\PH\V\V$}{HVV} interactions} \label{sec:Pheno}
The production and decay of $\PH$ is described by its interactions with a pair of vector bosons $\V\V$,
such as $\Z\Z, \Z \gamma, \gamma\gamma,\PW\PW$, and $\Pg\Pg$, or with a fermion-antifermion pair.
The relevant phenomenology for the interactions of a spin-zero, -one, and -two boson, as motivated by earlier
studies~\cite{Gao:2010qx, Bolognesi:2012mm, Gainer:2013rxa, Anderson:2013afp, Chen:2013waa, Chen:2014pia},
is presented below. In the following, the spin-parity state is generically denoted as $J^P$, with $J=0,1,$ or $2$,
while the quantum numbers of the SM Higgs boson are expected to be $J^{PC}=0^{++}$. However, the interactions
of the observed state do not necessarily conserve $C$-parity or $CP$-parity, and the general scattering amplitudes
describe the spin-parity properties of the new boson and its anomalous couplings with a pair of vector bosons
or fermions.

\subsection{Decay of a spin-zero resonance} \label{sec:spin0}

The scattering amplitude describing the interaction between a spin-zero $\PH$ and two spin-one gauge bosons $\V\V$,
such as $\Z\Z, \Z \gamma, \gamma\gamma, \PW\PW$, or $\Pg\Pg$, includes only three independent invariant tensor structures
with the coupling parameters $a_{i}^{\V\V}$ that can have both real and imaginary parts and in general are form factors
which can depend on the squared Lorentz invariant four-momenta of $\V_1$ and $\V_2$, $q_{\sss\V1}^2$ and $q_{\sss\V2}^2$.
In the following, the terms up to $q_{\sss\V}^2$  are kept in the expansion under the assumption of small contributions from
anomalous couplings
\ifthenelse{\boolean{cms@external}}{
\begin{multline}
A(\PH\V\V) \sim
\left[ a_{1}^{\V\V}
+ \frac{\kappa_1^{\V\V}q_{\sss\V1}^2 + \kappa_2^{\V\V} q_{\sss\V2}^{2}}{\left(\Lambda_{1}^{\V\V} \right)^{2}} \right]
m_{\sss\V1}^2 \epsilon_{\sss\V1}^* \epsilon_{\sss\V2}^*\\
+ a_{2}^{\V\V}  f_{\mu \nu}^{*(1)}f^{*(2),\mu\nu}
+ a_{3}^{\V\V}   f^{*(1)}_{\mu \nu} {\tilde f}^{*(2),\mu\nu},
\label{eq:formfact-fullampl-spin0}
\end{multline}
}{
\begin{equation}
A(\PH\V\V) \sim
\left[ a_{1}^{\V\V}
+ \frac{\kappa_1^{\V\V}q_{\sss\V1}^2 + \kappa_2^{\V\V} q_{\sss\V2}^{2}}{\left(\Lambda_{1}^{\V\V} \right)^{2}} \right]
m_{\sss\V1}^2 \epsilon_{\sss\V1}^* \epsilon_{\sss\V2}^*
+ a_{2}^{\V\V}  f_{\mu \nu}^{*(1)}f^{*(2),\mu\nu}
+ a_{3}^{\V\V}   f^{*(1)}_{\mu \nu} {\tilde f}^{*(2),\mu\nu},
\label{eq:formfact-fullampl-spin0}
\end{equation}
}
where $f^{(i){\mu \nu}} = \epsilon_{{\sss\V}i}^{\mu}q_{{\sss\V}i}^{\nu} - \epsilon_{{\sss\V}i}^\nu q_{{\sss\V}i}^{\mu} $
is the field strength tensor of a gauge boson with momentum $q_{{\sss\V}i}$ and polarization
vector $\epsilon_{{\sss\V}i}$, ${\tilde f}^{(i)}_{\mu \nu} = \frac{1}{2} \epsilon_{\mu\nu\rho\sigma} f^{(i),\rho\sigma}$
is the dual field strength tensor, the superscript~$^*$ designates a complex conjugate,
$m_{\sss\V1}$ is the pole mass of the $\Z$ or $\PW$ vector boson, while the cases with the massless vector bosons
are discussed below, and $\Lambda_{1}$ is the scale of BSM physics and is a free parameter
of the model~\cite{Anderson:2013afp}.
A different coupling in the scattering amplitude in Eq.~(\ref{eq:formfact-fullampl-spin0})
typically leads to changes of both the observed rate and the kinematic distributions of the process.
However, the analysis presented in this paper does not rely on any prediction of the overall rate
and studies only the relative contributions of different tensor structures.

In Eq.~(\ref{eq:formfact-fullampl-spin0}), $\V\V$ stands for $\Z\Z, \Z \gamma, \gamma\gamma, \PW\PW$, and $\Pg\Pg$.
The tree-level SM-like contribution corresponds to $a_{1}^{\Z\Z}\ne 0$ and
$a_{1}^{\PW\PW} \ne 0$, while there is no tree-level coupling to massless gauge bosons,
that is $a_{1}^{\V\V}= 0$ for $\Z \gamma, \gamma\gamma$, and $\Pg\Pg$.
Small values of the other couplings can be generated through loop effects in the SM, but their SM
values are not accessible experimentally with the available data.
Therefore, the other terms can be ascribed to anomalous couplings which are listed
for $\PH\Z\Z, \PH\PW\PW, \PH\Z\gamma,$ and $\PH\gamma\gamma$ in Table~\ref{tab:xsec_ratio} .
Among those, considerations of symmetry and gauge invariance require
$\kappa_1^{\Z\Z}=\kappa_2^{\Z\Z}=-\exp({i\phi^{\Z\Z}_{\Lambda{1}}})$,
$\kappa_1^{\gamma\gamma}=\kappa_2^{\gamma\gamma}=0$,
$\kappa_1^{\Pg\Pg}=\kappa_2^{\Pg\Pg}=0$,
$\kappa_1^{\Z\gamma}=0$ and $\kappa_2^{\Z\gamma}=-\exp({i\phi^{\Z\gamma}_{\Lambda{1}}})$.
While not strictly required, the same symmetry is considered in the $\PW\PW$ case
$\kappa_1^{\PW\PW}=\kappa_2^{\PW\PW}=-\exp({i\phi^{\PW\PW}_{\Lambda{1}}})$.
In the above, $\phi^{\V\V}_{\Lambda{1}}$ is the phase of the anomalous coupling
with $\Lambda_{1}^{\V\V}$, which is either 0 or $\pi$ for real couplings.
In the following, the $\Z\Z$ labels for the $\Z\Z$ interactions will be omitted,
and therefore the couplings $a_{1}$, $a_{2}$, $a_{3}$, and $\Lambda_{1}$ are not labeled explicitly
with a $\Z\Z$ superscript, while the superscript is kept for the other $\V\V$ states.

The parity-conserving interaction of a pseudoscalar ($CP$-odd state) corresponds to the $a_{3}^{\V\V}$ terms,
while the other terms describe the parity-conserving interaction of a scalar ($CP$-even state).
The $a_{3}^{\V\V}$ terms appear in the SM only at a three-loop level and receive a small contribution.
The $a_{2}^{\V\V}$ and $\Lambda_{1}^{\V\V}$ terms appear in loop-induced
processes and give small contributions $O(10^{-3}\text{--}10^{-2})$.
The dominant contributions to the SM expectation of the $\PH\to\Z\gamma$ and $\gamma\gamma$
decays are $a_2^{\Z\gamma}$ and $a_2^{\gamma\gamma}$, which are predicted to be
$a_2^{\Z\gamma}\simeq-0.007$ and $a_2^{\gamma\gamma}\simeq0.004$~\cite{Low:2012rj}.
The $a_i^{\Z\gamma}$ and $a_i^{\gamma\gamma}$ coupling terms contribute to the $\PH\to4\ell$
process through the $\PH\to\Z\gamma^*$ and $\gamma^*\gamma^*\to4\ell$ decays with off-shell
intermediate photons.
Anomalous couplings may be enhanced with BSM contributions and generally
acquire a non-trivial dependence on Lorentz invariant quantities and become complex.
The different contributions to the amplitude can therefore be tested without making
assumptions about the complex phase between different contributions.
When the particles in the loops responsible for these couplings are heavy in comparison
to the Higgs boson mass parameters, the couplings are real.

Under the assumption that the couplings are constant and real,
the above formulation is equivalent to an effective Lagrangian notation
for the $\PH\Z\Z$, $\PH\PW\PW$, $\PH\Z\gamma$, and $\PH\gamma\gamma$ interactions
\ifthenelse{\boolean{cms@external}}{
\begin{multline}
{L}(\PH\V\V) \sim
a_{1}\frac{m_{\sss\Z}^2}{2} \PH \Z^{\mu}\Z_{\mu} - \frac{\kappa_1}{\left(\Lambda_{1}\right)^{2}} m_{\sss\Z}^2 \PH  \Z_{\mu} \Box \Z^{\mu} \\
			  - \frac{1}{2}a_{2} \PH  \Z^{\mu\nu}\Z_{\mu\nu} - \frac{1}{2}a_{3} \PH  \Z^{\mu\nu}{\tilde \Z}_{\mu\nu} \\
+ a_{1}^{\PW\PW}{m_{\sss\PW}^2} \PH \PW^{+\mu} \PW^{-}_{\mu} \\
- \frac{1}{\left(\Lambda_{1}^{\PW\PW}\right)^{2}} m_{\sss\PW}^2 \PH
  \left(  \kappa_1^{\PW\PW} \PW^{-}_{\mu} \Box \PW^{+\mu} + \kappa_2^{\PW\PW} \PW^{+}_{\mu} \Box \PW^{-\mu} \right) \\
- a_{2}^{\PW\PW} \PH \PW^{+\mu\nu}\PW^{-}_{\mu\nu}
- a_{3}^{\PW\PW} \PH \PW^{+\mu\nu}{\tilde \PW}^{-}_{\mu\nu} \\
                          + \frac{\kappa_2^{\Z\gamma}}{\left(\Lambda_{1}^{\Z\gamma} \right)^{2}} m_{\sss\Z}^2 \PH  \Z_\mu \partial_\nu \F^{\mu\nu}
                          - a_{2}^{\Z\gamma} \PH \F^{\mu\nu} \Z_{\mu\nu} - a_{3}^{\Z\gamma} \PH  \F^{\mu\nu}{\tilde \Z}_{\mu\nu} \\
                           - \frac{1}{2}a_{2}^{\gamma\gamma} \PH  \F^{\mu\nu}\F_{\mu\nu} - \frac{1}{2}a_{3}^{\gamma\gamma}\PH  \F^{\mu\nu}{\tilde F}_{\mu\nu},
\label{eq:fullLag-spin0}
\end{multline}
}{
\begin{multline}
{L}(\PH\V\V) \sim
a_{1}\frac{m_{\sss\Z}^2}{2} \PH \Z^{\mu}\Z_{\mu} - \frac{\kappa_1}{\left(\Lambda_{1}\right)^{2}} m_{\sss\Z}^2 \PH  \Z_{\mu} \Box \Z^{\mu}
			  - \frac{1}{2}a_{2} \PH  \Z^{\mu\nu}\Z_{\mu\nu} - \frac{1}{2}a_{3} \PH  \Z^{\mu\nu}{\tilde \Z}_{\mu\nu} \\
+ a_{1}^{\PW\PW}{m_{\sss\PW}^2} \PH \PW^{+\mu} \PW^{-}_{\mu}
- \frac{1}{\left(\Lambda_{1}^{\PW\PW}\right)^{2}} m_{\sss\PW}^2 \PH
  \left(  \kappa_1^{\PW\PW} \PW^{-}_{\mu} \Box \PW^{+\mu} + \kappa_2^{\PW\PW} \PW^{+}_{\mu} \Box \PW^{-\mu} \right)\\
- a_{2}^{\PW\PW} \PH \PW^{+\mu\nu}\PW^{-}_{\mu\nu}
- a_{3}^{\PW\PW} \PH \PW^{+\mu\nu}{\tilde \PW}^{-}_{\mu\nu} \\
                          + \frac{\kappa_2^{\Z\gamma}}{\left(\Lambda_{1}^{\Z\gamma} \right)^{2}} m_{\sss\Z}^2 \PH  \Z_\mu \partial_\nu \F^{\mu\nu}
                          - a_{2}^{\Z\gamma} \PH \F^{\mu\nu} \Z_{\mu\nu} - a_{3}^{\Z\gamma} \PH  \F^{\mu\nu}{\tilde \Z}_{\mu\nu}
                           - \frac{1}{2}a_{2}^{\gamma\gamma} \PH  \F^{\mu\nu}\F_{\mu\nu} - \frac{1}{2}a_{3}^{\gamma\gamma}\PH  \F^{\mu\nu}{\tilde F}_{\mu\nu},
\label{eq:fullLag-spin0}
\end{multline}
}
where the notations are the same as in Eq.~(\ref{eq:formfact-fullampl-spin0}) and
$\PH$ is the real Higgs field, $\Z_{\mu}$ is the $\Z$ field, $\PW_{\mu}$ is the $\PW$ field,
$\F_{\mu}$ is the $\gamma^*$ field, $\V_{\mu\nu} = \partial_{\mu}\V_{\nu} - \partial_{\nu}\V_{\mu}$
is the bosonic field strength, the dual field strengths are defined as
${\tilde \V}_{\mu\nu} = \frac{1}{2}\epsilon_{\mu\nu\rho\sigma}\V^{\rho\sigma}$,
and $\Box$ is the D\,'\!Alembert operator.
The SM-like terms with tree-level couplings $a_{1}$ and $a_{1}^{\PW\PW}$ are associated with dimension-three operators,
and the rest of the terms tested correspond to operators of dimension five. Operators of higher dimension
are neglected in this study.

\begin{table*}[!bht]
\centering
\topcaption{
List of anomalous $\PH\V\V$ couplings considered in the measurements assuming a spin-zero Higgs boson.
The definition of the effective fractions is discussed in the text and the translation constant is given in each case.
The effective cross sections correspond to the processes
$\PH\to\V\V\to2\Pe2\mu$ and $\PH\to\PW\PW\to\ell\nu\ell\nu$
and the Higgs boson mass $m_{\sss\PH}=125.6\GeV$
using the \textsc{JHUGen}~\cite{Gao:2010qx,Bolognesi:2012mm,Anderson:2013afp} calculation.
The cross-section ratios for the $\PH\Z\gamma$ and $\PH\gamma\gamma$ couplings include
the requirement $\sqrt{q^2_{\sss\V}} \ge 4\GeV$.
}
\label{tab:xsec_ratio}
\begin{scotch}{lllll}
\multicolumn{1}{c}{Interaction} & \multicolumn{1}{c}{Anomalous}& \multicolumn{1}{c}{Coupling} & \multicolumn{1}{c}{Effective} & \multicolumn{1}{c}{Translation} \\
 & \multicolumn{1}{c}{Coupling}    & \multicolumn{1}{c}{Phase} & \multicolumn{1}{c}{Fraction}   & \multicolumn{1}{c}{Constant} \\
\hline
\multirow{3}{*}{$\PH\Z\Z$}& $\Lambda_{1}$ & $\phi_{\Lambda1}$ & $f_{\Lambda1}$ & $\sigma_{1}/{\tilde\sigma_{\Lambda1}}=1.45\times 10^{4}\TeV^{-4}$ \\
& $a_2$ & $\phi_{a2}$ & $f_{a2}$ & $\sigma_{1}/\sigma_{2}=2.68$ \\
& $a_3$ & $\phi_{a3}$ & $f_{a3}$ & $\sigma_{1}/\sigma_{3}=6.36$ \\[2ex]
\multirow{3}{*}{$\PH\PW\PW$}& $\Lambda_{1}^{\PW\PW}$ & $\phi_{\Lambda1}^{\PW\PW}$ & $f_{\Lambda1}^{\PW\PW}$ & $\sigma_{1}^{\PW\PW}/{\tilde\sigma_{\Lambda1}^{\PW\PW}}=1.87\times 10^{4}\TeV^{-4}$ \\

 & $a_2^{\PW\PW}$ & $\phi_{a2}^{\PW\PW}$ & $f_{a2}^{\PW\PW}$ & $\sigma^{\PW\PW}_{1}/\sigma^{\PW\PW}_{2}=1.25$ \\

& $a_3^{\PW\PW}$ & $\phi_{a3}^{\PW\PW}$ & $f_{a3}^{\PW\PW}$ & $\sigma^{\PW\PW}_{1}/\sigma^{\PW\PW}_{3}=3.01$ \\[2ex]

\multirow{3}{*}{$\PH\Z\gamma$} & $\Lambda_{1}^{\Z\gamma}$ & $\phi_{\Lambda1}^{\Z\gamma}$ & $f_{\Lambda1}^{\Z\gamma}$ & $\sigma_{1}^\prime/{\tilde\sigma_{\Lambda1}^{\Z\gamma}}=5.76\times 10^{3}\TeV^{-4}$ \\
 & $a_2^{\Z\gamma}$ & $\phi_{a2}^{\Z\gamma}$ & $f_{a2}^{\Z\gamma}$ & $\sigma_{1}^\prime/\sigma^{Z\gamma}_{2}=2.24\times10^{-3}$ \\

& $a_3^{\Z\gamma}$ & $\phi_{a3}^{\Z\gamma}$ & $f_{a3}^{\Z\gamma}$ & $\sigma_{1}^\prime/\sigma_{3}^{Z\gamma}=2.72\times10^{-3}$ \\[2ex]
\multirow{2}{*}{$\PH\gamma\gamma$} & $a_2^{\gamma\gamma}$ & $\phi_{a2}^{\gamma\gamma}$ & $f_{a2}^{\gamma\gamma}$  & $\sigma_{1}^\prime/\sigma^{\gamma\gamma}_{2}=2.82\times10^{-3}$ \\

& $a_3^{\gamma\gamma}$ & $\phi_{a3}^{\gamma\gamma}$ & $f_{a3}^{\gamma\gamma}$ & $\sigma_{1}^\prime/\sigma_{3}^{\gamma\gamma}=2.88\times10^{-3}$ \\
\end{scotch}

\end{table*}

In the analysis, the physics effects of the eleven anomalous couplings listed in Table~\ref{tab:xsec_ratio} are described,
where the hypothesis of the Higgs boson mass $m_{\sss\PH}=125.6\GeV$ is used, which is the best-fit value
in the study of the $\PH\to\V\V\to 4\ell$ and $\PH\to\PW\PW\to\ell\nu\ell\nu$ channels~\cite{Chatrchyan:2013mxa, Chatrchyan:2013iaa}.
The scenarios are parameterized in terms of the effective fractional cross sections $f_{ai}$ and their phases $\phi_{ai}$
with respect to the two dominant tree-level couplings $a_1$ and $a_1^{\PW\PW}$
in the $\PH\to \V\V\to 4\ell$ and $\PH\to \PW\PW\to \ell\nu\ell\nu$ processes, respectively.
In the $\PH\to\V\V$ decay the $q_{\sss\V}^2$ range does not exceed approximately 100\GeV due to the kinematic bound,
supporting the expansion up to $q_{\sss\V}^2$ in Eq.~(\ref{eq:formfact-fullampl-spin0}).
Even though the expansion with only three anomalous contributions in Eq.~(\ref{eq:formfact-fullampl-spin0})
may become formally incomplete when large values of $f_{ai}\sim1$ are considered, this remains a valuable test of
the consistency of the data with the SM. Moreover, certain models, such as models with a pseudoscalar Higgs boson state,
do not require sizable contribution of higher terms in the $q_{\sss\V}^2$ expansion even for $f_{ai}\sim1$.
Therefore, the full range $0\le f_{ai}\le1$ is considered in this study.

The effective fractional $\Z\Z$ cross sections $f_{ai}$ and phases $\phi_{ai}$ are defined as follows
\ifthenelse{\boolean{cms@external}}{
\begin{equation}\begin{aligned}
f_{\Lambda1} =& \frac{\tilde{\sigma}_{\Lambda1}/\left(\Lambda_{1}\right)^{4}}{\abs{a_1}^2 \sigma_{1} + \abs{a_2}^2 \sigma_{2} + \abs{a_3}^2 \sigma_{3} + \tilde{\sigma}_{\Lambda{1}}/\left(\Lambda_{1}\right)^{4} +\ldots},\\
 &\phi_{\Lambda1}, \\
f_{a2} =& \frac{\abs{a_2}^2 \sigma_{2}}{\abs{a_1}^2 \sigma_{1} + \abs{a_2}^2 \sigma_{2} + \abs{a_3}^2 \sigma_{3} + \tilde{\sigma}_{\Lambda{1}}/\left(\Lambda_{1}\right)^{4} +\ldots},\\
 &\phi_{a2} = \text{arg}\left(\frac{a_{2}}{a_{1}}\right), \\
f_{a3} =& \frac{\abs{a_3}^2 \sigma_{3}}{\abs{a_1}^2 \sigma_{1} + \abs{a_2}^2 \sigma_{2} + \abs{a_3}^2 \sigma_{3} + \tilde{\sigma}_{\Lambda{1}}/\left(\Lambda_{1}\right)^{4} +\ldots},\\
 &\phi_{a3} = \text{arg}\left(\frac{a_{3}}{a_{1}}\right),
\label{eq:fa_definitions}
\end{aligned}\end{equation}
}{
\begin{equation}\begin{aligned}
f_{\Lambda1} &= \frac{\tilde{\sigma}_{\Lambda1}/\left(\Lambda_{1}\right)^{4}}{\abs{a_1}^2 \sigma_{1} + \abs{a_2}^2 \sigma_{2} + \abs{a_3}^2 \sigma_{3} + \tilde{\sigma}_{\Lambda{1}}/\left(\Lambda_{1}\right)^{4} +\ldots}, \qquad \phi_{\Lambda1},    \\
f_{a2} &= \frac{\abs{a_2}^2 \sigma_{2}}{\abs{a_1}^2 \sigma_{1} + \abs{a_2}^2 \sigma_{2} + \abs{a_3}^2 \sigma_{3} + \tilde{\sigma}_{\Lambda{1}}/\left(\Lambda_{1}\right)^{4} +\ldots}, \qquad \phi_{a2} = \text{arg}\left(\frac{a_{2}}{a_{1}}\right),  \\
f_{a3} &= \frac{\abs{a_3}^2 \sigma_{3}}{\abs{a_1}^2 \sigma_{1} + \abs{a_2}^2 \sigma_{2} + \abs{a_3}^2 \sigma_{3} + \tilde{\sigma}_{\Lambda{1}}/\left(\Lambda_{1}\right)^{4} +\ldots}, \qquad \phi_{a3} = \text{arg}\left(\frac{a_{3}}{a_{1}}\right),
\label{eq:fa_definitions}
\end{aligned}\end{equation}
}
where $\sigma_{i}$ is the cross section of the process corresponding to $a_{i} = 1$, $a_{j \neq i} = 0$,
while $\tilde{\sigma}_{\Lambda{1}}$ is the effective cross section of the process corresponding to
$\Lambda_{1} =1\TeV$, given in units of $\mathrm{fb}\cdot\TeV^4$.
The effective fractional $\PW\PW$ cross sections are defined in complete analogy with the definitions for $\Z\Z$
as shown in Eq.~(\ref{eq:fa_definitions}).
The definition in Eq.~(\ref{eq:fa_definitions}) is independent of the collider energy because only the decay
rates of the processes $\PH\to \V\V\to 4\ell$ and $\PH\to \PW\PW\to \ell\nu\ell\nu$ affect the ratio.
It also has the advantage of the $f_{ai}$ parameters being bounded between 0 and 1,
and being uniquely defined, regardless of the convention used for the coupling constants.
In the four-lepton final state, the cross section of the $\PH\to\V\V\to2\Pe2\mu$ final state is used,
as this final state is not affected by the interference between same-flavor leptons in the final state.

In an analogous way, the effective fractional cross sections and phases of
$\Z\gamma$ and $\gamma\gamma$, generically denoted as $\V\gamma$ below,
in the $\PH\to\V\V\to2\Pe2\mu$ process are defined as
\ifthenelse{\boolean{cms@external}}{
\begin{multline}
f^{\V\gamma}_{ai} = \frac{\abs{a^{\V\gamma}_{i}}^2 \sigma^{\V\gamma}_{i}}{\abs{a_1}^2 \sigma_{1}^\prime + \abs{a^{\V\gamma}_{i}}^2 \sigma^{\V\gamma}_{i} +\ldots}, \\
 \phi^{\V\gamma}_{ai} = \text{arg}\left(\frac{a^{\V\gamma}_{ai}}{a_{1}}\right),
\label{eq:fa_definitions_zg}
\end{multline}
}{
\begin{equation}
f^{\V\gamma}_{ai} = \frac{\abs{a^{\V\gamma}_{i}}^2 \sigma^{\V\gamma}_{i}}{\abs{a_1}^2 \sigma_{1}^\prime + \abs{a^{\V\gamma}_{i}}^2 \sigma^{\V\gamma}_{i} +\ldots}, \qquad \phi^{\V\gamma}_{ai} = \text{arg}\left(\frac{a^{\V\gamma}_{ai}}{a_{1}}\right),
\label{eq:fa_definitions_zg}
\end{equation}
}
where the requirement $\sqrt{q^2_{{\sss\V}i}} \ge 4\GeV$ is used in the cross-section
calculations for all processes, including the $\Z\Z$ tree-level process with $a_1$ as indicated with $\sigma^\prime_1$.
This requirement on $q^2_{{\sss\V}i}$ is introduced to restrict the definition to a region without
infrared divergence and to define the fractions within the empirically relevant range.
The ellipsis $(...)$ in Eqs.~(\ref{eq:fa_definitions}) and~(\ref{eq:fa_definitions_zg}) indicates any other contribution
not listed explicitly.

Given the measured values of the effective fractions, it is possible to extract the ratios of the coupling constants
$a_i/a_1$, the scale of BSM physics $\Lambda_{1}$, or the ratios of the $Z\gamma^*$ ($\gamma^*\gamma^*$)
cross sections with respect to the SM predictions in any parameterization.
Following Eq.~(\ref{eq:formfact-fullampl-spin0}) the translation of the $f_{ai}$ measurements can be performed as
\ifthenelse{\boolean{cms@external}}{
\begin{equation}\begin{aligned}
\frac{\abs{a_{i}}}{\abs{a_1}}=&\sqrt{{f_{ai}}/{f_{a1}}}\times \sqrt{{\sigma_{1}}/{\sigma_{i}}},\\
\Lambda_1\sqrt{\abs{a_1}}=&\sqrt[4]{{f_{a1}}/{f_{\Lambda{1}}}}\times \sqrt[4]{\tilde{\sigma}_{\Lambda{1}}/{\sigma_{1}}},
\label{eq:fa_conversion}
\end{aligned}\end{equation}
}{
\begin{equation}
\frac{\abs{a_{i}}}{\abs{a_1}}=\sqrt{{f_{ai}}/{f_{a1}}}\times \sqrt{{\sigma_{1}}/{\sigma_{i}}},~~~~~~~~~
\Lambda_1\sqrt{\abs{a_1}}=\sqrt[4]{{f_{a1}}/{f_{\Lambda{1}}}}\times \sqrt[4]{\tilde{\sigma}_{\Lambda{1}}/{\sigma_{1}}},
\label{eq:fa_conversion}
\end{equation}
}
where the cross-section ratios for a 125.6\GeV Higgs boson are given in Table~\ref{tab:xsec_ratio},
and the fraction $f_{a1} = (1-{f_{\Lambda1}}-{f_{a2}}-{f_{a3}}-\ldots)$ corresponds to
the effective SM tree-level contribution, which is expected to dominate.
The ellipsis in the $f_{a1}$ definition indicates any other contribution,
such as $\Z\gamma^*$ and $\gamma^*\gamma^*$, where relevant.

The couplings of the Higgs boson to $\Z\gamma$ and $\gamma\gamma$ are generally much
better measured in the decays with the on-shell gauge bosons $\PH\to\Z\gamma$ and
$\gamma\gamma$~\cite{Chatrchyan:2013vaa, Aad:2013wqa, Aad:2014fia, Khachatryan:2014ira}.
Therefore, the measurements of the $\PH\Z\gamma$ and $\PH\gamma\gamma$
anomalous couplings are provided mostly as a feasibility study without going
into detailed measurements of correlations of parameters.
Once a sufficient number of events is accumulated for the discovery of these modes in the
$\PH\to\V\V\to 4\ell$ channel with a high-luminosity LHC, the study of $CP$ properties
can be performed with the $\PH\Z\gamma$ and $\PH\gamma\gamma$
couplings~\cite{Dawson:2013bba,Chen:2014gka}.

The couplings of a spin-zero particle to $\PW$ and $\Z$ bosons can be related given the assumption of certain symmetries.
For example, in the case of the custodial singlet Higgs boson, the relation is $a_1^{\PW\PW} = a_1$~\cite{Low:2010jp,Sikivie:1980hm}.
Generally, these couplings could have a different relationship and the $\PH\V\V$ couplings are controlled by two free parameters.
When one parameter is expressed as the $f_{ai}$ fraction in the $\PH\Z\Z$ coupling, the other parameter can be chosen as a ratio
of anomalous couplings in the $\PH\to\Z\Z$ and $\PH\to\PW\PW$ channels
\begin{eqnarray}
r_{ai} = \frac{a_i^{\PW\PW} / a_1^{\PW\PW}  }{  a_i / a_1}, \text{  or  } R_{ai} = \frac{ r_{ai} \abs{r_{ai}} }{  1 + r_{ai}^2 }.
\label{eq:ratio_ww_zz}
\end{eqnarray}
Using Eq.~(\ref{eq:fa_conversion}) the effective fractions $f_{ai}^{\PW\PW}$ and $f_{ai}$ can be related as
\begin{equation}
f_{ai} = \left[ 1+r_{ai}^2(1/f_{ai}^{\PW\PW}-1)\sigma_{i}^{\PW\PW}\sigma_{1}/(\sigma_{1}^{\PW\PW}\sigma_{i}) \right]^{-1}.
\label{eq:a2_conversion}
\end{equation}
In this way, the measurement of $f_{ai}^{\PW\PW}$ can be converted to $f_{ai}$ and vice versa,
and the combination of the results in the $\Z\Z$ and $\PW\PW$ channels can be achieved.

\subsection{Decay of a spin-one resonance} \label{sec:spin1}

In the case of a spin-one resonance, the amplitude of its interaction with a pair of massive gauge bosons,
$\Z\Z$ or $\PW\PW$, consists of two independent terms, which can be written as
\ifthenelse{\boolean{cms@external}}{
\begin{multline}
A(\X_{J=1} \V\V) \sim b_{1}^{\V\V}  \big[ \left(\epsilon_{\sss\V1}^{*}q\right)\left(\epsilon_{\sss\V2}^{*}\epsilon_{\sss\X}\right) + \left(\epsilon_{\sss\V2}^{*}q\right)\left(\epsilon_{\sss\V1}^{*}\epsilon_{\sss\X}\right) \big] \\+ b_{2}^{\V\V}  \epsilon_{\alpha\mu\nu\beta}\epsilon_{\sss\X}^{\alpha}\epsilon_{\sss\V1}^{*\mu}\epsilon_{\sss\V2}^{*\nu}{\tilde q}^{\beta},
\label{eq:ampl-spin1}
\end{multline}

}{
\begin{equation}
A(\X_{J=1} \V\V) \sim b_{1}^{\V\V}  \left[ \left(\epsilon_{\sss\V1}^{*}q\right)\left(\epsilon_{\sss\V2}^{*}\epsilon_{\sss\X}\right) + \left(\epsilon_{\sss\V2}^{*}q\right)\left(\epsilon_{\sss\V1}^{*}\epsilon_{\sss\X}\right) \right] + b_{2}^{\V\V}  \epsilon_{\alpha\mu\nu\beta}\epsilon_{\sss\X}^{\alpha}\epsilon_{\sss\V1}^{*\mu}\epsilon_{\sss\V2}^{*\nu}{\tilde q}^{\beta},
\label{eq:ampl-spin1}
\end{equation}
}
where $\epsilon_{\sss\X}$ is the polarization vector of the boson $\X$ with spin one,
${q}=q_{\sss\V1}+q_{\sss\V2}$ and ${\tilde q}=q_{\sss\V1}-q_{\sss\V2}$~\cite{Gao:2010qx, Bolognesi:2012mm}.
Here the $b_{1}^{\V\V}  \neq 0$ coupling corresponds to a vector particle, while
the $b_{2}^{\V\V}\neq 0$ coupling corresponds to a pseudovector.
The $\Z\gamma$ interactions of the spin-one particle are not considered, while
the $\gamma\gamma$  and $\Pg\Pg$ interactions are forbidden by the Landau-Yang theorem~\cite{Landau, Yang},
where the $\Pg\Pg$ case is justified by the assumption that the state $\X$ is color-neutral.
Here, and throughout this paper, a boson with an exotic spin is denoted as $\X$ to distinguish it from a spin-zero Higgs boson $\PH$.

Similarly, the lowest order terms in the scattering amplitudes can be mapped to the corresponding terms in the effective Lagrangian
\ifthenelse{\boolean{cms@external}}{
\begin{multline}
{L}({\X_{J=1} \V\V}) \sim
b_1  \partial_{\mu}\X_{\nu}\Z^{\mu}\Z^{\nu} + b_2  \epsilon_{\alpha\mu\nu\beta}\X^{\alpha}\Z^{\mu} \partial^{\beta}\Z^{\nu} \\
+ b_1^{\PW\PW}  \partial_{\mu}\X_{\nu}\left( \PW^{+\mu}\PW^{-\nu} + \PW^{-\mu}\PW^{+\nu}   \right) \\
+ b_2^{\PW\PW}  \epsilon_{\alpha\mu\nu\beta}\X^{\alpha}
\left( \PW^{-\mu} \partial^{\beta}\PW^{+\nu} + \PW^{+\mu} \partial^{\beta}\PW^{-\nu} \right)
\,.
\label{eq:lagr-spin1}
\end{multline}
}{
\begin{multline}
{L}({\X_{J=1} \V\V}) \sim
b_1  \partial_{\mu}\X_{\nu}\Z^{\mu}\Z^{\nu} + b_2  \epsilon_{\alpha\mu\nu\beta}\X^{\alpha}\Z^{\mu} \partial^{\beta}\Z^{\nu} \\
+ b_1^{\PW\PW}  \partial_{\mu}\X_{\nu}\left( \PW^{+\mu}\PW^{-\nu} + \PW^{-\mu}\PW^{+\nu}   \right)
+ b_2^{\PW\PW}  \epsilon_{\alpha\mu\nu\beta}\X^{\alpha}
\left( \PW^{-\mu} \partial^{\beta}\PW^{+\nu} + \PW^{+\mu} \partial^{\beta}\PW^{-\nu} \right)
\,.
\label{eq:lagr-spin1}
\end{multline}
}

Despite the fact that the experimental observation~\cite{Aad:2012tfa,Chatrchyan:2012ufa,Chatrchyan:2013lba}
of the $\PH \to \gamma \gamma$ decay channel prevents
the observed boson from being a spin-one particle, it is still important to experimentally
study the spin-one models in the decay to massive vector bosons in case that the observed state is a different one.
The CMS and ATLAS experiments have already tested the compatibility of the observed boson
with the $J^P=1^{+}$ and $1^{-}$ hypotheses~\cite{Aad:2013xqa,Chatrchyan:2013mxa},
where CMS has tested this using both production-independent and production-dependent methods.
The compatibility of the data with the hypothesis of the boson being a mixture of the $1^{+}$ and $1^{-}$ states is now tested
allowing for the presence of each of the terms in the scattering amplitude in Eq.~(\ref{eq:ampl-spin1}).
A continuous parameter that uniquely describes the presence of the corresponding terms $b_{1}^{\V\V} $ and
$b_{2}^{\V\V}$ is defined as an effective fractional cross section
\begin{equation}
f_{b2}^{\V\V}  = \frac{\abs{b_{2}^{\V\V}}^2 \sigma_{b2}}{\abs{b_{1}^{\V\V}}^2 \sigma_{b1} + \abs{b_{2}^{\V\V}}^2 \sigma_{b2}},
\label{eq:fa_definitions_spin1}
\end{equation}
where $\sigma_{bi}$ is the cross section of the process corresponding
to $b_{i}^{\V\V}  = 1, b^{\V\V}_{j \neq i} = 0$ in the
$\X\to\Z\Z\to2\Pe2\mu$ or $\PW\PW\to\ell\nu\ell\nu$ final state and $\sigma_{b1} = \sigma_{b2}$.
This effective fraction is used in the analysis to test if the data favor the SM Higgs boson
scalar hypothesis or some particular mixture of the vector and pseudovector states.

\subsection{Decay of a spin-two resonance} \label{sec:spin2}

In the case of a general spin-two resonance, its decay to a pair of massive vector bosons,
$\Z\Z$ or $\PW\PW$, is considered in their sequential decay to four leptons, but not with $Z\gamma^*$
and $\gamma^*\gamma^*$, as those are generally suppressed by the $\gamma^*\to\ell^+\ell^-$ selection.
The decay to two on-shell photons $\X\to \gamma\gamma$ is also considered.
The corresponding $\X\V\V$ amplitude is used to describe the $\X\to\Z\Z$ and $\PW\PW$, as well as $\Pg\Pg\to \X$, processes
\ifthenelse{\boolean{cms@external}}{
\begin{multline}
\label{eq:ampl-spin2-a}
A(\X_{J=2} \V\V)  \sim \Lambda^{-1} \Big [
2 c_{1}^{\V\V} t_{\mu \nu} f^{*1,\mu \alpha} f^{*2,\nu}_{~~~~~\alpha}
\\+ 2 c_{2}^{\V\V} t_{\mu \nu} \frac{q_\alpha q_\beta }{\Lambda^2} f^{*1,\mu \alpha}  f^{*2,\nu \beta}
   \\
+ c_{3}^{\V\V} t_{\beta \nu} \frac{{\tilde q}^\beta {\tilde q}^{\alpha}}{\Lambda^2}
 ( f^{*1,\mu \nu} f^{*2}_{\mu \alpha} + f^{*2,\mu \nu} f^{*1}_{\mu \alpha} )
 \\+ c_{ 4}^{\V\V}t_{\mu \nu} \frac{{\tilde q}^{\nu} {\tilde q}^\mu}{{\Lambda^2} }  f^{*1,\alpha \beta} f^{*2}_{\alpha \beta}
\\
 + m_{\sss\V}^2  \big (
2 c_{ 5}^{\V\V}  t_{\mu\nu} \epsilon_{\sss\V1}^{*\mu} \epsilon_{\sss\V2}^{*\nu}
+2 c_{ 6}^{\V\V}  t_{\mu \nu} \frac{{\tilde q}^\mu q_\alpha}{\Lambda^2}
\big ( \epsilon_{\sss\V1}^{*\nu} \epsilon_{\sss\V2}^{*\alpha} -
\epsilon_{\sss\V1}^{*\alpha} \epsilon_{\sss\V2}^{*\nu} \big )
\\+c_{ 7}^{\V\V} t_{\mu \nu}  \frac{{\tilde q}^\mu {\tilde q}^\nu}{\Lambda^2}  \epsilon^*_{\sss\V1} \epsilon^*_{\sss\V2}
\Big)
+c_{ 8}^{\V\V} t_{\mu \nu} \frac{{\tilde q}^{\mu} {\tilde q}^{\nu}}{\Lambda^2}
  f^{*1,\alpha \beta} {\tilde f}^{*2}_{\alpha \beta}
   \\
+  m_{\sss\V}^2  \Big(
c_{ 9}^{\V\V} t^{\mu \alpha}
\frac{
{\tilde q}_{\alpha} \epsilon_{\mu \nu \rho \sigma} \epsilon_{\sss\V1}^{*\nu} \epsilon_{\sss\V2}^{*\rho} q^{\sigma}
}{\Lambda^2}
\\+c_{ 10}^{\V\V} t^{\mu \alpha}
\frac{
{\tilde q}_{\alpha} \epsilon_{\mu \nu \rho \sigma} q^\rho {\tilde q}^{\sigma}
\big ( \epsilon_{\sss\V1}^{*\nu}(q\epsilon_{\sss\V2}^*)+
\epsilon_{\sss\V2}^{*\nu}(q\epsilon_{\sss\V1}^*) \big )
}{\Lambda^4}
\Big)
\Big],
\end{multline}
}{
\begin{multline}
\label{eq:ampl-spin2-a}
A(\X_{J=2} \V\V)  \sim \Lambda^{-1} \left [
2 c_{1}^{\V\V} t_{\mu \nu} f^{*1,\mu \alpha} f^{*2,\nu}_{~~~~~\alpha}
+ 2 c_{2}^{\V\V} t_{\mu \nu} \frac{q_\alpha q_\beta }{\Lambda^2} f^{*1,\mu \alpha}  f^{*2,\nu \beta}
\right.   \\
\left.
+ c_{3}^{\V\V} t_{\beta \nu} \frac{{\tilde q}^\beta {\tilde q}^{\alpha}}{\Lambda^2}
 ( f^{*1,\mu \nu} f^{*2}_{\mu \alpha} + f^{*2,\mu \nu} f^{*1}_{\mu \alpha} )
 + c_{ 4}^{\V\V}t_{\mu \nu} \frac{{\tilde q}^{\nu} {\tilde q}^\mu}{{\Lambda^2} }  f^{*1,\alpha \beta} f^{*2}_{\alpha \beta}
\right.\\
\left. + m_{\sss\V}^2  \left (
2 c_{ 5}^{\V\V}  t_{\mu\nu} \epsilon_{\sss\V1}^{*\mu} \epsilon_{\sss\V2}^{*\nu}
+2 c_{ 6}^{\V\V}  t_{\mu \nu} \frac{{\tilde q}^\mu q_\alpha}{\Lambda^2}
\left ( \epsilon_{\sss\V1}^{*\nu} \epsilon_{\sss\V2}^{*\alpha} -
\epsilon_{\sss\V1}^{*\alpha} \epsilon_{\sss\V2}^{*\nu} \right )
+c_{ 7}^{\V\V} t_{\mu \nu}  \frac{{\tilde q}^\mu {\tilde q}^\nu}{\Lambda^2}  \epsilon^*_{\sss\V1} \epsilon^*_{\sss\V2}
\right)
\right.  \\
\left.
+c_{ 8}^{\V\V} t_{\mu \nu} \frac{{\tilde q}^{\mu} {\tilde q}^{\nu}}{\Lambda^2}
  f^{*1,\alpha \beta} {\tilde f}^{*2}_{\alpha \beta}
\right.   \\
 \left.
+  m_{\sss\V}^2  \left (
c_{ 9}^{\V\V} t^{\mu \alpha}
\frac{
{\tilde q}_{\alpha} \epsilon_{\mu \nu \rho \sigma} \epsilon_{\sss\V1}^{*\nu} \epsilon_{\sss\V2}^{*\rho} q^{\sigma}
}{\Lambda^2}
+c_{ 10}^{\V\V} t^{\mu \alpha}
\frac{
{\tilde q}_{\alpha} \epsilon_{\mu \nu \rho \sigma} q^\rho {\tilde q}^{\sigma}
\left ( \epsilon_{\sss\V1}^{*\nu}(q\epsilon_{\sss\V2}^*)+
\epsilon_{\sss\V2}^{*\nu}(q\epsilon_{\sss\V1}^*) \right )
}{\Lambda^4}
\right )
\right ],
\end{multline}
}
where $t^{\mu\nu}$ is the wavefunction of a spin-two particle $\X$ given by a symmetric traceless tensor,
$m_{\sss\V}$ is the mass of the considered gauge boson,
and $\Lambda$ is the scale of BSM physics~\cite{Gao:2010qx, Bolognesi:2012mm}.
The couplings $c_{1}^{\V\V}$ and $c_{5}^{\V\V}$ correspond to the parity-conserving interaction
of a spin-two tensor with minimal gravity-like couplings. As in the spin-zero case,
the couplings $c_{i}^{\V\V}$ are in general momentum-dependent form factors. In this
analysis it is assumed that the form factors are momentum-independent constants and, thus,
only the lowest $q_i^2$  order terms in the scattering amplitude are considered.

The terms in Eq.~(\ref{eq:ampl-spin2-a}) can be mapped to the corresponding
terms (operators up to dimension seven) in the effective Lagrangian
\ifthenelse{\boolean{cms@external}}{
\begin{multline}
{L}({\X_{J=2} \Z\Z}) \sim
 \Lambda^{-1} \left (
- c_{1}^{}  \X_{\mu \nu} \Z^{\mu \alpha} \Z^{\nu}_{~\alpha}
\right.  \\ \left. %
+ \frac{c_{2}^{} }{\Lambda^2} \left(\partial_\alpha \partial_\beta \X_{\mu \nu} \right )\Z^{\mu \alpha} \Z^{\nu \beta}
\right.  \\ \left. %
+\frac{c_{3}^{} }{\Lambda^2} \X_{\beta \nu}  \left[ \partial^\alpha, \left[ \partial^\beta, \Z^{\mu \nu} \right] \right] \Z_{\mu \alpha}
\right.  \\ \left.
+ \frac{c_{4}^{} }{2\Lambda^2} \X_{\mu \nu}  \left[ \partial^\mu, \left[ \partial^\nu, \Z^{\alpha \beta} \right] \right] \Z_{\alpha \beta}
\right.  \\ \left. %
+ c_{5}^{}  m_{\sss\Z}^2 X_{\mu \nu} \Z^{\mu} \Z^{\nu}
 +  \frac{2c_{6}^{}m_{\sss\Z}^2}{\Lambda^2} \partial_{\alpha} \X_{\mu \nu} \left[ \partial^{\mu}, \Z^{\nu} \right] \Z^{\alpha}
\right.  \\ \left.
  -   \frac{c_{7}^{} m_{\sss\Z}^2}{2\Lambda^2} \X_{\mu \nu}  \left[ \partial^\mu, \left[ \partial^\nu, \Z_{\alpha} \right] \right] \Z^{\alpha}
\right.  \\ \left. %
+ \frac{c_{8}^{} }{2\Lambda^2} \X_{\mu \nu}  \left[ \partial^\mu, \left[ \partial^\nu, {\Z}^{\alpha \beta} \right] \right] \tilde{\Z}_{\alpha \beta}
\right.  \\ \left.
  - \frac{c_{9}^{} m_{\sss\Z}^2}{\Lambda^2}  \epsilon_{\mu \nu \rho \sigma} \partial^{\sigma} \X^{\mu \alpha} \Z_{\nu}  \partial_{\alpha} \Z^{\rho}
\right.  \\ \left. %
+ \frac{c_{10}^{} m_{\sss\Z}^2}{\Lambda^4} \epsilon_{\mu \nu \rho \sigma} \partial^{\rho} \partial^{\beta} \X^{\mu \alpha} \left[ \partial^\sigma, \left[ \partial_\alpha, \Z^{\nu} \right] \right] \Z_{\beta}
 \right ).
 \end{multline}
}{
\begin{multline}
{L}({\X_{J=2} \Z\Z}) \sim
 \Lambda^{-1} \left (
- c_{1}^{}  \X_{\mu \nu} \Z^{\mu \alpha} \Z^{\nu}_{~\alpha}
+ \frac{c_{2}^{} }{\Lambda^2} \left(\partial_\alpha \partial_\beta \X_{\mu \nu} \right )\Z^{\mu \alpha} \Z^{\nu \beta} +
 \frac{c_{3}^{} }{\Lambda^2} \X_{\beta \nu}  \left[ \partial^\alpha, \left[ \partial^\beta, \Z^{\mu \nu} \right] \right] \Z_{\mu \alpha}
 \right.  \\ \left.
  + \frac{c_{4}^{} }{2\Lambda^2} \X_{\mu \nu}  \left[ \partial^\mu, \left[ \partial^\nu, \Z^{\alpha \beta} \right] \right] \Z_{\alpha \beta}
 + c_{5}^{}  m_{\sss\Z}^2 X_{\mu \nu} \Z^{\mu} \Z^{\nu}
 +  \frac{2c_{6}^{}m_{\sss\Z}^2}{\Lambda^2} \partial_{\alpha} \X_{\mu \nu} \left[ \partial^{\mu}, \Z^{\nu} \right] \Z^{\alpha}
 \right.  \\ \left.
  -   \frac{c_{7}^{} m_{\sss\Z}^2}{2\Lambda^2} \X_{\mu \nu}  \left[ \partial^\mu, \left[ \partial^\nu, \Z_{\alpha} \right] \right] \Z^{\alpha}
 + \frac{c_{8}^{} }{2\Lambda^2} \X_{\mu \nu}  \left[ \partial^\mu, \left[ \partial^\nu, {\Z}^{\alpha \beta} \right] \right] \tilde{\Z}_{\alpha \beta}
 \right.  \\ \left.
  - \frac{c_{9}^{} m_{\sss\Z}^2}{\Lambda^2}  \epsilon_{\mu \nu \rho \sigma} \partial^{\sigma} \X^{\mu \alpha} \Z_{\nu}  \partial_{\alpha} \Z^{\rho}
  + \frac{c_{10}^{} m_{\sss\Z}^2}{\Lambda^4} \epsilon_{\mu \nu \rho \sigma} \partial^{\rho} \partial^{\beta} \X^{\mu \alpha} \left[ \partial^\sigma, \left[ \partial_\alpha, \Z^{\nu} \right] \right] \Z_{\beta}
\right ).
 \end{multline}
}
The study of a subset of these ten terms in $\X\to \Z\Z$, $\PW\PW$, and $\gamma\gamma$ decays and $gg\to\X$ production
has already been performed by the CMS and ATLAS
experiments~\cite{Chatrchyan:2012jja,Aad:2013xqa,Chatrchyan:2013iaa,Chatrchyan:2013mxa,Khachatryan:2014ira}.
In this analysis the study of spin-two hypotheses is completed by considering the remaining terms
in the spin-two $\V\V$ scattering amplitude in Eq.~(\ref{eq:ampl-spin2-a}) and different production scenarios.
Ten spin-two scenarios are listed in Table~\ref{table-scenarios}.
Both $\qqbar$ production, discussed in Section~\ref{sec:production},
and gluon fusion, described by Eq.~(\ref{eq:ampl-spin2-a}), of a spin-two state are considered.
In the $\X\to \gamma\gamma$ decay channel, the full list of models is not analyzed.

The spin-two model with minimal couplings, which is common to $\X\to \Z\Z$, $\PW\PW$, and $\gamma\gamma$ ,
represents a massive graviton-like boson as suggested in models with warped extra dimensions (ED)~\cite{Randall:1999vf,Randall:1999ee}.
The individual results for the $2^+_{{m}}$ model were presented for the $\X\to \Z\Z$, $\PW\PW$, and $\gamma\gamma$ decays
earlier~\cite{Chatrchyan:2013iaa,Chatrchyan:2013mxa,Khachatryan:2014ira}. A combination is reported here.

A modified minimal coupling model $2^+_{{b}}$ is also considered,
where the SM fields are allowed to propagate in the bulk of the ED~\cite{Agashe:2007zd},
corresponding to $c_1^{\V\V} \ll c_5^{\V\V}$ in the $\X\Z\Z$ or $\X\PW\PW$ couplings only.
Moreover, eight spin-two models with higher-dimension operators are considered for
the $\X\Z\Z$ and $\X\PW\PW$ couplings. The above list of ten spin-two decay models and several
production mechanisms does not exhaust all the possible scenarios with mixed amplitudes,
but it does provide a comprehensive sample of spin-two alternatives to test the validity
of the SM-like $J^P=0^+$ hypothesis.

\begin{table*}[htbp]
\topcaption{
List of spin-two models with the production and decay couplings of an exotic $\X$ particle.
The subscripts $m$ (minimal couplings), $h$ (couplings with higher-dimension operators),
and $b$ (bulk) distinguish different scenarios.
}
\centering
\begin{scotch}{cccc}
$\jp$ Model  &  $\Pg\Pg\to$ $\X$ Couplings   & $\qqbar\to\X$ Couplings & $\X\to\V\V$ Couplings \\

\hline

$2_m^+$  &   $c_{ 1}^{\Pg\Pg}\ne0$  &  $\rho_1\ne0$  & $c_{ 1}^{\V\V}=c_{ 5}^{\V\V}\ne0$     \\

$2_{h2}^+$      &   $c_{ 2}^{\Pg\Pg}\ne0$  &  $\rho_1\ne0$     & $c_{ 2}^{\V\V}\ne0$  \\

$2_{h3}^+$     &   $c_{ 3}^{\Pg\Pg}\ne0$  &  $\rho_1\ne0$ & $c_{ 3}^{\V\V}\ne0$  \\

$2_h^+$  &   $c_{ 4}^{\Pg\Pg}\ne0$ &  $\rho_1\ne0$ & $c_{ 4}^{\V\V}\ne0$  \\

$2_b^+$   &   $c_{ 1}^{\Pg\Pg}\ne0$  &  $\rho_1\ne0$ &  $c_{ 1}^{\V\V}\ll c_{ 5}^{\V\V}\ne0$ \\

$2_{h6}^+$  &   $c_{ 1}^{\Pg\Pg}\ne0$ &  $\rho_1\ne0$ & $c_{ 6}^{\V\V}\ne0$  \\

$2_{h7}^+$ &   $c_{ 1}^{\Pg\Pg}\ne0$ &  $\rho_1\ne0$ & $c_{ 7}^{\V\V}\ne0$ \\

$2_h^-$  &    $c_{ 8}^{\Pg\Pg}\ne0$ &  $\rho_2\ne0$ & $c_{ 8}^{\V\V}\ne0$  \\

$2_{h9}^-$ &    $c_{ 8}^{\Pg\Pg}\ne0$ &  $\rho_2\ne0$ & $c_{ 9}^{\V\V}\ne0$ \\

$2_{h10}^-$ &    $c_{ 8}^{\Pg\Pg}\ne0$ &  $\rho_2\ne0$ & $c_{ 10}^{\V\V}\ne0$ \\
\end{scotch}

\label{table-scenarios}
\end{table*}

\subsection{Production of a resonance} \label{sec:production}

While the above discussion of Eqs.~(\ref{eq:formfact-fullampl-spin0}), (\ref{eq:ampl-spin1}), and~(\ref{eq:ampl-spin2-a})
is focussed on the decay $\PH\to\V\V$, these amplitudes also describe production of a resonance via gluon fusion,
weak vector boson fusion (VBF) with associated jets, or associated production with a weak vector boson $\V\PH$.
All these mechanisms, along with the $\mathrm{t}\mathrm{\bar{t}}\PH$ production, are considered
in the analysis of the spin-zero hypothesis of the $\PH$ boson, where the gluon fusion production dominates.
It is possible to study $\PH\V\V$ interactions using the kinematics of particles produced in
association with the Higgs boson, such as VBF jets or vector boson daughters in $\V\PH$ production.
While the $q_{\sss\V}^2$ range in the $\PH\to\V\V$ process does not exceed approximately 100\GeV due to
the kinematic bound, in the associated production no such bound exists, and therefore consideration of more restricted
ranges of $q_{\sss\V}^2$ might be required~\cite{Anderson:2013afp}, bringing an additional uncertainty to such a study.
Instead, the analysis presented here is designed to minimize the dependence on the production mechanism
focusing on the study of the $\PH\to\V\V$ decay kinematics.
In the case of a spin-zero particle, there is no spin correlation between the production process
and decay, which allows for production-independent studies. In the case of a non-zero spin particle,
it is possible to study decay information only without dependence on the polarization of the resonance,
and therefore without dependence on the production mechanism.

The production of on-shell Higgs boson is considered in this analysis. In gluon fusion, about 10\%
of $\PH\to\Z\Z$ and $\PH\to\PW\PW$  events are produced off-shell, with a Higgs boson invariant
mass above 150 GeV~\cite{Kauer:2012hd}. A similar effect appears in VBF production~\cite{CMS-HIG-14-002},
while it is further suppressed for other production mechanisms. However, this off-shell contribution
depends on the width of the Higgs boson~\cite{CaolaMelnikov:1307.4935}. A relative enhancement of the off-shell
with respect to the on-shell production is expected in models with anomalous $\PH\V\V$ couplings~\cite{CMS-HIG-14-002, Gainer:2014hha}.
Therefore, it is possible to study anomalous $\PH\V\V$ interactions using the kinematics of the Higgs boson produced
off-shell, including relative off-shell enhancement. However, such a study requires additional assumptions about the width of
the Higgs boson, its production mechanisms, and the extrapolation of the coupling constants
in Eqs.~(\ref{eq:formfact-fullampl-spin0}), (\ref{eq:ampl-spin1}), and  (\ref{eq:ampl-spin2-a}) to $q_\PH^2$ values
significantly larger than 100\GeV. Therefore, the study of anomalous $\PH\V\V$ couplings with the off-shell
Higgs boson is left for a future work. Instead, the $\PH\to\Z\Z$ events with an invariant $\Z\Z$ mass above 140\GeV
are not considered, effectively removing off-shell effects and associated model dependence.
In the $\PH\to\PW\PW$ analysis, the event selection discussed below reduces the off-shell contribution to less than 2\%.
Even though this contribution may increase with anomalous $\PH\V\V$ couplings, no such enhancement has been
observed in the $\PH\to\Z\Z$ study~\cite{CMS-HIG-14-002}, which limits it to be less than five times the SM expectation
at a 95\% \CLend This constraint is expected to be further improved with more data and additional final states studied. In the present analysis, any off-shell contribution in the study of the on-shell production and $\PH\to\PW\PW$ decay is neglected.

Since the production of a color-neutral spin-one resonance is forbidden in gluon fusion, its dominant
production mechanism is expected to be quark-antiquark annihilation. The production mechanisms
of a spin-two boson are expected to be gluon fusion and quark-antiquark annihilation, as for example
in the ED models in Refs.~\cite{Randall:1999vf,Randall:1999ee,Agashe:2007zd}.
While gluon fusion is expected to dominate over the $\qqbar$ production of a spin-two state,
the latter is a possibility in the effective scattering amplitude with form factors.
Therefore, the $\qqbar\to\X$ production of both spin-one and spin-two resonances
and the $gg\to\X$ production of a spin-two resonance are considered.
The fractional contribution of the $\qqbar$ process to the production of a spin-two resonance is denoted as $f(\qqbar)$,
and can be interpreted as the fraction of events produced with $J_z=\pm1$.
For both spin-one and spin-two states, the analysis of the $\X\to \Z\Z\to 4\ell$ decay mode is also performed
without dependence on the production mechanism, allowing coverage of other mechanisms including
associated production.

For the analysis of the $\qqbar\to\X$ production, the general scattering amplitudes
are considered for the interaction of the spin-one and spin-two bosons with fermions
\ifthenelse{\boolean{cms@external}}{
\begin{multline}
 A(\X_{J=1} f\bar f) = \epsilon_{\sss\X}^\mu
\bar u_2
\big (
\gamma_\mu  \left (  \rho_{ 1} + \rho_{ 2} \gamma_5 \right )
\\+ \frac{m_f {\tilde q}_\mu}{\Lambda^2} \left ( \rho_{ 3} +
\rho_{ 4} \gamma_5
\right ) \big ) u_1,
\label{eq:ampl-spin1-qq}
\end{multline}
\begin{multline}
 A(\X_{J=2} f\bar f) = \frac{1}{\Lambda} t^{\mu \nu}
\bar u_2 \big (
 \gamma_\mu {\tilde q}_\nu \left ( \rho_{ 1} + \rho_{ 2}
\gamma_5 \right )
\\+ \frac{ m_f {\tilde q}_\mu {\tilde q}_\nu}{\Lambda^2}
\left (
 \rho_{ 3} + \rho_{ 4} \gamma_5
\right )
\big ) u_1,
\label{eq:ampl-spin2-qq}
\end{multline}
}{
\begin{equation}
A(\X_{J=1} f\bar f) = \epsilon_{\sss\X}^\mu
\bar u_2
\left (
\gamma_\mu  \left (  \rho_{ 1} + \rho_{ 2} \gamma_5 \right )
+ \frac{m_f {\tilde q}_\mu}{\Lambda^2} \left ( \rho_{ 3} +
\rho_{ 4} \gamma_5
\right ) \right ) u_1,
\label{eq:ampl-spin1-qq}
\end{equation}
\begin{equation}
A(\X_{J=2} f\bar f) = \frac{1}{\Lambda} t^{\mu \nu}
\bar u_2 \left (
 \gamma_\mu {\tilde q}_\nu \left ( \rho_{ 1} + \rho_{ 2}
\gamma_5 \right )
+ \frac{ m_f {\tilde q}_\mu {\tilde q}_\nu}{\Lambda^2}
\left (
 \rho_{ 3} + \rho_{ 4} \gamma_5
\right )
\right ) u_1,
\label{eq:ampl-spin2-qq}
\end{equation}
}
where $m_f$ is the fermion mass, $u_i$ is the Dirac spinor,
and $\Lambda$ is the scale of BSM physics~\cite{Gao:2010qx, Bolognesi:2012mm}.
The couplings $\rho_{i}$ are assumed to be the same for all quark flavors.
This assumption, along with the choice of $\rho_{i}$ couplings in general,
has little effect on the analysis since this affects only the expected longitudinal boost of the resonance
from different mixtures of parton production processes without affecting its polarization,
whose projection on the parton collision axis is always $J_z=\pm1$, since
the $\rho_{3}$ and $\rho_{4}$ terms are suppressed in the annihilation of light quarks.
Therefore, $\qqbar$ production leads to a resonance with polarization $J_z=\pm1$
along the parton collision axis, while gluon fusion leads to $J_z=0$ or $\pm2$.
In the case of minimal $c_{1}^{\Pg\Pg}$ coupling, only $J_z=\pm2$ is possible.
The terms proportional to $m_{\sss\V}^2$ in Eq.~(\ref{eq:ampl-spin2-a}) are absent
for couplings to massless vector bosons, either $\Pg\Pg\to\X$ in production or $\X\to \gamma\gamma$ in decay.
Therefore, the list of models in Table~\ref{table-scenarios} covers ten decay couplings to massive vector
bosons but only five couplings for the massless gluons.

The presence of an additional resonance can be inferred from the kinematics of the decay products
when separation in invariant mass alone is not sufficient.
For example, composite particles can have multiple narrow states
with different spin-parity quantum numbers and nearly degenerate masses.
Some examples of this phenomenon include ortho/para-positronia, $\chi_b$ and $\chi_c$ particles
where the mass splitting between the different $J^{P}$ states is orders of magnitude smaller than
their mass~\cite{Agashe:2014kda, Cawlfield:2005ra, Karshenboim:2003vs}.

In an approach common to both the spin-one and spin-two scenarios,
the production of a second resonance with different $J^{P}$ quantum numbers
but close in mass to the SM Higgs-like state can be probed.
The two states are assumed to be sufficiently separated in mass or produced by different
mechanisms, so that they do not interfere, but still to be closer than the experimental mass resolution
\begin{eqnarray}
\Gamma_{J^{P}}\text{ and }\Gamma_{0^{+}} \ll | m_{J^P} - m_{0^{+}} | \ll \delta_m \sim 1\GeV .
\label{eq:Non-interf-limits}
\end{eqnarray}
The fractional cross section $f\!\left(J^{P}\right)$ is defined as follows
\begin{eqnarray}
f\!\left(J^{P}\right) = \frac{ \sigma_{J^{P}}}{ \sigma_{0^{+}} + \sigma_{J^{P}} },
\label{eq:fJCP_definition}
\end{eqnarray}
where $\sigma_{J^{P}}$ ($\sigma_{0^{+}}$) is the cross section of the process
corresponding to the $J^{P}$ ($0^{+}$) model defined at the LHC energy of 8\TeV and,
in the case of the $\Z\Z$ channel, for the $\X\to \Z\Z \to 2\Pe2\mu$ decay mode.
In this case the notation $J^{P}$ refers to a model name and in practice should reflect all
relevant model properties, including spin, parity, production, and decay modes.
It should be noted that the effective fractions $f_{ai}$ and $f\!\left(J^{P}\right)$ have a distinct nature.
The fractions $f_{ai}$ denote the effective fractions related to the corresponding $a_i$ terms within
the scattering amplitude of a given state, and are used in measurements that consider interference
effects between different parts of the amplitude.

\section{The CMS detector, simulation, and reconstruction} \label{sec:CMS}
The central feature of the CMS apparatus is a superconducting solenoid of 6\unit{m}
internal diameter, providing a magnetic field of 3.8\unit{T}. Within the superconducting
solenoid volume are a silicon pixel and strip tracker, a lead tungstate crystal electromagnetic
calorimeter (ECAL), and a brass/scintillator hadron calorimeter, each composed
of a barrel and two endcap sections. Muons are measured in gas-ionization detectors
embedded in the steel flux-return yoke outside the solenoid. Extensive forward calorimetry
complements the coverage provided by the barrel and endcap detectors.  A more detailed
description of the CMS detector, together with a definition of the coordinate system used
and the relevant kinematic variables, can be found in Ref.~\cite{Chatrchyan:2008zzk}.

The data samples used in this analysis are the same as those described in
Refs.~\cite{Chatrchyan:2013iaa,Chatrchyan:2013mxa,CMS-HIG-14-002,Khachatryan:2014ira},
corresponding to an integrated luminosity of 5.1,(4.9)\fbinv collected in 2011 at a center-of-mass energy
of 7\TeV and 19.7\,(19.4)\fbinv in 2012 at 8\TeV in the case of the
$\PH\to\V\V\to 4\ell$ and $\PH\to\gamma\gamma$ ($\PH\to\PW\PW\to\ell\nu\ell\nu$) channels.
The integrated luminosity is measured using data from the CMS hadron forward calorimeter system
and the pixel detector~\cite{Chatrchyan:2014mua, CMS-PAS-LUM-13-001}.
The uncertainties in the integrated luminosity measurement are 2.2\% and 2.6\% in the 2011 and 2012 datasets, respectively.

\subsection{Monte Carlo simulation} \label{sec:mc}

The simulation of the signal process is essential for the study of anomalous couplings in $\PH\V\V$ interactions,
and all the relevant Monte Carlo (MC) samples are generated following the description in Section~\ref{sec:Pheno}.
A dedicated simulation program, \textsc{JHUGen}~4.8.1~\cite{Gao:2010qx,Bolognesi:2012mm,Anderson:2013afp},
is used to describe anomalous couplings in the production and decay to two vector bosons of spin-zero,
spin-one, and spin-two resonances in hadron-hadron collisions, including all the models listed in
Tables~\ref{tab:xsec_ratio}~and~\ref{table-scenarios}. For the spin-zero and spin-one studies, interference effects
are included by generating mixed samples produced with either of the different tensor structures shown
in Eqs.~(\ref{eq:formfact-fullampl-spin0}) and (\ref{eq:ampl-spin1}).

For gluon fusion production of a spin-zero state, the kinematics of the Higgs boson decay products
and of an associated jet are not affected by the anomalous $\PH\Pg\Pg$ interactions,
and therefore the next-to-leading-order (NLO) QCD effects are introduced in production with the SM couplings
through the \POWHEG~\cite{Frixione:2007vw,Bagnaschi:2011tu,Nason:2009ai} event generator.
It is also found that the NLO QCD effects that are relevant for the analysis of a spin-zero state
are well approximated with the combination of leading-order (LO) QCD matrix elements and parton
showering~\cite{Anderson:2013afp}. Therefore, \textsc{JHUGen} at LO QCD is adopted for the simulation
of anomalous interactions in all the other production processes where it is important to model the correlations
between production and the kinematics of the final-state particles, such as in VBF and $\V\PH$ production
of a spin-zero state, $\qqbar\to\X$ production of a spin-one state, and $\qqbar$ and $\Pg\Pg\to\X$ production
of a spin-two state. In the case of a spin-two $\X$ boson, the LO QCD modeling of production
avoids potentially problematic $\PT$ spectrum of the $\X$ boson appearing at NLO with non-universal
$\X\qqbar$ and $\X\Pg\Pg$ couplings~\cite{Artoisenet:2013puc} allowed in this study.
In all cases, the decays $\PH/\X\to \Z\Z$ / $\Z\gamma^*$ / $\gamma^*\gamma^*\to4\ell$,
$\PH/\X \to \PW\PW \to \ell\nu\ell\nu$, and $\PH/\X\to\gamma\gamma$  are simulated with \textsc{JHUGen},
including all spin correlations in the production and decay processes and interference effects between
all contributing amplitudes.

To increase the number of events in the simulated samples for each hypothesis studied,
the \textsc{mela} package~\cite{Chatrchyan:2012ufa,Gao:2010qx,Bolognesi:2012mm,Anderson:2013afp}
is adopted to apply weights to events in any $\PH\to\V\V\to4\ell$ or $\PH\to\PW\PW\to\ell\nu\ell\nu$ spin-zero sample
to model any other spin-zero sample. The same re-weighting technique has also been used in the study of the
$\qqbar$ and $\Pg\Pg\to\Z\Z/\Z\gamma^*$ backgrounds.

All MC samples are interfaced with \PYTHIA~6.4.24~\cite{Sjostrand:2006za} for parton showering and further
processing through a dedicated simulation of the CMS detector based on \GEANTfour~\cite{Agostinelli2003250}.
The simulation includes overlapping pp interactions (pileup) matching the distribution of the number
of interactions per LHC beam crossing observed in data.

Most of the background event simulation is unchanged since
Refs.~\cite{Chatrchyan:2013iaa,Chatrchyan:2013mxa,CMS-HIG-14-002,Khachatryan:2014ira}.
In the $\PH\to\V\V\to 4\ell$ analysis, the $\qqbar\to\Z\Z/\Z\gamma^*$ process is simulated with
 \POWHEG.
The $\Pg\Pg\to\Z\Z/\Z\gamma^*$ process is simulated with both \textsc{gg2zz}  3.1.5~\cite{Binoth:2008pr}
and \MCFM~6.7~\cite{MCFM,Campbell:2011bn,Campbell:2013una},
where the Higgs boson production K-factor is applied to the LO cross section~\cite{CMS-HIG-14-002}.
In the $\PH\to\PW\PW\to\ell\nu\ell\nu$ analysis,
the $\PW\Z$, $\Z\Z$, $\V\V\V$, Drell--Yan (DY) production of $\Z/\gamma^*$, $\Wjets$,  $\wgamma^{*}$,
and $\qqbar \to\PW\PW$ processes are generated using the \MADGRAPH 5.1 event generator~\cite{madgraph},
the $\Pg\Pg \to \PW\PW$ process using the \textsc{gg2ww} 3.1 generator~\cite{ggww}, and the $\ttbar$
and $\tw$ processes are generated with \POWHEG. The electroweak production of the non-resonant
$\PW\PW$ + 2~jets process, which is not part of the inclusive $\PW\PW$ + jets sample, has been generated
using the \textsc{phantom} 1.1 event generator~\cite{phantom}  including terms of order $(\alpha_{EM}^{6})$.
The \TAUOLA package~\cite{tauola} is used in the simulation
of $\tau$-lepton decays to account for $\Pgt$-polarization effects.

\subsection{Event reconstruction} \label{sec:Reco}

The analysis uses the same event reconstruction and selection as in the previous measurements
of the properties of the Higgs boson in the
$\PH\to\V\V\to4\ell$~\cite{Chatrchyan:2013mxa,CMS-HIG-14-002},
$\PH\to\PW\PW\to\ell\nu\ell\nu$~\cite{Chatrchyan:2013iaa},
and $\PH\to\gamma\gamma$~\cite{Khachatryan:2014ira} channels.
The data from the CMS detector and the simulated samples are reconstructed using the same algorithms.

For the $\PH\to\V\V\to4\ell$ and $\PH\to\PW\PW\to\ell\nu\ell\nu$ analyses described in this paper, events are triggered by requiring
the presence of two leptons, electrons or muons, with asymmetric requirements on their transverse momenta, $\pt$.
Several single-lepton triggers with relatively tight lepton identification are used for the $\PH\to\PW\PW$ analysis.
A triple-electron trigger is also used for the $\PH\to\V\V\to4\ell$ analysis.
For the $\PH\to\gamma\gamma$ analysis, the events are selected by diphoton triggers with asymmetric transverse
energy thresholds and complementary photon selections.
The particle-flow (PF) algorithm~\cite{CMS-PAS-PFT-09-001, CMS-PAS-PFT-10-001} is used to reconstruct the observable
particles in the event. The PF event reconstruction consists of reconstructing and identifying each single particle
with an optimized combination of all sub-detector information.

The $\PH\to\V\V\to4\ell$ and $\PH\to\PW\PW\to\ell\nu\ell\nu$ analyses require four and two lepton candidates
(electrons or muons), respectively, originating from a vertex with the largest $\sum \pt^2$ of all tracks associated with it.
Electron candidates are defined by a reconstructed charged-particle track in the
tracking detector pointing to an energy deposition in the ECAL.
The electron energy is measured primarily from the ECAL cluster energy.
Muon candidates are identified by signals of charged-particle tracks in the muon system
that are compatible with a track reconstructed in the central tracking system.
Electrons and muons are required to be isolated. Electrons are reconstructed within the geometrical acceptance, $\abs{\eta} < 2.5$,
and for $\PT > 7\GeV$. Muons are reconstructed within
$\abs{\eta} < 2.4$ and $\PT > 5\GeV$~\cite{Chatrchyan:2012xi}.

Photons, used in the $\PH\to\gamma\gamma$ analysis, are identified as ECAL energy clusters
not linked to the extrapolation of any charged-particle trajectory to the ECAL.
Jets, used in the $\PH\to\PW\PW$ analysis, are reconstructed from the PF candidates,
clustered with the anti-$k_t$ algorithm~\cite{Cacciari:2008gp, Cacciari:2011ma} with a size parameter of 0.5.
The jet momentum is determined as the vector sum of all particle momenta in the jet.
Identification of b-quark decays is used to reject backgrounds containing top quarks that subsequently decay to a b quark and a $\PW$ boson
in the $\PH\to\PW\PW$ analysis.
The missing transverse energy vector $\VEtmiss$ is defined as
the negative vector sum of the transverse momenta of all reconstructed
particles (charged or neutral) in the event, with $\ETm = |\VEtmiss|$.

\subsection{Four-lepton event selection}

To study the $\PH\to\V\V\to4\ell$ decay, events are selected with at least four identified and isolated electrons or muons.
A $\V\to \ell^+\ell^-$ candidate originating from a pair of leptons of the same flavor and
opposite charge is required.
The $\ell^+\ell^-$ pair with an invariant mass, $m_1$, nearest to the nominal $\Z$ boson mass is retained
and is denoted $\Z_{1}$ if it is in the range $40 \le m_1 \le 120 \GeV$.
A second $\ell^{+}\ell^{-}$ pair, denoted $\Z_{2}$, is required to have $12 \le m_2 \le 120 \GeV$.
If more than one Z$_{2}$ candidate satisfies all criteria, the pair of
leptons with the highest scalar $\PT$ sum is chosen.
At least one lepton should have $\PT \ge 20 \GeV$, another one
$\PT \ge 10 \GeV$ and any oppositely charged pair of leptons among the four selected must satisfy
$m_{\ell\ell} \ge 4 \GeV$. Events are restricted to a window around the observed
125.6\GeV resonance, $105.6 \le m_{4\ell} \le 140.6 \GeV$.

After the selection, the dominant background for $\PH\to\V\V\to4\ell$ originates from the
$\Pq\Paq \to \Z\Z/\Z\gamma^*$ and $\Pg\Pg \to \Z\Z/\Z\gamma^*$ processes
and is evaluated from simulation, following Refs.~\cite{Chatrchyan:2013mxa,CMS-HIG-14-002}.
The other backgrounds come from the production of $\cPZ$ and $\PW\cPZ$ bosons in association with jets,
as well as $\ttbar$, with one or two jets misidentified as an electron or a muon.
The $\Z + X$ background is evaluated using a tight-to-loose misidentification rate method~\cite{Chatrchyan:2013mxa}.
The number of estimated background and signal events, and the number of
observed candidates after the final selection in data in the
narrow mass region around 125.6\GeV  is given in Table~\ref{tab:EventYieldsLowMass}.

\begin{table*}[!hbt]
\centering
\topcaption{
Number of background (Bkg.) and signal events expected in the SM, and number of
observed candidates, for the $\PH\to\V\V\to4\ell$ analysis after the final selection in the
mass region $105.6 < m_{4\ell} < 140.6 \GeV$. The signal and $\Z\Z$
background are estimated from MC simulation, while the $\Z +X$ background
is estimated from data. Only systematic uncertainties are quoted.
\label{tab:EventYieldsLowMass}}
\begin{scotch}{lcccccc}
Channel & \multicolumn{2}{c}{$4\Pe$} &  \multicolumn{2}{c}{$4\Pgm$} &  \multicolumn{2}{c}{$2\Pe2\Pgm$} \\
Energy & 7\TeV & 8\TeV & 7\TeV & 8\TeV & 7\TeV & 8\TeV  \\
\hline
$\Pq\Paq \to \Z\Z$ &  0.84  $\pm$  0.10  &  2.94 $\pm$  0.33  &  1.80  $\pm$  0.11  & 7.65  $\pm$  0.49  &  2.24  $\pm$  0.28 & 8.86  $\pm$  0.68  \\
$\Pg\Pg \to \Z\Z$ &  0.03  $\pm$  0.01 &  0.20  $\pm$  0.05    &  0.06  $\pm$  0.02 &  0.41  $\pm$  0.10    &  0.07  $\pm$  0.02   &  0.50  $\pm$  0.13 \\
$\Z + X$ & 0.62 $\pm$ 0.14 & 2.77 $\pm$ 0.62 & 0.22 $\pm$ 0.09 & 1.19 $\pm$ 0.48 & 1.06 $\pm$ 0.29 & 4.29 $\pm$ 1.10\\
\hline
Bkg. & 1.49 $\pm$ 0.17 & 5.91 $\pm$ 0.71  & 2.08 $\pm$ 0.14  & 9.25 $\pm$ 0.69 &  3.37 $\pm$ 0.40 & 13.65 $\pm$ 1.30 \\
Signal &  0.70  $\pm$  0.11  &  3.09 $\pm$  0.47  &  1.24  $\pm$  0.14 &  5.95  $\pm$  0.71  & 1.67  $\pm$  0.26  &  7.68  $\pm$  0.98 \\
Observed  & 1 & 9 & 3 & 15 & 6 & 16 \\
\end{scotch}

\end{table*}

\subsection{Two-lepton event selection}

In the case of the $\PH\to\PW\PW\to\ell\nu\ell\nu$  analysis, events with exactly one electron and one muon are selected.
The leptons must have opposite charge and pass the full identification and isolation criteria
presented in detail in Ref.~\cite{Chatrchyan:2013iaa}.
The highest-$\pt$ (leading) lepton should have $\pt> 20\GeV$, and the second one $\pt > 10\GeV$.
Events are classified according to the number of selected jets that satisfy $\et>30\GeV$ and $\abs{\eta}<4.7$.
Two categories of events with exactly zero and exactly one jet are selected, in which the signal is produced mostly
by the gluon fusion process.
The $\Pe\mu$ pair is required to have an invariant mass above $12\GeV$,
and a \pt above $30\GeV$. Events are also required to have \textit{projected}~$\MET$ above 20~$\GeV$,
as defined in Ref.~\cite{Chatrchyan:2013iaa}.

The main background processes from non-resonant $\PW\PW$ production and
from top-quark production, including top-quark pair ($\ttbar$) and
single-top-quark (mainly $\tw$) processes, are estimated using data.
Instrumental backgrounds arising from
misi\-denti\-fi\-ed (``nonprompt") leptons in $\PW$+jets production and
mismeasurement of $\VEtmiss$ in $\Z/\gamma^*$+jets events are also estimated from data.
The contributions from other sub-dominant diboson ($\PW\Z$ and $\Z\Z$)
and triboson ($\VVV$) production
processes are estimated from simulation.
The $\wgamma^*$ cross section is measured from data.
The shapes of the discriminant variables used in the signal extraction
for the $\wgamma$ process are also obtained from data.
The $\dytt$ background process is estimated using $\dymm$ events selected in
data where the muons are replaced with simulated $\tau$-lepton decays.
To suppress the background from top-quark production,
events that are identified as coming from top decays are rejected based on soft-muon and b-jet
identification.
The number of estimated background and signal events, and number of
observed candidates after the final selection are given in Table~\ref{tab:hww01j_yields}.
After all selection criteria are applied, the contribution from other Higgs boson
decay channels is negligible.

\begin{table*}[htbp]
\centering
\topcaption{
Number of background and signal events expected in the SM, and number of
observed candidates, for the $\PH\to\PW\PW$ analysis after final selection. The signal and
background are estimated from MC simulation and from data control regions,
as discussed in the text. Only systematic uncertainties are quoted.
}
\label{tab:hww01j_yields}
 \begin{scotch}{lcccc}
 Channel & \multicolumn{2}{c}{ 0-jet} &  \multicolumn{2}{c}{1-jet} \\
Energy & 7\TeV & 8\TeV & 7\TeV & 8\TeV  \\
\hline
 $\PW\PW$ 					& $861\pm12$ 	& $4185\pm63$	& $249.9\pm4.0$ & $1268\pm21$  \\
$\PW\Z+\Z\Z+\Z/\gamma^*$ & $22.7\pm1.2$ & $178.3\pm9.5$ 	& $26.4\pm1.4$ & $193\pm11$\\
$\ttbar+\tw$ 				& $91\pm20$ 	& $500\pm96$	&  $226\pm14$ & $1443\pm46$\\
$\Wjets$ 					& $150\pm39$ 	&  $620\pm160$ 	& $60\pm16$ & $283\pm72$\\
$\wgamma^{(*)}$ 			& $68\pm20$	& $282\pm76$	& $10.1\pm2.8$ & $55\pm14$  \\

\hline

Background              		& $1193\pm50$ & $5760\pm210$	& $573\pm22$ & $3242\pm90$  \\
Signal $\Pg\Pg\to\PH$ 				& $50\pm10$ 	& $227\pm46$ 	& $17.1\pm5.5$ & $88\pm28$\\
Signal VBF+$\V\PH$ 				& $0.44\pm0.03$ & $10.27\pm0.41$	& $2.09\pm0.12$ & $19.83\pm0.81$\\
Observed 					& 1207 		& 5747 		& 589 &  3281\\
\end{scotch}

\end{table*}

\subsection{Two-photon event selection} \label{sec:hgg_select}

In the $\PH\to\gamma\gamma$ analysis,
the energy of photons used in the global event reconstruction is directly obtained from the ECAL measurement.
The selection requires a loose calorimetric identification based on the shape
of the electromagnetic shower and loose isolation requirements on the photon candidates.
For the spin-parity studies, the ``cut-based'' analysis described in Ref.~\cite{Khachatryan:2014ira}
is used. This analysis does not use multivariate techniques for selection or classification of events,
which allows for a categorization better suited for the study of the Higgs boson decay kinematics.
The cosine of the scattering angle in the Collins--Soper frame, $\cos\theta^*$, is used to discriminate
between the spin hypotheses.
The angle is defined in the diphoton rest frame as that between the collinear photons
and the line that bisects the acute angle between the colliding protons.
To increase the sensitivity, the events are categorized using
the same four diphoton event classes used in the cut-based analysis
but without the additional classification based on $\PT$ used there.
Within each diphoton class, the events are binned in $|\cos\theta^*|$ to discriminate between the different spin hypotheses.
The events are thus split into 20 event classes, four $(\eta, R_9)$~\cite{Khachatryan:2014ira}
diphoton classes with five $|\cos\theta^*|$ bins each, for both the 7 and 8\TeV datasets, giving a total of 40 event classes.
In Table~\ref{tab:hgg_yields},
the number of estimated background and signal events, and number of observed candidates are given after the final selection
in an $m_{\gamma\gamma}$ range centered at $m_\PH=125$ GeV and corresponding to the full width at half maximum
for the signal distribution for each of the four $(\eta, R_9)$ categories.
The total expected number of selected signal events, summed over all categories and integrated over the full signal distribution, is 421 (94) at 8\TeV (7\TeV).

\begin{table*}[htbp]
\centering
\topcaption{
Number of background and signal events expected in the SM, and number of
observed candidates, for the $\PH\to\gamma\gamma$  analysis after final selection.
The four categories are defined as follows~\cite{Khachatryan:2014ira}:
low $\abs{\eta}$ indicates that both photons are in the barrel with $\abs{\eta} < 1.5$ and  high $\abs{\eta}$ otherwise,
high $R_9$ indicates that both photons have $R_9>0.94$ and low $R_9$ otherwise.
The $m_{\gamma\gamma}$ range (\GeVns{}) centered at $m_\PH=125$ GeV
corresponds to the full width at half maximum for the signal distribution in each category. Only systematic uncertainties are quoted, which include uncertainty from the background
$m_{\gamma\gamma}$ parameterization in the background estimates.
}
\label{tab:hgg_yields}
 \begin{scotch}{lcccccccc}
 Channel & \multicolumn{2}{c}{(low $\abs{\eta}$, high $R_9$)} & \multicolumn{2}{c}{(low $\abs{\eta}$, low $R_9$)}
               & \multicolumn{2}{c}{(high $\abs{\eta}$, high $R_9$)} & \multicolumn{2}{c}{(high $\abs{\eta}$, low $R_9$)} \\
Energy & 7\TeV & 8\TeV & 7\TeV & 8\TeV & 7\TeV & 8\TeV & 7\TeV  & 8\TeV\\
Range $m_{\gamma\gamma}$   &  2.44 & 2.30   & 3.34 & 2.94  & 4.72 & 4.58  & 5.48 & 5.57\\
\hline
Background              		&  230.1$\pm$2.5& 875$\pm$5  & 604$\pm$4 & 2210$\pm$8  & 456$\pm$8 & 1685$\pm$9 & 911$\pm$6 & 2045$\pm$11\\
Signal				 & 18.6$\pm$2.3 & 74$\pm$9  & 23.5$\pm$3.0 & 103$\pm$13  & 9.3$\pm$1.3 & 38$\pm$5  & 12.0$\pm$1.6 & 57$\pm$8 \\
Observed 					 & 263 & 11047 & 647 & 1963  & 459 & 1638   & 913& 1988\\
\end{scotch}
\end{table*}

\section{Analysis techniques} \label{sec:AnalysisStrategyIntro}
\begin{figure}[htbp]
\centering
\includegraphics[width=\cmsFigWidth]{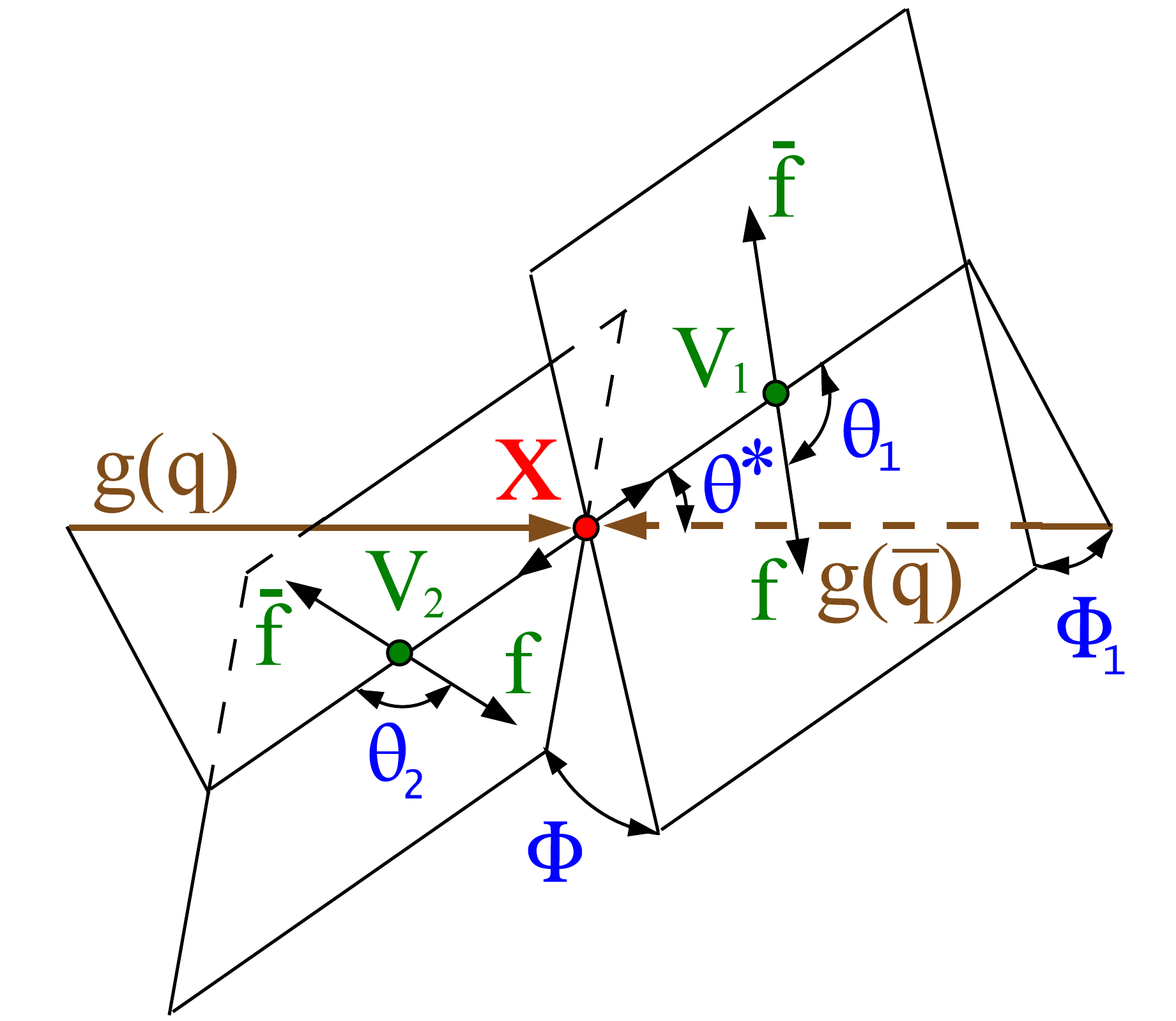}
\caption{
Illustration of the production of a system $\X$ in a parton collision and its decay to two vector bosons
$\Pg\Pg$ or  $\qqbar \to \X\to \Z\Z, \PW\PW, \Z\gamma,$ and $\gamma\gamma$ either with or without
sequential decay of each vector boson to a fermion-antifermion pair~\cite{Gao:2010qx,Cabibbo:1965zz}.
The two production angles $\theta^*$ and $\Phi_1$ are shown in the $\X$ rest
frame and the three decay angles $\theta_1$, $\theta_2$, and $\Phi$ are shown in the $\V$ rest frames.
Here $\X$ stands either for a Higgs boson, an exotic particle, or, in general, the genuine or misidentified $\V\V$ system, including background.}
\label{fig:decay}
\end{figure}

The kinematics of the Higgs boson decay to four charged leptons, two charged leptons and two neutrinos,
or two photons, and their application to the study of the properties of the Higgs boson have been extensively
studied in the literature~\cite{Soni:1993jc,Barger:1993wt,Choi:2002jk,Choi:2002jk,
Buszello:2002uu,Godbole:2007cn,Keung:2008ve,Antipin:2008hj,Hagiwara:2009wt,
Gao:2010qx,DeRujula:2010ys,Gainer:2011xz, Bolognesi:2012mm, Chen:2012jy,Anderson:2013afp,Gainer:2013rxa}.
The schematic view of the production and decay information can be seen
in Fig.~\ref{fig:decay}~\cite{Gao:2010qx,Cabibbo:1965zz}.

If the resonance has a non-zero spin, its polarization depends on the production mechanism.
As a result, a non-trivial correlation of the kinematic distributions of production
and sequential decay is observed for a resonance with non-zero spin,
while there is no such direct correlation due to polarization for a spin-zero resonance.
Furthermore, the kinematics and polarization of the vector bosons in the $\PH\to\V\V$ process depend
on the initial polarization of the resonance and the tensor structure of the $\PH\V\V$ interactions and
this affects the kinematics of the leptons in the  $\V\V\to 4\ell$ or $\ell\nu\ell\nu$ decay.

The analysis of the $\PH\V\V$ interactions requires the study of the kinematic distributions
of the Higgs boson decay products comparing to the prediction of the corresponding models.
In the case of the $\PH\to\V\V\to4\ell$ decay, the full kinematic information can be reconstructed with small
experimental uncertainties. In the case of the $\PH\to\PW\PW\to\ell\nu\ell\nu$
decay, the two missing neutrinos lead to a loss of kinematic information, but in some cases the
V-A nature of the $\PW\to\ell\nu$ coupling, compared to a different V-A coupling in the $\Z\to\ell\ell$ decay,
leads to more pronounced kinematic effects. In the following, the partial kinematic information used
in the analysis of this decay mode is also introduced.

The spin-parity analysis of the $\PH\to\gamma\gamma$ decay is also possible
and this channel is studied  in the context of the exotic spin-two hypothesis tested with respect
to the SM hypothesis.
However, only one angle $\theta^*$ out of the five identified in Fig.~\ref{fig:decay}
is observable in this case. Its distribution is isotropic in the boson frame for any spin-zero model,
and therefore such models cannot be distinguished in this way.
Details of the reconstruction and analysis of the $\cos\theta^*$ distribution in
the $\PH\to\gamma\gamma$ channel are discussed in Ref.~\cite{Khachatryan:2014ira}.

The rest of this section is organized as follows. The kinematic observables reconstructed in
the $\PH\to\V\V\to4\ell$ and $\PH\to\PW\PW\to\ell\nu\ell\nu$ channels are discussed first. A matrix
element likelihood approach is introduced next. Its goal is to reduce the number of observables
to be manageable in the following analysis, while retaining full information for the measurements
of interest. A maximum likelihood fit employs the template parameterization of the probability
distribution of the kinematic observables using full simulation of the processes in the detector.
This method is validated with the analytic parameterization of some of the multidimensional
distributions using a simplified modeling of the detector response in the $\PH\to\Z\Z\to4\ell$ channel.
Systematic uncertainties and validation tests are also discussed.

\subsection{Observables in the \texorpdfstring{$\PH\to\V\V\to 4\ell$}{HVV to 4 l} analysis}\label{sec:Observables}

The four-momenta of the $\PH \to 4\mathrm{\ell}$ decay products carry eight independent degrees of freedom,
which fully describe the kinematic configuration of a four-lepton system in its center-of-mass frame, except for an arbitrary
rotation around the beam axis. These can be conveniently expressed in terms of the five angles
$\vec\Omega\equiv(\theta^*, \Phi_1, \theta_1, \theta_2, \Phi)$ defined in Fig.~\ref{fig:decay}, the invariant masses
of the dilepton pairs, $m_{1}$ and $m_{2}$, and of the four-lepton system, $m_{4\ell}$.
The boost of the $\PH$ boson system in the laboratory frame, expressed as $\PT$ and rapidity, depends on the production
mechanism and generally carries some but limited discrimination power between either signal or background
hypotheses originating from different production processes. These observables are not used in the analysis
to remove the dependence of the results on the production model. For the same reason, information about particles
produced in association with $\PH$ boson is not used either. This approach differs from the study reported in
Ref.~\cite{Chatrchyan:2013mxa} where such observables were used to investigate the production mechanisms
of the Higgs boson.

\begin{figure*}[htbp]
\centering
\includegraphics[width=0.32\textwidth]{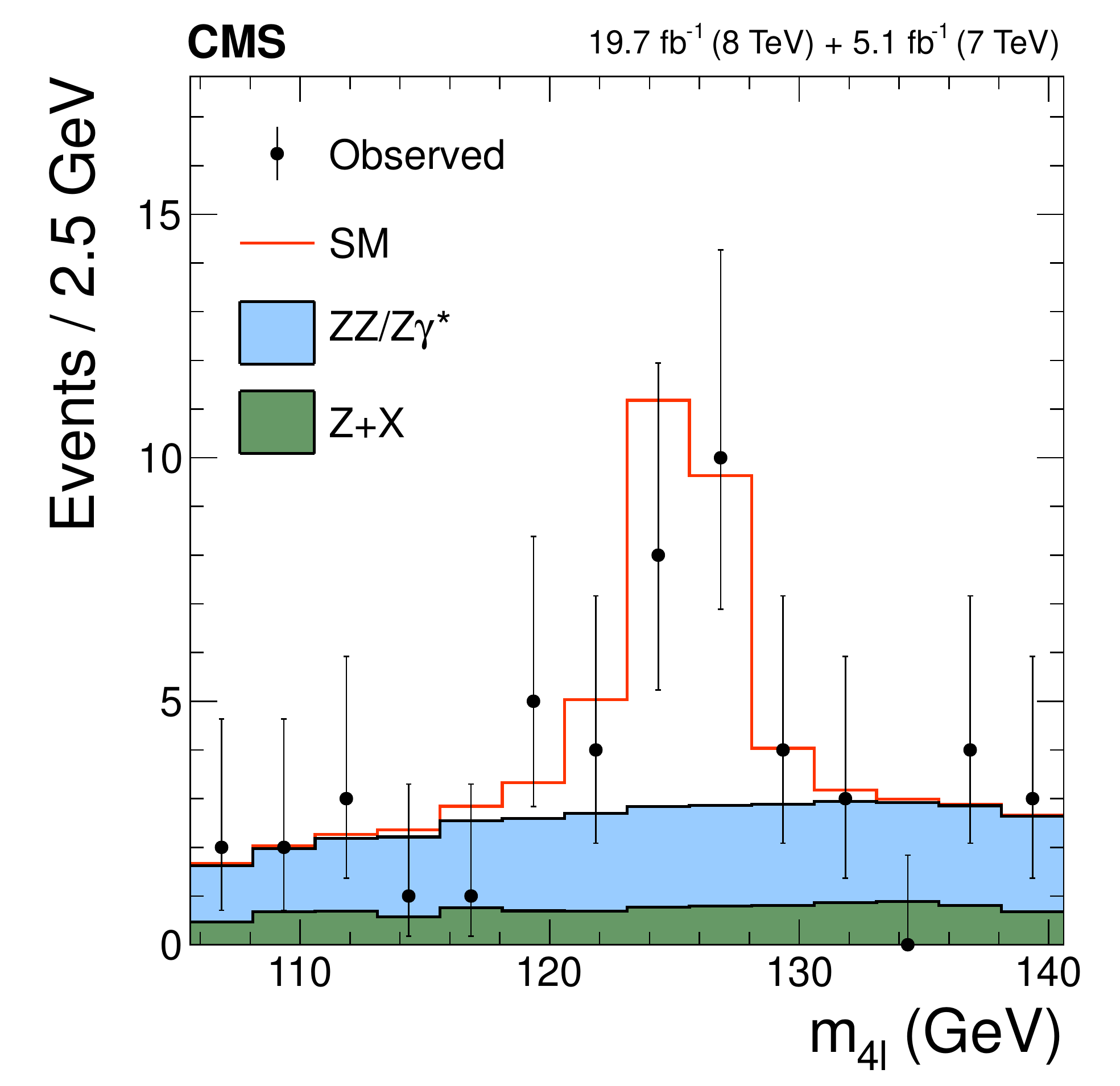}
\includegraphics[width=0.32\textwidth]{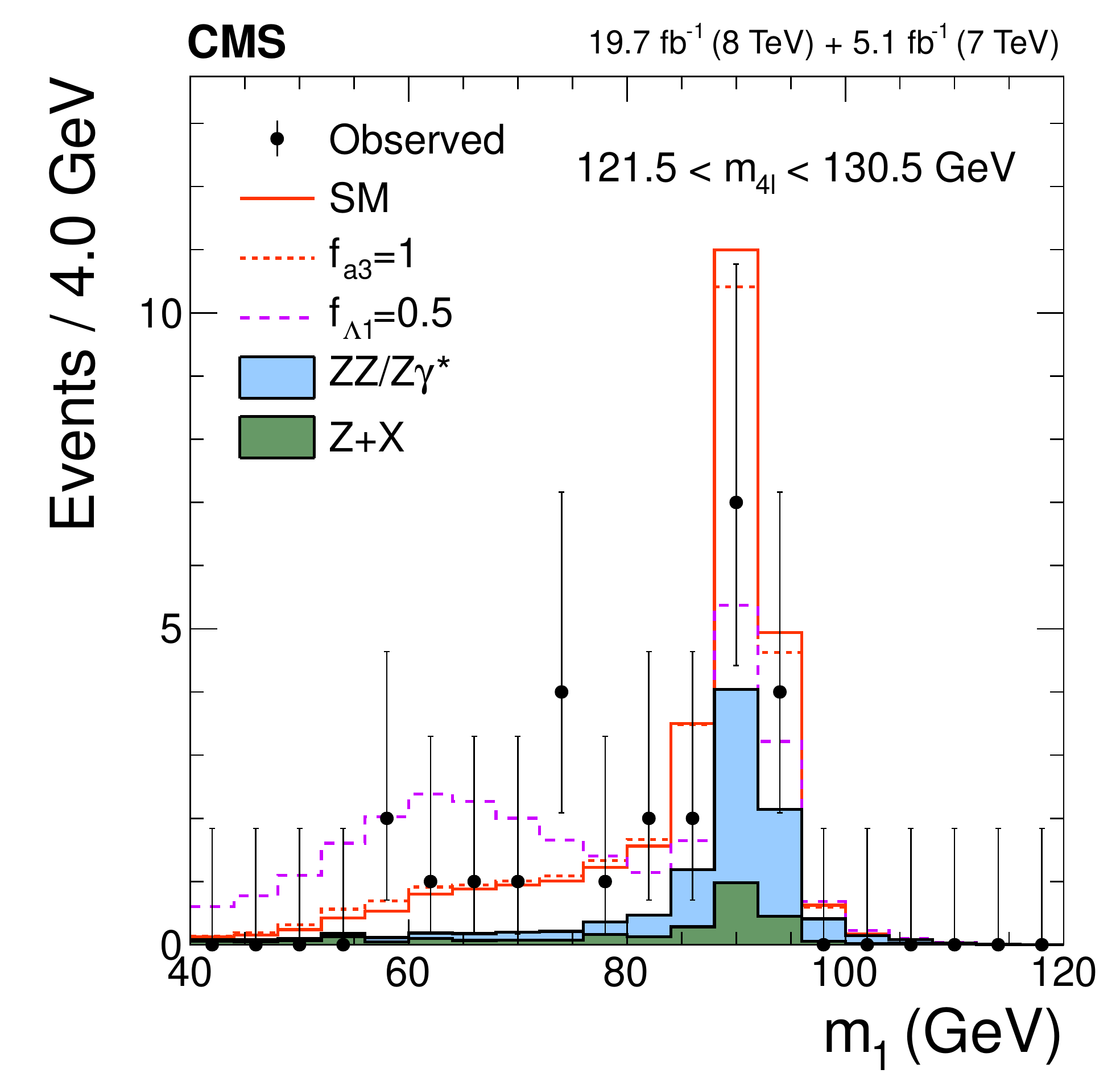}
\includegraphics[width=0.32\textwidth]{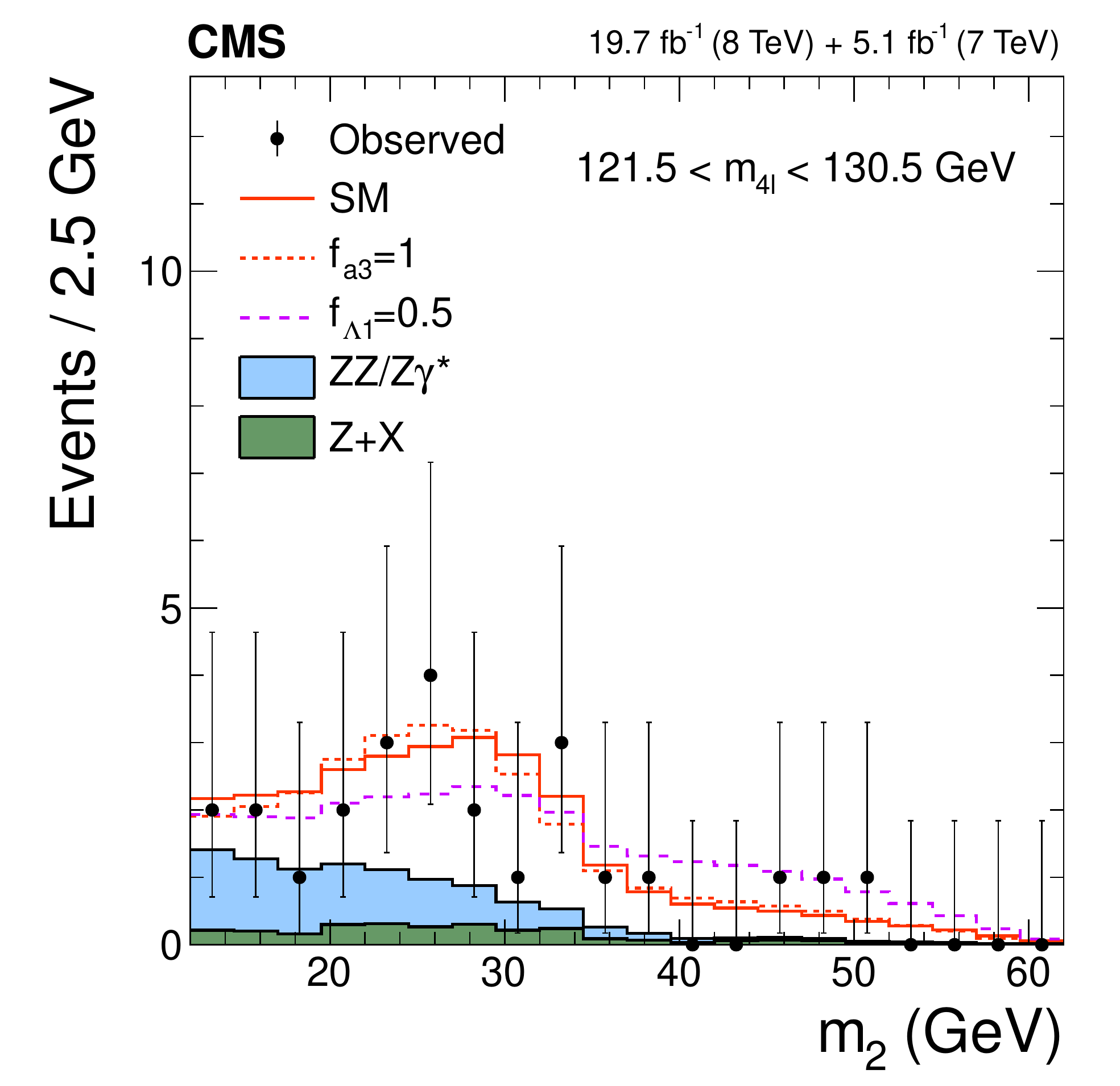}
\includegraphics[width=0.32\textwidth]{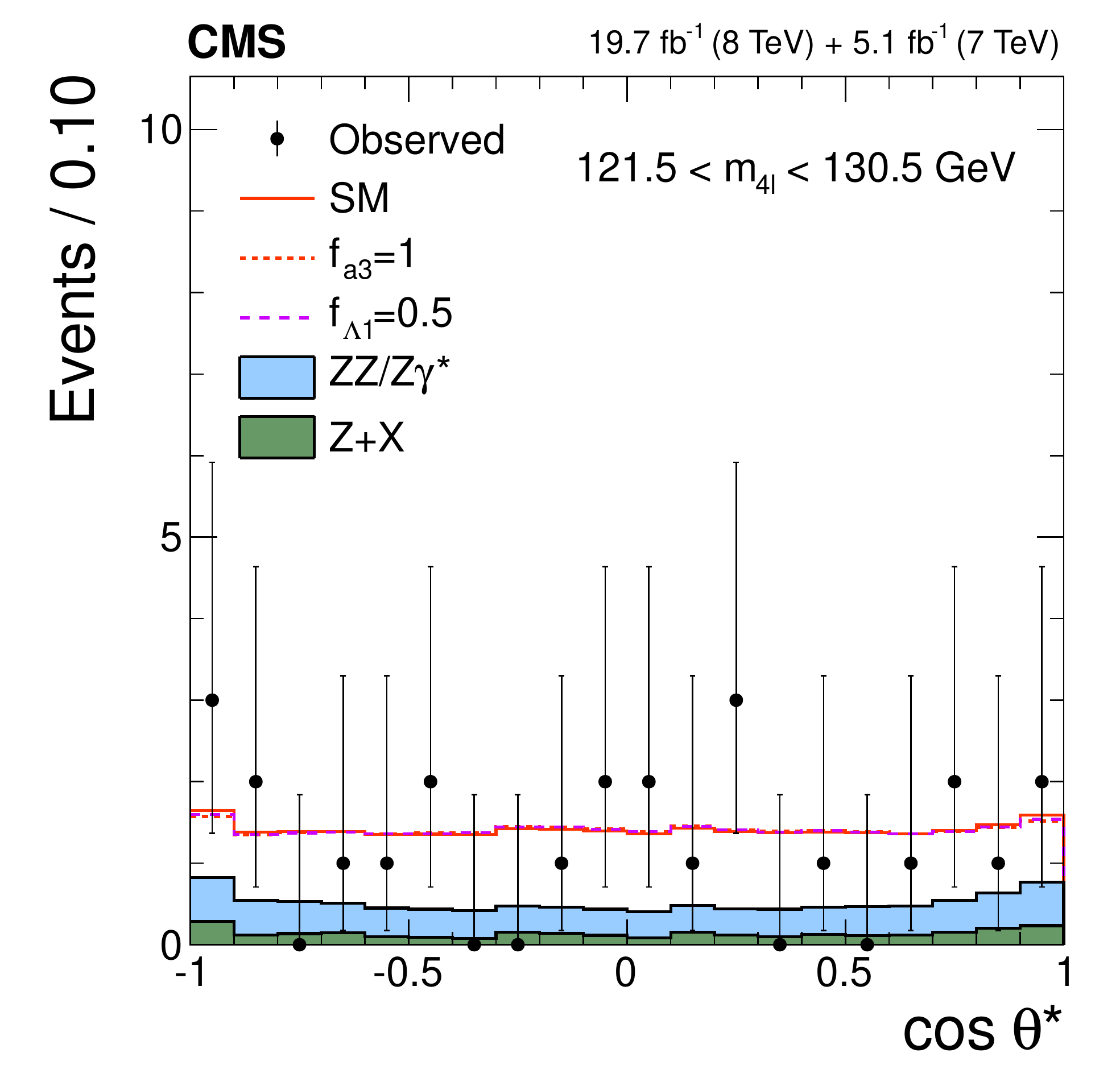}
\includegraphics[width=0.32\textwidth]{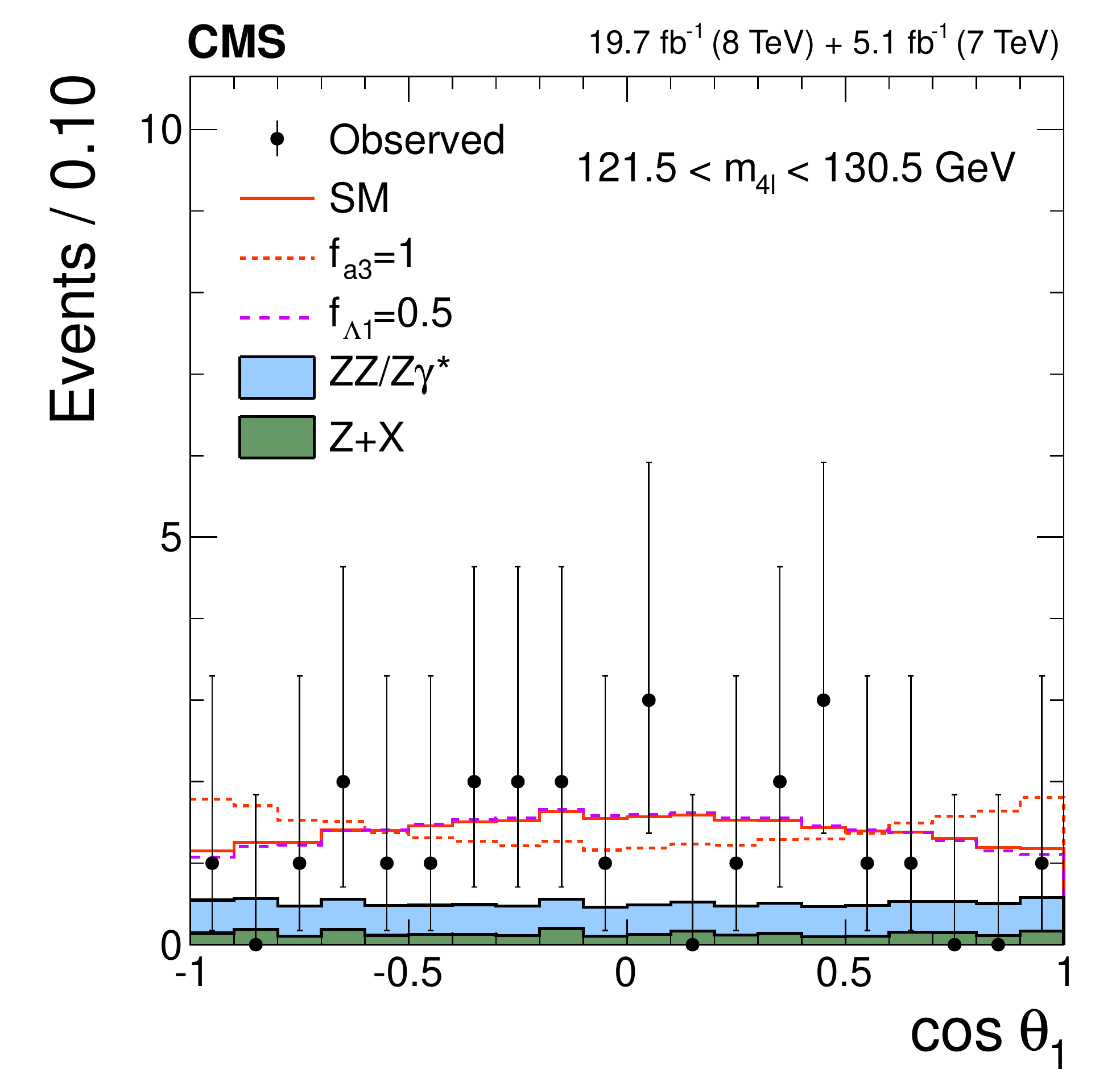}
\includegraphics[width=0.32\textwidth]{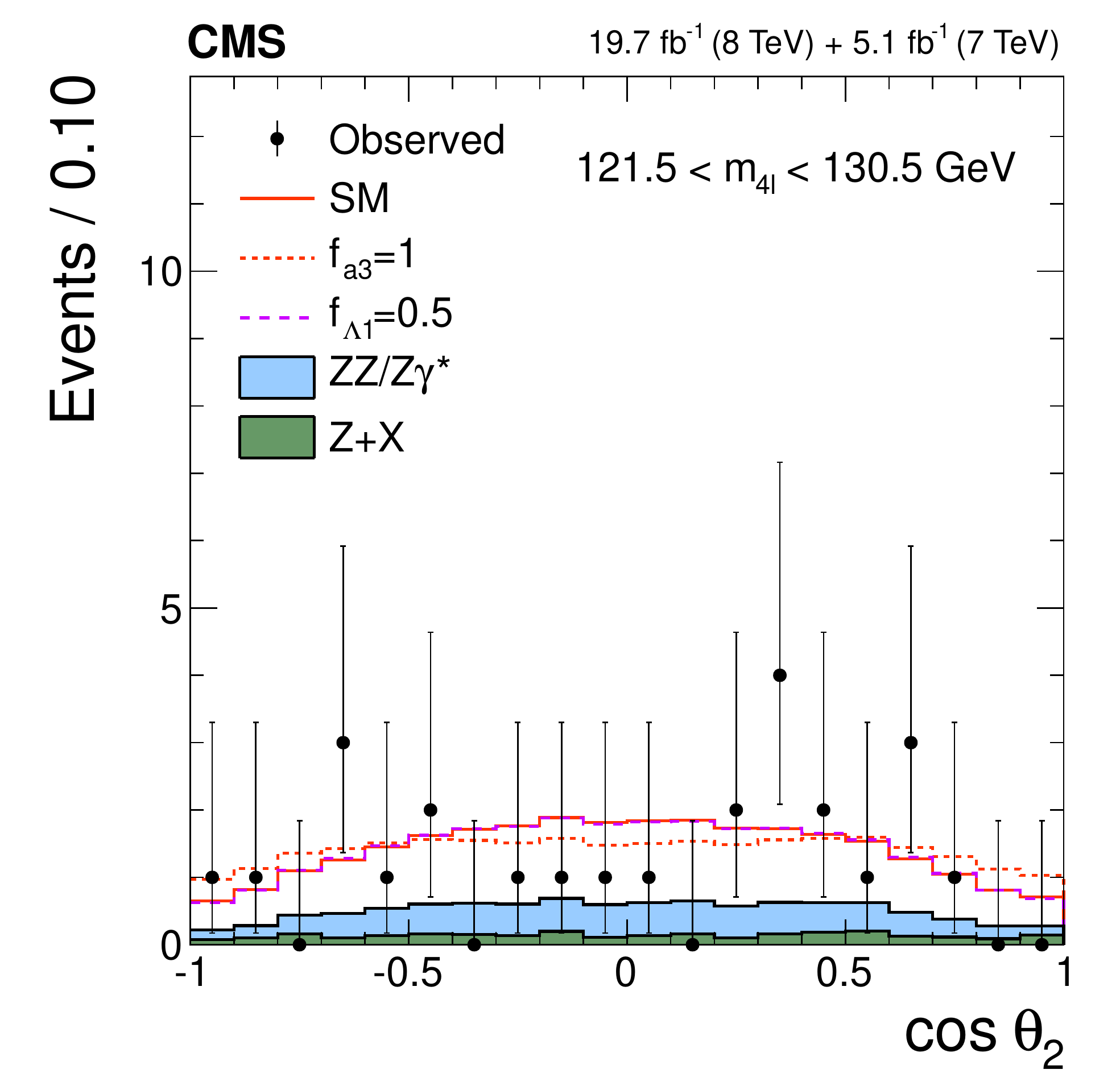}
\includegraphics[width=0.32\textwidth]{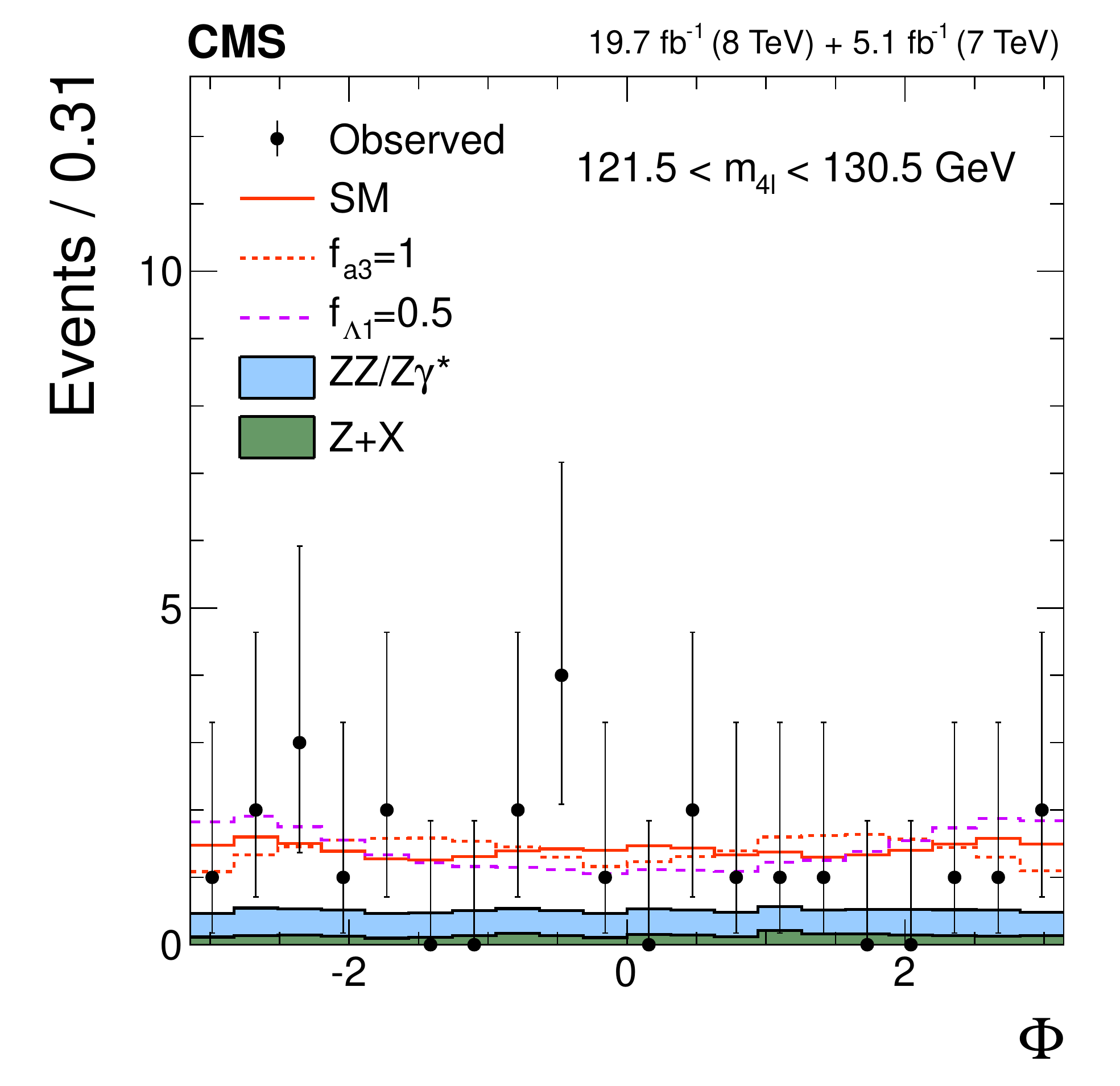}
\includegraphics[width=0.32\textwidth]{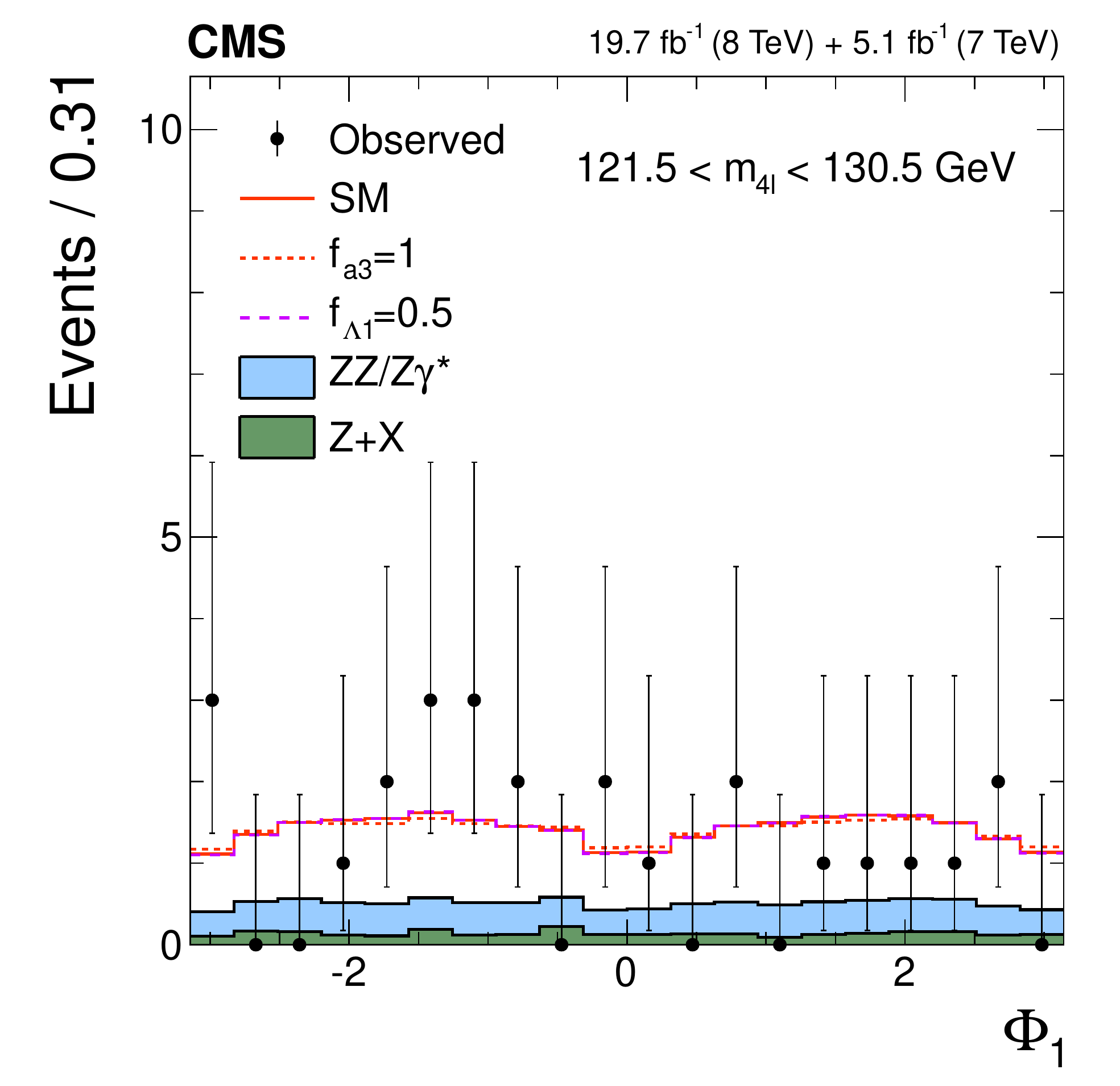}
\caption{
Distributions of the eight kinematic observables used in the $\PH\to\V\V\to4\ell$ analysis:
$m_{4\ell}$,
$m_1$,
$m_2$,
$\cos\theta^*$,
$\cos\theta_{1}$,
$\cos\theta_{2}$,
$\Phi$, and
$\Phi_{1}$.
The observed data (points with error bars), the expectations for the SM background (shaded areas),
the SM Higgs boson signal (open areas under the solid histogram),
and the alternative spin-zero resonances (open areas under the dashed histograms) are shown,
as indicated in the legend.
The mass of the resonance is taken to be 125.6\GeV and the SM cross section is used.
All distributions, with the exception of $m_{4\ell}$, are presented with the requirement
$121.5 < m_{4\ell} < 130.5$\GeV.
}
\label{fig:kinematics}

\end{figure*}

The distributions of the eight kinematic observables $(m_1, m_2, m_{4\ell}, \vec\Omega)$ in data,
as well as the expectations for the SM background, the Higgs boson signal, and some characteristic alternative
spin-zero scenarios are shown in Fig.~\ref{fig:kinematics}. All distributions in Fig.~\ref{fig:kinematics},
with the exception of the $m_{4\ell}$ distribution, are presented using events in the $m_{4\ell}$ range of
$121.5 - 130.5$\GeV to enhance the signal purity.
The observables with their correlations are used in the analysis to establish the consistency of the spin
and parity quantum numbers and tensor structure of interactions with respect to the SM predictions.
These observables also permit a further discrimination of signal from background,
increasing the signal sensitivity and reducing the statistical uncertainty in the measurements.

\subsection{Observables in the \texorpdfstring{$\PH\to\PW\PW\to\ell\nu\ell\nu$}{H to WW to l nu l nu} analysis} \label{sec:Observables_ww}

Only partial reconstruction is possible in the $\PH \to \PW\PW \to \ell\nu\ell\nu$ decay.
This channel features two isolated, high-$\PT$, charged leptons and $\ETm$
due to the presence of neutrinos in the final state.
The kinematic distributions of the decay products exhibit the characteristic properties of the parent boson.
There are three main observables in this channel: the azimuthal opening angle between
the two leptons ($\delphill$), which is correlated with the spin of the Higgs boson;
the dilepton mass ($\mll$), which is one of the most discriminating kinematic variables
for a Higgs boson with low mass, and it is also correlated to the spin and to $\delphill$;
and the transverse mass ($\mt$) of the final state objects,
which scales with the Higgs boson mass.
The transverse mass is defined as $\mt^2 = 2 \ptll \ETm (1-\cos\delphillmet)$,
where $\ptll$ is the dilepton transverse momentum and
$\delphillmet$ is the azimuthal angle between the dilepton momentum and~$\VEtmiss$.

\begin{figure*}[htbp]
\centering
\includegraphics[width=0.45\textwidth]{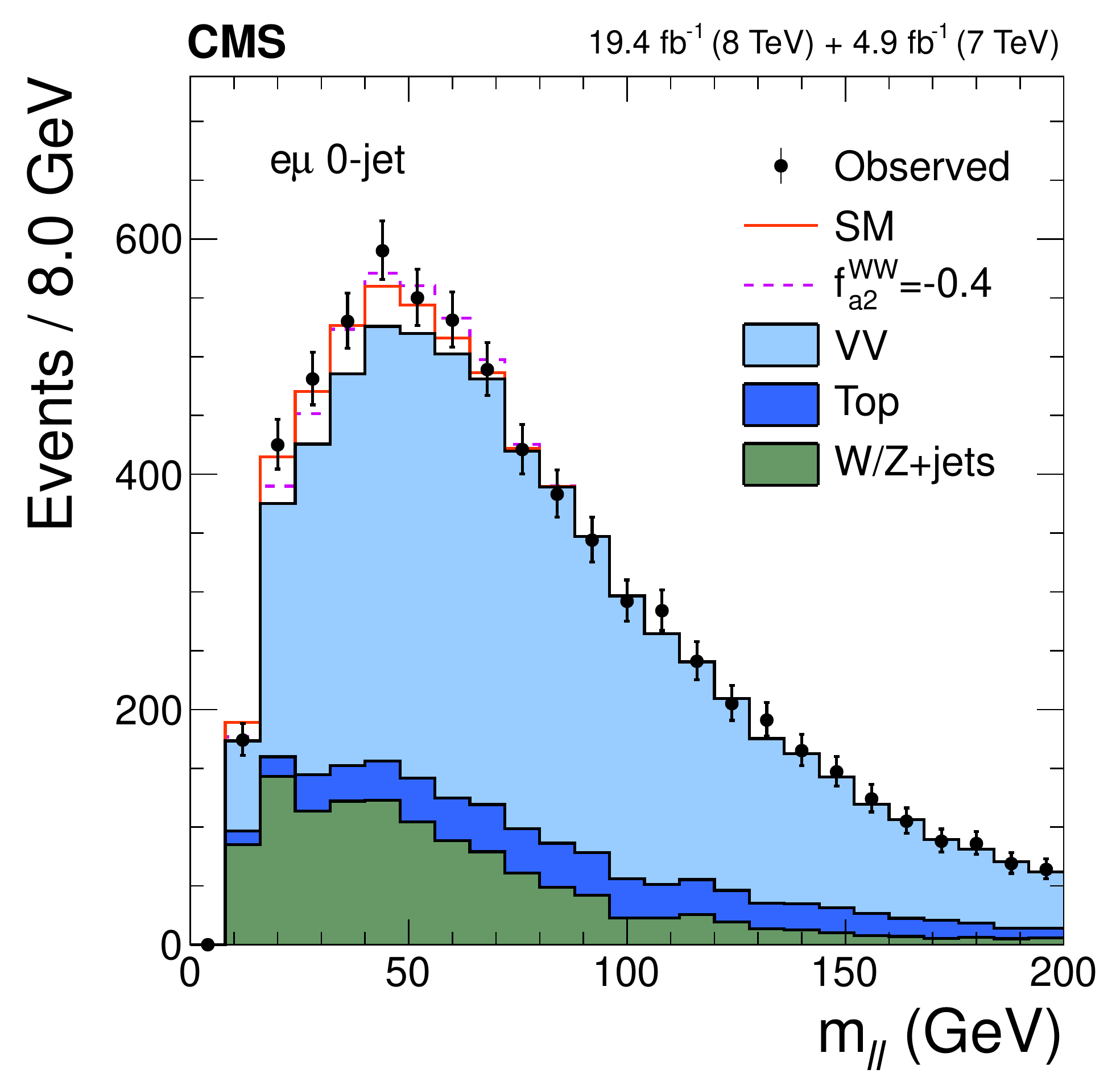}
\includegraphics[width=0.45\textwidth]{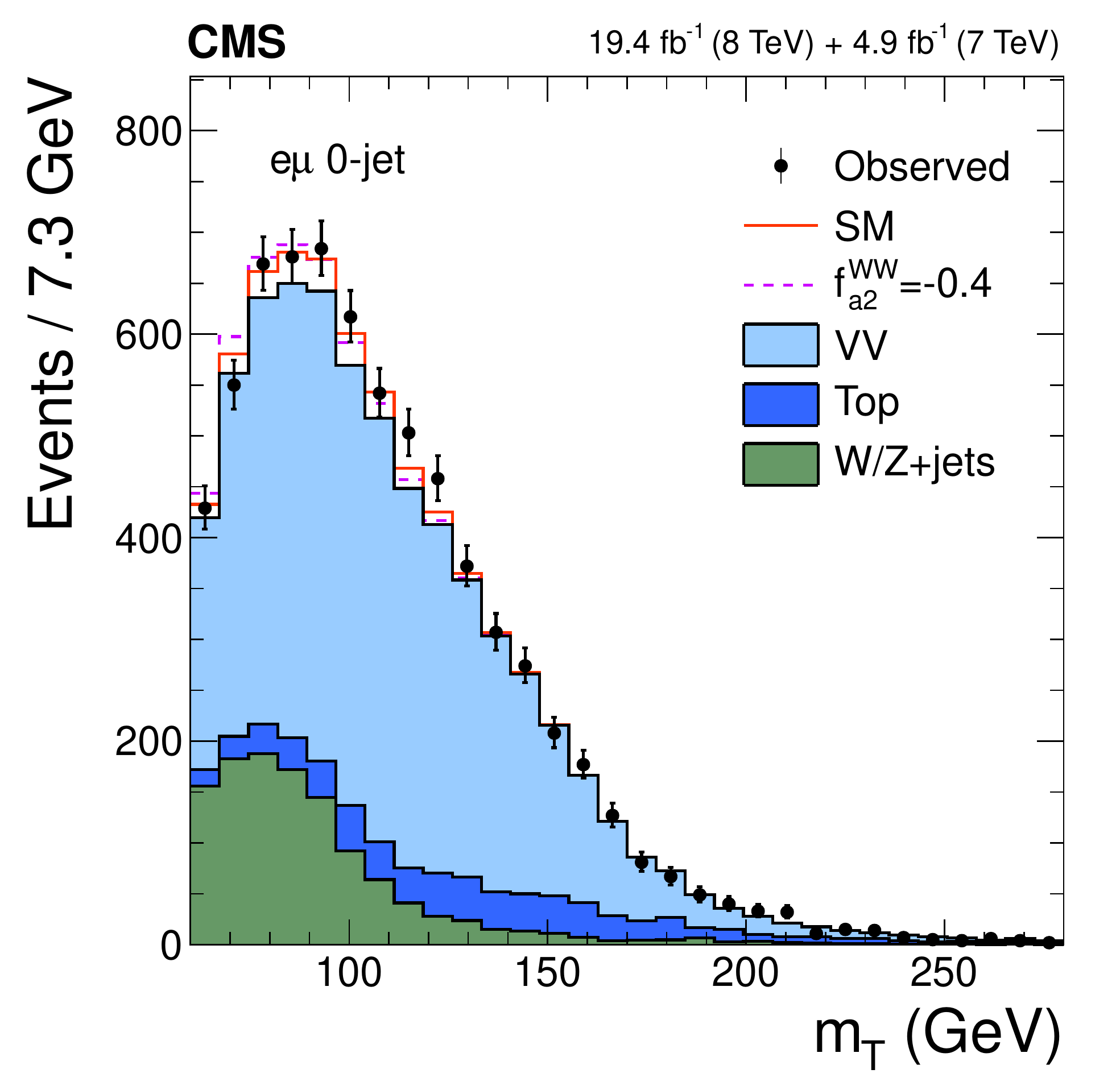}
\includegraphics[width=0.45\textwidth]{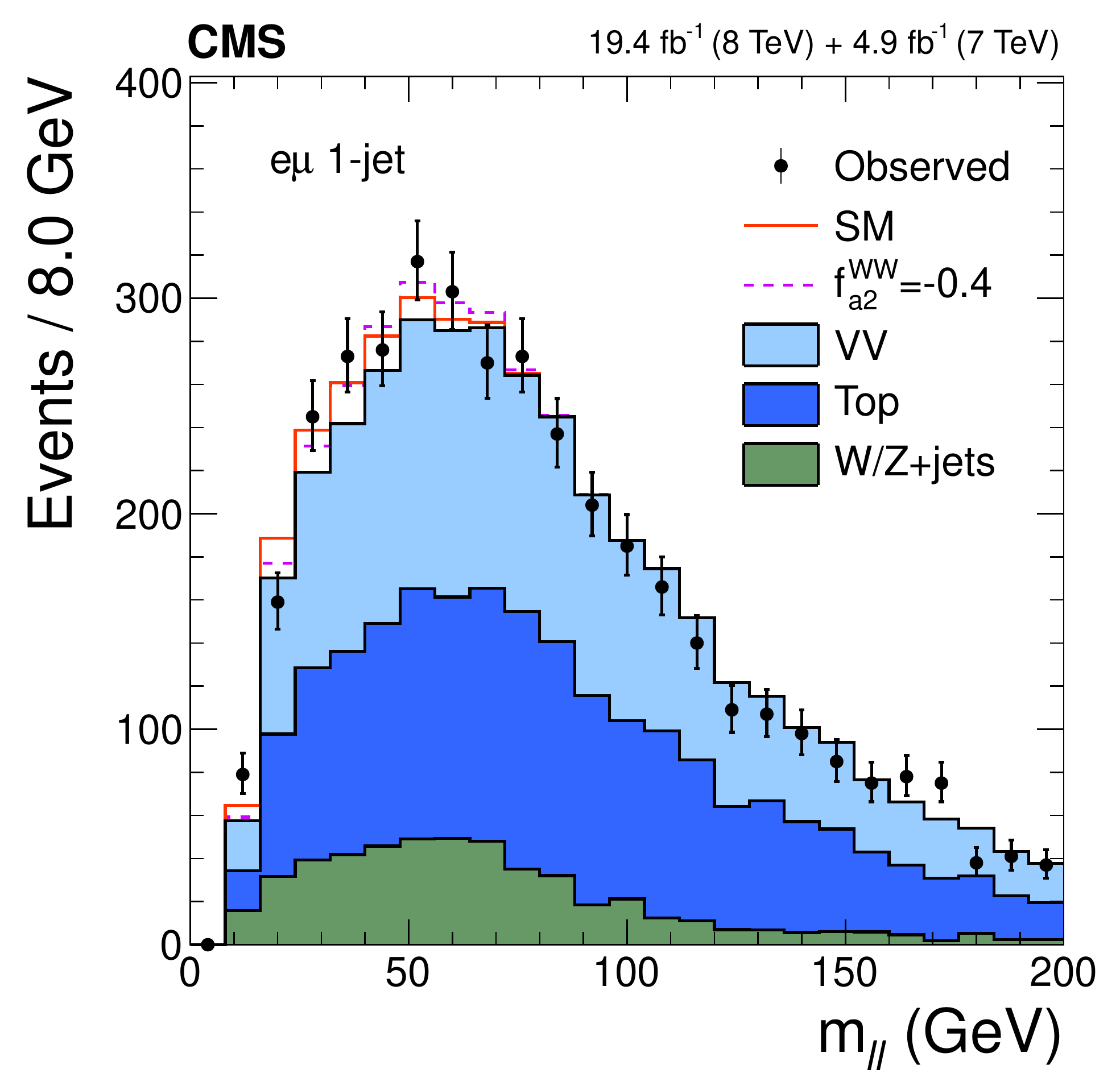}
\includegraphics[width=0.45\textwidth]{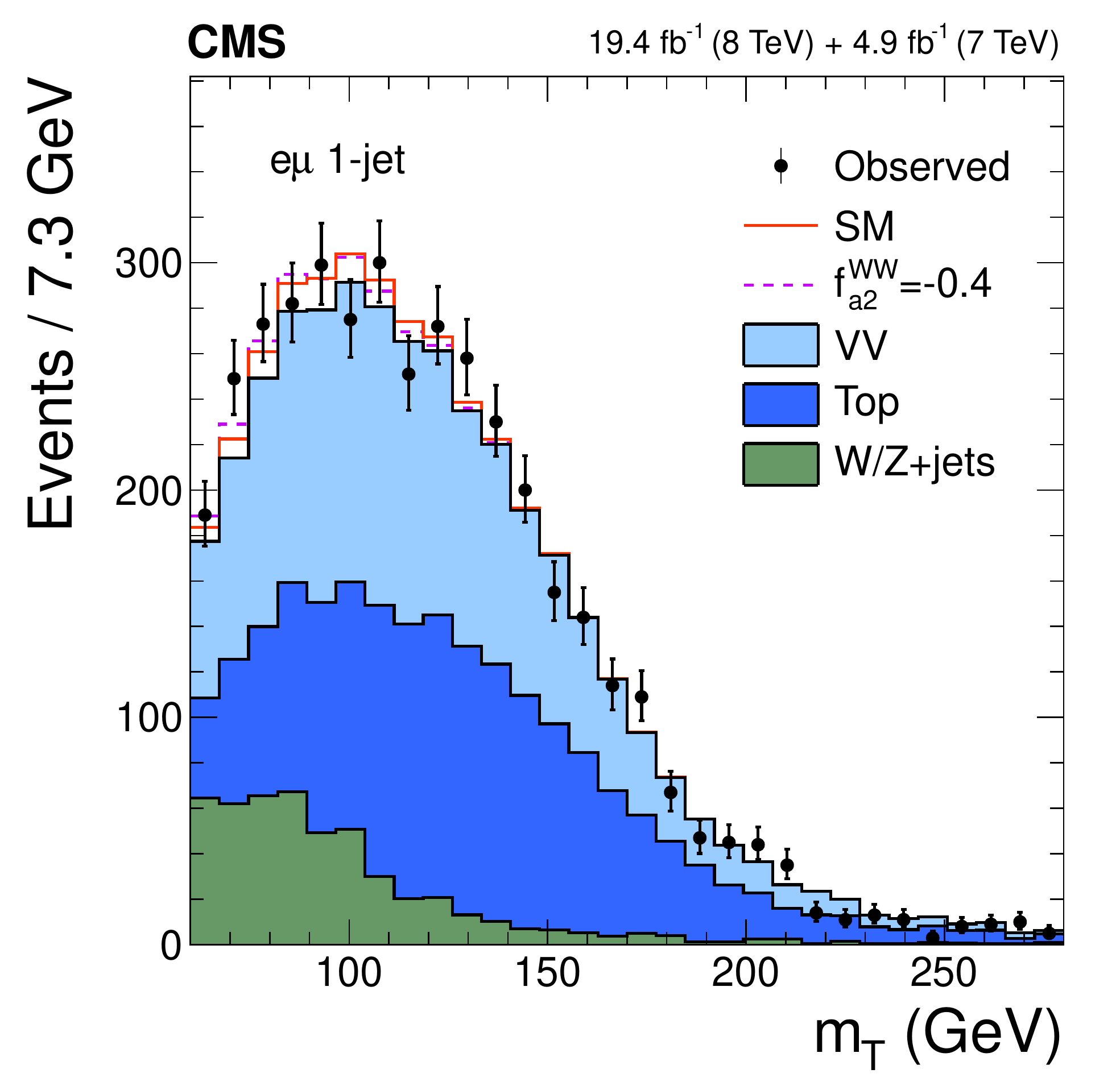}
\caption{
Distributions of $\mll$ (left) and $\mt$ (right) for events with 0 jets (upper row) and 1 jet (lower row)
in the $\PH\to\PW\PW\to\ell\nu\ell\nu$ analysis.
The observed data (points with error bars), the expectations for the SM background (shaded areas),
the SM Higgs boson signal (open areas under the solid histogram),
and the alternative spin-zero resonance (open areas under the dashed histograms) are shown,
as indicated in the legend.
The mass of the resonance is taken to be 125.6\GeV and the SM cross section is used.
}
    \label{fig:data1}

\end{figure*}

Two observables are used in the final analysis, $\mll$ and $\mt$. These two kinematic observables are independent
quantities that effectively discriminate the signal against most of the backgrounds and between different
signal models in the dilepton analysis in the 0-jet and 1-jet categories and have already been used in Ref.~\cite{Chatrchyan:2013iaa}.
The signal region is defined by $\mll < 200$\GeV, and $60 \leq \mt \leq 280\GeV$.
The distributions of these observables for data, an expected SM Higgs signal, an alternative signal model with $f_{a3}^{WW}=-0.4$,
and backgrounds are presented in Fig.~\ref{fig:data1}.

\subsection{Observables in the matrix element likelihood approach} \label{sec:KDMethod}

A comprehensive analysis of the kinematics of the decay of a Higgs boson would include up to eight observables,
as discussed above. In such an analysis, it is required to have a parameterization of the multidimensional
distributions as a function of the parameters of interest. However, it becomes challenging to describe all
the correlations of the observables and detector effects. It is possible to reduce the number of observables
and keep the necessary information using the matrix element likelihood approach.
In this approach, the kinematic information is stored in a discriminant designed for the separation of
either background, the alternative signal components, or interference between those components.
The parameterization of up to three observables can be performed with full simulation or data from
the control regions. This approach is adopted in the $\PH \to \V\V \to 4\ell$ analysis.
A similar approach is also possible in the $\PH \to \PW\PW \to \ell\nu\ell\nu$ channel,
but the construction of the discriminants is more challenging because of the presence of unobserved neutrinos.
Therefore, a simpler approach with the two observables defined above is used in this case.

The use of kinematic discriminants in Higgs boson studies was introduced in
previous CMS analyses~\cite{Chatrchyan:2012ufa,Chatrchyan:2012jja, Chatrchyan:2013mxa,CMS-HIG-14-002}
and feasibility studies~\cite{Bolognesi:2012mm,Anderson:2013afp},
and here it is extended both to a number of new models and to new techniques.
The construction of the kinematic discriminants follows the matrix element likelihood approach,
where the probabilities for an event are calculated using the LO matrix elements as a function
of angular and mass observables. In this way, the kinematic information, which fully characterizes
the $4\ell$ event topology of a certain process in its center-of-mass frame, is condensed to a
reduced number of observables.

The kinematic discriminants used in this study are computed using the {\sc MELA}
package~\cite{Chatrchyan:2012ufa,Gao:2010qx,Bolognesi:2012mm,Anderson:2013afp},
which provides the full set of processes studied in this paper and uses \textsc{JHUGen} matrix elements
for the  signal,  $\Pg\Pg$ or $\qqbar\to \X\to \Z\Z$ / $\Z\gamma^*$ / $\gamma^*\gamma^*\to4\ell$,
and \MCFM matrix elements for the background,
$\Pg\Pg$ or $\qqbar\to\Z\Z$ / $\Z\gamma^*$ / $\gamma^*\gamma^*$ / $\Z\to 4\ell$.
This library of processes is also consistent with the MC simulation used,
as discussed in Section~\ref{sec:CMS}, and also includes other production and decay mechanisms.
Within the {\sc MELA} framework, an analytic parameterization of the matrix elements
for signal~\cite{Gao:2010qx,Bolognesi:2012mm} and background~\cite{Chen:2012jy} was adopted in the previous CMS
analyses, reported in Refs.~\cite{Chatrchyan:2012ufa,Chatrchyan:2013lba,Chatrchyan:2012jja}.
The above matrix element calculations are validated against each other and tested
with the \textsc{mekd} package~\cite{Avery:2012um}, which is based on \MADGRAPH
and \textsc{FeynRules}~\cite{Christensen:2008py}, for a subset of processes implemented in common.
The analytic parameterizations of the spin-zero signal and $\qqbar\to\Z\Z$ / $\Z\gamma^*\to 4\ell$ background
processes are available from an independent implementation~\cite{Chen:2012jy, Chen:2014pia, Chen:2014gka}
and are used in a multidimensional distribution parameterization without the calculation of discriminants.

Given several signal hypotheses defined for $\Pg\Pg$ or $\qqbar\to X\to \Z\Z$ / $\Z\gamma^*$ / $\gamma^*\gamma^*\to4\ell$,
and the main background hypotheses $\Pg\Pg$ or $\qqbar\to\Z\Z$ / $\Z\gamma^*$ / $\gamma^*\gamma^*$ / $\Z\to 4\ell$,
the effective probabilities are defined for each event using a set of kinematic observables~$(m_1, m_2, m_{4\ell}, \vec\Omega)$
\ifthenelse{\boolean{cms@external}}{
\begin{equation}\begin{aligned}
 \mathcal{P}_\mathrm{SM} =& \mathcal{P}^\text{kin}_\mathrm{SM} (m_1, m_2, \vec\Omega|m_{4\ell}) \times \mathcal{P}^\text{mass}_\text{sig} (m_{4\ell}|m_{\sss\PH}), \\
 \mathcal{P}_{J^P} =& \mathcal{P}^\text{kin}_{J^P} (m_1, m_2, \vec\Omega|m_{4\ell}) \times \mathcal{P}^\text{mass}_\text{sig} (m_{4\ell}|m_{\sss\PH}) ,\\
 \mathcal{P}_{J^P} ^\text{int} =& \big(\mathcal{P}^\text{kin}_\mathrm{SM + J^P}(m_1, m_2, \vec\Omega|m_{4\ell}) \\&- \mathcal{P}^\text{kin}_{J^P} (m_1, m_2, \vec\Omega|m_{4\ell}) - \mathcal{P}^\text{kin}_\mathrm{SM}(m_1, m_2, \vec\Omega|m_{4\ell}) \big) , \\
 \mathcal{P}_{J^P} ^{\text{int} \perp} =& \big(\mathcal{P}^\text{kin}_{\mathrm{SM + J^P}\perp}(m_1, m_2, \vec\Omega|m_{4\ell}) \\&- \mathcal{P}^\text{kin}_{J^P} (m_1, m_2, \vec\Omega|m_{4\ell}) - \mathcal{P}^\text{kin}_\mathrm{SM}(m_1, m_2, \vec\Omega|m_{4\ell}) \big) , \\
 \mathcal{P}_{\qqbar\Z\Z} =& \mathcal{P}^\text{kin}_{\qqbar\Z\Z} (m_1, m_2, \vec\Omega|m_{4\ell}) \times \mathcal{P}^\text{mass}_{\qqbar\Z\Z} (m_{4\ell}) , \\
 \mathcal{P}_{\Pg\Pg\Z\Z} =& \mathcal{P}^\text{kin}_{\Pg\Pg\Z\Z} (m_1, m_2, \vec\Omega|m_{4\ell}) \times \mathcal{P}^\text{mass}_{\Pg\Pg\Z\Z} (m_{4\ell}) ,
\label{eq:kd-prob-gg}
\end{aligned}\end{equation}
}{
\begin{equation}\begin{aligned}
 \mathcal{P}_\mathrm{SM} =& \mathcal{P}^\text{kin}_\mathrm{SM} (m_1, m_2, \vec\Omega|m_{4\ell}) \times \mathcal{P}^\text{mass}_\text{sig} (m_{4\ell}|m_{\sss\PH}), \\
 \mathcal{P}_{J^P} =& \mathcal{P}^\text{kin}_{J^P} (m_1, m_2, \vec\Omega|m_{4\ell}) \times \mathcal{P}^\text{mass}_\text{sig} (m_{4\ell}|m_{\sss\PH}) ,\\
 \mathcal{P}_{J^P} ^\text{int} =& \left(\mathcal{P}^\text{kin}_\mathrm{SM + J^P}(m_1, m_2, \vec\Omega|m_{4\ell}) - \mathcal{P}^\text{kin}_{J^P} (m_1, m_2, \vec\Omega|m_{4\ell}) - \mathcal{P}^\text{kin}_\mathrm{SM}(m_1, m_2, \vec\Omega|m_{4\ell}) \right) , \\
 \mathcal{P}_{J^P} ^{\text{int} \perp} =& \left(\mathcal{P}^\text{kin}_{\mathrm{SM + J^P}\perp}(m_1, m_2, \vec\Omega|m_{4\ell}) - \mathcal{P}^\text{kin}_{J^P} (m_1, m_2, \vec\Omega|m_{4\ell}) - \mathcal{P}^\text{kin}_\mathrm{SM}(m_1, m_2, \vec\Omega|m_{4\ell}) \right) , \\
 \mathcal{P}_{\qqbar\Z\Z} =& \mathcal{P}^\text{kin}_{\qqbar\Z\Z} (m_1, m_2, \vec\Omega|m_{4\ell}) \times \mathcal{P}^\text{mass}_{\qqbar\Z\Z} (m_{4\ell}) , \\
 \mathcal{P}_{\Pg\Pg\Z\Z} =& \mathcal{P}^\text{kin}_{\Pg\Pg\Z\Z} (m_1, m_2, \vec\Omega|m_{4\ell}) \times \mathcal{P}^\text{mass}_{\Pg\Pg\Z\Z} (m_{4\ell}) ,
\label{eq:kd-prob-gg}
\end{aligned}\end{equation}
}
where $\mathcal{P}^\text{kin}(m_1, m_2, \vec\Omega|m_{\sss\PH}) = |{A}(m_1, m_2, \vec\Omega|m_{\sss\PH})|^2$
are the probabilities computed from the LO matrix elements and generally are not normalized.
The variable $\mathcal{P}^\text{mass}(m_{4\ell}|m_{\sss\PH})$ is the probability as a function of the four-lepton reconstructed mass
and is calculated using the $m_{4\ell}$ parameterization described in Refs.~\cite{Chatrchyan:2012jja, Chatrchyan:2013mxa}
including the $m_{\sss\PH}=125.6\GeV$ hypothesis for signal.
The probabilities $\mathcal{P}_{J^P} ^\text{int}$ parameterize interference between contributions from the SM and
anomalous couplings, where $J^P$ refers to a spin-zero tensor structure of interest, and are allowed to have both positive
and negative values. In the calculation of the mixed amplitude used for $\mathcal{P}^\text{kin}_\mathrm{SM + J^P}$,
the same coupling strengths are used as in the individual probabilities $\mathcal{P}^\text{kin}_\mathrm{SM}$ and $\mathcal{P}^\text{kin}_{J^P}$,
and these couplings are required to provide equal cross sections for the two individual processes.
The quantity $\mathcal{P}_{J^P} ^{\text{int}\perp}$ is constructed in the same way
as $\mathcal{P}_{J^P} ^\text{int}$ except that the phase of the $J^P$ amplitude is changed by $\pi/2$.
The matrix element calculations in Eq.~(\ref{eq:kd-prob-gg}) are also used for the re-weighting of simulated samples,
as discussed in Section~\ref{sec:CMS}.

Several kinematic discriminants are constructed for the main signal and background processes
from the set of probabilities described above
\ifthenelse{\boolean{cms@external}}{\begin{equation}\begin{aligned}
 \mathcal{D}_\text{bkg} =& \frac{\mathcal{P}_\mathrm{SM} }{\mathcal{P}_\mathrm{SM} +c\times\mathcal{P}_{\qqbar\Z\Z} }=
\big[1+c(m_{4\ell})
\\ &\times\frac{\mathcal{P}^\text{kin}_{\qqbar\Z\Z} (m_1, m_2, \vec\Omega | m_{4\ell})\times \mathcal{P}^\text{mass}_{\qqbar\Z\Z} (m_{4\ell})  }
{\mathcal{P}^\text{kin}_\mathrm{SM} (m_1, m_2, \vec\Omega | m_{4\ell}) \times \mathcal{P}^\text{mass}_\text{sig} (m_{4\ell}|m_{\sss\PH}) } \big]^{-1} ,
 \\
\mathcal{D}_{J^P} =& \frac{\mathcal{P}_\mathrm{SM} }{\mathcal{P}_\mathrm{SM} + \mathcal{P}_{J^P} }=
\left[1+ \frac{\mathcal{P}^\text{kin}_{J^P} (m_1, m_2, \vec\Omega | m_{4\ell}) }
{\mathcal{P}^\text{kin}_\mathrm{SM} (m_1, m_2, \vec\Omega | m_{4\ell}) } \right]^{-1} ,
 \\
\mathcal{D}_\text{int} =& \frac{ \mathcal{P}_{J^P}^\text{int}(m_1, m_2, \vec\Omega | m_{4\ell})}
{\mathcal{P}^\text{kin}_\mathrm{SM} +\mathcal{P}^\text{kin}_{J^P} }  .
\end{aligned}\end{equation}
}{
\begin{equation}\begin{aligned}
 \mathcal{D}_\text{bkg} =& \frac{\mathcal{P}_\mathrm{SM} }{\mathcal{P}_\mathrm{SM} +c\times\mathcal{P}_{\qqbar\Z\Z} }=
\left[1+c(m_{4\ell})\times\frac{\mathcal{P}^\text{kin}_{\qqbar\Z\Z} (m_1, m_2, \vec\Omega | m_{4\ell})\times \mathcal{P}^\text{mass}_{\qqbar\Z\Z} (m_{4\ell})  }
{\mathcal{P}^\text{kin}_\mathrm{SM} (m_1, m_2, \vec\Omega | m_{4\ell}) \times \mathcal{P}^\text{mass}_\text{sig} (m_{4\ell}|m_{\sss\PH}) } \right]^{-1} ,
 \\
\mathcal{D}_{J^P} =& \frac{\mathcal{P}_\mathrm{SM} }{\mathcal{P}_\mathrm{SM} + \mathcal{P}_{J^P} }=
\left[1+ \frac{\mathcal{P}^\text{kin}_{J^P} (m_1, m_2, \vec\Omega | m_{4\ell}) }
{\mathcal{P}^\text{kin}_\mathrm{SM} (m_1, m_2, \vec\Omega | m_{4\ell}) } \right]^{-1} ,
 \\
\mathcal{D}_\text{int} =& \frac{ \mathcal{P}_{J^P}^\text{int}(m_1, m_2, \vec\Omega | m_{4\ell})}
{\mathcal{P}^\text{kin}_\mathrm{SM} +\mathcal{P}^\text{kin}_{J^P} }  .
\end{aligned}\end{equation}
}
Here, the coefficient $c(m_{4\ell})$ is tuned to adjust the relative normalization of the signal and background probabilities for a given value of $m_{4\ell}$.
The observable $\mathcal{D}_\text{bkg}$ is used to separate signal from $\qqbar \to \Z\Z$, $\Pg\Pg \to \Z\Z$, and $\Z+X$ backgrounds,
using the $m_{4\ell}$ probability in addition to $\mathcal{P}^\text{kin}$. The discriminant $\mathcal{D}_{J^P}$ is created to separate
the SM signal from an alternative $J^P$ state. The discriminant $\mathcal{D}_\text{int}$ is created to isolate interference between
the SM and anomalous coupling contributions. Since the analysis is designed to probe small anomalous couplings, interference
between different anomalous contributions is a negligible effect and dedicated discriminants for those contributions are not
considered. The variable $\mathcal{D}_\text{int}$ is denoted as $\mathcal{D}_{C\!P}$ for interference between the $a_{1}$ and $a_{3}$ contributions
because it is sensitive to $CP$ violation~\cite{Anderson:2013afp}.

To remove the dependence of the spin-one and spin-two discriminants on the production model,
the probability $\mathcal{P}^\text{kin}$ is averaged over the two production angles $\cos\theta^*$ and $\Phi_1$,
defined in Fig.~\ref{fig:decay}, or equivalently the signal matrix element squared is averaged over the
polarization of the resonance~\cite{Anderson:2013afp}. The production independent discriminants
are defined as
\ifthenelse{\boolean{cms@external}}{
\begin{equation}\begin{aligned}
\label{eq:melaSigProd}
\mathcal{D}^\text{dec}_\text{bkg} =&
\Big[1+\frac{c(m_{4\ell})}{4\pi}\\&\times\frac{\mathcal{P}^\text{mass}_{\qqbar\Z\Z} (m_{4\ell})}{\mathcal{P}^\text{kin}_\mathrm{SM} (m_1, m_2, \vec\Omega | m_{4\ell}) \times \mathcal{P}^\text{mass}_\text{sig} (m_{4\ell}|m_{\sss\PH}) }\\ &\times\int \rd\Phi_1  \rd\cos\theta^{*}
\mathcal{P}^\text{kin}_{\qqbar\Z\Z} (m_1, m_2, \vec\Omega | m_{4\ell})
 \Big]^{-1} ,
 \\
\mathcal{D}^\text{dec}_{J^P} =&
\left[1+\frac{\frac{1}{4\pi}\int \rd\Phi_1  \rd\cos\theta^{*}
\mathcal{P}^\text{kin}_{J^P} (m_1, m_2, \vec\Omega | m_{4\ell}) }
{\mathcal{P}^\text{kin}_\mathrm{SM} (m_1, m_2, \vec\Omega | m_{4\ell}) } \right]^{-1} .
\end{aligned}\end{equation}
}{
\begin{equation}\begin{aligned}
\label{eq:melaSigProd}
\mathcal{D}^\text{dec}_\text{bkg} =&
\Bigg[1+c(m_{4\ell})\times\frac{\frac{1}{4\pi}\int \rd\Phi_1  \rd\cos\theta^{*}
\mathcal{P}^\text{kin}_{\qqbar\Z\Z} (m_1, m_2, \vec\Omega | m_{4\ell})\times \mathcal{P}^\text{mass}_{\qqbar\Z\Z} (m_{4\ell})  }
{\mathcal{P}^\text{kin}_\mathrm{SM} (m_1, m_2, \vec\Omega | m_{4\ell}) \times \mathcal{P}^\text{mass}_\text{sig} (m_{4\ell}|m_{\sss\PH}) } \Bigg]^{-1} ,
 \\
\mathcal{D}^\text{dec}_{J^P} =&
\left[1+\frac{\frac{1}{4\pi}\int \rd\Phi_1  \rd\cos\theta^{*}
\mathcal{P}^\text{kin}_{J^P} (m_1, m_2, \vec\Omega | m_{4\ell}) }
{\mathcal{P}^\text{kin}_\mathrm{SM} (m_1, m_2, \vec\Omega | m_{4\ell}) } \right]^{-1} .
\end{aligned}\end{equation}
}
The decay kinematics of a spin-zero resonance are already independent of the production
mechanism, due to the lack of spin correlations for any spin-zero particle.
The small differences in the distributions of the production-independent discriminants
with the different production mechanisms are due to detector acceptance effects
and are treated as systematic uncertainties.

\begin{figure*}[htbp]
\centering
\includegraphics[width=0.32\textwidth]{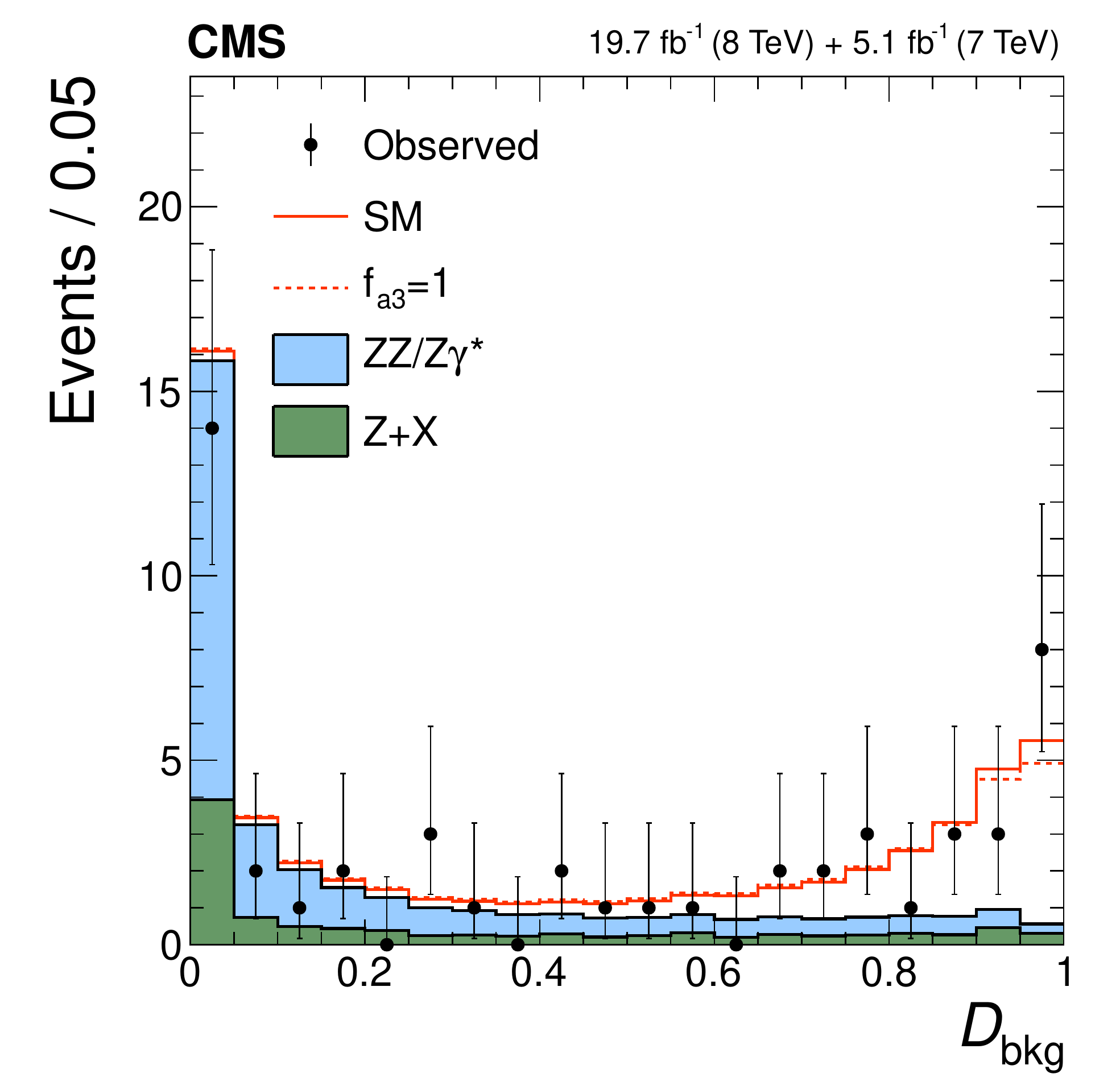}
\includegraphics[width=0.32\textwidth]{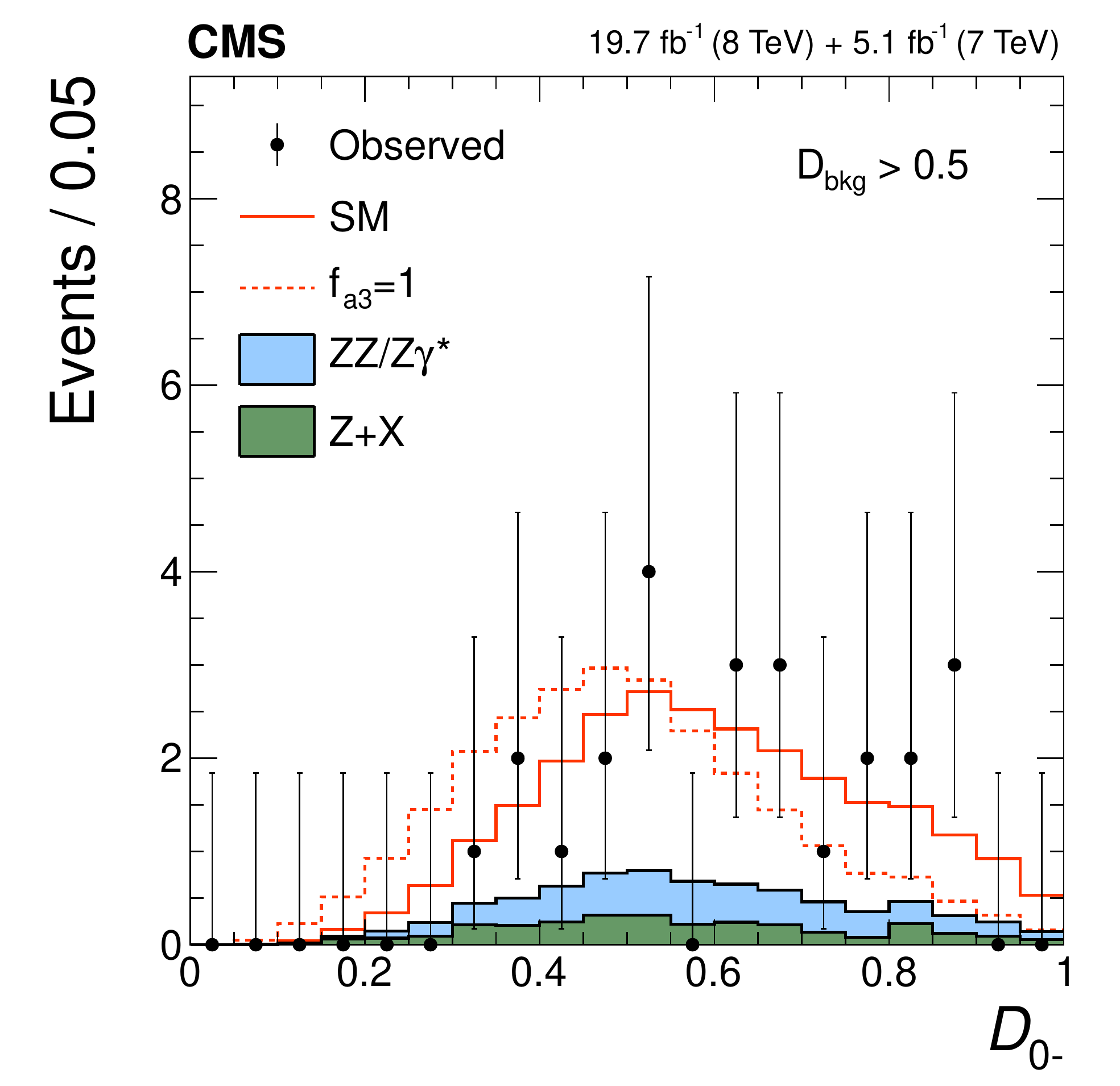}
\includegraphics[width=0.32\textwidth]{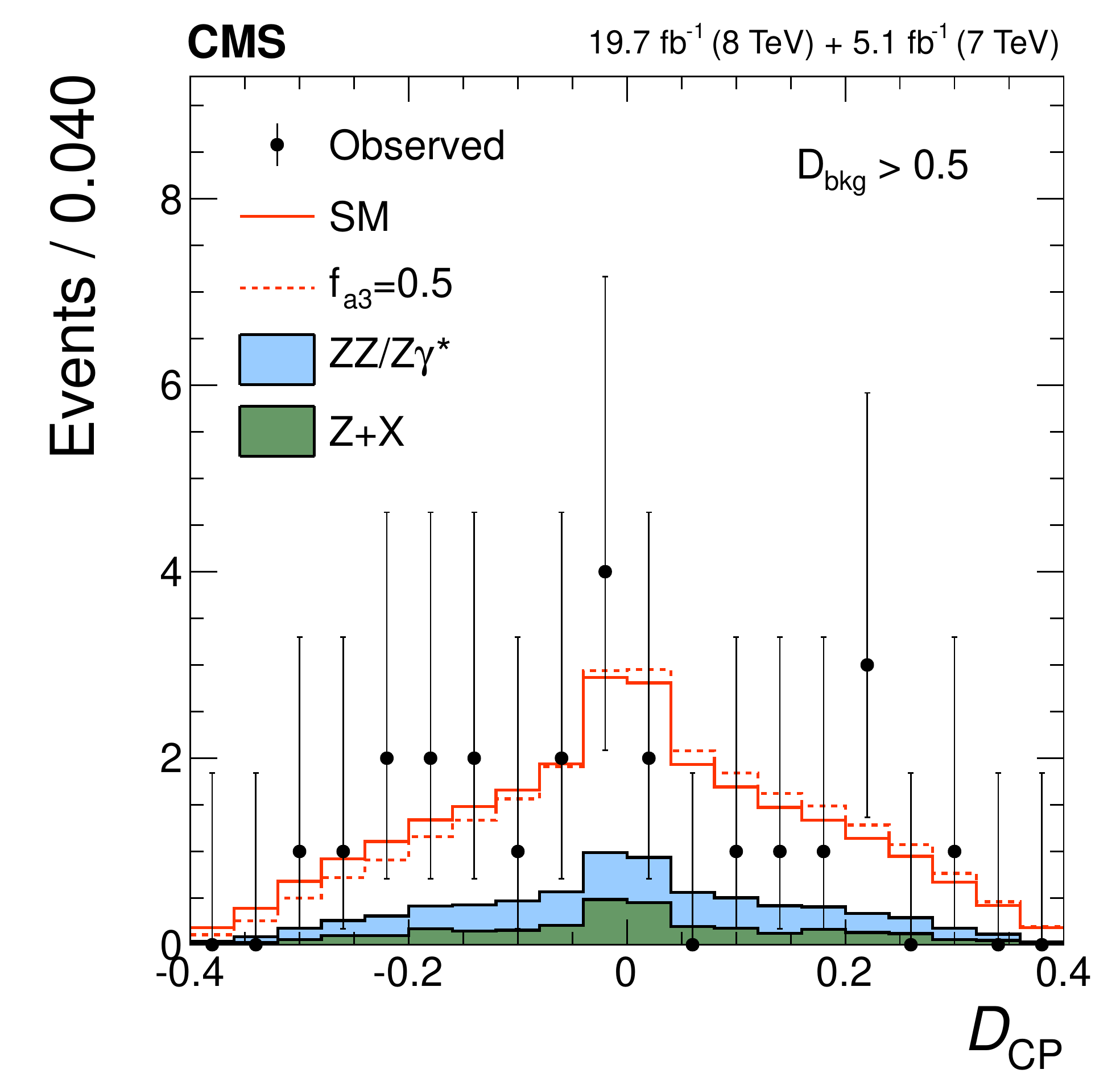}
\includegraphics[width=0.32\textwidth]{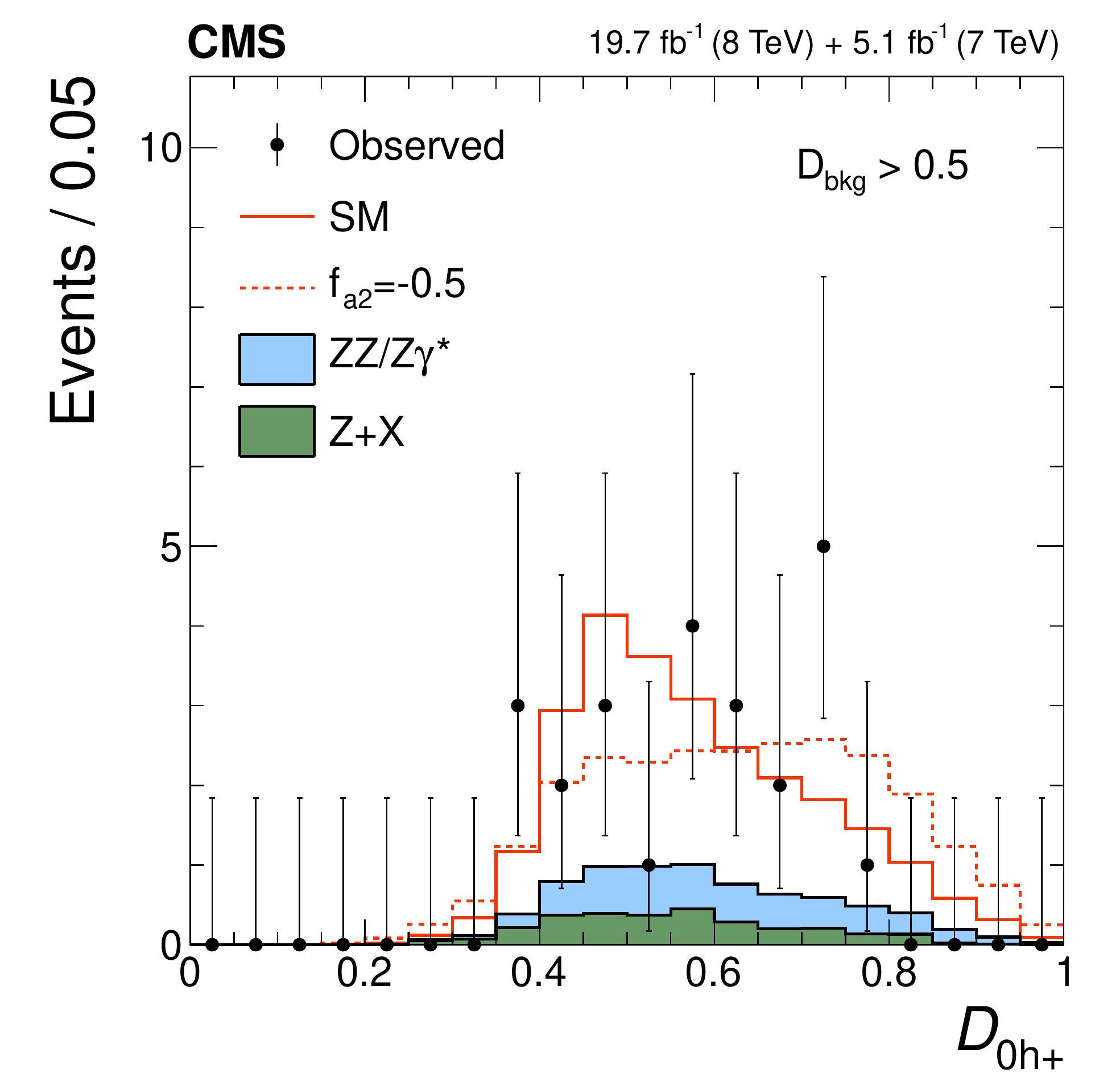}
\includegraphics[width=0.32\textwidth]{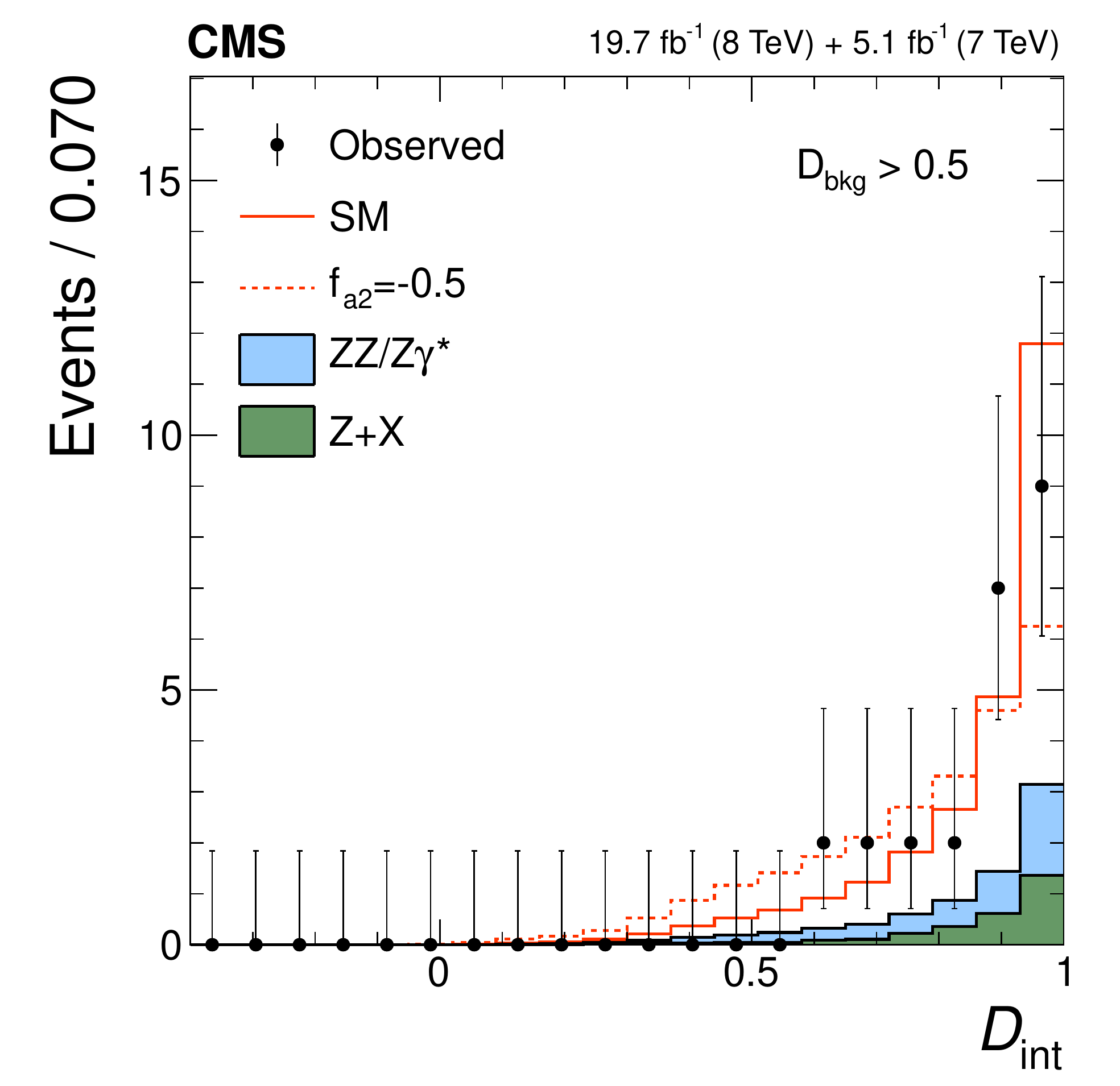}
\includegraphics[width=0.32\textwidth]{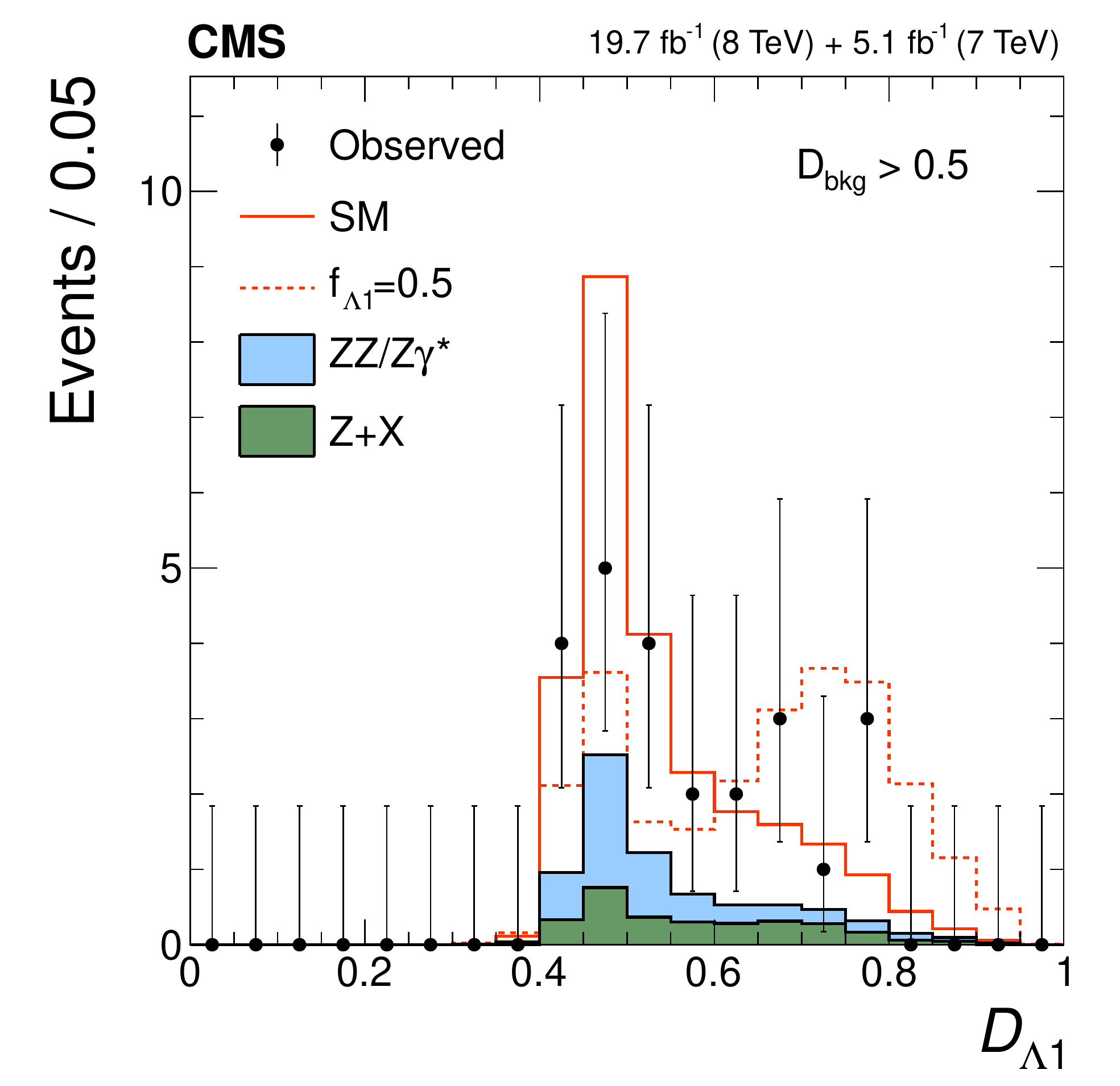}
\caption{
Distributions of the kinematic discriminants for
the observed data (points with error bars), the expectations for the SM background (shaded areas),
the SM Higgs boson signal (open areas under the solid histogram),
and the alternative spin-zero resonances (open areas under the dashed histograms) are shown,
as indicated in the legend.
The mass of the resonance is taken to be 125.6\GeV and the SM cross section is used.
Top row from left to right:
$\mathcal{D}_\text{bkg}$,
$\mathcal{D}_{0-}$,
$\mathcal{D}_{C\!P}$;
bottom row from left to right:
$\mathcal{D}_{0h+}$,
$\mathcal{D}_\text{int}$,
$\mathcal{D}_{\Lambda1}$.
All distributions, with the exception of $\mathcal{D}_\text{bkg}$, are shown
with the requirement $\mathcal{D}_\text{bkg}>0.5$ to enhance signal purity.
}
\label{fig:discriminants}

\end{figure*}

A complete list of all the discriminants used in the analysis is presented in Table~\ref{tab:kdlist}.
Some examples of the distributions as expected from simulation and as observed in data can be seen
in Fig.~\ref{fig:discriminants} for all the discriminants used in the study of the spin-zero $\PH\Z\Z$ couplings.
A complete list of the measurements performed and observables used is discussed in Sections~\ref{sec:ResultsExotic}
and~\ref{sec:ResultsSpinZero}.

\begin{table}[htbp]
\centering
\topcaption{
List of observables $\vec{x}$ used in the analysis of the $\PH\V\V$ couplings.
The $J^P$ notation for spin-two refers to the ten scenarios defined in Table~\ref{table-scenarios}.
The $\PH\to\gamma\gamma$ channel is illustrated with two main observables,
where $\cos\theta^*$ represents categories constructed from the angular and other observables,
and more details are given in Section~\ref{sec:hgg_select} and Ref.~\cite{Khachatryan:2014ira}.
}
\label{tab:kdlist}
\begin{scotch}{cccc}
 Measurement  & \multicolumn{3}{c}{Observables $\vec{x}$} \\
\hline
 $f_{\Lambda1}$ & $\mathcal{D}_\text{bkg}$ & $\mathcal{D}_{\Lambda1}$ & \\
 $f_{a2}$ &  $\mathcal{D}_\text{bkg}$ & $\mathcal{D}_{0h+}$ & $\mathcal{D}_\text{int}$ \\
  $f_{a3}$ &  $\mathcal{D}_\text{bkg}$ & $\mathcal{D}_{0-}$ & $\mathcal{D}_{C\!P}$ \\
 $f_{\Lambda1}^{\PW\PW}$ & $\mt$ & $\mll$ & \\
 $f_{a2}^{\PW\PW}$ &  $\mt$ & $\mll$ & \\
  $f_{a3}^{\PW\PW}$ &  $\mt$ & $\mll$ & \\
  $f_{\Lambda1}^{\Z\gamma}$ &  $\mathcal{D}_\text{bkg}$ & $\mathcal{D}^{\Z\gamma}_{\Lambda1}$  & $\mathcal{D}^{\Z\gamma,\Lambda1}_\text{int}$  \\
  $f_{a2}^{\Z\gamma}$ &  $\mathcal{D}_\text{bkg}$ & $\mathcal{D}^{\Z\gamma}_{a2}$  & $\mathcal{D}_\text{int}^{\Z\gamma}$ \\
  $f_{a3}^{\Z\gamma}$ &  $\mathcal{D}_\text{bkg}$ & $\mathcal{D}^{\Z\gamma}_{a3}$  & $\mathcal{D}_{C\!P}^{\Z\gamma}$ \\
 $f_{a2}^{\gamma\gamma}$  &  $\mathcal{D}_\text{bkg}$ & $\mathcal{D}^{\gamma\gamma}_{a2}$  & $\mathcal{D}_\text{int}^{\gamma\gamma}$ \\
  $f_{a3}^{\gamma\gamma}$ & $\mathcal{D}_\text{bkg}$ & $\mathcal{D}^{\gamma\gamma}_{a3}$ & $\mathcal{D}_{C\!P}^{\gamma\gamma}$ \\
 \multicolumn{1}{l}{spin-one $\qqbar\to X (f_{b2})\to \Z\Z$} &  $\mathcal{D}_\text{bkg}$ & $\mathcal{D}_{1-}$ & $\mathcal{D}_{1+}$ \\

 \multicolumn{1}{l}{spin-one decay  $X(f_{b2})\to \Z\Z$} &  $\mathcal{D}_\text{bkg}^\text{dec}$ & $\mathcal{D}_{1-}^\text{dec}$ & $\mathcal{D}_{1+}^\text{dec}$ \\
 \multicolumn{1}{l}{spin-two $\qqbar\to X (J^P)\to \Z\Z$} &  $\mathcal{D}_\text{bkg}$ & $\mathcal{D}_{J^P}^{\qqbar}$ & \\

 \multicolumn{1}{l}{spin-two $\Pg\Pg\to X (J^P)\to \Z\Z$} &  $\mathcal{D}_\text{bkg}$ & $\mathcal{D}_{J^P}^{\Pg\Pg}$ & \\

 \multicolumn{1}{l}{spin-two decay $X (J^P)\to \Z\Z$}&  $\mathcal{D}_\text{bkg}^\text{dec}$ & $\mathcal{D}_{J^P}^\text{dec}$ & \\
 \multicolumn{1}{l}{spin-one $\qqbar\to X (f_{b2}^{\PW\PW}) \to \PW\PW$} & $\mt$ & $\mll$ & \\

 \multicolumn{1}{l}{spin-two $\Pg\Pg$ or $\qqbar\to X (J^P) \to \PW\PW$} & $\mt$ & $\mll$ & \\
 \multicolumn{1}{l}{spin-two $\Pg\Pg$ or $\qqbar\to X (2^+_m) \to \gamma\gamma$} & $m_{\gamma\gamma}$ & $\cos\theta^*$ & \\
\end{scotch}

\end{table}

\subsection{Maximum likelihood fit with the template method}  \label{sec:MLfit}

The goal of the analysis is to determine if a set of anomalous coupling parameters $\vec{\zeta}$,
defined both for the production and decay of a resonance with either spin zero, one, or two
is consistent, for a given set of observables $\vec{x}$, with the data.
The coupling parameters $\vec{\zeta}$ are discussed in detail in Section~\ref{sec:Pheno}.
They are summarized in Eqs.~(\ref{eq:formfact-fullampl-spin0}), (\ref{eq:fa_definitions}), (\ref{eq:fa_definitions_zg})
and Table~\ref{tab:xsec_ratio} for spin-zero,
in Eqs.~(\ref{eq:ampl-spin1}) and (\ref{eq:fa_definitions_spin1}) for spin-one, and
in Eqs.~(\ref{eq:ampl-spin2-a}), (\ref{eq:ampl-spin2-qq}) and Table~\ref{table-scenarios} for spin-two.
The observables $\vec{x}_i$ are defined for each event $i$, listed in Table~\ref{tab:kdlist},
and discussed above. The extended likelihood function is defined for $N$ candidate events as
\ifthenelse{\boolean{cms@external}}{
\begin{multline}
\mathcal{L} =  \exp\Big( - n_\text{sig}-\sum_k n_\text{bkg}^k  \Big)
\\ \times\prod_i^{N} \Big( n_\text{sig} \times\mathcal{P}_\text{sig}(\vec{x}_{i};~\vec{\zeta})
+\sum_k n_\text{bkg}^{k} \times\mathcal{P}_\text{bkg}^k(\vec{x}_{i})
\Big),
\label{eq:likelihood}
\end{multline}
}{
\begin{equation}
\mathcal{L} =  \exp\Big( - n_\text{sig}-\sum_k n_\text{bkg}^k  \Big)
\prod_i^{N} \Big( n_\text{sig} \times\mathcal{P}_\text{sig}(\vec{x}_{i};~\vec{\zeta})
+\sum_k n_\text{bkg}^{k} \times\mathcal{P}_\text{bkg}^k(\vec{x}_{i})
\Big),
\label{eq:likelihood}
\end{equation}
}
where $n_\text{sig}$ is the number of signal events and $n_\text{bkg}^k$ is the number of background events of type $k$.
The probability density functions $\mathcal{P}_\text{sig}(\vec{x}_{i};\vec{\zeta})$ and $\mathcal{P}_\text{bkg}^k(\vec{x}_{i})$
are defined for the signal and background, respectively.

There are several event categories, such as $4\Pe$, $4\mu$, and $2\Pe2\mu$ in the
$\PH\to \V\V \to 4\ell$ analysis,  0 and 1-jet in the
$\PH\to \PW\PW \to\ell\nu\ell\nu$ analysis, or the 7\TeV and 8\TeV categories, and several types of background.
The total signal yield $n_\text{sig}$ is a free parameter to avoid
using the overall signal event yield as a part of the discrimination between alternative hypotheses.
However, when several channels are used in the same decay,
such as $\PH \to \V\V \to 4\Pe$, $2\Pe2\mu$, and $4\mu$,
the relative yields between the channels depend on the terms considered in the tensor structure
due to interference effects in the presence of identical leptons,
and this information is exploited in the analysis.

The method adopted for all the measurements presented in this paper is a template method.
The probability density functions $\mathcal{P}_\text{sig}$ and  $\mathcal{P}_\text{bkg}^k$ are described
as histograms (templates) with two or three dimensions, see observables in Table~\ref{tab:kdlist},
and with up to 50 bins in each dimension. The number of dimensions used is limited by the number
of simulated events that can be generated or the number of events in the control regions in data.
However, an optimal construction of observables allows for the retention of all the necessary information for
the measurement with up to three observables. The templates are built for signal and background from
histograms of fully simulated events, or from control regions in data. In the $\PH\to\V\V\to 4\ell$ analyses,
where the number of bins is larger than in the $\PH\to\PW\PW\to\ell\nu\ell\nu$ analysis, statistical fluctuations
are removed using a smoothing algorithm~\cite{rosenblatt1956, parzen1962}.

The signal probability density functions $\mathcal{P}_\text{sig}$ depend on the coupling parameters $\vec{\zeta}$.
For spin-zero, these functions can be parameterized as a linear combination of the terms originating from
the SM-like and anomalous amplitudes and their interference~\cite{Anderson:2013afp}
\ifthenelse{\boolean{cms@external}}{
\begin{multline}
\mathcal{P}_\text{sig}\left(\vec{x}; \vec{\zeta}=\{f_{ai},\phi_{ai}\}\right) = \\ \left(1-\sum_{ai} f_{ai}\right) \, \mathcal{P}_{0^+}\left(\vec{x}\right)
 + \sum_{ai} f_{ai} \, \mathcal{P}_{ai}\left(\vec{x}\right)  \\
+ \sum_{ai} \sqrt{f_{ai}\left(1-\sum_{aj} f_{aj}\right)}\, \mathcal{P}^\text{int}_{ai,0^+}\left(\vec{x}; \phi_{ai}\right) \\
+ \sum_{ai<aj} \sqrt{f_{ai}f_{aj}} \, \mathcal{P}^\text{int}_{ai,aj}\left(\vec{x}; \phi_{ai}-\phi_{aj}\right),
\label{eq:fractions-general}
\end{multline}
}{
\begin{multline}
\mathcal{P}_\text{sig}\left(\vec{x}; \vec{\zeta}=\{f_{ai},\phi_{ai}\}\right) =  \left(1-\sum_{ai} f_{ai}\right) \, \mathcal{P}_{0^+}\left(\vec{x}\right)
 + \sum_{ai} f_{ai} \, \mathcal{P}_{ai}\left(\vec{x}\right)  \\
+ \sum_{ai} \sqrt{f_{ai}\left(1-\sum_{aj} f_{aj}\right)}\, \mathcal{P}^\text{int}_{ai,0^+}\left(\vec{x}; \phi_{ai}\right) \\
+ \sum_{ai<aj} \sqrt{f_{ai}f_{aj}} \, \mathcal{P}^\text{int}_{ai,aj}\left(\vec{x}; \phi_{ai}-\phi_{aj}\right),
\label{eq:fractions-general}
\end{multline}
}
where $\mathcal{P}_{ai}$ is the probability of a pure $a_i$ term and $\mathcal{P}^\text{int}_{ai, aj}$ describes
the interference between the two terms, each parameterized as a template.
Each term in Eq.~(\ref{eq:fractions-general}) is extracted from the dedicated simulation
and includes proper normalization.
For spin-one or spin-two, in the case of a study of non-interfering states there is only one fraction $f(J^P)$
and no interference contribution.

The likelihood in Eq.~(\ref{eq:likelihood}) can be used in two different ways.
In both approaches, the likelihood is maximized with respect to the nuisance parameters which include the signal yield
and constrained parameters describing the systematic uncertainties discussed in Section~\ref{sec:Systematics}.
In one approach the likelihood is maximized to estimate the values of anomalous couplings, and the confidence intervals are
determined from profile likelihood scans of the respective parameters. This is used for the measurement of anomalous
couplings under the spin-zero hypothesis, as well as for the $f(J^P)$ measurements of the spin-one and spin-two hypotheses.
The allowed 68\% and 95\% \CL intervals are defined using the profile likelihood function, $-2\,\Delta \ln{\mathcal L} = 1.00$ and 3.84,
for which exact coverage is expected in the asymptotic limit~\cite{wilks1938}. The approximate coverage
has been tested with generated samples for several true parameter values and the quoted results have been found to be conservative.

The other approach is used to distinguish an alternative spin-one or spin-two signal hypothesis from the SM Higgs boson.
In this case, the test statistic $q=-2{\ln(\mathcal{L}_{J^P}/\mathcal{L}_{0^+})}$ is defined using the ratio
of signal plus background likelihoods for two signal hypotheses. To quantify the consistency of the observed test statistic
$q_\text{obs}$ with respect to the SM Higgs boson hypothesis ($0^+$), the probability $p = P( q \leq q_\text{obs} \, | \, 0^+ +\text{bkg} )$
is assessed and converted into a number of standard deviations via the Gaussian one-sided tail integral.
The consistency of the observed data with the alternative signal hypothesis ($J^P$) is assessed from
$P( q \geq q_\text{obs} \, | \, J^P + \text{bkg} )$. The \CLs~criterion~\cite{0954-3899-28-10-313,Junk:1999kv}, defined
as $\CLs = { P( q \geq q_\text{obs} \, | \, J^P + \text{bkg} ) }$ / ${ P( q \geq q_\text{obs} \, | \, 0^+ + \text{bkg} ) } < \alpha,$
is used for the final inference of whether a particular alternative signal
hypothesis is excluded or not at a given confidence level $(1 -\alpha)$.
The following quantities are used to characterize the expected and observed results:
(i) separation, defined as the tail area $A_\text{tail}$ calculated
at the value of q where the tails of the two distributions have identical area,
(ii) the probability of each hypothesis to fluctuate beyond $q_\text{obs}$, and
(iii) the expected and observed $\CLs$ value.
Option (i) is used to characterize the expected results as this quantity is symmetric between the two hypotheses,
and it is expressed as the number of standard deviations multiplied by two.
Options (ii) and (iii) are used to characterized the observed results for exclusion of a particular hypothesis.
The observed separation (ii) is also expressed as the number of standard deviations, and the sign
is positive if the tail extends away or negative if it extends towards the median of the other hypothesis.

\subsection{Analysis validation and systematic uncertainties} \label{sec:Systematics}

The validation of this analysis and the assignment of systematic uncertainties follows
various aspects of the parameterization in Eq.~(\ref{eq:likelihood}). Estimates of the expected
background yields and shapes of the probability distributions for signal and background are
investigated.
The performance of the fit has been tested using events from full simulation discussed in
Section~\ref{sec:mc} and using events generated directly from probability distributions.
Both approaches are found to give consistent expected results and unbiased parameter
estimates in the fit for anomalous couplings for the full spectrum of measurements listed
in Table~\ref{tab:kdlist}.
These tests rely on the proper simulation of the signal and background processes
and further studies propagate any systematic uncertainties in the simulation to the final results,
which are specific to each final state. The overall signal yield is left unconstrained in the fit
and therefore the associated theoretical uncertainties do not affect the constraints on anomalous couplings.

The statistical uncertainties dominate over the systematic ones for all the results quoted in this paper.
The systematic uncertainties in the $\PH\to\V\V\to4\ell$ channel are generally the same as the ones
investigated in Ref.~\cite{Chatrchyan:2013mxa}.
Among the yield uncertainties, experimental systematic uncertainties are evaluated from data for the
lepton trigger efficiency and combined object reconstruction, identification, and isolation efficiencies.
The theoretical uncertainties on the $\Z\Z$ background are described in Ref.~\cite{Chatrchyan:2013mxa},
but the calculations have been updated using the recommendations in Ref.~\cite{LHCHiggsCrossSectionWorkingGroup:2013tqa}
and the treatment of the $\Pg\Pg\to\Z\Z/\Z\gamma^*$ process follows Ref.~\cite{CMS-HIG-14-002}.
The $\Z+X$ uncertainties include the effects on both the expected yields and on the shape.
The yield uncertainties are estimated to be 20\%, 25\%, and 40\% for the $4\Pe$, $2\Pe2\mu$, and $4\mu$
decay channels, respectively. The shape uncertainty is taken into account by considering the difference
between the $\Z+X$ and $\qqbar\to\Z\Z$ distributions for a particular final state, which was found to cover
any potential biases in $\Z+X$ parameterization.
To account for the lepton momentum scale and resolution uncertainty in
the $m_{4\ell}$ distribution, the alternative signal shapes are taken from the variations of both of these
contributions, following~Ref.~\cite{Chatrchyan:2013mxa}.

In the $\PH\to\PW\PW\to\ell\nu\ell\nu$ analysis, the same treatment of the systematic uncertainties as in Ref.~\cite{Chatrchyan:2013iaa}
has been performed.
The uncertainty related to the size of the simulated samples is such that it is at least a factor of two smaller
than the rest of the systematic uncertainties and varies from 1.0\% for Higgs boson signal to 20\% for some of the backgrounds
( $\dyll$, $\Wjets$, and $\V\gamma^{(*)}$).
Systematic uncertainties are represented by individual nuisance parameters
with log-normal distributions. An exception is applied to the $\qqbar \to \PW\PW$
normalization, which is an unconstrained parameter in the fit.

The analysis is optimized for the $\Pg\Pg\to\PH$ production mode, which has the largest cross section,
as verified experimentally~\cite{Chatrchyan:2013mxa,Chatrchyan:2013iaa,Khachatryan:2014ira},
and is characterized by low hadronic activity in the final state.
Other production modes such as VBF, $\V\PH$, and $\mathrm{t\bar{t}}\PH$ are considered in the
analysis, representing a small or negligible fraction of the signal.
In the $\PH\to\V\V\to 4\ell$ analysis, only the exclusive four-lepton final state is reconstructed and it has been verified that all
observables are similar for all production mechanisms of a spin-zero particle. For the spin-one and spin-two
models using decay-only observables, any residual dependence on the production mechanism is small
and enters only through the difference in detector acceptance effects. Uncertainties in this approach are accounted for
with alternative parameterization of the observable distributions, covering the difference between the gluon fusion and
$\qqbar$ production mechanisms of a spin-two particle, or an equivalent variation for a spin-one particle production which
reflects the difference in the boost of the resonance.

In the $\PH\to\PW\PW\to\ell\nu\ell\nu$ analysis, the VBF contribution, which has similar kinematics as
$\Pg\Pg\to\PH$, represents 5\% of the total Higgs boson signal in the 1-jet category,
where it is the second-largest mode in terms of rate after $\Pg\Pg\to\PH$,
and less than 0.5\% in the 0-jet bin, where it is highly suppressed.
The associated production $\V\PH$, and in particular $\Z\PH$, shows some differences in the observables
compared to $\Pg\Pg\to\PH$ because of the additional vector bosons present in the final state, but
contributes less than 1\% to the total signal yield in the 0- and 1-jet categories.
There is no expected $\mathrm{t\bar{t}}\PH$ contribution in the signal region after all selection requirements.
For the measurements presented in Section~\ref{sec:Resultsexotichww}, a full combination of all Higgs boson
production mechanisms is considered in the parameterization, while the alternative exotic-spin hypotheses
are produced via $\Pg\Pg$, $\qqbar$, or a combination of the two.
For the measurements presented in Section~\ref{sec:Resultspinzerohww}, the $\Pg\Pg\to\PH$ model is used to create
the templates and the full variation of the distributions after the inclusion of all the production mechanisms according
to the SM expectation is used for the evaluation of the systematic uncertainties. This approach is taken because a priori
the fraction of various production mechanisms is not known for an arbitrary BSM model.
However, those fractions have been experimentally constrained to be consistent with the SM
expectations~\cite{Chatrchyan:2013mxa,Chatrchyan:2013iaa,Khachatryan:2014ira,Khachatryan:2014jba}.

The correlations between the systematic uncertainties in the different categories and final states
are taken into account. In particular, the main sources of correlated systematic
uncertainties are those related to the experimental measurements
such as the integrated luminosity, lepton and trigger selection efficiencies, lepton momentum scale,
and the theoretical uncertainties affecting the background processes.
Uncertainties in the background normalization or background model parameters
from control regions and uncertainties of a statistical nature are uncorrelated.

It is instructive to validate the matrix element method with the study of spin-parity and anomalous
interactions of the $\Z$ boson, which has already established SM properties~\cite{Agashe:2014kda}.
An earlier CMS analysis tested the $\Z$ boson couplings to fermions in the two-body decay
 $\qqbar\to \Z/\gamma^* \to \ell^+\ell^-$~\cite{Chatrchyan:2011ya}, using a matrix element formalism similar
 to the one used in the Higgs boson studies~\cite{Gao:2010qx} and established consistency with the SM.
Here the study is extended to the four-lepton decay of the $\Z$ boson in the topology
$\qqbar\to \Z/\gamma^* \to \ell^+\ell^-\gamma^*\to 4\ell$~\cite{CMS:2012bw}.
A hypothesis test is performed between the SM $\Z$ boson and an exotic Higgs-like resonance
$\Pg\Pg\to\PH(91.2)\to \Z\Z\to 4\ell$ with the same mass and width as the $\Z$ boson.
The mass window $80<m_{4\ell}<100$\GeV is used, just below the mass range used in the Higgs boson analysis.
In addition, the non-resonant $\qqbar\to \Z\Z / \Z\gamma^*/\gamma^*\gamma^* \to 4\ell$
contribution is parameterized including interference with $\qqbar\to \Z/\gamma^* \to \ell^+\ell^-\gamma^*\to 4\ell$
following the formalism of Eq.~(\ref{eq:fractions-general}) and its effective fractional cross section
is fitted in data in analogy with the $f_{ai}$ parameters.
The results show that the SM $\Z$ boson hypothesis is highly preferred in the data
and the small contribution of the production $\qqbar\to \Z\Z / \Z\gamma^*/\gamma^*\gamma^* \to 4\ell$,
including its phase, is consistent with the SM expectation.
The alternative Higgs-like hypothesis $\PH(91.2)$ has been excluded with a $\CL >99.99\%$.

\subsection{Analysis validation with analytic parameterization of kinematic distributions} \label{sec:8DMethod}

In the $\PH\to\V\V\to 4\ell$ channel, the template method discussed above can be extended to the complete
set of eight kinematic observables $\vec{x}=(m_{1},m_{2},m_{4\ell},\vec\Omega)$ described in
Section~\ref{sec:Observables} and shown in Fig.~\ref{fig:kinematics}.
Such an approach would allow us to parameterize the data distributions directly without constructing the
dedicated discriminants. However, the parameterization of templates in eight dimensions using full simulation
is nearly impossible to perform because of the large number of events required.
Therefore, a simplified approach is performed with parametrization of eight-dimensional distributions
to cross-check a subset of results, specifically measurements
of the $f_{a2}$ and $f_{a3}$ parameters in the spin-zero studies (see Section~\ref{sec:ResultsSpinZero}).
The signal and the dominant $\qqbar\to \Z\Z / \Z\gamma^*$ background are parameterized analytically
and reconstruction effects are incorporated in the probability function numerically.
About one third of the background events coming from the $\Z+X$ and $\Pg\Pg\to \Z\Z / \Z\gamma^*$ processes
is parameterized with the template approach in eight dimensions using generated events with detector
effects incorporated using the same approximate numerical parameterization.

The likelihood construction follows Eq.~(\ref{eq:likelihood}) and the probability distribution
is equivalent to Eq.~(\ref{eq:fractions-general}).
The normalization of the probability distributions in eight dimensions is one of the main
computational challenges in this approach and is performed with MC integration.
The final state with $2\Pe$ and $2\mu$ is split into $2\Pe2\mu$ and $2\mu2\Pe$
sub-categories where the distinction between them is determined by the flavor of the leptons from the $\Z_{1}$ decay.
Additionally, a narrower mass window ($115 - 135\GeV$) is used compared
to the template method.

The analytic parameterization is the product of the differential decay cross section,
$\rd\sigma_{4\ell}$, and the production spectrum, $W_{\text{prod}}$, written as
\ifthenelse{\boolean{cms@external}}{
\begin{multline}
\label{eqn:gen_pdf}
\mathcal{P}(\ptvec,Y,\Phi^*, \vec{x} | \vec{\zeta}) =\\
W_{\mathrm{prod}}(\ptvec,Y,\Phi^*,\hat{s}) \times
\frac{\rd\sigma_{4\ell}(m_{4\ell},m_1, m_2, \vec\Omega | \vec{\zeta})}{\rd{}m_1^2\rd{}m_2^2\rd\vec\Omega},
\end{multline}
}{
\begin{equation}
\label{eqn:gen_pdf}
\mathcal{P}(\ptvec,Y,\Phi^*, \vec{x} | \vec{\zeta}) =
W_{\mathrm{prod}}(\ptvec,Y,\Phi^*,\hat{s}) \times
\frac{\rd\sigma_{4\ell}(m_{4\ell},m_1, m_2, \vec\Omega | \vec{\zeta})}{\rd{}m_1^2\rd{}m_2^2\rd\vec\Omega},
\end{equation}
}
where $\ptvec$, $Y$, and $\Phi^*$ are the transverse momentum, rapidity, and azimuthal orientation
of the four-lepton system illustrated in Fig.~\ref{fig:decay},
and $\hat{s}=m_{4\ell}^2$ is the center-of-mass energy of the parton-parton system.
In order to convert the above probability to an expression in terms of detector-level reconstructed
observables, it is convoluted with a transfer function $T({\vec{x}^{\prime\mathrm{R}}} | \vec{x}^{\prime\mathrm{G}})$ describing the
detector response to produced leptons
\begin{equation}
\label{eqn:reco_pdf1}
\mathcal{P}(\vec{x}^{\prime\mathrm{R}} | \vec{\zeta} ) = \int \mathcal{P}(\vec{x}^{\prime\mathrm{G}} | \vec{\zeta} )
T({\vec{x}^{\prime\mathrm{R}}} | \vec{x}^{\prime\mathrm{G}}) \rd\vec{x}^{\prime\mathrm{G}},
\end{equation}
where $\vec{x}^\prime = (\ptvec,Y,\Phi^*,m_{1},m_{2},m_{4\ell},\vec\Omega)$ and the superscripts
R and G denote reconstruction and generator level, respectively.

It is important to model accurately the lepton momentum response and the dependence of the efficiency on $\PT$ and $\eta$,
which can all significantly affect the shape of the distributions of the eight observables used in the likelihood function.
The transfer  functions are constructed from the fully simulated samples for both signal and background.
Because of the excellent angular resolution of the CMS tracker, for the purpose of
this measurement, the effect of the resolution on the direction of each lepton is
negligible compared with the effect of the momentum resolution.
As a result, the effect of the direction is neglected, and only the $\PT$ response of the leptons is modeled.
It is also assumed that the detector response for each lepton is
independent of the other leptons so that the transfer function can be written
as a product of the transfer functions for each individual lepton.
Furthermore, an overall efficiency factor to account for inefficiencies in the lepton selection requirements
is applied. The transfer functions are validated by comparing the full detector simulation
with the generator-level samples, where track parameters are convolved with these functions.

The production spectrum $W_{\text{prod}}$ in Eq.~(\ref{eqn:gen_pdf})
is obtained empirically using simulation. The observables (\ptvec, $Y$, $\Phi^*$)
are found to be uncorrelated to a good approximation, and their distribution is modeled
as a product of three one-dimensional distributions. Then these observables are integrated out
to keep the parameterization with the eight main kinematic observables $\vec{x}^{\mathrm{R}}$.
For the main background, $\qqbar\to \Z\Z / \Z\gamma^*$, the four-lepton mass spectrum
$m_{4\ell}$ is also modeled empirically.
To construct the $m_{4\ell}$ model, the mass spectrum is parameterized with an empirical
exponential function in several bins of rapidity using MC simulation. These distributions are
interpolated between different bins in rapidity. The reconstructed $m_{4\ell}$ spectrum is parameterized
between 115 and 135\GeV, while the generator-level spectrum is wider to model smearing into
and out of this region.

There is no explicit analytic form for the differential cross section for the $\Z+X$ and
$\Pg\Pg \to \Z\Z / \Z\gamma^*$ backgrounds. Instead, the likelihood is calculated
by filling a multidimensional template histogram using very large
samples of generator-level \PYTHIA and \MCFM events, respectively,
with parton showering modeled by \PYTHIA.
These samples are smeared with transfer functions to account for detector effects.
This approach is validated using the $\qqbar\to \Z\Z / \Z\gamma^*$ analytic description
and the corresponding templates, which have been confirmed to have
a sufficient accuracy for the description of these backgrounds.
The remaining discrepancies observed between the $\Z+X$ background templates and
the control regions used in the template analysis~\cite{Chatrchyan:2013mxa}
are covered by assigning a corresponding systematic uncertainty.
The systematic uncertainties in the lepton
momentum scale and resolution are propagated using alternative parameterizations generated through variations
of the transfer function for both signal and background. The sizes of these variations were determined to be consistent
with the size of the lepton momentum and resolution systematic uncertainty in Ref.~\cite{Chatrchyan:2013mxa}.
A systematic uncertainty in the production spectrum of the signal is included using variation
of the $\PT$ spectrum of the four-lepton system when averaging over the production spectrum.
The parameterization of the $\Pg\Pg\to\Z\Z/\Z\gamma^*$ and $\Z+X$ background shape
is varied using the alternative parameterization from the $\qqbar \to \Z\Z/\Z\gamma^*$ background process.

\section{Study of exotic spin-one and spin-two scenarios} \label{sec:ResultsExotic}

The study of the exotic-spin $J^P$ hypotheses of the observed boson with mass around 125\GeV
using the $\X\to\Z\Z$ and $\PW\PW$ channels that have not been presented in previous
publications~\cite{Chatrchyan:2013iaa, Chatrchyan:2013mxa}, is summarized in this Section.
Mixed spin-one state hypotheses, as well as the spin-two models listed in Table~\ref{tab:kdlist} are examined.
In addition, the fractional presence of $J^P$ models of a state nearly degenerate in mass with the SM
state are tested. In all cases, the template method is employed as discussed in Section~\ref{sec:MLfit}.
The $\X\to\gamma\gamma$ decay channel is also studied in the context of the exotic spin-two
scenarios and the results presented in Ref.~\cite{Khachatryan:2014ira} are combined with those
obtained in the $\X\to\Z\Z$ and $\PW\PW$ channels~\cite{Chatrchyan:2013mxa,Chatrchyan:2013iaa}.
All spin-one and spin-two scenarios studied are excluded, which motivates the detailed study
of the spin-zero scenario in Section~\ref{sec:ResultsSpinZero}.
All studies in this paper are presented under the hypothesis of a boson mass of
$m_{\sss\PH}=125.6\GeV$, which is the combined value in
the $\PH\to\Z\Z$ and $\PW\PW$ channels~\cite{Chatrchyan:2013mxa,Chatrchyan:2013iaa}.
The only exception is the analysis of the $\X\to \Z\Z$, $\PW\PW$, and $\gamma\gamma$ channels combined
that is performed with the $m_{\sss\PH}=125.0$ $\GeV$ hypothesis, which is the combined value for the three
channels~\cite{Chatrchyan:2013mxa,Chatrchyan:2013iaa,Khachatryan:2014ira}.
This mass difference has little effect on the results and it is in the same range as the
systematic uncertainties assigned to the energy scale in the mass reconstruction.

\subsection{Exotic-spin study with the \texorpdfstring{$\PH\to\Z\Z\to 4\ell$}{H to ZZ to 4l} channel}  \label{sec:Resultsexotichzz}

In the case of the spin-one studies, the hypothesis testing is performed for a discrete set of values of the
parameter $f_{b2}$. The input observables are $(\superKD,\psvectorKD,\vectorKD)$.
It has been demonstrated in the context of this study
that the distributions of these observables are not sensitive to the phase between
the $b_1$ and $b_2$ coupling parameters in Eq.~(\ref{eq:ampl-spin1}) and therefore the results of the $f_{b2}$ scan
are valid for any value of the phase term in the interference. The spin-one hypothesis is tested for
two scenarios, $\qqbar$ production and using only decay information. The latter requires the input
observables $(\superKD^\text{dec},\psvectorKD^\text{dec},\vectorKD^\text{dec})$.

\begin{figure}[tbhp]
  \centering
    \includegraphics[width=0.45\textwidth]{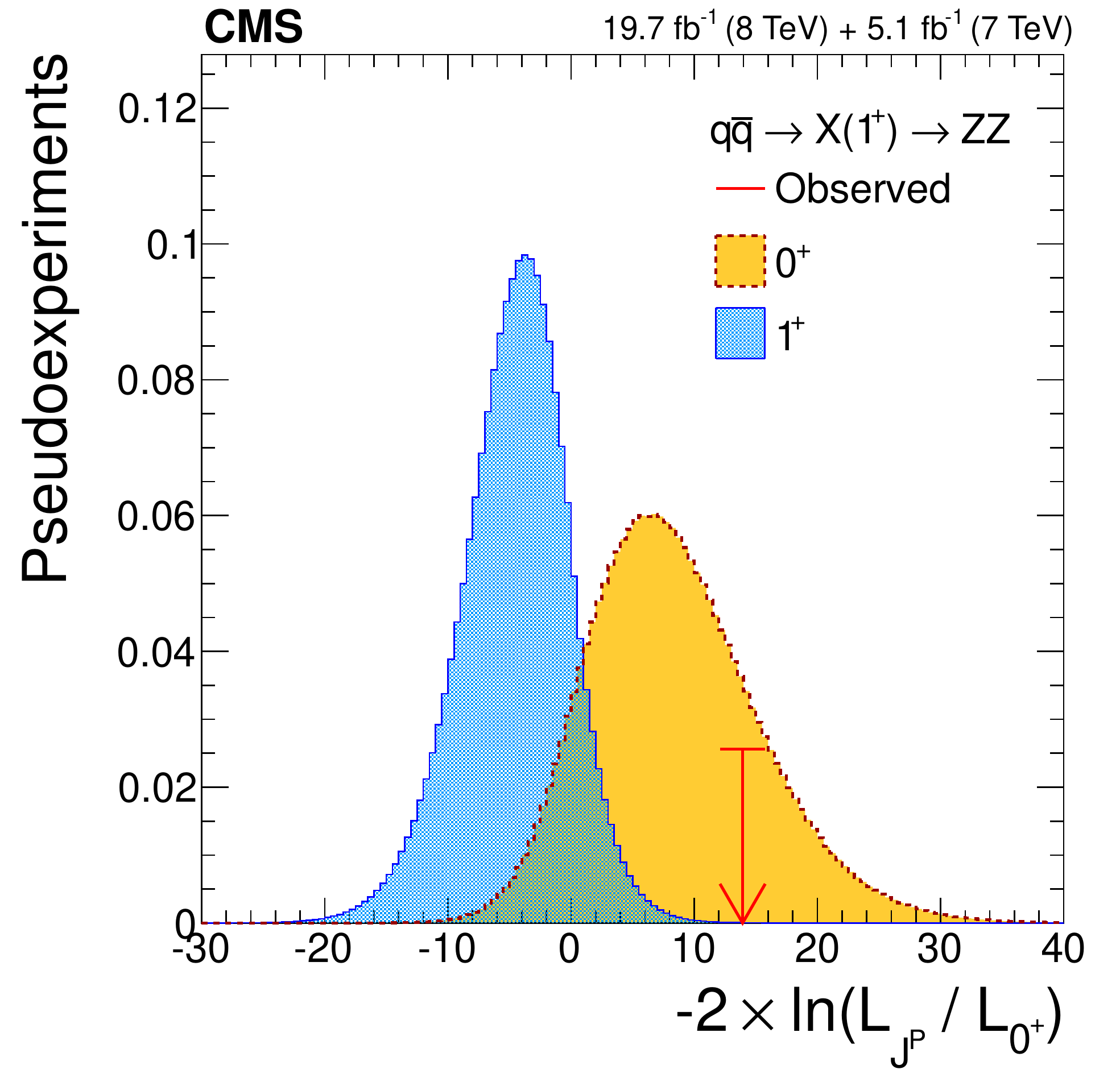}
    \includegraphics[width=0.45\textwidth]{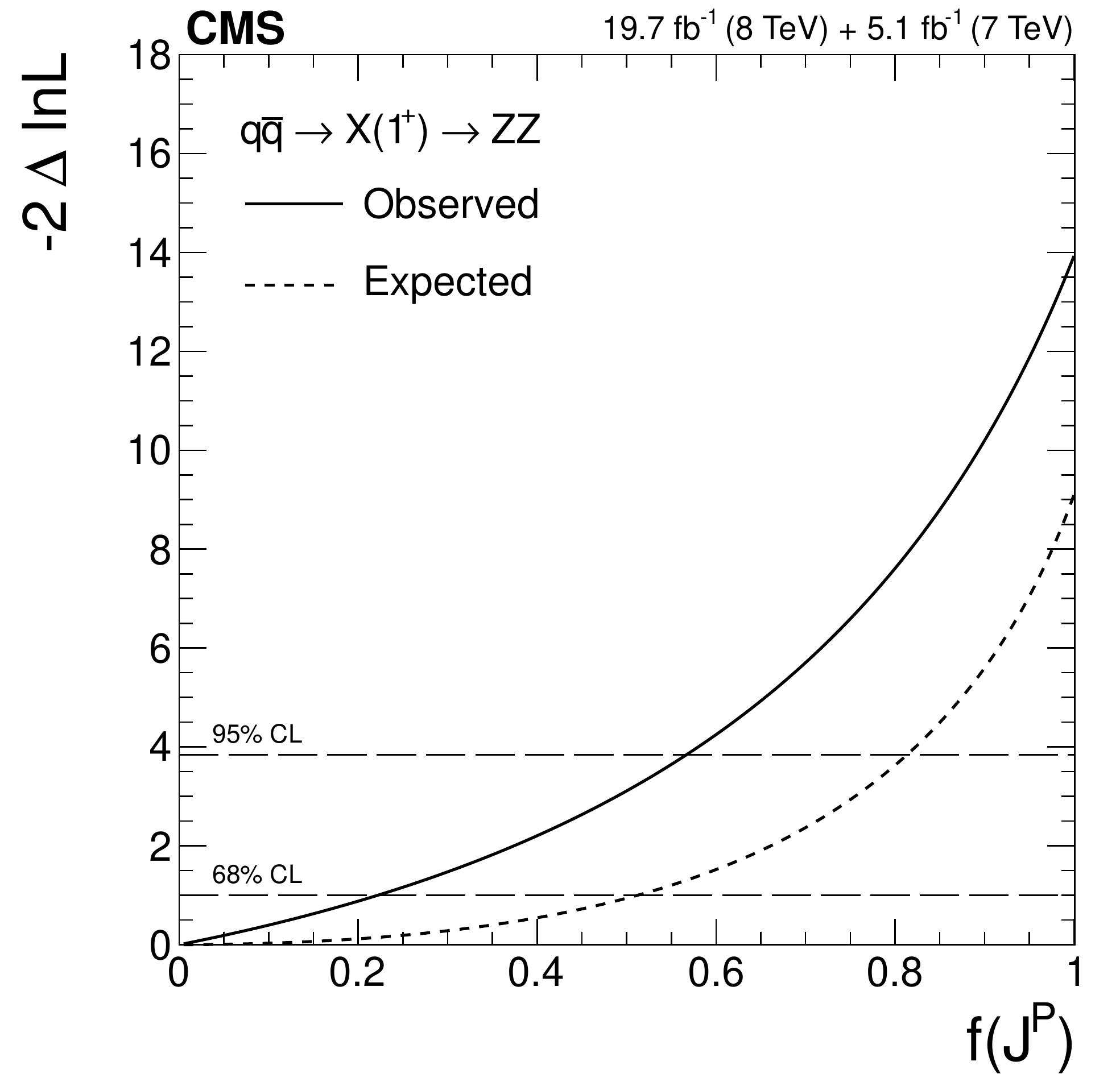}
    \caption{(\cmsLeft) Distributions of the test statistic $q=-2\ln(\mathcal{L}_{J^P}/\mathcal{L}_{0^+})$
       for the $J^P=1^+$ hypothesis of $\qqbar\to\X(1^+)\to \Z\Z$
      tested against the SM Higgs boson hypothesis ($0^+$).
      The expectation for the SM Higgs boson is represented by the yellow histogram on the right and the alternative $J^P$ hypothesis by the
      blue histogram on the left. The red arrow indicates the observed $q$ value.
      (\cmsRight) Observed value of $-2\Delta\ln\mathcal{L}$ as a function of $f(J^P)$ and the expectation in the SM
      for the $\qqbar\to\X(1^+)\to \Z\Z$ alternative $J^P$ model.
      \label{fig:sepPD_1+}
      }

\end{figure}

\begin{figure*}[htbp]
  \centering
    \includegraphics[width=0.45\textwidth]{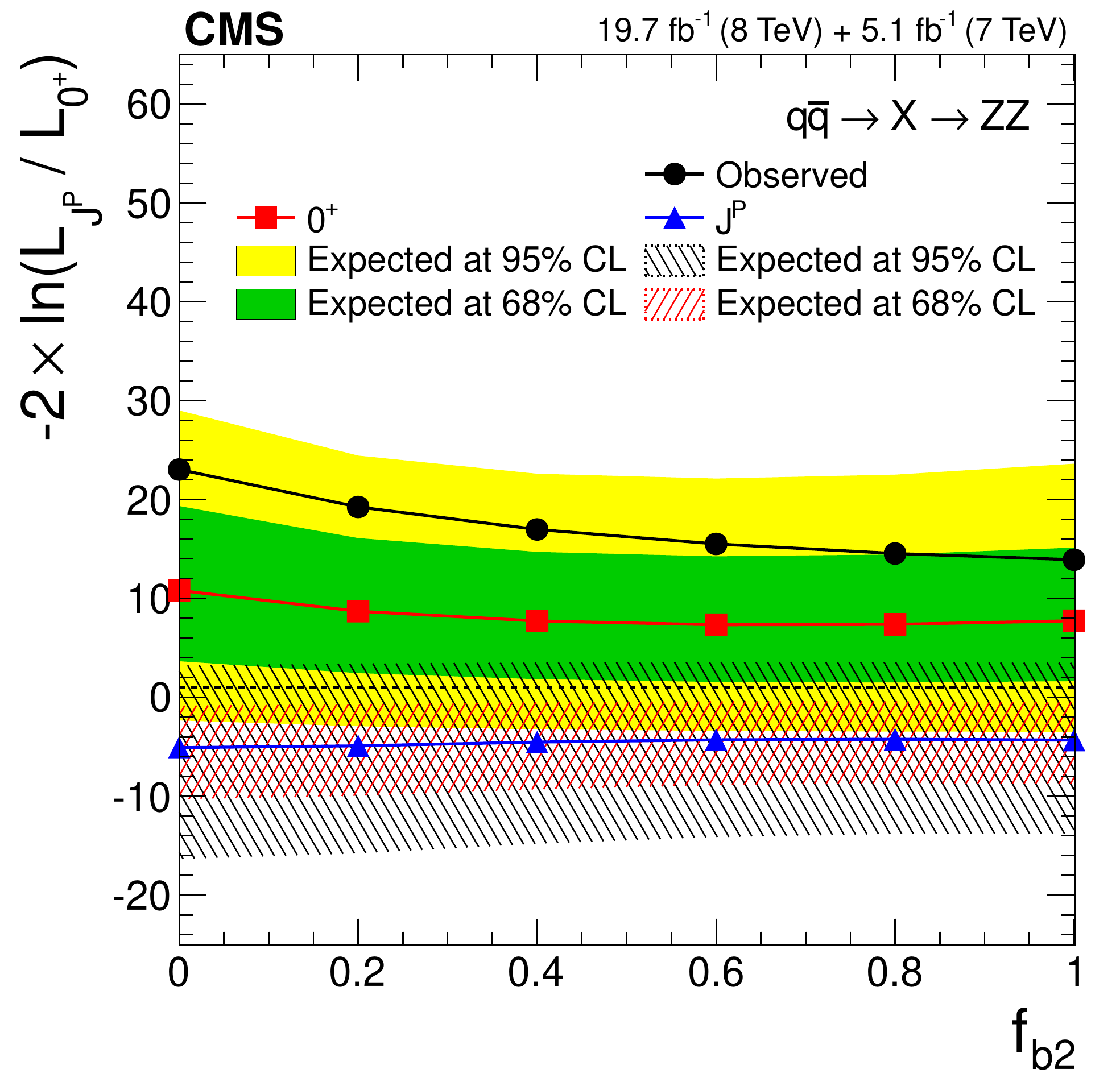}
    \includegraphics[width=0.45\textwidth]{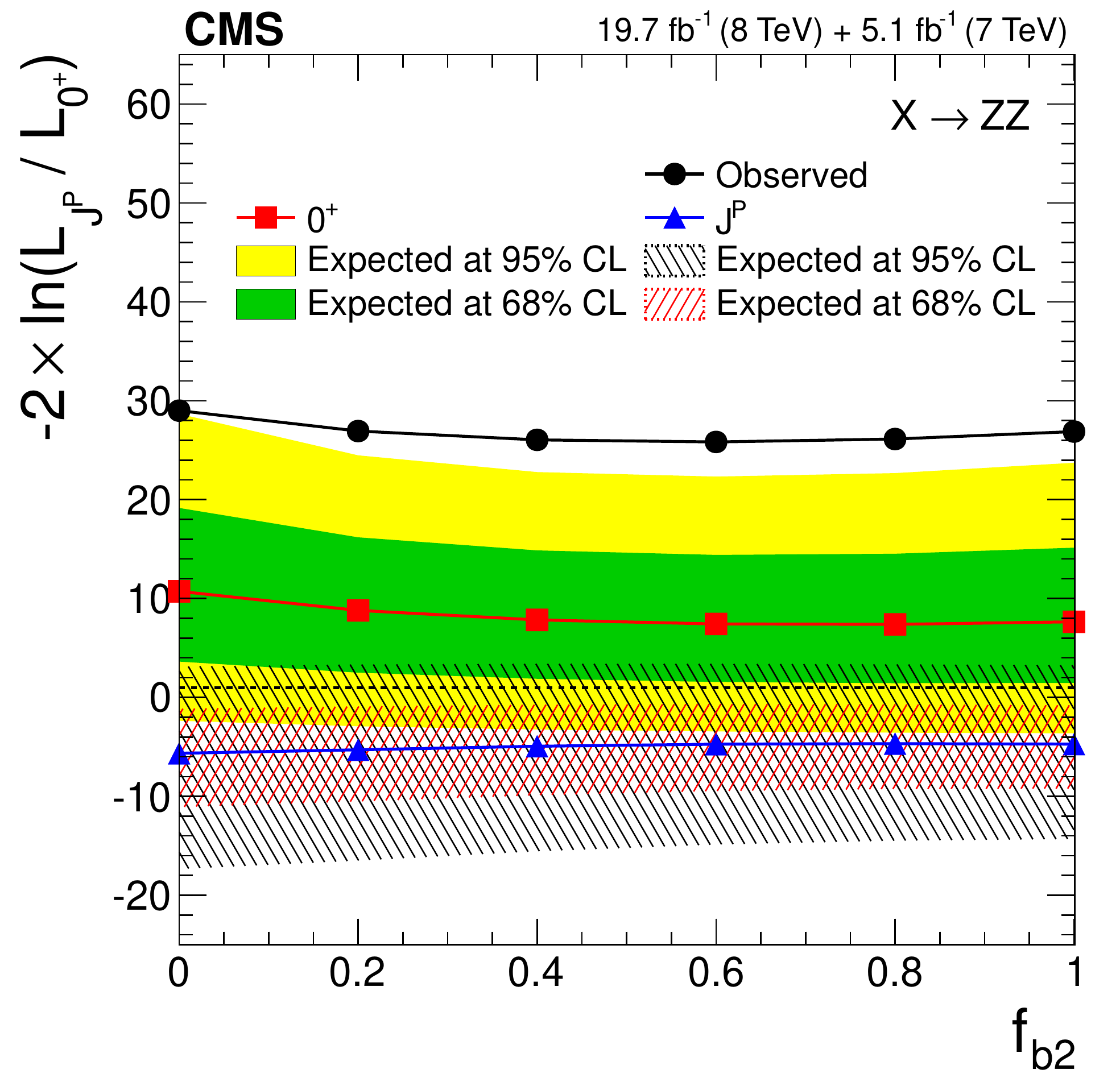}
    \caption{Distributions of the test statistic $q=-2\ln(\mathcal{L}_{J^P}/\mathcal{L}_{0^+})$
    as a function of $f_{b2}$
      for the spin-one $J^{P}$ models tested against the SM Higgs boson hypothesis
      in the $\qqbar\to \X\to\Z\Z$ (left) and decay-only $\X\to\Z\Z$ (right) analyses.
      The median expectation for the SM Higgs boson is represented
      by the red squares with the green (68\% \CL) and yellow (95\% \CL) solid color regions and
       for the alternative $J^P$ hypotheses by the blue triangles
       with the red (68\% \CL) and blue (95\% \CL) hatched regions.
     The observed values are indicated by the black dots.
      \label{fig:jp_summary_1}
      }

\end{figure*}

Figure~\ref{fig:sepPD_1+} (\cmsLeft) shows the distribution of the test statistic
$q=-2\ln(\mathcal{L}_{J^P}/\mathcal{L}_{0^+})$ for a SM Higgs boson and for the $J^P=1^+$ hypothesis.
The expected and observed separations of spin-one models from the test statistic distributions
are summarized in Table~\ref{tab:jpmodels_1} and in Fig.~\ref{fig:jp_summary_1}.
The expected separation between the alternative signal hypotheses is
quoted for two cases. In the first case, the expected SM Higgs boson
signal strength and the alternative signal cross section are
the ones obtained in the fit to the data.
The second case assumes the nominal SM Higgs boson signal strength
(defined as $\mu=1$), while the cross section for the alternative signal hypothesis is
taken to be the same as for the SM Higgs boson (the $2\Pe2\Pgm$
channel at 8\TeV is taken as a reference). Since the observed signal strength
is very close to unity, the two results for the expected separations are also similar.

\begin{figure*}[htbp]
  \centering
    \includegraphics[width=0.95\textwidth]{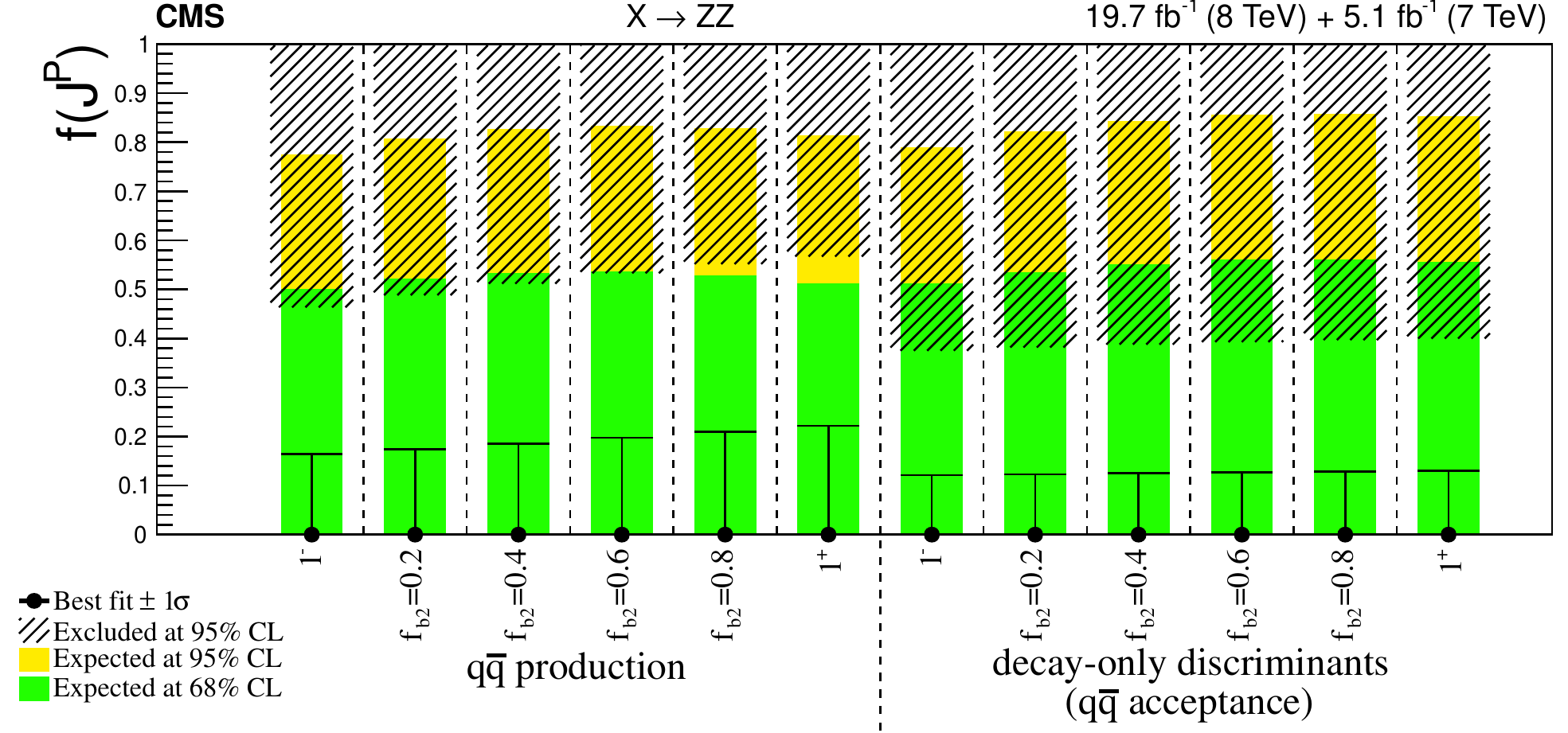}
    \caption{
Summary of the $f(J^P)$ constraints as a function of $f_{b2}$ from Table~\ref{tab:jpmodels_1},
where the decay-only measurements are performed using the efficiency of the $\qqbar \to \X\to \Z\Z$ selection.
The expected 68\% and 95\% \CL regions are shown as green and yellow bands.
The observed constraints at 68\% and 95\% \CL are shown as the points with error bars
and the excluded hatched regions.
       \label{fig:non_interf_summary_spin1}
       }

\end{figure*}

Figure~\ref{fig:sepPD_1+} (\cmsRight) also shows an example of the likelihood scan, $-2\Delta\ln \mathcal{L}$
as a function of $f(J^P)$ for the $\qqbar$ produced $1^+$ model, where the fractional cross section
of a second overlapping but non interfering resonance $f(J^P)$ is defined in Eq.~(\ref{eq:fJCP_definition}).
The expected and observed measurements of the non-interfering fractions are also summarized
in Table~\ref{tab:jpmodels_1} and in Fig.~\ref{fig:non_interf_summary_spin1}.
The production cross section fractions are represented by $f(J^P)$ and therefore require
knowledge of the reconstruction efficiency for the interpretation of the measured yields.
In the case of the production independent scenarios of spin-one models, the $f(J^P)$ results
are extracted using the reconstruction efficiency of the $\qqbar \to\X$ process.
The values of  $-2\Delta \ln \mathcal{L} = 1$ and 3.84 represent the 68\% and 95\% \CL, respectively.

All spin-one tests are consistent with the expectation for the SM Higgs boson.
While the decay-only analysis uses less information and is expected to provide weaker constraints,
the fluctuations in the observed data lead to stronger constraints for spin-one models.
The least restrictive result corresponds to the $1^+$ model in the $\qqbar$ production test
with a \CLs value of 0.031\%.

Any arbitrary spin-one model for the resonance observed in the $\X\to \Z\Z\to 4\ell$ decay mode with any
mixture of parity-even and parity-odd interactions and any production mechanism is excluded at a \CL of
99.97\% or higher.

\begin{table*}[htbp]
  \centering
\topcaption{
List of spin-one models tested in the $\X\to \Z\Z$ analysis.
The expected separation is quoted for two scenarios, for the signal production cross section
obtained from the fit to data for each hypothesis and using the SM expectation ($\mu=1$).
The observed separation shows the consistency of the observation with the SM Higgs boson model
or the alternative $J^{P}$ model, from which the \CLs value is derived.
The $f(J^P)$ constraints are quoted, where the decay-only measurements are valid for any
production (Prod.) mechanism and are performed using the efficiency of the $\qqbar \to \X\to \Z\Z$
selection.
\label{tab:jpmodels_1}
}
\begin{scotch}{lccccccc}
$f_{b2} (J^P)$                 & $J^P$        &        Expected  &                       &                    &               &        $f(J^P)$  95\% \CL & $f(J^P)$ \\
Model                          & Prod. &        ($\mu$=1) &  Obs. $0^+$  & Obs. $J^P$ & \CLs &        Obs. (Exp.)                 & Best Fit \\
    \hline
     $0.0 (1^-)$             & $$\qqbar$$    &  2.9$\sigma$ (2.8$\sigma$)   & $-$1.4$\sigma$ & $+$5.0$\sigma$ & $<$0.001\% & $<$0.46 (0.78) & $0.00^{+0.16}_{-0.00}$ \\
    0.2      & $$\qqbar$$    &  2.6$\sigma$ (2.6$\sigma$)   & $-$1.4$\sigma$ & $+$4.6$\sigma$ & 0.002\% & $<$0.49 (0.81) & $0.00^{+0.17}_{-0.00}$ \\
    0.4      & $$\qqbar$$    &  2.5$\sigma$ (2.4$\sigma$)   & $-$1.3$\sigma$ & $+$4.4$\sigma$ & 0.005\% & $<$0.51 (0.83) & $0.00^{+0.19}_{-0.00}$ \\
    0.6      & $$\qqbar$$    &  2.4$\sigma$ (2.4$\sigma$)   & $-$1.2$\sigma$ & $+$4.1$\sigma$ &  0.015\% & $<$0.53 (0.83) & $0.00^{+0.20}_{-0.00}$ \\
    0.8      & $$\qqbar$$    &  2.4$\sigma$ (2.4$\sigma$)   & $-$1.0$\sigma$ & $+$4.0$\sigma$ &  0.021\% & $<$0.55 (0.83) & $0.00^{+0.21}_{-0.00}$ \\
    $1.0 (1^+)$             & $$\qqbar$$    &  2.4$\sigma$ (2.4$\sigma$)   & $-$0.8$\sigma$ & $+$3.8$\sigma$ &  0.031\% & $<$0.57 (0.81) & $0.00^{+0.22}_{-0.00}$ \\
   \hline
    $0.0 (1^-)$             & any               &  2.9$\sigma$ (2.7$\sigma$)   & $-$2.0$\sigma$ & $>$5.0$\sigma$ & $<$0.001\% & $<$0.37 (0.79) & $0.00^{+0.12}_{-0.00}$ \\
    0.2      & any               &  2.7$\sigma$ (2.5$\sigma$)   & $-$2.2$\sigma$ & $>$5.0$\sigma$ & $<$0.001\% & $<$0.38 (0.82) & $0.00^{+0.12}_{-0.00}$ \\
    0.4      & any               &  2.5$\sigma$ (2.4$\sigma$)   & $-$2.3$\sigma$ & $>$5.0$\sigma$ & $<$0.001\% & $<$0.39 (0.84) & $0.00^{+0.13}_{-0.00}$ \\
    0.6      & any               &  2.5$\sigma$ (2.3$\sigma$)   & $-$2.4$\sigma$ & $>$5.0$\sigma$ & $<$0.001\% & $<$0.39 (0.86) & $0.00^{+0.13}_{-0.00}$ \\
    0.8      & any               &  2.4$\sigma$ (2.3$\sigma$)   & $-$2.3$\sigma$ & $>$5.0$\sigma$ & $<$0.001\% & $<$0.40 (0.86) & $0.00^{+0.13}_{-0.00}$ \\
    $1.0 (1^+)$             & any               &  2.5$\sigma$ (2.3$\sigma$)   & $-$2.3$\sigma$ & $>$5.0$\sigma$ & $<$0.001\% & $<$0.41 (0.85) & $0.00^{+0.13}_{-0.00}$ \\
  \end{scotch}
\end{table*}

\begin{figure*}[htbp]
\centering
\includegraphics[width=0.48\textwidth]{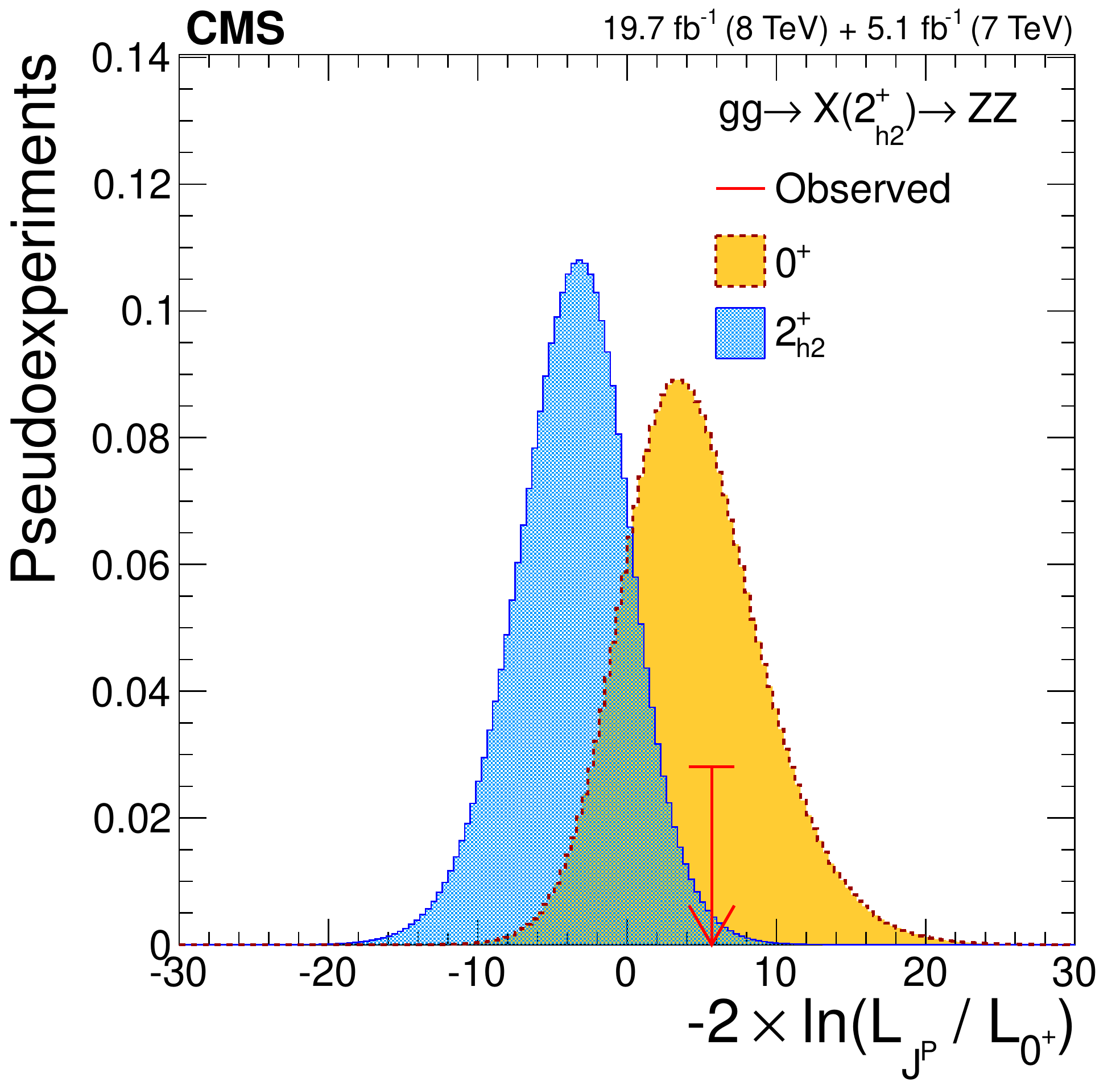}
\includegraphics[width=0.48\textwidth]{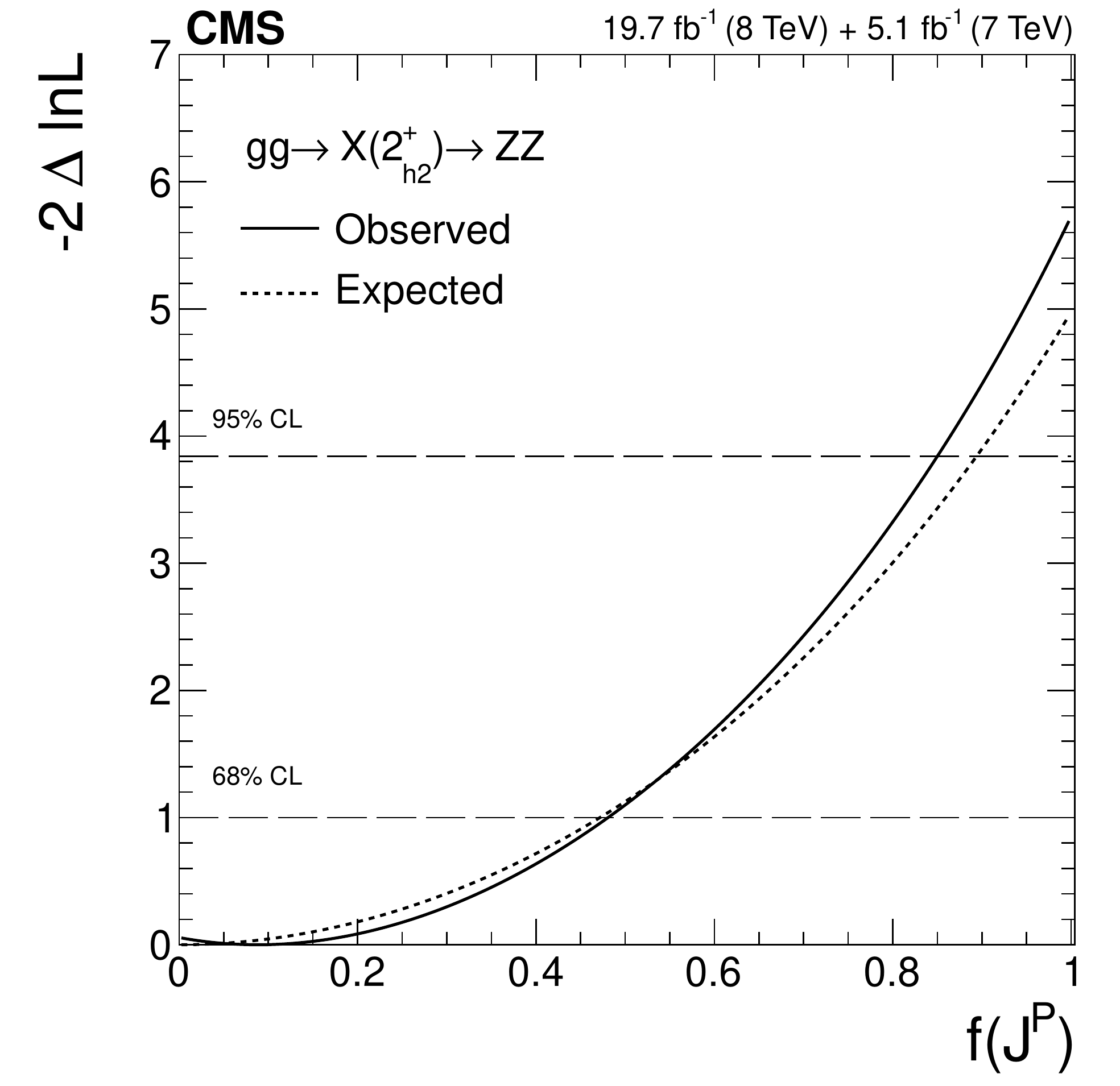}
\caption{(left) Distributions of the test statistic $q=-2\ln(\mathcal{L}_{J^P}/\mathcal{L}_{0^+})$
       for the $J^P=2^+_{h2}$ hypothesis of $\Pg\Pg\to\X(2^+_{h2})\to \Z\Z$
      tested against the SM Higgs boson hypothesis ($0^+$).
            The expectation for the SM Higgs boson is represented by the yellow histogram on the right and the alternative $J^P$ hypothesis by the
      blue histogram on the left. The red arrow indicates the observed $q$ value.
      (right) Observed value of $-2\Delta\ln\mathcal{L}$ as a function of $f(J^P)$ and the expectation in the SM
      for the $\Pg\Pg\to\X(2^+_{h2})\to \Z\Z$ alternative $J^P$ model.
  }
\label{fig:sep_2h10-}

\end{figure*}

In the case of the spin-two studies, hypothesis testing is performed for ten models and three
scenarios: $\Pg\Pg$, $\qqbar$ production, and using only decay information. Two input observables
are used since interference between the different amplitude components is not considered.
Several models have been tested in Ref.~\cite{Chatrchyan:2013mxa} and here those results
are repeated for completeness. They cover all the lowest order terms in the amplitude without considering mixing of different contributions.

\begin{figure*}[htbp]
  \centering
    \includegraphics[width=\textwidth]{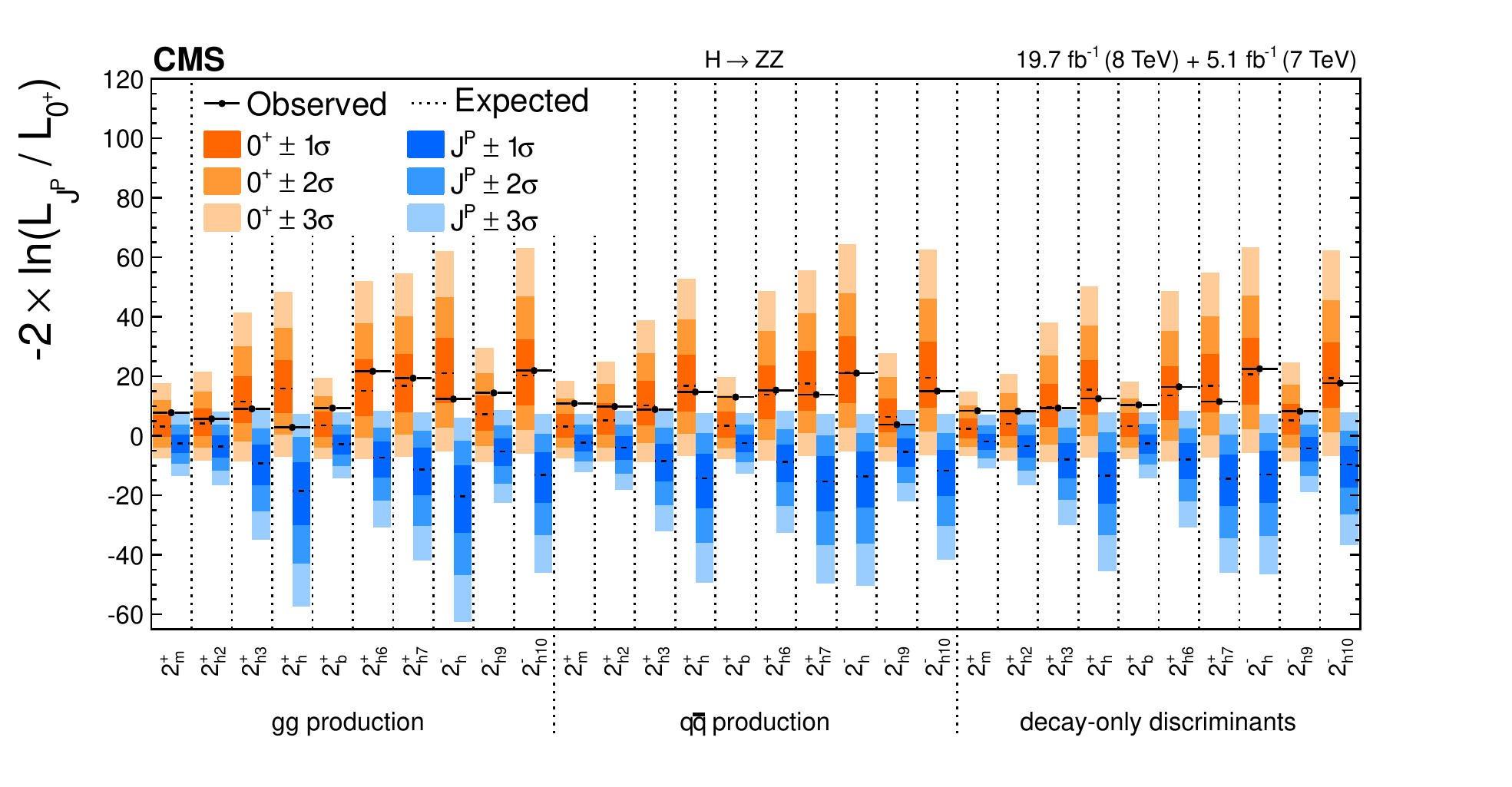}
    \caption{
        Distributions of the test statistic $q=-2\ln(\mathcal{L}_{J^P}/\mathcal{L}_{0^+})$
       for the spin-two $J^{P}$ models tested against the SM Higgs boson hypothesis
      in the $\X\to\Z\Z$ analyses.
      The expected median and the 68.3\%, 95.4\%, and 99.7\% \CL regions for the SM Higgs boson (orange, the left for each model)
      and for the alternative $J^P$ hypotheses (blue, right) are shown.
     The observed $q$ values are indicated by the black dots.
      \label{fig:jp_summary}
      }

\end{figure*}

\begin{figure*}[htbp]
  \centering
    \includegraphics[width=\textwidth]{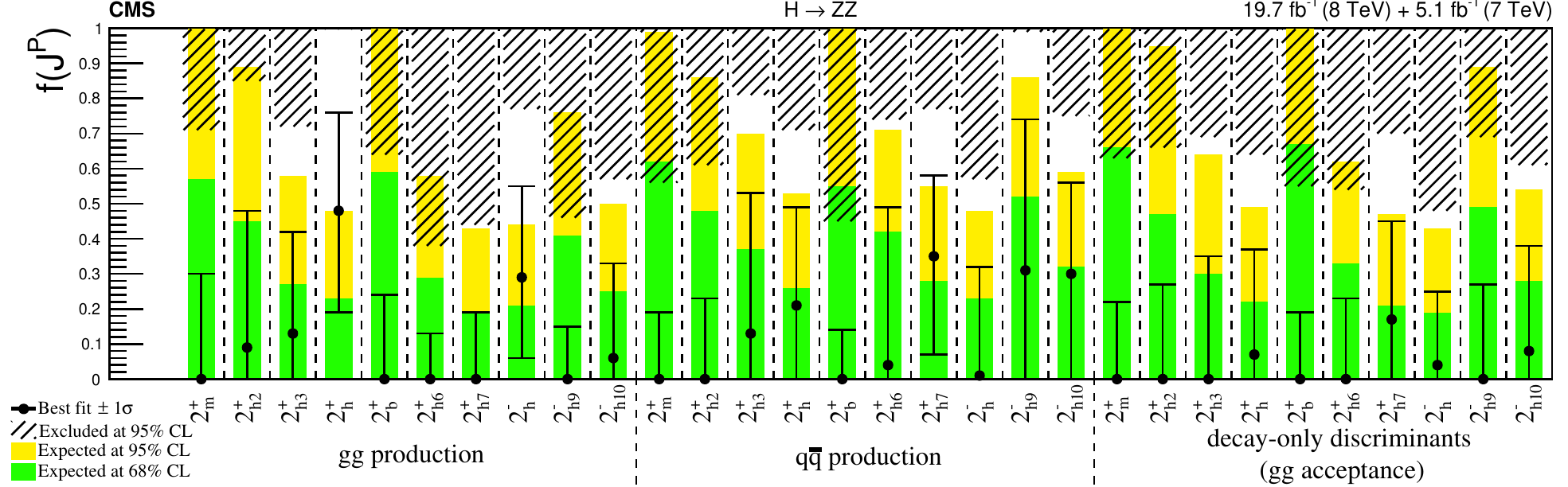}
    \caption{
    Summary of the $f(J^P)$ constraints for the spin-two models from Table~\ref{tab:jpmodels},
where the decay-only measurements are performed using the efficiency of the $\Pg\Pg \to \X\to \Z\Z$ selection.
The expected 68\% and 95\% \CL regions are shown as the green and yellow bands.
The observed constraints at 68\% and 95\% \CL are shown as the points with error bars
and the excluded hatched regions.
      \label{fig:non_interf_summary}}

\end{figure*}

An example distribution of the test statistic and observed value in the case of the SM Higgs boson
and the spin-two hypothesis $2_{h2}^{+}$ is shown in Fig.~\ref{fig:sep_2h10-}~(left).
The expected and observed separation from the test statistic distributions for all the spin-two
models considered is summarized in Table~\ref{tab:jpmodels} and in Fig.~\ref{fig:jp_summary}.
The $2_{h2}^{+}$ model is the least restricted one, see Table~\ref{tab:jpmodels}: \CLs = 0.74\% for any production mechanism.
The observed non-interfering fraction measurements are summarized in Table~\ref{tab:jpmodels} and in Fig.~\ref{fig:non_interf_summary}.
In the case of production-independent scenarios the $f(J^P)$ results are extracted using the $\Pg\Pg \to\X$ efficiency. Figure~\ref{fig:sep_2h10-}~(right) shows the likelihood scan for the $2_{h2}^{+}$ hypothesis as a function of $f(J^P)$.

The data disfavor all the spin-two $\X\to \Z\Z\to 4\ell$ hypotheses tested in favor of the SM hypothesis
$J^P=0^+$ with $1-\CLs$ values larger than 99\% \CL when only decay information is used (Table~\ref{tab:jpmodels}).

\begin{table*}[htbp]
\topcaption{
List of spin-two models tested in the $\X\to \Z\Z$ analysis.
The expected separation is quoted for two scenarios, for the signal production cross section
obtained from the fit to data for each hypothesis, and using the SM expectation ($\mu=1$).
The observed separation shows the consistency of the observation with the SM Higgs boson
or an alternative $J^{P}$ model, from which the \CLs value is derived.
The $f(J^P)$ constraints are quoted, where the decay-only measurements are valid for any
production (Prod.) mechanism and are performed using the efficiency of the $\Pg\Pg \to \X\to \Z\Z$ selection.
Results from Ref.~\cite{Chatrchyan:2013mxa} are explicitly noted.
}
\centering
\renewcommand{\arraystretch}{1.25}
\begin{scotch}{lccccccccc}
$J^P$                           & $J^P$        &        Expected  &                       &                    &               &        $f(J^P)$  95\% \CL & $f(J^P)$ \\
Model                           & Prod. &        ($\mu$=1) &  Obs. $0^+$  & Obs. $J^P$ & \CLs &        Obs. (Exp.)                 & Best Fit \\
\hline
$2_{m}^+$~\cite{Chatrchyan:2013mxa}     & $\Pg\Pg$       &  1.9$\sigma$ (1.8$\sigma$)  & $-$1.1$\sigma$  & +3.0$\sigma$  &  0.90\%
       & $<$0.71 (1.00) & $0.00^{+0.30}_{-0.00}$ \\
$2^+_{h2}$ & $\Pg\Pg$         & 2.0$\sigma$ (2.1$\sigma$) & $-$0.3$\sigma$ & $+$2.4$\sigma$ & 2.0\% & $<$0.85 (0.89) & $0.09^{+0.39}_{-0.09}$ \\
$2^+_{h3}$ & $\Pg\Pg$         & 3.2$\sigma$ (3.4$\sigma$) & $+$0.3$\sigma$ & $+$3.0$\sigma$ & 0.17\% & $<$0.72 (0.58) & $0.13^{+0.29}_{-0.13}$ \\
$2_{h}^+$~\cite{Chatrchyan:2013mxa}      & $\Pg\Pg$       &  3.8$\sigma$ (4.0$\sigma$)  & +1.8$\sigma$    & +2.0$\sigma$  &  2.3\%
         & $<$1.00 (0.48)  & $0.48^{+0.28}_{-0.29}$ \\
$2_{b}^+$~\cite{Chatrchyan:2013mxa}     & $\Pg\Pg$       &  1.6$\sigma$ (1.8$\sigma$)  & $-$1.4$\sigma$  & +3.4$\sigma$  &  0.50\%
     & $<$0.64 (1.00) & $0.00^{+0.24}_{-0.00}$ \\
$2^+_{h6}$ & $\Pg\Pg$         & 3.4$\sigma$ (3.7$\sigma$) & $-$0.6$\sigma$ & $+$4.9$\sigma$ & $<$0.001\% & $<$0.38 (0.58) & $0.00^{+0.13}_{-0.00}$ \\
$2^+_{h7}$ & $\Pg\Pg$         & 3.8$\sigma$ (4.5$\sigma$) & $-$0.3$\sigma$ & $+$4.5$\sigma$ & $<$0.001\% & $<$0.44 (0.43) & $0.00^{+0.19}_{-0.00}$ \\
 $2^-_{h}$~\cite{Chatrchyan:2013mxa}     & $\Pg\Pg$       &  4.2$\sigma$ (4.5$\sigma$)  & +1.0$\sigma$    & +3.2$\sigma$  &  0.090\%
      & $<$0.77 (0.44) & $0.29^{+0.26}_{-0.23}$  \\
$2^-_{h9}$ & $\Pg\Pg$          & 2.5$\sigma$ (2.6$\sigma$) & $-$1.1$\sigma$ & $+$4.0$\sigma$ & 0.029\% & $<$0.46 (0.76) & $0.00^{+0.15}_{-0.00}$ \\
$2^-_{h10}$ & $\Pg\Pg$         & 4.2$\sigma$ (4.3$\sigma$) & $-$0.1$\sigma$ & $+$4.8$\sigma$ & $<$0.001\% & $<$0.57 (0.50) & $0.06^{+0.27}_{-0.06}$ \\
\hline
$2_{m}^+$~\cite{Chatrchyan:2013mxa}     & $\Pq\Paq$       &  1.7$\sigma$ (1.7$\sigma$)  & $-$1.7$\sigma$  & +3.8$\sigma$  &  0.17\%
      & $<$0.56 (0.99) & $0.00^{+0.19}_{-0.00}$ \\
$2^+_{h2}$ & $\Pq\Paq$ & 2.2$\sigma$ (2.2$\sigma$) & $-$0.8$\sigma$ & $+$3.3$\sigma$ & 0.26\% & $<$0.61 (0.86) & $0.00^{+0.23}_{-0.00}$ \\
$2^+_{h3}$ & $\Pq\Paq$ & 3.1$\sigma$ (3.0$\sigma$) & $+$0.2$\sigma$ & $+$3.0$\sigma$ & 0.21\% & $<$0.81 (0.70) & $0.13^{+0.40}_{-0.13}$ \\
$2^+_{h}$  & $\Pq\Paq$ & 4.0$\sigma$ (3.9$\sigma$) & $+$0.2$\sigma$ & $+$3.9$\sigma$ & 0.008\% & $<$0.71 (0.53) & $0.21^{+0.28}_{-0.21}$ \\
$2^+_{b}$  & $\Pq\Paq$ & 1.7$\sigma$ (1.7$\sigma$) & $-$1.9$\sigma$ & $+$4.1$\sigma$ & 0.062\% & $<$0.45 (1.00) & $0.00^{+0.14}_{-0.00}$ \\
$2^+_{h6}$ & $\Pq\Paq$ & 3.4$\sigma$ (3.3$\sigma$) & $-$0.2$\sigma$ & $+$4.0$\sigma$ & 0.008\% & $<$0.74 (0.71) & $0.04^{+0.45}_{-0.04}$ \\
$2^+_{h7}$ & $\Pq\Paq$ & 4.1$\sigma$ (3.9$\sigma$) & $+$0.4$\sigma$ & $+$3.8$\sigma$ & 0.010\% & $<$0.77 (0.55) & $0.35^{+0.23}_{-0.28}$ \\
$2^-_{h}$  & $\Pq\Paq$  & 4.3$\sigma$ (4.4$\sigma$) & $+$0.0$\sigma$ & $+$4.6$\sigma$ & $<$0.001\% & $<$0.57 (0.48) & $0.01^{+0.31}_{-0.01}$ \\
$2^-_{h9}$ & $\Pq\Paq$  & 2.4$\sigma$ (2.2$\sigma$) & $+$0.5$\sigma$ & $+$2.0$\sigma$ & 3.1\% & $<$0.99 (0.86) & $0.31^{+0.43}_{-0.31}$ \\
$2^-_{h10}$ & $\Pq\Paq$ & 4.0$\sigma$ (3.9$\sigma$) & $+$0.4$\sigma$ & $+$4.0$\sigma$ & 0.006\% & $<$0.75 (0.59) & $0.30^{+0.26}_{-0.30}$ \\
\hline
$2_{m}^+$~\cite{Chatrchyan:2013mxa}      & any                   &  1.5$\sigma$ (1.5$\sigma$)  & $-$1.6$\sigma$  & +3.4$\sigma$  &  0.71\%
    & $<$0.63 (1.00) & $0.00^{+0.22}_{-0.00}$ \\
$2^+_{h2}$ & any                                 & 1.9$\sigma$ (2.0$\sigma$) & $-$0.9$\sigma$ & $+$3.0$\sigma$ & 0.74\% & $<$0.66 (0.95) & $0.00^{+0.27}_{-0.00}$ \\
$2^+_{h3}$ & any                                 & 3.0$\sigma$ (3.1$\sigma$) & $+$0.0$\sigma$ & $+$3.1$\sigma$ & 0.18\% & $<$0.69 (0.64) & $0.00^{+0.35}_{-0.00}$ \\
$2^+_{h}$  & any                                 & 3.8$\sigma$ (4.0$\sigma$) & $+$0.3$\sigma$ & $+$3.6$\sigma$ & 0.025\% & $<$0.64 (0.49) & $0.07^{+0.30}_{-0.07}$ \\
$2^+_{b}$  & any                                 & 1.7$\sigma$ (1.7$\sigma$) & $-$1.6$\sigma$ & $+$3.6$\sigma$ & 0.29\% & $<$0.55 (1.00) & $0.00^{+0.19}_{-0.00}$ \\
$2^+_{h6}$ & any                                 & 3.3$\sigma$ (3.4$\sigma$) & $-$0.3$\sigma$ & $+$4.2$\sigma$ & 0.003\% & $<$0.54 (0.62) & $0.00^{+0.23}_{-0.00}$ \\
$2^+_{h7}$ & any                                 & 4.0$\sigma$ (4.2$\sigma$) & $+$0.6$\sigma$ & $+$3.5$\sigma$ & 0.032\% & $<$0.70 (0.47) & $0.17^{+0.28}_{-0.17}$ \\
$2^-_{h}$  & any                                  & 4.2$\sigma$ (4.6$\sigma$) & $-$0.2$\sigma$ & $+$4.8$\sigma$ & $<$0.001\% & $<$0.48 (0.43) &  $0.04^{+0.21}_{-0.04}$ \\
$2^-_{h9}$ & any                                  & 2.2$\sigma$ (2.1$\sigma$) & $-$0.6$\sigma$ & $+$2.9$\sigma$ & 0.57\% & $<$0.69 (0.89) & $0.00^{+0.27}_{-0.00}$ \\
$2^-_{h10}$ & any                                 & 3.9$\sigma$ (4.0$\sigma$) & $+$0.1$\sigma$ & $+$4.3$\sigma$ & 0.002\% & $<$0.61 (0.54) & $0.08^{+0.30}_{-0.08}$ \\
\end{scotch}

\label{tab:jpmodels}
\end{table*}

\subsection{Exotic-spin study with the \texorpdfstring{$\PH\to\PW\PW\to\ell\nu\ell\nu$}{H to WW to l nu l nu} channel} \label{sec:Resultsexotichww}

\begin{figure}[thbp]
  \centering
      \includegraphics[width=0.45\textwidth]{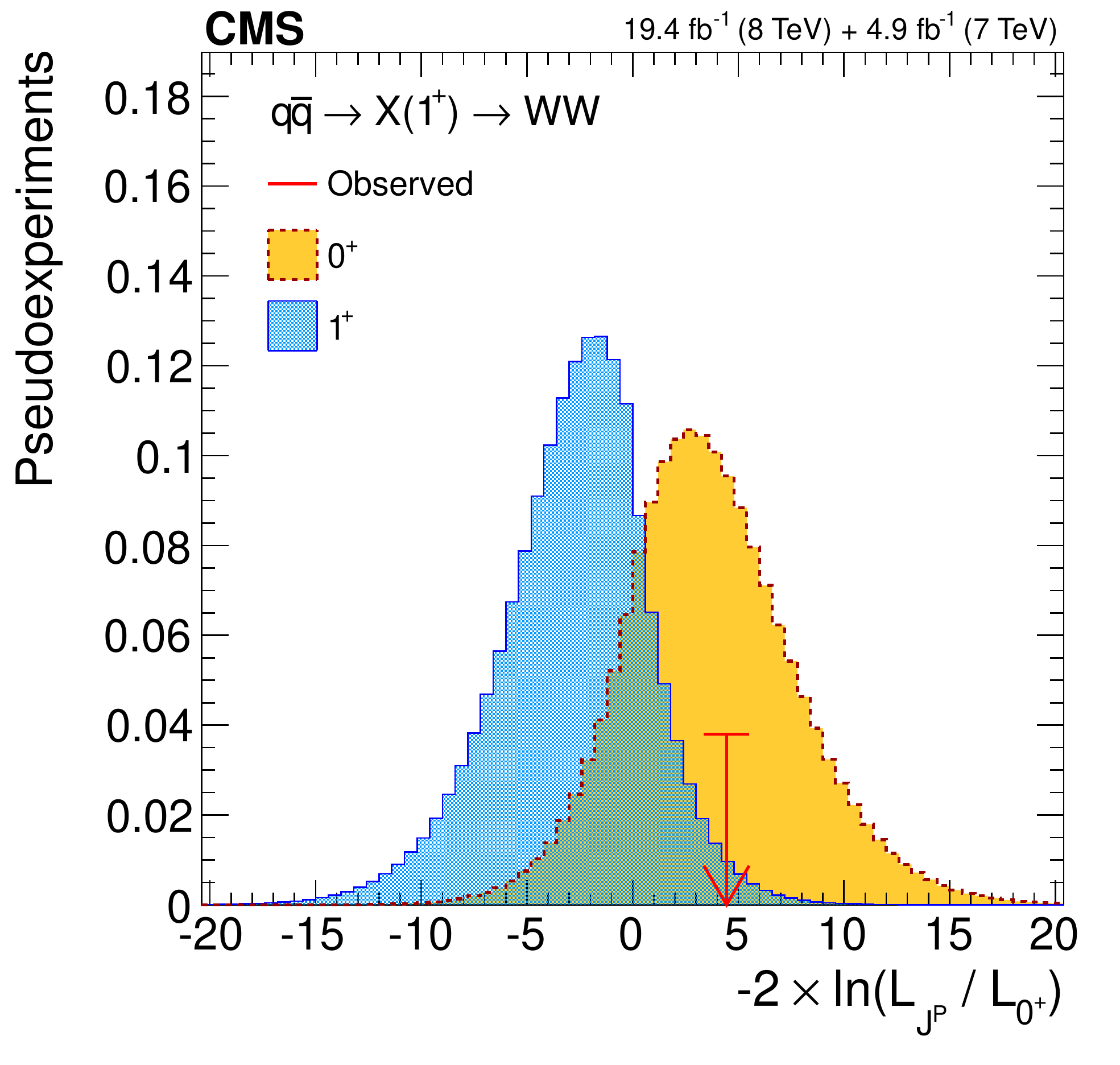}
           \includegraphics[width=0.45\textwidth]{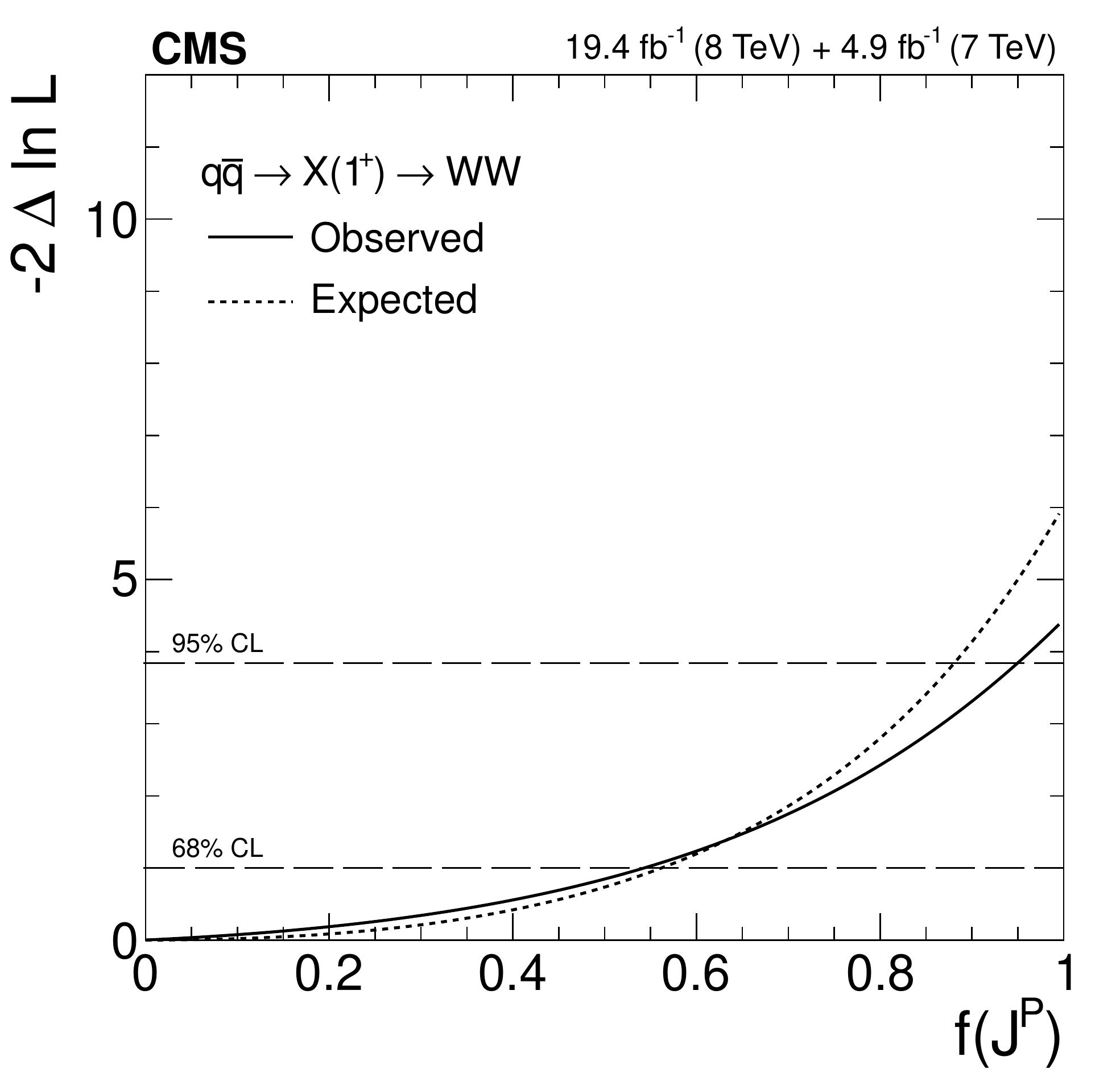}
         \caption{
           (\cmsLeft) Distributions of the test statistic $q=-2\ln(\mathcal{L}_{J^P}/\mathcal{L}_{0^+})$
       for the $J^P=1^+$ hypothesis of $\qqbar\to\X(1^+)\to \PW\PW$
      against the SM Higgs boson hypothesis ($0^+$).
      The expectation for the SM Higgs boson is represented by the yellow histogram on the right and the alternative $J^P$ hypothesis by the
    blue histogram on the left. The red arrow indicates the observed $q$ value.
      (\cmsRight) Observed value of $-2\Delta\ln\mathcal{L}$ as a function of $f(J^P)$ and the expectation in the SM
      for the $\qqbar\to\X(1^+)\to \PW\PW$ alternative $J^P$ model.
}
	    \label{fig:1jphww}

\end{figure}

\begin{figure}[htbp]
  \centering
  \includegraphics[width=0.45\textwidth]{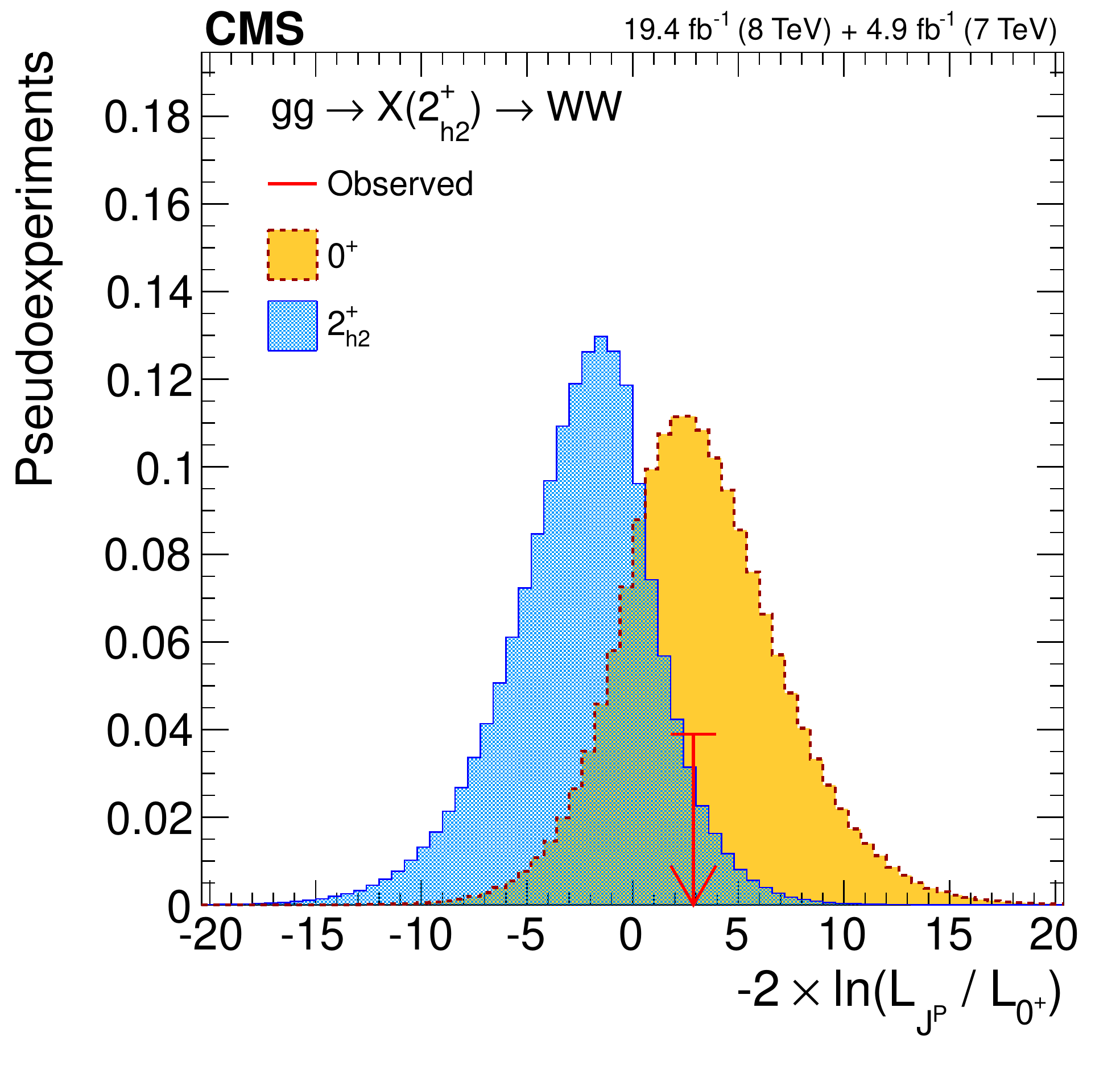}
  \includegraphics[width=0.45\textwidth]{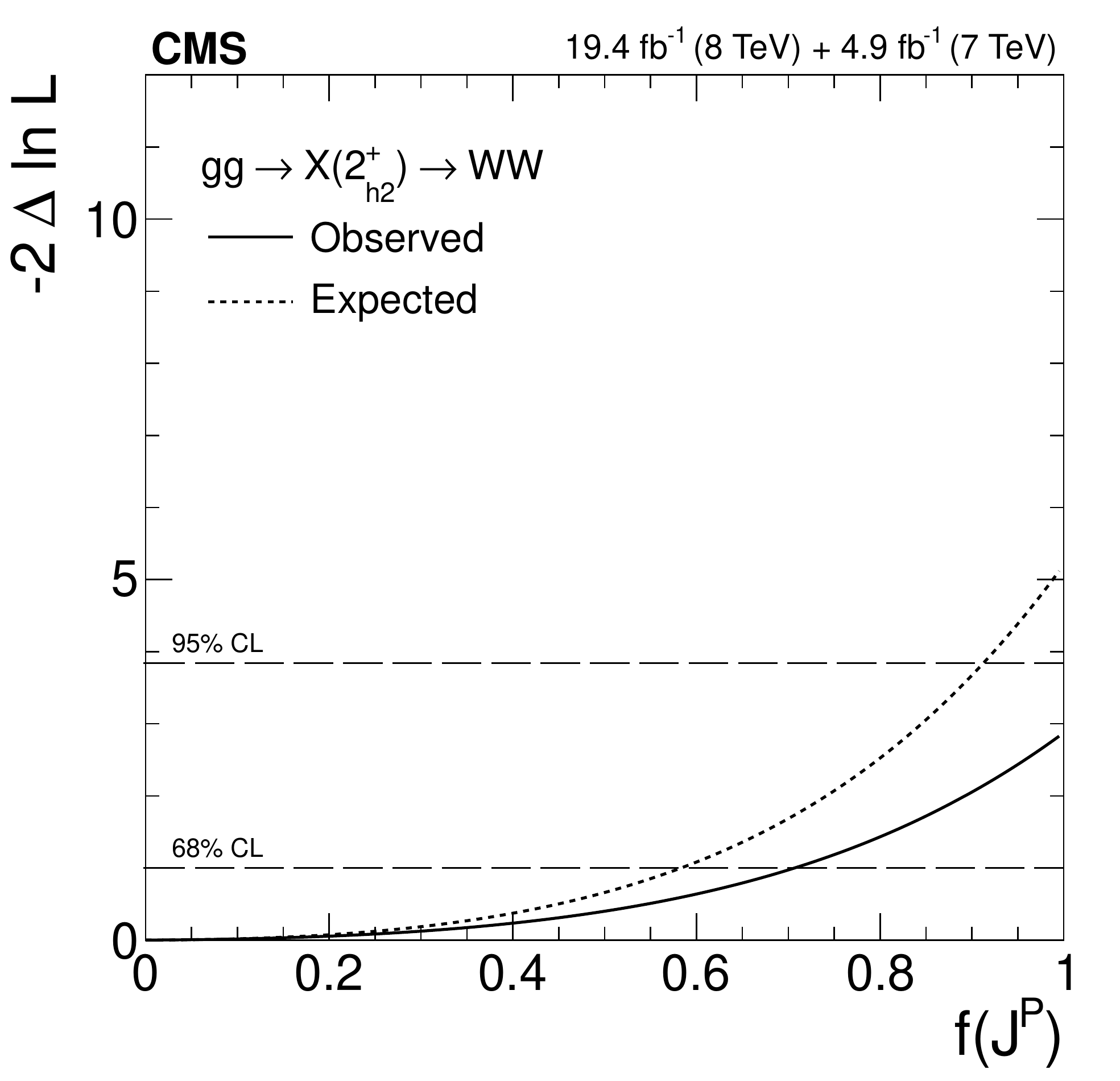}
                      \caption{
       (\cmsLeft) Distributions of the test statistic $q=-2\ln(\mathcal{L}_{J^P}/\mathcal{L}_{0^+})$
       for the $J^P=2^+_{h2}$ hypothesis of $\Pg\Pg\to\X(2^+_{h2})\to \PW\PW$
     against the SM Higgs boson hypothesis ($0^+$).
      The expectation for the SM Higgs boson is represented by the yellow histogram on the right and the alternative $J^P$ hypothesis by the
    blue histogram on the left. The red arrow indicates the observed $q$ value.
      (\cmsRight) Observed value of $-2\Delta\ln\mathcal{L}$ as a function of $f(J^P)$ and the expectation in the SM
      for the $\Pg\Pg\to\X(2^+_{h2})\to \PW\PW$ alternative $J^P$ model.
}
   \label{fig:2jphww}

\end{figure}

Similar to the $\X\to \Z\Z\to 4\ell$ study above, ten spin-two hypotheses, listed in Table~\ref{table-scenarios},
and three spin-one hypotheses, including a mixed case with $f_{b2}^{\PW\PW} = 0.5$, are tested using the
$\X\to \PW\PW\to \ell\nu\ell\nu$ decay.
Examples of distributions of the test statistic, $q=-2\ln(\mathcal{L}_{J^P}/\mathcal{L}_{0^+})$, for the SM Higgs
boson and alternative spin-one and spin-two models are shown in Figs.~\ref{fig:1jphww} and~\ref{fig:2jphww} (\cmsLeft).
Examples of the likelihood scans, $-2\Delta\ln \mathcal{L}$, as a function of $f(J^P)$ are also shown
in Figs.~\ref{fig:1jphww} and~\ref{fig:2jphww} (\cmsRight).

The expected and observed separation of the test statistic for the various models are summarized
in Table~\ref{tab:summaryhtspin1hww} for spin-one and
in Table~\ref{tab:summaryhtspin2} for spin-two.
The expected separation between the SM Higgs boson and
each alternative spin-one or spin-two hypothesis is larger than one standard deviation in most cases,
reaching three standard deviation for several models.

\begin{figure*}[htbp]
  \centering
   \includegraphics[width=0.45\textwidth]{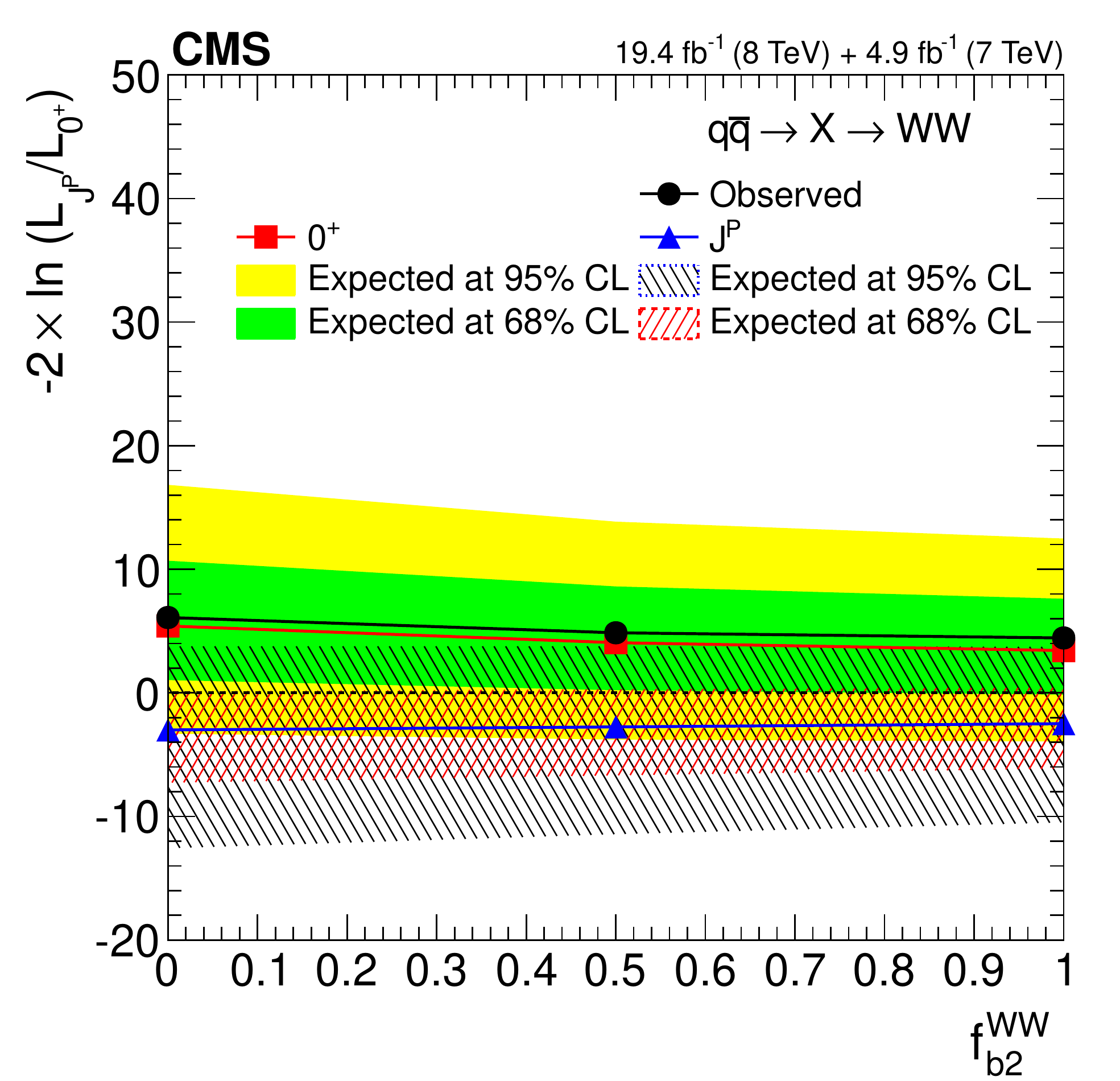}
   \includegraphics[width=0.45\textwidth]{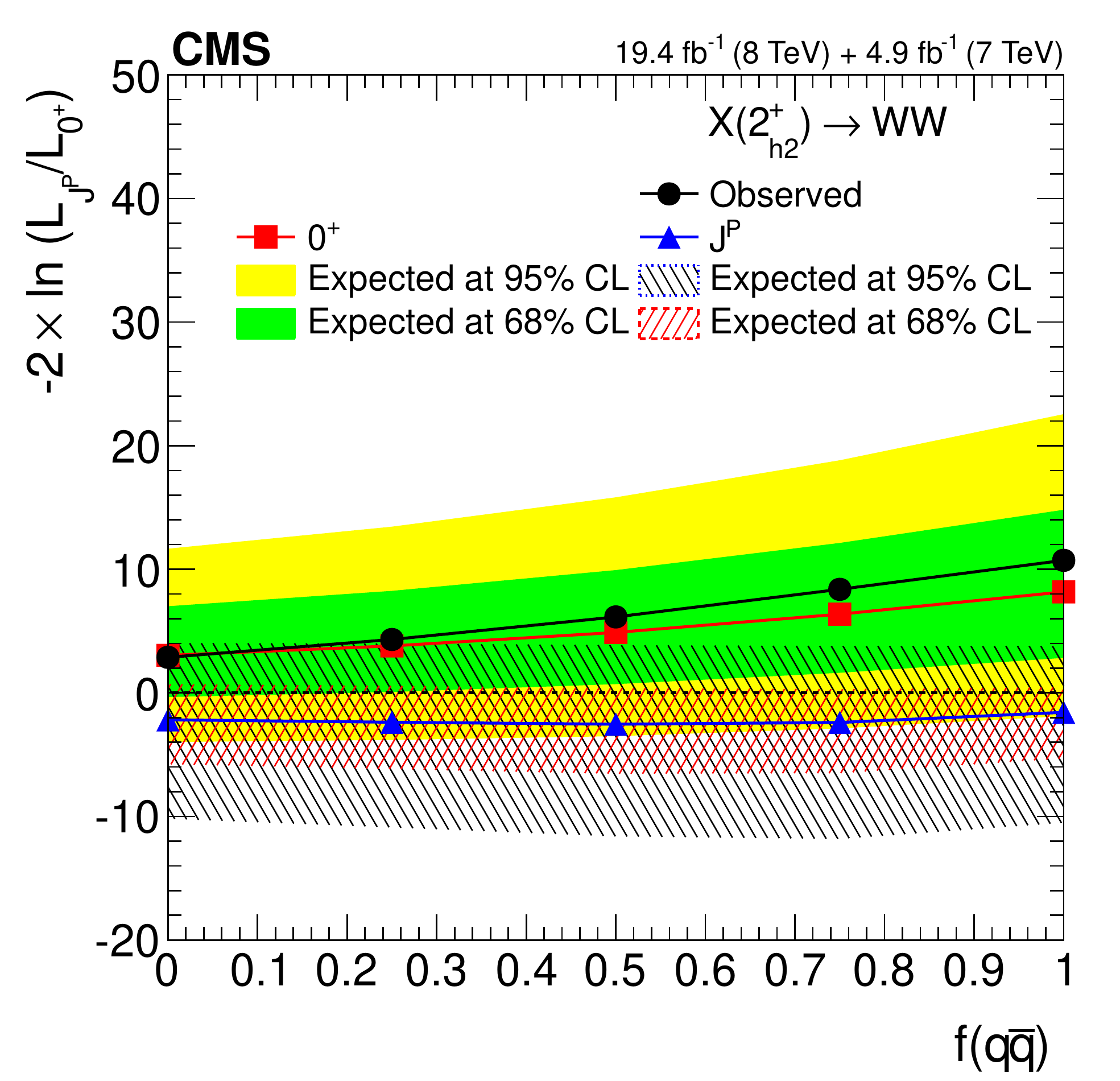}
   \caption{
      (left) Distributions of the test statistic $q=-2\ln(\mathcal{L}_{J^P}/\mathcal{L}_{0^+})$
      as a function of $f_{b2}^{\PW\PW}$
      for the hypothesis of the spin-one $J^{P}$ models against the SM Higgs boson hypothesis
      in the $\qqbar\to \X\to\PW\PW$ analysis.
      (right) Distribution of the test statistic $q=-2\ln(\mathcal{L}_{J^P}/\mathcal{L}_{0^+})$
     as a function of $f(\qqbar)$
      for the $2^+_{h2}$ hypothesis against the SM Higgs boson hypothesis
      in the $\PH\to\PW\PW$ analysis.
      The median expectation for the SM Higgs boson is represented
      by the red squares with the green (68\% \CL) and yellow (95\% \CL) solid color regions and
       for the alternative $J^P$ hypotheses by the blue triangles
       with the red (68\% \CL) and blue (95\% \CL) hatched regions.
     The observed values are indicated by the black dots.
     }
   \label{fig:2jpfqqhww}
\end{figure*}

The  spin-one $J^{P}$ hypothesis is tested against the SM Higgs boson
for several values of $f_{b2}^{\PW\PW}$. The results are shown in Fig.~\ref{fig:2jpfqqhww} (left) and
summarized in Table~\ref{tab:summaryhtspin1hww}. As in the $\X\to \Z\Z\to 4\ell$ study,
the $1^+$ model is found to be the least restricted.

\begin{table*}[htbp]
\centering
\topcaption{
List of spin-one models tested in the $\X\to \PW\PW$ analysis.
The expected separation is quoted for two scenarios, for the signal production cross section
obtained from the fit to data for each hypothesis, and using the SM expectation ($\mu=1$).
The observed separation shows the consistency of the observation with the SM Higgs boson model
or the alternative $J^{P}$ model, from which a \CLs value is derived. The constraints on the non-interfering $J^{P}$ fraction are quoted in the last two columns.
\label{tab:summaryhtspin1hww}
}
\renewcommand{\arraystretch}{1.25}
\begin{scotch}{@{}lccccccc}
$f_{b2}^{\PW\PW} (J^P)$                 & $J^P$        	&        Expected  &                       &                    &               &        $f(J^P)$  95\% \CL & $f(J^P)$ \\
Model                          		  & Prod. 		&        ($\mu$=1) &  Obs. $0^+$  & Obs. $J^P$ & \CLs &        Obs. (Exp.)                 & Best Fit \\
\hline
$0.0 (1^{-})$ & $\qqbar$ & 2.2$\sigma$ (3.3$\sigma$) & $-$0.1$\sigma$ & $+$2.5$\sigma$ & 1.5\% &$<$0.88 (0.81) & 0.00$^{+0.55}_{-0.00}$ \\
0.5  & $\qqbar$ 	 & 2.0$\sigma$ (3.0$\sigma$) & $-$0.2$\sigma$ & $+$2.2$\sigma$ & 3.1\% &$<$0.93 (0.86) & 0.00$^{+0.57}_{-0.00}$ \\
$1.0 (1^{+})$  & $\qqbar$& 1.8$\sigma$ (2.7$\sigma$) & $-$0.3$\sigma$ & $+$2.1$\sigma$ & 4.1\% &$<$0.95 (0.88) & 0.00$^{+0.54}_{-0.00}$ \\
\end{scotch}

  \end{table*}

\begin{figure*}[htbp]
  \centering
   \includegraphics[width=0.9\textwidth]{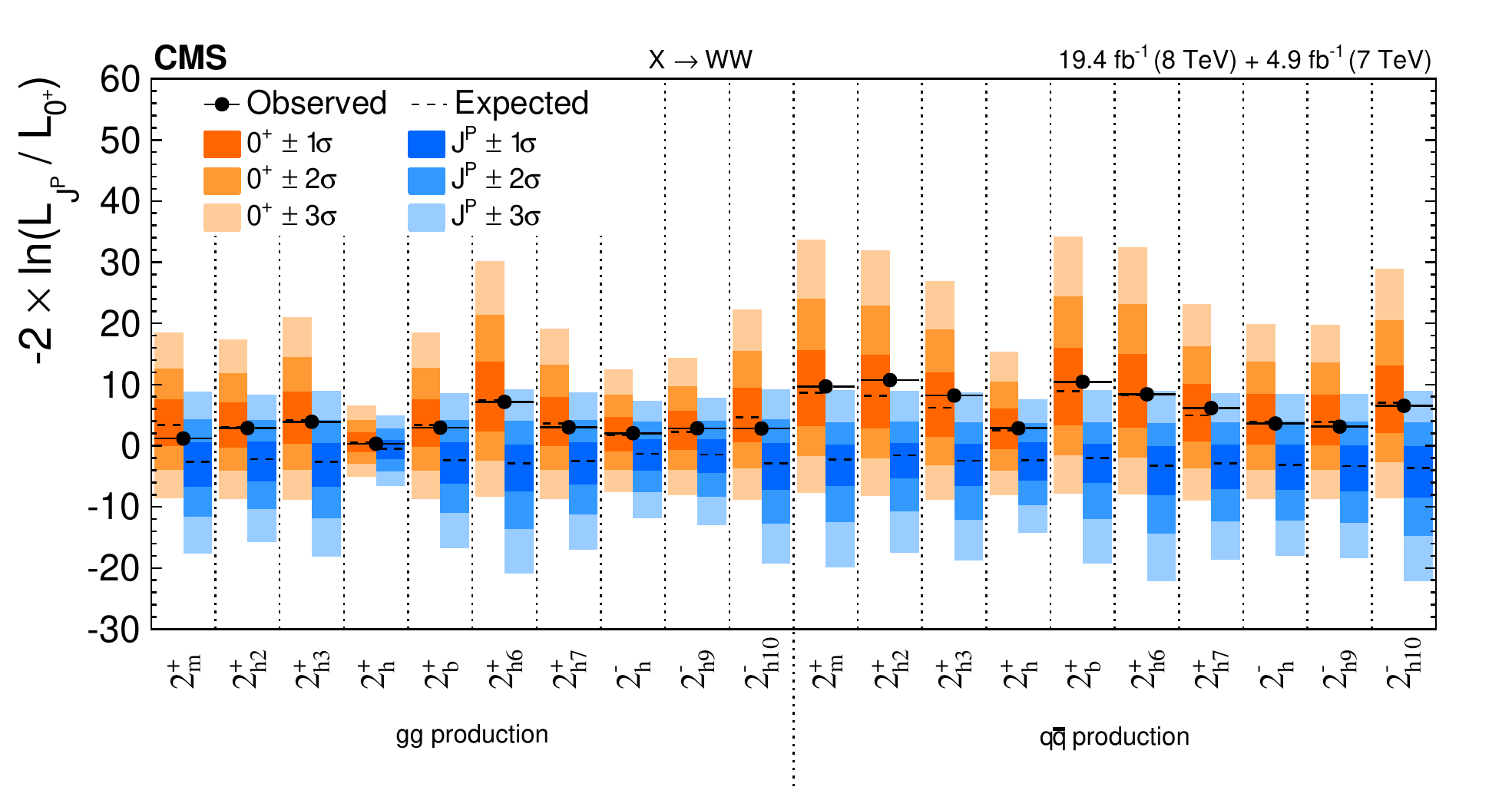}
         \caption{
                Distribution of the test statistic $q=-2\ln(\mathcal{L}_{J^P}/\mathcal{L}_{0^+})$
       for the spin-two $J^{P}$ models tested against the SM Higgs boson hypothesis
      in the $\X\to\PW\PW$ analyses.
      The expected median and the 68.3\%, 95.4\%, and 99.7\% \CL regions for the SM Higgs boson (orange, the left for each model)
      and for the alternative $J^P$ hypotheses (blue, right) are shown.
     The observed $q$ values are indicated by the black dots.
}
   \label{fig:2jpsummaryplot}

\end{figure*}

\begin{figure*}[htbp]
  \centering
    \includegraphics[width=1.0\textwidth]{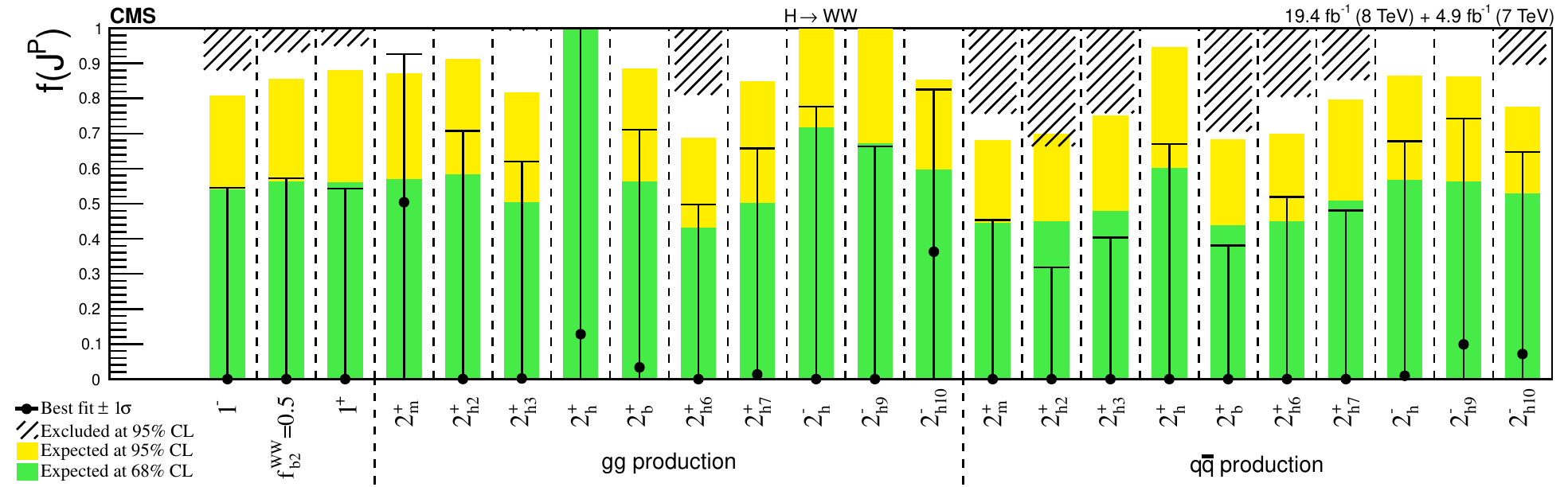}
    \caption{
    Summary of the $f(J^P)$ constraints for the spin-one and spin-two models from Tables~\ref{tab:summaryhtspin1hww}
    and \ref{tab:summaryhtspin2} in the $\X\to\PW\PW$ analyses.
The expected 68\% and 95\% \CL regions are shown as the green and yellow bands.
The observed constraints at 68\% and 95\% \CL are shown as the points with error bars
and the excluded hatched regions.
      \label{fig:non_interf_summary_ww}}

\end{figure*}

The summary of the spin-two results is presented in Fig.~\ref{fig:2jpsummaryplot} and Table~\ref{tab:summaryhtspin2}.
In the case of the spin-two studies, the results for the different scenarios are estimated assuming
different production fractions from $f(\qqbar)=0$, representing the pure $\Pg\Pg\to \X$ process, to $f(\qqbar)=1$, representing the pure $\Pq\Paq\to \X$ process.
A scan of the $f(\qqbar)$ fraction is performed in each case,
with an example of the scan for the $2^+_{h2}$ model shown in Fig.~\ref{fig:2jpfqqhww} (\cmsRight).
The results with pure gluon fusion production $f(\qqbar)=0$ are found to be the least restricted in each case.
The observed non-interfering fraction measurements are summarized in Fig.~\ref{fig:non_interf_summary_ww}.

In all cases the data favor the SM hypothesis over the alternative spin-one or spin-two hypotheses.

\begin{table*}[htbp]
\centering
\topcaption{
List of spin-two models tested in the $\X\to \PW\PW$ analysis.
The expected separation is quoted for two scenarios, for the signal production cross section
obtained from the fit to data for each hypothesis, and using the SM expectation ($\mu=1$).
The observed separation shows the consistency of the observation with the SM Higgs boson
or an alternative $J^{P}$ model, from which the \CLs value is derived. The constraints on the non-interfering $J^{P}$ fraction are quoted in the last two columns.
Results from Ref.~\cite{Chatrchyan:2013iaa} are explicitly noted.
\label{tab:summaryhtspin2}
}
\renewcommand{\arraystretch}{1.25}
\begin{scotch}{@{}lccccccc}
$J^P$                           & $J^P$        &        Expected  &                       &                    &               &        $f(J^P)$  95\% \CL & $f(J^P)$ \\
Model                           & Mrod. &        ($\mu$=1) &  Obs. $0^+$  & Obs. $J^P$ & \CLs &        Obs. (Exp.)                 & Best Fit \\
\hline
$2^{+}_{m}$~\cite{Chatrchyan:2013iaa}   & $\Pg\Pg$	  & 1.8$\sigma$ (2.9$\sigma$) & $+$0.6$\sigma$ & $+$1.2$\sigma$ & 16\% &$<$1.00 (0.87) & 0.50$^{+0.42}_{-0.50}$ \\
$2^{+}_{h2}$ & $\Pg\Pg$  	  & 1.7$\sigma$ (2.6$\sigma$) &       0.0$\sigma$ & $+$1.6$\sigma$ & 10\% &$<$1.00 (0.91) & 0.00$^{+0.71}_{-0.00}$\\
$2^{+}_{h3}$ & $\Pg\Pg$  	  & 1.9$\sigma$ (2.8$\sigma$) & $+$0.1$\sigma$ & $+$1.9$\sigma$ & 5.2\% &$<$0.99 (0.82) & 0.00$^{+0.62}_{-0.00}$ \\
$2^{+}_{h}$ & $\Pg\Pg$  	  & 0.7$\sigma$ (1.3$\sigma$) & $+$0.1$\sigma$ & $+$0.6$\sigma$ & 52\% &$<$1.00 (1.00) & 0.13$^{+0.87}_{-0.13}$ \\
$2^{+}_{b}$ & $\Pg\Pg$ 	  & 1.8$\sigma$ (2.7$\sigma$) & $+$0.1$\sigma$ & $+$1.7$\sigma$ & 8.6\% &$<$1.00 (0.89) & 0.03$^{+0.68}_{-0.03}$ \\
$2^{+}_{h6}$ & $\Pg\Pg$  	  & 2.5$\sigma$ (3.4$\sigma$) & $+$0.0$\sigma$ & $+$2.6$\sigma$ & 0.88\% &$<$0.81 (0.69) & 0.00$^{+0.50}_{-0.00}$  \\
$2^{+}_{h7}$ & $\Pg\Pg$  	  & 1.8$\sigma$ (2.5$\sigma$) & $+$0.2$\sigma$ & $+$1.7$\sigma$ & 8.1\% &$<$1.00 (0.85) & 0.01$^{+0.64}_{-0.01}$\\
$2^{-}_{h}$ & $\Pg\Pg$ 		  & 1.2$\sigma$ (2.3$\sigma$) & $-$0.1$\sigma$ & $+$1.4$\sigma$ & 19\% &$<$1.00 (1.00) & 0.00$^{+0.78}_{-0.00}$\\
$2^{-}_{h9}$ & $\Pg\Pg$  	  & 1.4$\sigma$ (2.5$\sigma$) & $-$0.2$\sigma$ & $+$1.6$\sigma$ & 12\% &$<$1.00 (1.00) & 0.00$^{+0.66}_{-0.00}$\\
$2^{-}_{h10}$ & $\Pg\Pg$	  & 2.0$\sigma$ (3.3$\sigma$) & $+$0.4$\sigma$ & $+$1.6$\sigma$ & 7.8\% &$<$1.00 (0.85) & 0.36$^{+0.46}_{-0.36}$ \\
\hline
$2^{+}_{m}$~\cite{Chatrchyan:2013iaa}  & $\Pq\Paq$	     & 2.7$\sigma$ (3.9$\sigma$) & $-$0.2$\sigma$ & $+$3.1$\sigma$ & 0.25\% &$<$0.76 (0.68) & 0.00$^{+0.45}_{-0.00}$ \\
$2^{+}_{h2}$ & $\Pq\Paq$	     & 2.6$\sigma$ (3.7$\sigma$) & $-$0.4$\sigma$ & $+$3.3$\sigma$ & 0.16\% &$<$0.66 (0.70) & 0.00$^{+0.32}_{-0.00}$ \\
$2^{+}_{h3}$ & $\Pq\Paq$	     & 2.3$\sigma$ (3.3$\sigma$) & $-$0.4$\sigma$ & $+$2.9$\sigma$ & 0.56\% &$<$0.76 (0.75) & 0.00$^{+0.40}_{-0.00}$ \\
$2^{+}_{h}$ & $\Pq\Paq$ 	     & 1.6$\sigma$ (2.3$\sigma$) & $-$0.1$\sigma$ & $+$1.7$\sigma$ & 8.8\% &$<$1.00 (0.95) & 0.00$^{+0.67}_{-0.00}$ \\
$2^{+}_{b}$ & $\Pq\Paq$ 	     & 2.8$\sigma$ (3.8$\sigma$) & $-$0.2$\sigma$ & $+$3.2$\sigma$ & 0.18\% &$<$0.71 (0.68) & 0.00$^{+0.38}_{-0.00}$ \\
$2^{+}_{h6}$ & $\Pq\Paq$	     & 2.8$\sigma$ (3.7$\sigma$) & 0.0$\sigma$ & $+$2.9$\sigma$ & 0.41\% &$<$0.80 (0.70) & 0.00$^{+0.52}_{-0.00}$ \\
$2^{+}_{h7}$ & $\Pq\Paq$	     & 2.2$\sigma$ (3.1$\sigma$) & $-$0.2$\sigma$ & $+$2.5$\sigma$ & 1.6\% &$<$0.85 (0.80) & 0.00$^{+0.48}_{-0.00}$ \\
$2^{-}_{h}$ & $\Pq\Paq$ 	     & 2.0$\sigma$ (2.9$\sigma$) & $+$0.1$\sigma$ & $+$1.9$\sigma$ & 5.1\% &$<$1.00 (0.87) & 0.01$^{+0.67}_{-0.01}$ \\
$2^{-}_{h9}$ & $\Pq\Paq$	     & 2.0$\sigma$ (2.9$\sigma$) & $+$0.2$\sigma$ & $+$1.8$\sigma$ & 6.2\% &$<$1.00 (0.86) & 0.10$^{+0.64}_{-0.10}$ \\
$2^{-}_{h10}$ & $\Pq\Paq$	     & 2.6$\sigma$ (3.6$\sigma$) & $+$0.1$\sigma$ & $+$2.5$\sigma$ & 1.1\% &$<$0.90 (0.78) & 0.07$^{+0.58}_{-0.07}$ \\
  \end{scotch}

  \end{table*}

\subsection{Combined exotic-spin results with the \texorpdfstring{$\PH\to\Z\Z$}{H to ZZ} and \texorpdfstring{$\PW\PW$}{WW} channels} \label{sec:Combination_Exotic}

The results of testing the spin-one and spin-two hypotheses obtained
by considering the $\X\to\Z\Z\to4\ell$ and $\X\to\PW\PW\to\ell\nu\ell\nu$ decay channels together are presented in this section.
The assumption made is that the same tensor structure for the interactions appears in both
$\X\Z\Z$ and $\X\PW\PW$ couplings, as outlined for spin-two models in Table~\ref{table-scenarios}.

Since only isolated tensor structure terms, and not the interference between them, are tested, the relationship between the absolute strengths of those couplings is
not important and is not used in the analysis. Therefore the combined spin-one exclusion of pure $1^-$ and $1^+$ states is tested, and for spin-two the ten hypotheses listed in
Table~\ref{table-scenarios} are tested.
The combination of the $f(J^P)$ results is not considered here because the relative strength between the
two channels is left unconstrained and the fractions remain independent measurements.

The $\Pq\Paq$ production mechanism is tested for spin-one
and spin-two models, and gluon fusion is tested for spin-two models.  The combination of an arbitrary admixture of the $\Pq\Paq$ and gluon production mechanisms is also tested.
These results are based on the channels presented in
Sections~\ref{sec:Resultsexotichzz} and ~\ref{sec:Resultsexotichww}. For several of the models
some production mechanisms have been tested already~\cite{Chatrchyan:2012jja,
Chatrchyan:2013mxa, Chatrchyan:2013iaa}.

\begin{figure}[htbp]
  \centering
    \includegraphics[width=0.45\textwidth]{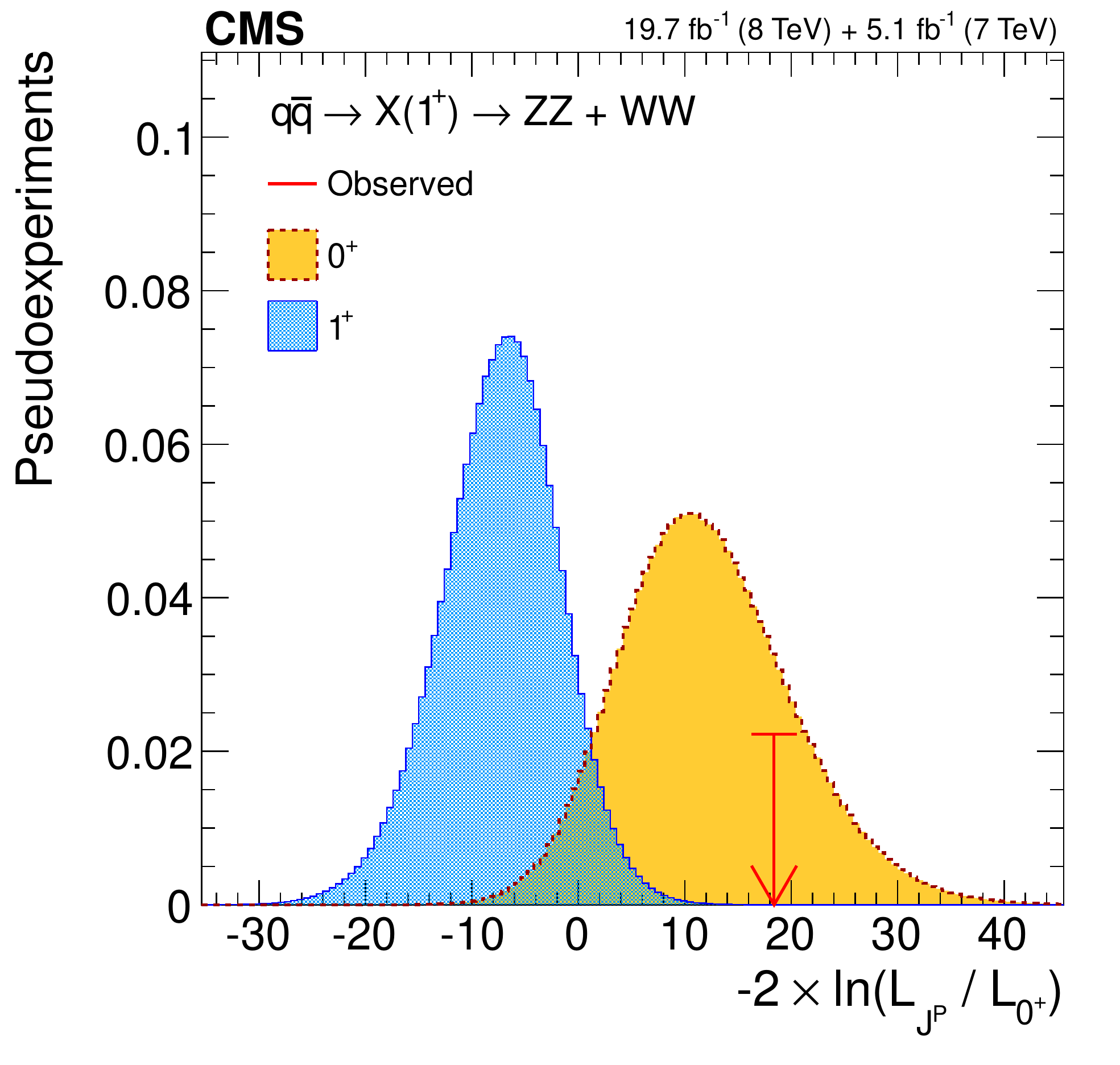}
    \includegraphics[width=0.45\textwidth]{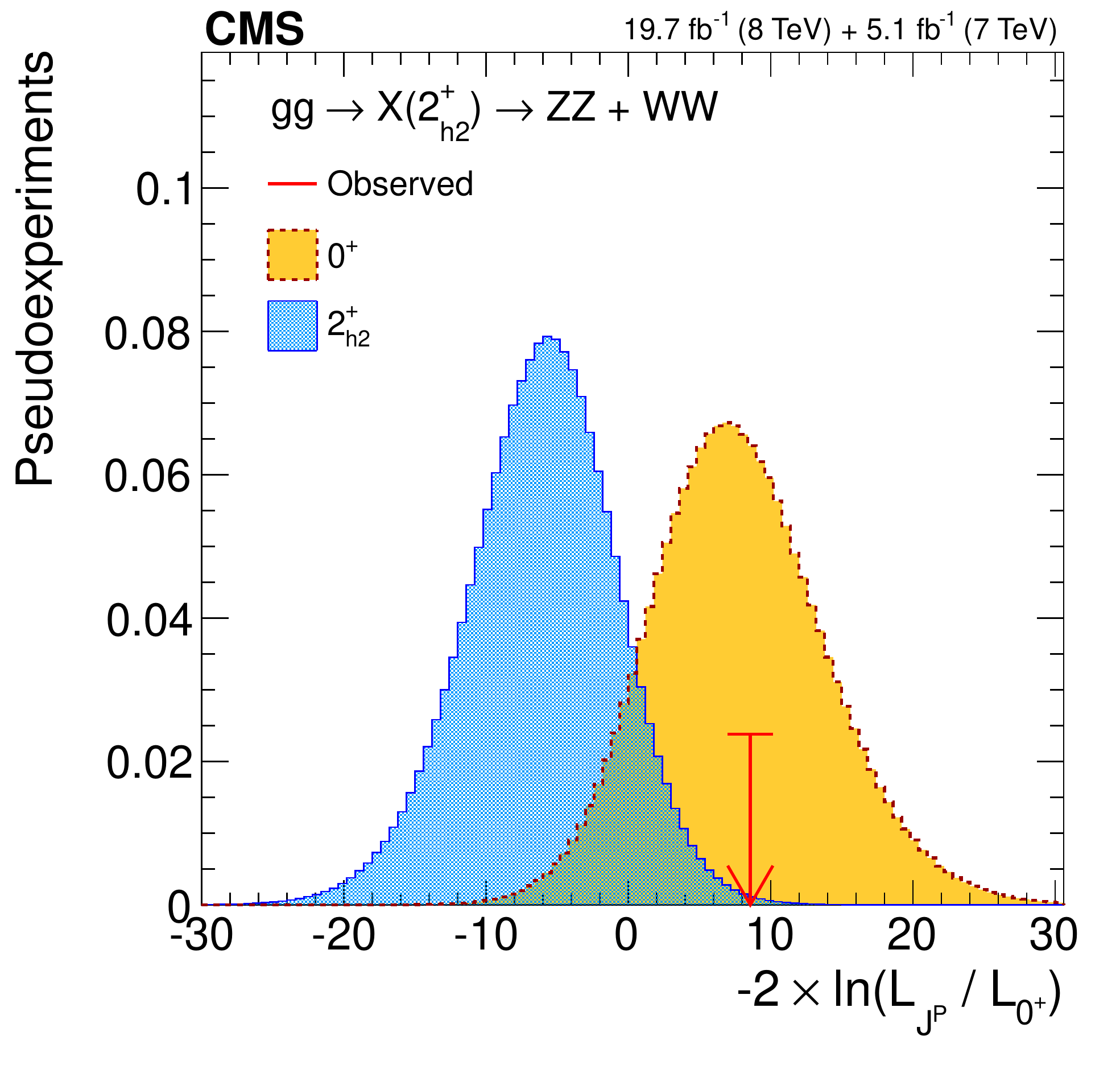}
    \caption{
    Distributions of the test statistic $q=-2\ln(\mathcal{L}_{J^P}/\mathcal{L}_{0^+})$ in the combination of the
    $\X\to\Z\Z$ and $\PW\PW$ channels
    for the hypotheses of $\qqbar\to\X(1^+)$ (\cmsLeft) and $\Pg\Pg\to\X(2^+_{h2})$ (\cmsRight)
      tested against the SM Higgs boson hypothesis.
       The expectation for the SM Higgs boson is represented by the
      yellow histogram on the right of each plot and the alternative $J^P$ hypothesis by the
      blue histogram on the left. The red arrow indicates the observed $q$ value.
      \label{fig:sepComb_1}
      }

\end{figure}

\begin{figure}[htbp]
  \centering
   \includegraphics[width=0.45\textwidth]{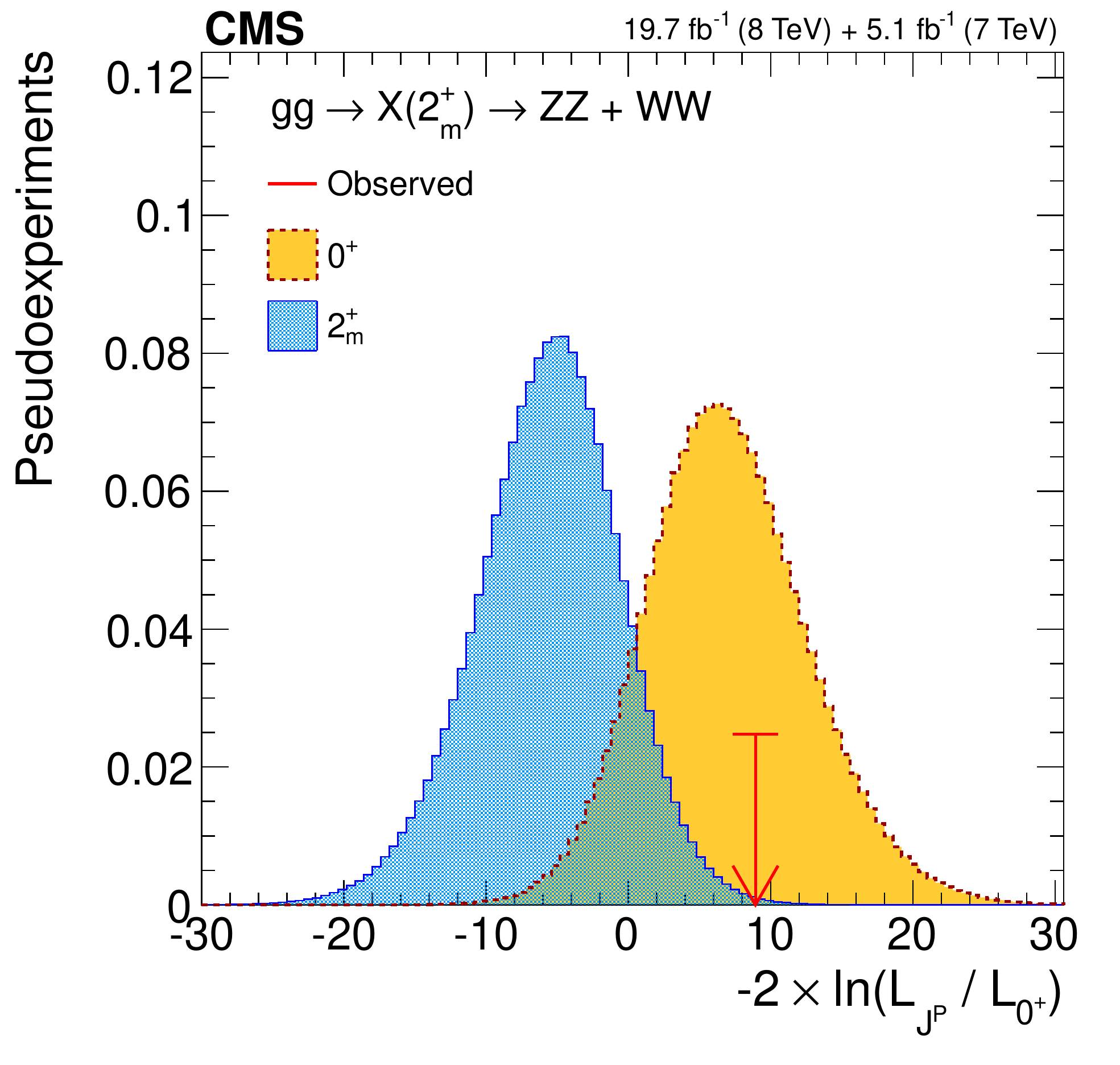}
    \includegraphics[width=0.45\textwidth]{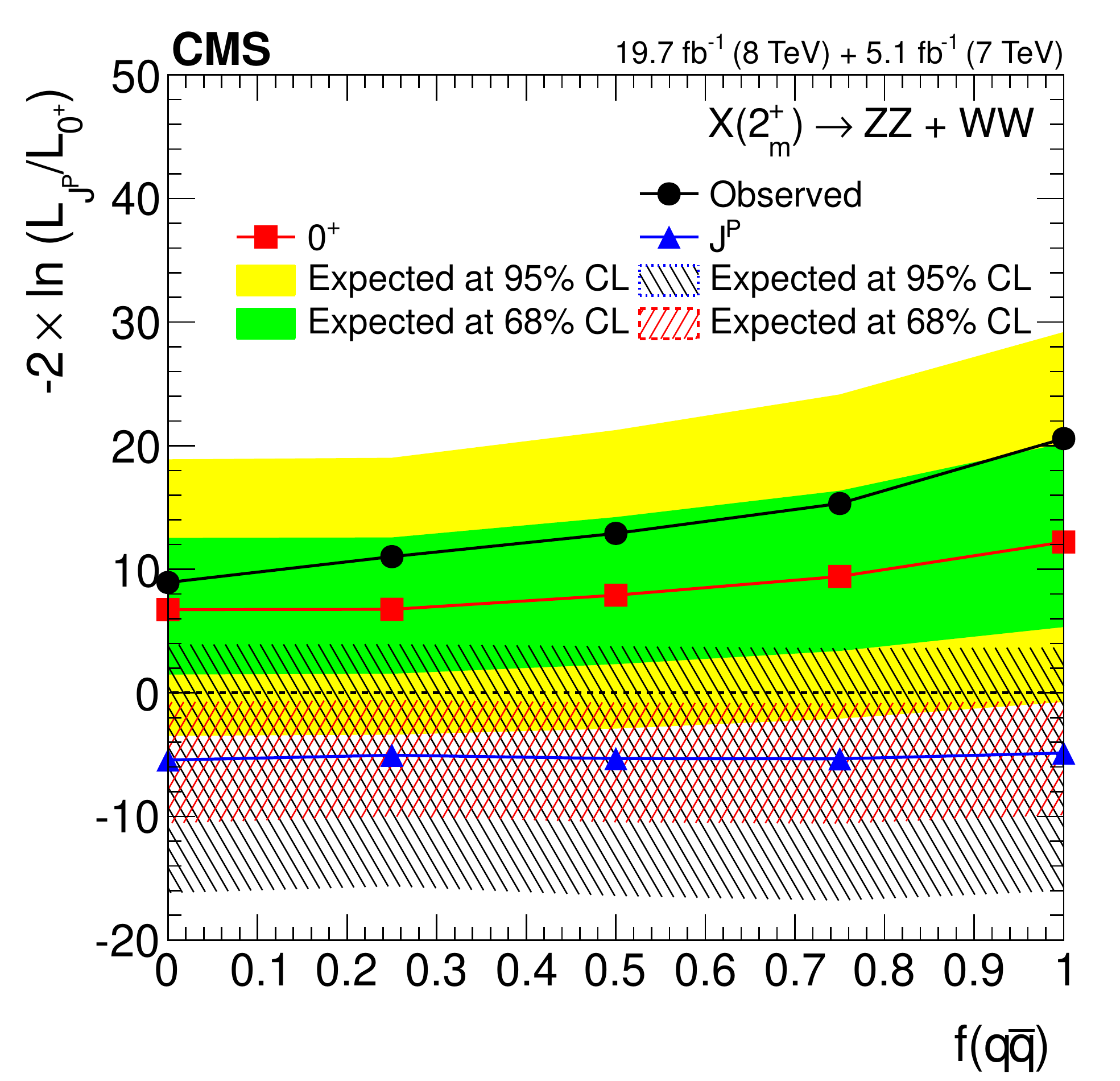}
     \caption{
     (\cmsLeft) Distributions of the test statistic $q=-2\ln(\mathcal{L}_{J^P}/\mathcal{L}_{0^+})$ in the combination of the
    $\X\to\Z\Z$ and $\PW\PW$ channels for the hypothesis of $\Pg\Pg\to\X(2^+_{m})$
      tested against the SM Higgs boson hypothesis.
       The expectation for the SM Higgs boson is represented by the
      yellow histogram on the right and the alternative $J^P$ hypothesis by the
      blue histogram on the left. The red arrow indicates the observed $q$ value.
      (\cmsRight) Distribution of $q$ as a function of $f(\qqbar)$
      for the $2^+_{m}$ hypothesis against the SM Higgs boson hypothesis
      in the $\X\to\Z\Z$ and $\PW\PW$ channels.
      The median expectation for the SM Higgs boson is represented  with the solid green (68\% \CL)
      and yellow (95\% \CL) regions. The alternative $2^+_{m}$ hypotheses are represented  by the blue triangles
       with the red (68\% \CL) and blue (95\% \CL) hatched regions.
     The observed values are indicated by the black dots.
       \label{fig:sepComb_2}}

\end{figure}

\begin{figure*}[htbp]
  \centering
    \includegraphics[width=0.9\textwidth]{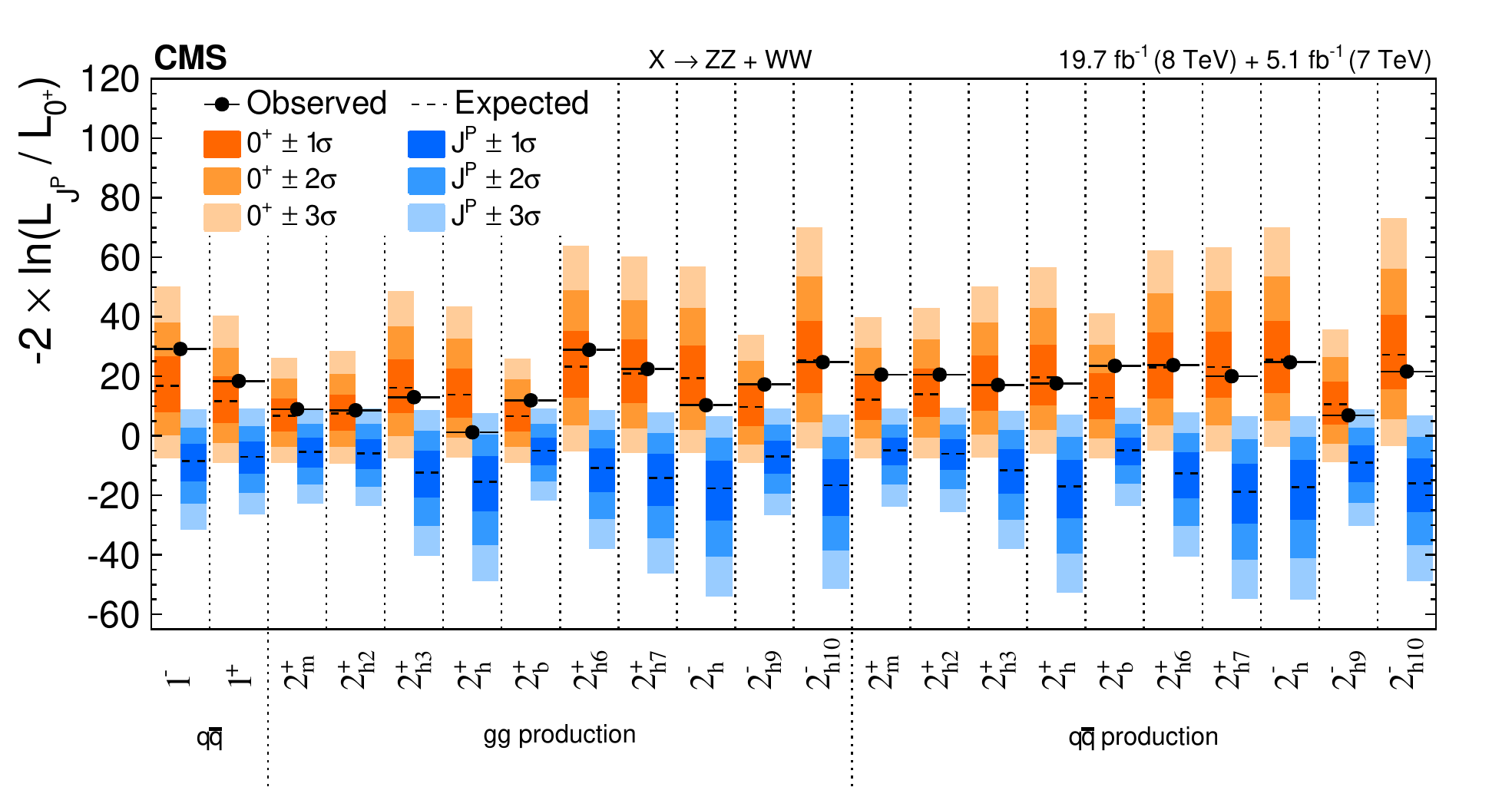}
    \caption{
       Distributions of the test statistic $q=-2\ln(\mathcal{L}_{J^P}/\mathcal{L}_{0^+})$
       for the spin-one and spin-two $J^{P}$ models tested against the SM Higgs boson hypothesis
      in the combined $\X\to\Z\Z$ and $\PW\PW$ analyses.
      The expected median and the 68.3\%, 95.4\%, and 99.7\% \CL regions for the SM Higgs boson (orange, the left for each model)
      and for the alternative $J^P$ hypotheses (blue, right) are shown.
     The observed $q$ values are indicated by the black dots.
      \label{fig:jp_summaryComb}
      }

\end{figure*}

In the spin-one studies, an example of the distribution of the test statistic and observed value
in the case of the SM Higgs boson along with the spin-one $1^+$ hypothesis is shown in Fig.~\ref{fig:sepComb_1}.
The expected and observed separations from the test
statistic distributions are summarized in Table~\ref{tab:jpmodelsComb}.
In the case of the spin-two studies,
the distributions of the test statistic and observed value
in the case of the SM Higgs boson along with the
spin-two hypotheses $\Pg\Pg \to \X(2_{h2}^{+})$ and  $\Pg\Pg \to \X(2_{m}^{+})$
are shown in Figs.~\ref{fig:sepComb_1} (\cmsRight) and~\ref{fig:sepComb_2} (\cmsLeft).
All the spin-one and spin-two models tested in the combination are summarized in Fig.~\ref{fig:jp_summaryComb}.

The expected separations between the test statistic distributions for all the models considered
are summarized in Table~\ref{tab:jpmodelsComb}.
In all cases, the expected separation between the alternative signal hypotheses is
quoted for the case where the expected SM Higgs boson
signal strength and the alternative signal cross sections are obtained in the fit of the data. The signal strengths in
the $\X\to\Z\Z$ and $\X\to\PW\PW$ channels are fit independently.
The expected separation is also quoted for the case where the events
are generated with the SM expectation for the signal cross section ($\mu$=1).

These tests are performed for several choices of the ratio of the two production rates $f(\qqbar)$.
The analysis, which uses information from the $\X \to \Z\Z \to 4\ell$
decay channel, is performed in a production-independent way, unless $f(\qqbar)=0$ or $1$.
Part of the analysis, which is based on the $\X \to \PW\PW \to \ell\nu\ell\nu$ decay channel,
tests several choices of the $f(\qqbar)$ ratio explicitly.
An example of such a test is shown in Fig.~\ref{fig:sepComb_2} (\cmsRight).
For the combined $\X\to\Z\Z$ and $\PW\PW$ analysis, as in the case of the $\X\to\PW\PW$ analysis,
the results with gluon fusion ($f(\qqbar) =0$) and with $\qqbar$ production ($f(\qqbar) =1$) exhibit the
largest and the smallest observed separation when compared to any other value in the scan of $0<f(\qqbar)<1$.
The data disfavor all the spin-one and spin-two hypotheses tested in favor of
the SM hypothesis $J^P=0^+$ with $1-\CLs$ values  larger than 98\% \CL (Table~\ref{tab:jpmodelsComb}).

\begin{table*}[htbp]
\centering
\topcaption{
List of spin-one and spin-two models tested in the combination of the $\X\to\Z\Z$ and $\X\to\PW\PW$ channels.
The combined expected separation is quoted for two scenarios, for the signal production cross section
obtained from the fit to data for each hypothesis and using the SM expectation ($\mu=1$).
For comparison, the former expectations are also quoted for the individual channels as in Tables~\ref{tab:jpmodels_1}-\ref{tab:summaryhtspin2}.
The observed separation shows the consistency of the observation with the SM Higgs boson model
or an alternative $J^{P}$ model, from which the \CLs value is derived.
\label{tab:jpmodelsComb}}
\begin{scotch}{lccccccc}
$J^P$                           & $J^P$        &        Expected    &       Expected  &       Expected  &                       &     &       \\
Model                           & Prod. &        $\X\to\Z\Z$ & $\X\to\PW\PW$ &($\mu$=1) &  Obs. $0^+$  & Obs. $J^P$ & \CLs\\
\hline
 $1^-$             & $\Pq\Paq$ & 2.9$\sigma$ & 2.2$\sigma$ & 3.6$\sigma$ (4.6$\sigma$) & $-$1.2$\sigma$ & $+$4.9$\sigma$ & $<$0.001\%\\
$1^+$             & $\Pq\Paq$ &  2.4$\sigma$ & 1.8$\sigma$ &  3.0$\sigma$ (3.8$\sigma$) & $-$0.8$\sigma$ & $+$4.3$\sigma$ & 0.004\%\\
\hline
$2^+_{m}$  & $\Pg\Pg $       &  1.9$\sigma$ & 1.8$\sigma$ & 2.4$\sigma$ (3.4$\sigma$) & $-$0.4$\sigma$ & $+$2.9$\sigma$ & 0.53\%\\
$2^+_{h2}$ & $\Pg\Pg $       &  2.0$\sigma$ & 1.7$\sigma$ &  2.5$\sigma$ (3.3$\sigma$) & $-$0.2$\sigma$ & $+$2.8$\sigma$ & 0.52\% \\
$2^+_{h3}$ & $\Pg\Pg $       &  3.2$\sigma$ & 1.6$\sigma$ & 3.7$\sigma$ (4.3$\sigma$) & $+$0.4$\sigma$ & $+$3.5$\sigma$ & 0.031\% \\
$2^+_{h}$  & $\Pg\Pg $       &  3.8$\sigma$ & 0.7$\sigma$ & 3.8$\sigma$ (4.2$\sigma$) & $+$1.7$\sigma$ & $+$2.1$\sigma$ & 1.9\% \\
$2^+_{b}$  & $\Pg\Pg $       &  1.6$\sigma$ & 1.8$\sigma$ & 2.4$\sigma$ (3.2$\sigma$) & $-$0.9$\sigma$ & $+$3.4$\sigma$ & 0.16\%  \\
$2^+_{h6}$ & $\Pg\Pg $       &  3.4$\sigma$ & 2.5$\sigma$ & 4.2$\sigma$ (4.9$\sigma$) & $-$0.5$\sigma$ & $>$5$\sigma$ & $<$0.001\%  \\
$2^+_{h7}$ & $\Pg\Pg $       &  3.8$\sigma$ & 1.8$\sigma$ & 4.2$\sigma$ (5.0$\sigma$) & $-$0.1$\sigma$ & $+$4.7$\sigma$ & $<$0.001\%  \\
$2^-_{h}$  & $\Pg\Pg $       &  4.2$\sigma$ & 1.2$\sigma$ & 4.3$\sigma$ (5.0$\sigma$) &$+$1.0$\sigma$ & $+$3.4$\sigma$ & 0.039\%  \\
$2^-_{h9}$ & $\Pg\Pg $       &  2.5$\sigma$ & 1.4$\sigma$ & 2.8$\sigma$ (3.5$\sigma$) & $-$1.0$\sigma$ & $+$4.2$\sigma$ & 0.009\%  \\
$2^-_{h10}$& $\Pg\Pg $       &  4.2$\sigma$ & 2.0$\sigma$ & 4.6$\sigma$ (5.3$\sigma$) & $+$0.1$\sigma$ & $+$4.9$\sigma$ &  $<$0.001\% \\
\hline
$2^+_{m}$  & $\Pq\Paq$   &  1.7$\sigma$ & 2.7$\sigma$ & 3.1$\sigma$ (4.3$\sigma$) & $-$1.0$\sigma$ & $+$4.5$\sigma$ & 0.002\%  \\
$2^+_{h2}$ & $\Pq\Paq$ &  2.2$\sigma$ & 2.6$\sigma$ & 3.3$\sigma$ (4.3$\sigma$) & $-$0.8$\sigma$ & $+$4.4$\sigma$ & 0.002\%   \\
$2^+_{h3}$ & $\Pq\Paq$ &  3.1$\sigma$ & 2.6$\sigma$ & 3.8$\sigma$ (4.5$\sigma$) & 0.0$\sigma$ & $+$4.1$\sigma$ & 0.005\%  \\
$2^+_{h}$  & $\Pq\Paq$ &  4.0$\sigma$ & 1.6$\sigma$ & 4.3$\sigma$ (4.5$\sigma$) & $+$0.2$\sigma$ & $+$4.3$\sigma$ & 0.002\%  \\
$2^+_{b}$  & $\Pq\Paq$ &  1.7$\sigma$ & 2.8$\sigma$ &  3.1$\sigma$ (4.2$\sigma$) & $-$1.3$\sigma$ & $+$4.8$\sigma$ &  $<$0.001\% \\
$2^+_{h6}$ & $\Pq\Paq$ &  3.4$\sigma$ & 2.8$\sigma$ & 4.3$\sigma$ (5.0$\sigma$) & $-$0.1$\sigma$ & $+$4.8$\sigma$ & $<$0.001\% \\
$2^+_{h7}$ & $\Pq\Paq$ &  4.1$\sigma$ & 2.2$\sigma$ & 4.6$\sigma$ (5.0$\sigma$) & $+$0.3$\sigma$ & $+$4.5$\sigma$ & $<$0.001\%  \\
$2^-_{h}$  & $\Pq\Paq$ &  4.3$\sigma$ & 2.0$\sigma$ & 4.7$\sigma$ (5.2$\sigma$) & $+$0.1$\sigma$ & $+$5.0$\sigma$ & $<$0.001\%  \\
$2^-_{h9}$ & $\Pq\Paq$ &  2.4$\sigma$ & 2.0$\sigma$ & 3.1$\sigma$ (3.8$\sigma$) & $+$0.5$\sigma$ & $+$2.7$\sigma$ & 0.55\%   \\
$2^-_{h10}$ &  $\Pq\Paq$ &  4.0$\sigma$ & 2.6$\sigma$ & 4.7$\sigma$ (5.3$\sigma$) & $+$0.5$\sigma$ & $+$4.6$\sigma$ & $<$0.001\% \\
\end{scotch}
\end{table*}

\subsection{Combined exotic-spin results with the \texorpdfstring{$\PH\to\Z\Z, \PW\PW,$ and  $\gamma\gamma$}{H to ZZ, WW, and gamma gamma} channels}\label{sec:Combination_HGG} \label{combination_hgg}

In this analysis, the $\X\to\gamma\gamma$ decay channel is studied only in the context of the exotic
spin-two $2_m^+$ hypothesis.
Several spin-two scenarios in Table~\ref{table-scenarios} are only defined for couplings
to massive vector bosons and are not defined for $\X\to\gamma\gamma$.
Several of the remaining higher-dimension operators in the spin-two scenario are not considered here.
However, the direct model-independent analysis of the $\cos\theta^*$ distribution can be
performed~\cite{Khachatryan:2014ira,Bolognesi:2012mm}.
The spin-one scenario of a resonance decaying to a two-photon final state is forbidden~\cite{Landau, Yang},
and all spin-zero scenarios have an identical isotropic two-photon distribution in the rest frame of the boson. Therefore the spin-zero and spin-one scenarios are not considered.

\begin{figure}[htbp]
  \centering
   \includegraphics[width=0.45\textwidth]{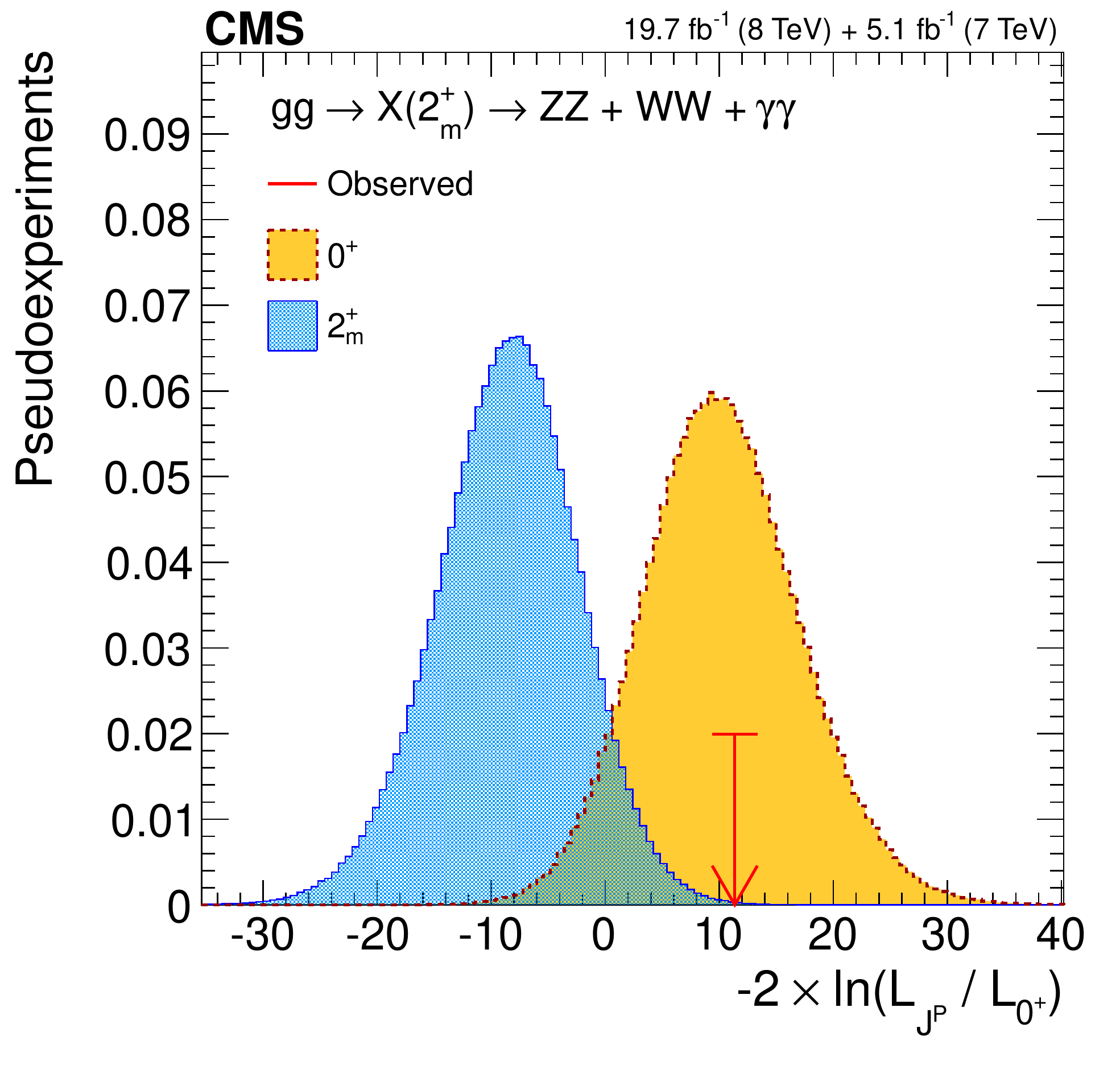}
   \includegraphics[width=0.45\textwidth]{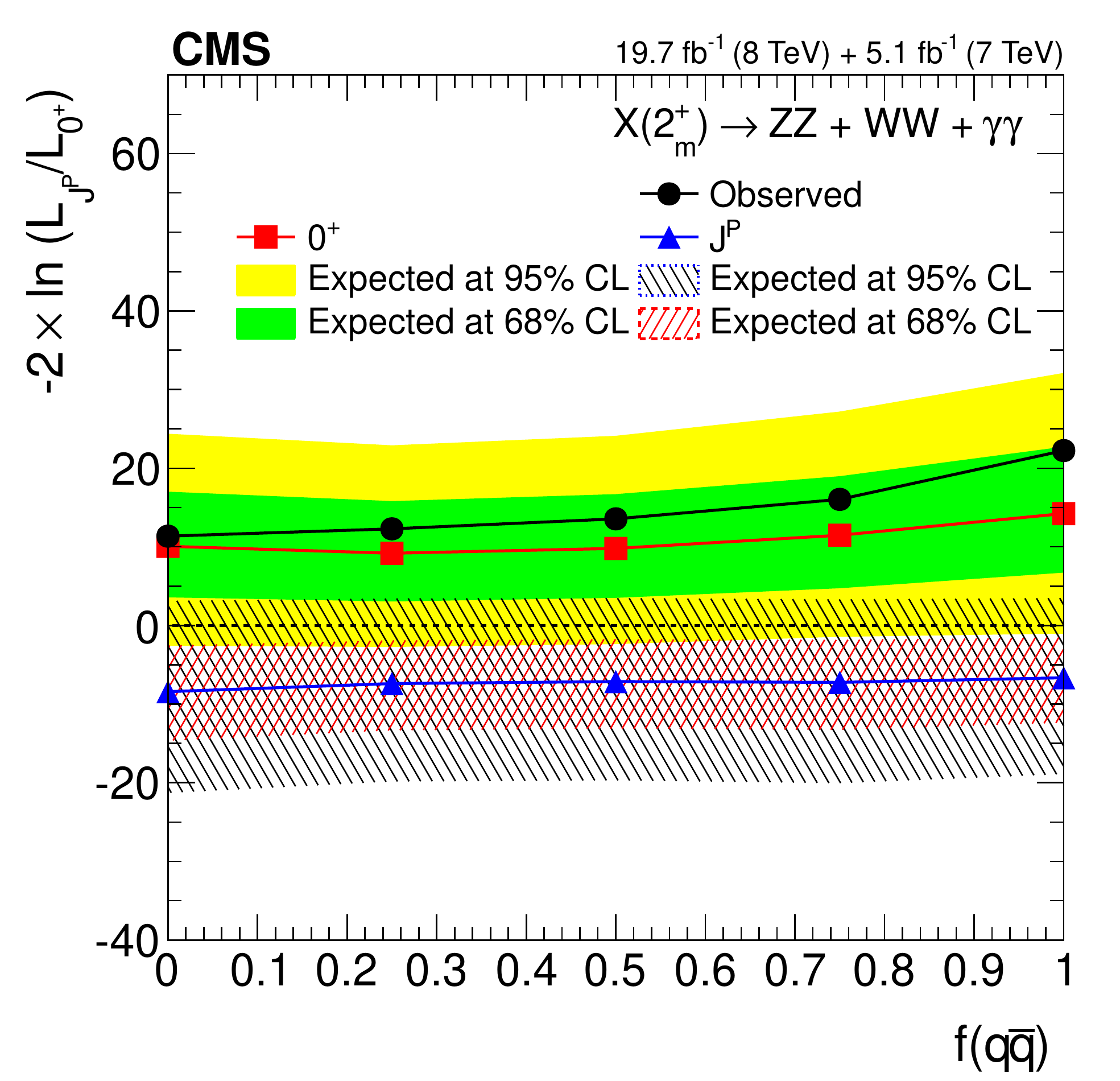}
     \caption{
      (\cmsLeft) Distributions of the test statistic $q=-2\ln(\mathcal{L}_{J^P}/\mathcal{L}_{0^+})$ in the combination of the
    $\X\to\Z\Z, \PW\PW,$ and $\gamma\gamma$ channels for the hypothesis of $\Pg\Pg\to X(2^+_{m})$
      tested against the SM Higgs boson hypothesis ($0^+$).
      The expectation for the SM Higgs boson is represented by the
      yellow histogram on the right and the alternative $J^P$ hypothesis by the
      blue histogram on the left. The red arrow indicates the observed $q$ value.
      (\cmsRight) Distributions of the test statistic $q=-2\ln(\mathcal{L}_{J^P}/\mathcal{L}_{0^+})$
     as a function of $f(\qqbar)$
      for the hypotheses of the $2^+_{m}$ model tested against the SM Higgs boson hypothesis
      in the $\X\to\Z\Z, \PW\PW,$ and $\gamma\gamma$ channels.
      The median expectation for the SM Higgs boson is represented  with the solid green (68\% \CL) and
      yellow (95\% \CL) regions. The alternative $2^+_{m}$ hypotheses are represented  by the blue triangles
       with the red (68\% \CL) and blue (95\% \CL) hatched regions. The observed values are indicated by the black dots.
       \label{fig:sepComb_all2}
       }

\end{figure}

The individual $2_m^+$ hypothesis test results in each channel
were presented earlier~\cite{Chatrchyan:2013iaa,Chatrchyan:2013mxa,Khachatryan:2014ira}
and the combined results are shown in Table~\ref{tab:jpmodelsComb-hgg}.
In Fig.~\ref{fig:sepComb_all2} examples of the test statistic, $q=-2\ln(\mathcal{L}_{J^P}/\mathcal{L}_{0^+})$,
are shown for various fractions of the $\qqbar$ production mechanism $f(\qqbar)$.
As a result, the $2_m^+$ model is excluded with a 99.87\% \CL or higher for any
combination of the $\Pg\Pg$ and $\qqbar$ production mechanisms.

\begin{table*}[htb]
\centering
\topcaption{
Results of the study of the $2^+_{m}$ model for the combination of the $\X\to\Z\Z$,
$\PW\PW$, and $\gamma\gamma$ decay channels.
The expected separation is quoted for the three channels separately and for the combination with
the signal strength for each hypothesis determined from the fit to data independently in each channel.
Also shown in parentheses is the expectation with the SM signal cross section ($\mu$=1).
The observed separation shows the consistency of the observation with the SM $0^+$ model or $J^P$
model and corresponds to the scenario where the signal strength
is floated in the fit to data.
\label{tab:jpmodelsComb-hgg}}
\begin{scotch}{lcccccccc}
$J^P$                           & $J^P$        &        Expected    &       Expected  &    Expected  &       Expected  &                       &      &      \\
Model                           & Prod. &        $\X\to\Z\Z$ & $\X\to\PW\PW$ & $\X\to\gamma\gamma$ &($\mu$=1) &  Obs. $0^+$  & Obs. $J^P$ & \CLs \\
\hline
$2^+_{m}$  & $\Pg\Pg$       &  1.9$\sigma$ & 1.8$\sigma$  & 1.6 $\sigma$ & 3.0$\sigma$ (3.7$\sigma$) & $-$0.2$\sigma$ & $+$3.3$\sigma$ & 0.13\%  \\
$2^+_{m}$  & $\qqbar$   &  1.7$\sigma$ & 2.7$\sigma$ & 1.2 $\sigma$ & 3.3$\sigma$ (4.4$\sigma$) & $-$0.9$\sigma$ & $+$4.7$\sigma$  & 0.001\%  \\
\end{scotch}
\end{table*}

\section{Study of spin-zero \texorpdfstring{$\PH\V\V$}{HVV} couplings} \label{sec:ResultsSpinZero}
Given the exclusion of the exotic spin-one and spin-two scenarios presented in Section~\ref{sec:ResultsExotic},
detailed studies of $\PH\V\V$ interactions under the assumption that the new boson is a spin-zero
resonance are performed. The results are obtained following the techniques
presented in Section~\ref{sec:AnalysisStrategyIntro}.

\begin{figure*}[htbp]
  \centering
    \includegraphics[width=\textwidth]{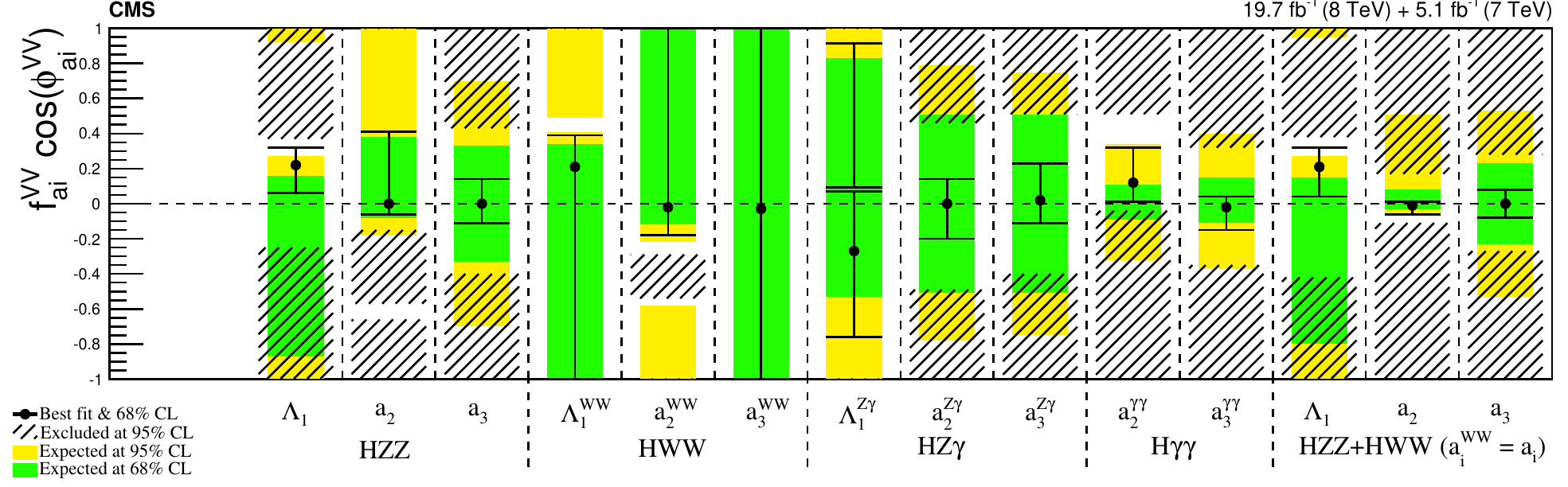}
    \caption{
Summary of allowed confidence level intervals on anomalous coupling parameters in $\PH\V\V$
interactions under the assumption that all the coupling ratios are real ($\phi_{ai}^{\V\V}=0$ or $\pi$).
The expected 68\% and 95\% \CL regions are shown as the green and yellow bands.
The observed constraints at 68\% and 95\% \CL are shown as the points with errors
and the excluded hatched regions.
In the case of the $f_{\Lambda1}^{Z\gamma}$ measurement, there are two minima and two
68\%~\CL intervals, while only one global minimum is indicated with a point.
The combination of the  $\PH\Z\Z$ and $\PH\PW\PW$ measurements is presented,
assuming the symmetry $a_i = a_i^{\PW\PW}$, including $R_{ai}=0.5$.
      \label{fig:spin0_summary}}

\end{figure*}

First, constraints are applied on the presence of only one anomalous term in the $\PH\V\V$
amplitude where the couplings are considered to be real. A summary of such results is presented
in Table~\ref{tab:summary_spin0} and Fig.~\ref{fig:spin0_summary}.
The details of these and other measurements are presented in the following subsections, with further
measurements considering simultaneously up to four fractions and phase parameters in several cases.
The combination of the $\PH\Z\Z$ and $\PH\PW\PW$ coupling measurements provides further constraints
on the $\PH\V\V$ interactions. All results are obtained with the template method, and
the $f_{a2}$ and $f_{a3}$ measurements in $\PH\Z\Z$ interactions are also validated
with the multidimensional distribution method.

\begin{table*}[b!th]
\centering
\topcaption{
Summary of allowed 68\%~\CL (central values with uncertainties) and 95\%~\CL (ranges in square brackets)
intervals on anomalous coupling parameters in $\PH\V\V$ interactions under the assumption that all the coupling
ratios are real ($\phi_{ai}^{\V\V}=0$ or $\pi$).
The ranges are truncated at the physical boundaries of $f_{ai}^{\V\V}=1$.
The last column indicates the observed (expected) confidence level of a pure anomalous coupling
corresponding to $f_{ai}^{\V\V}=1$ when compared to the SM expectation $f_{ai}^{\V\V}=0$.
The expected results are quoted for the SM signal production cross section ($\mu=1$).
The results are obtained with the template method.
}
\renewcommand{\arraystretch}{1.25}
\begin{scotch}{ccccccc}
 & Parameter                                   &  \multicolumn{2}{c}{Observed} &  \multicolumn{2}{c}{Expected}  & $f_{ai}^{\V\V}=1$  \\
\hline
& $f_{\Lambda1}\cos(\phi_{\Lambda1})$        & \multicolumn{2}{c}{$0.22^{+0.10}_{-0.16}$ $ [-0.25,0.37] $}          & \multicolumn{2}{c}{$0.00^{+0.16}_{-0.87}$ $ [-1.00,0.27]$}
& 1.1\% (16\%)                                    \\
&   & \multicolumn{2}{c}{ }          & \multicolumn{2}{c}{~~~~~~~~~$\cup [0.92,1.00] $}       &            \\
& $f_{a2}\cos(\phi_{a2})$         & \multicolumn{2}{c}{$0.00^{+0.41}_{-0.06}$ $ [-0.66, -0.57]$}     & \multicolumn{2}{c}{$0.00^{+0.38}_{-0.08}$ $ [-0.18,1.00]$}
 & 5.2\% (5.0\%)               \\
 &     & \multicolumn{2}{c}{~~~~~~~~~$ \cup [-0.15,1.00]$}     & \multicolumn{2}{c}{ }       &            \\
& $f_{a3}\cos(\phi_{a3})$         & \multicolumn{2}{c}{$0.00^{+0.14}_{-0.11}$ $ [-0.40,0.43] $} & \multicolumn{2}{c}{$0.00^{+0.33}_{-0.33}$ $ [-0.70,0.70] $}
& 0.02\% (0.41\%)         \\
& $f_{\Lambda1}^{\PW\PW}\cos(\phi_{\Lambda1}^{\PW\PW})$   & \multicolumn{2}{c}{$0.21^{+0.18}_{-1.21}$ $ [-1.00 ,1.00]$}     & \multicolumn{2}{c}{$0.00^{+0.34}_{-1.00}$ $[-1.00, 0.41]$}
& 78\% (67\%)        \\
& & \multicolumn{2}{c}{ }     & \multicolumn{2}{c}{~~~~~~~~~$\cup  [0.49,1.00]$}   &    \\
& $f_{a2}^{\PW\PW}\cos(\phi_{a2}^{\PW\PW})$         & \multicolumn{2}{c}{$-0.02^{+1.02}_{-0.16}$ $ [-1.00, -0.54]$}     & \multicolumn{2}{c}{$0.00^{+1.00}_{-0.12}$ $[-1.00, -0.58]$}
& 42\% (46\%)             \\
 &  & \multicolumn{2}{c}{~~~~~~~~~$\cup  [-0.29,1.00]$}     & \multicolumn{2}{c}{~~~~~~~~~$\cup  [-0.22,1.00]$}   &     \\
& $f_{a3}^{\PW\PW}\cos(\phi_{a3}^{\PW\PW})$         & \multicolumn{2}{c}{$-0.03^{+1.03}_{-0.97}$ $ [-1.00,1.00] $} & \multicolumn{2}{c}{$0.00^{+1.00}_{-1.00}$ $ [-1.00,1.00] $}
& 34\% (49\%)        \\
& $f_{\Lambda1}^{Z\gamma}\cos(\phi_{\Lambda1}^{Z\gamma})$        & \multicolumn{2}{c}{$-0.27^{+0.34}_{-0.49}$ $ [-1.00,1.00] $}          & \multicolumn{2}{c}{$0.00^{+0.83}_{-0.53}$ $[-1.00,1.00] $}
& 26\% (16\%)                                              \\
& $f_{a2}^{Z\gamma}\cos(\phi_{a2}^{Z\gamma})$          & \multicolumn{2}{c}{$0.00^{+0.14}_{-0.20}$ $[-0.49,0.46]$} &   \multicolumn{2}{c}{$0.00^{+0.51}_{-0.51}$ $[-0.78,0.79]$}
& $<$0.01\% (0.01\%)         \\
& $f_{a3}^{Z\gamma}\cos(\phi_{a3}^{Z\gamma})$           & \multicolumn{2}{c}{$0.02^{+0.21}_{-0.13}$ $[-0.40,0.51]$} &   \multicolumn{2}{c}{$0.00^{+0.51}_{-0.51}$ $[-0.75,0.75]$}
& $<$0.01\% ($<$0.01\%)         \\
 & $f_{a2}^{\gamma\gamma}\cos(\phi_{a2}^{\gamma\gamma})$    & \multicolumn{2}{c}{$0.12_{-0.11}^{+0.20}$ $[-0.04,+0.51]$} &   \multicolumn{2}{c}{$0.00_{-0.09}^{+0.11}$ $[-0.32,0.34]$}
& $<$0.01\% ($<$0.01\%)        \\
& $f_{a3}^{\gamma\gamma}\cos(\phi_{a3}^{\gamma\gamma})$    & \multicolumn{2}{c}{$-0.02_{-0.13}^{+0.06}$ $[-0.35,0.32]$} &   \multicolumn{2}{c}{$0.00_{-0.11}^{+0.15}$ $[-0.37,0.40]$}
& $<$0.01\% ($<$0.01\%)        \\
\end{scotch}
\label{tab:summary_spin0}
\end{table*}

\subsection{Study of \texorpdfstring{$\PH\Z\Z$}{HZZ} couplings with the \texorpdfstring{$\PH\to\Z\Z\to 4\ell$}{H to ZZ to 4l} channel} \label{sec:Resultspinzerohzz}

The study of the anomalous $\PH\V\V$ couplings starts with the test of three contributions to the $\PH\Z\Z$ interaction
as shown in Eq.~(\ref{eq:formfact-fullampl-spin0}). Only real couplings are considered in this test,
$\phi_{ai}=0$ or $\pi$, where $\phi_{ai}$ generically refers to the phase of the coupling in question,
such as $\phi_{\Lambda1}$, $\phi_{a2}$, or $\phi_{a3}$. Since the expansion of terms in
Eq.~(\ref{eq:formfact-fullampl-spin0}) is considered for small anomalous contributions,
all other parameters are set to zero when the anomalous couplings of interest are considered.
These constraints of real couplings and zero contribution from other terms are relaxed in further tests
discussed below.
In the template approach, the three sets of observables in each fit are given in Table~\ref{tab:kdlist}.
The only exception is in the $f_{\Lambda1}$ measurement, where the usual interference discriminant
does not provide additional information and instead the third observable is $\mathcal{D}_{0h+}$
to minimize the number of configurations also used for other studies. Since $\mathcal{D}_{0h+}$
does not bring additional information for this measurement, it is not reflected in Table~\ref{tab:kdlist}.

\begin{figure*}[htbp]
\centering
       \includegraphics[width=0.40\textwidth]{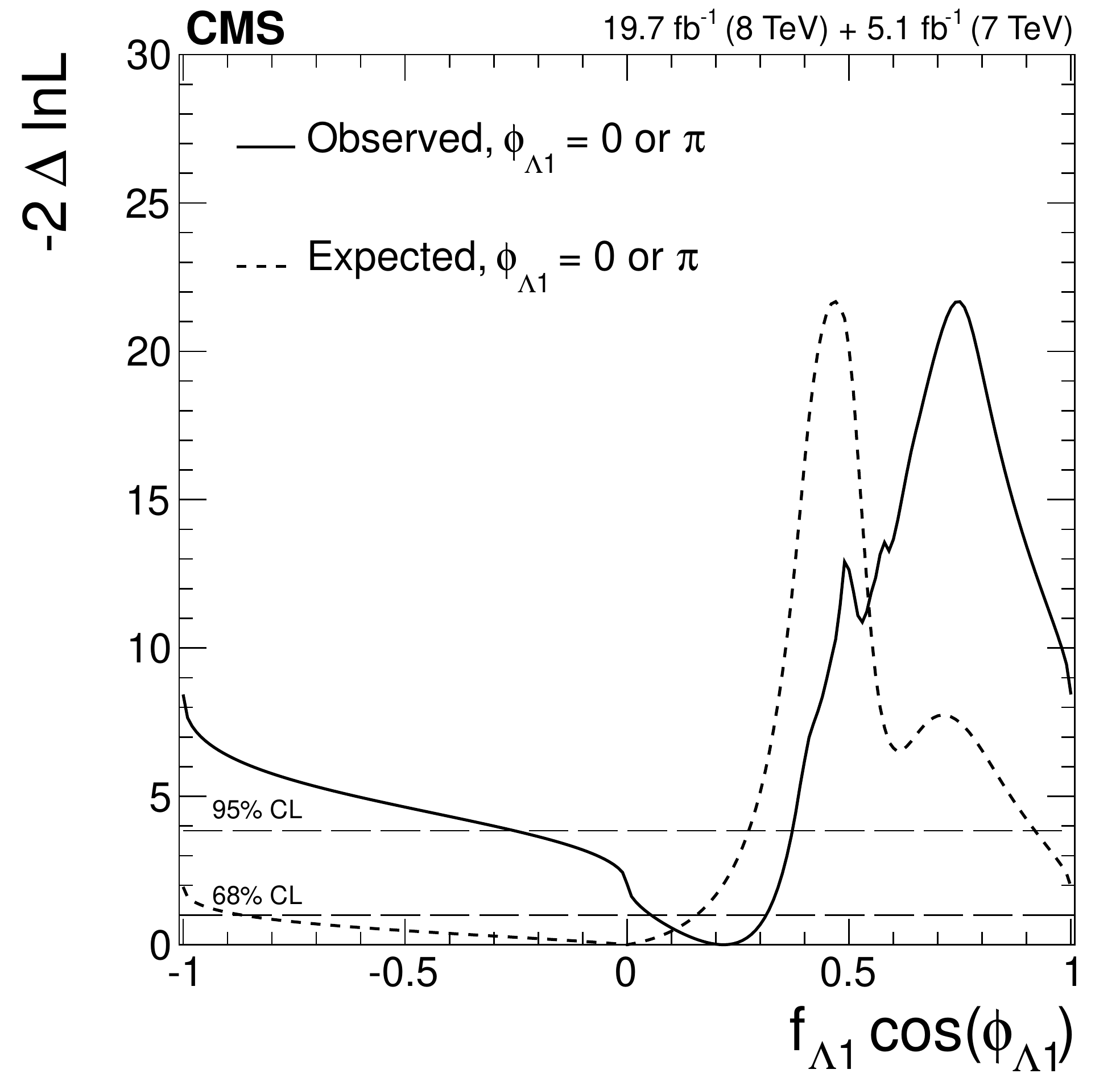}
       \includegraphics[width=0.40\textwidth]{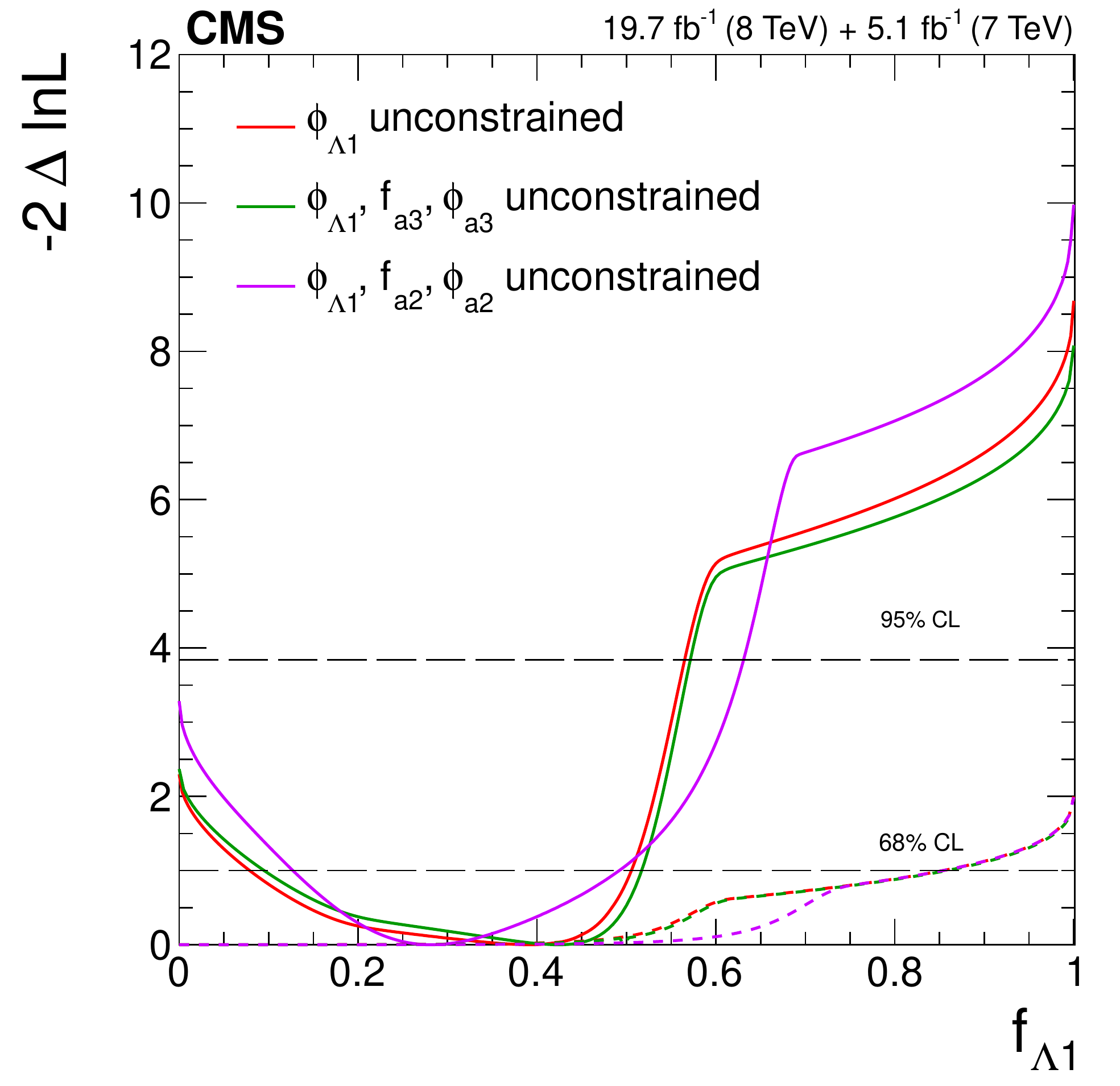} \\
       \includegraphics[width=0.40\textwidth]{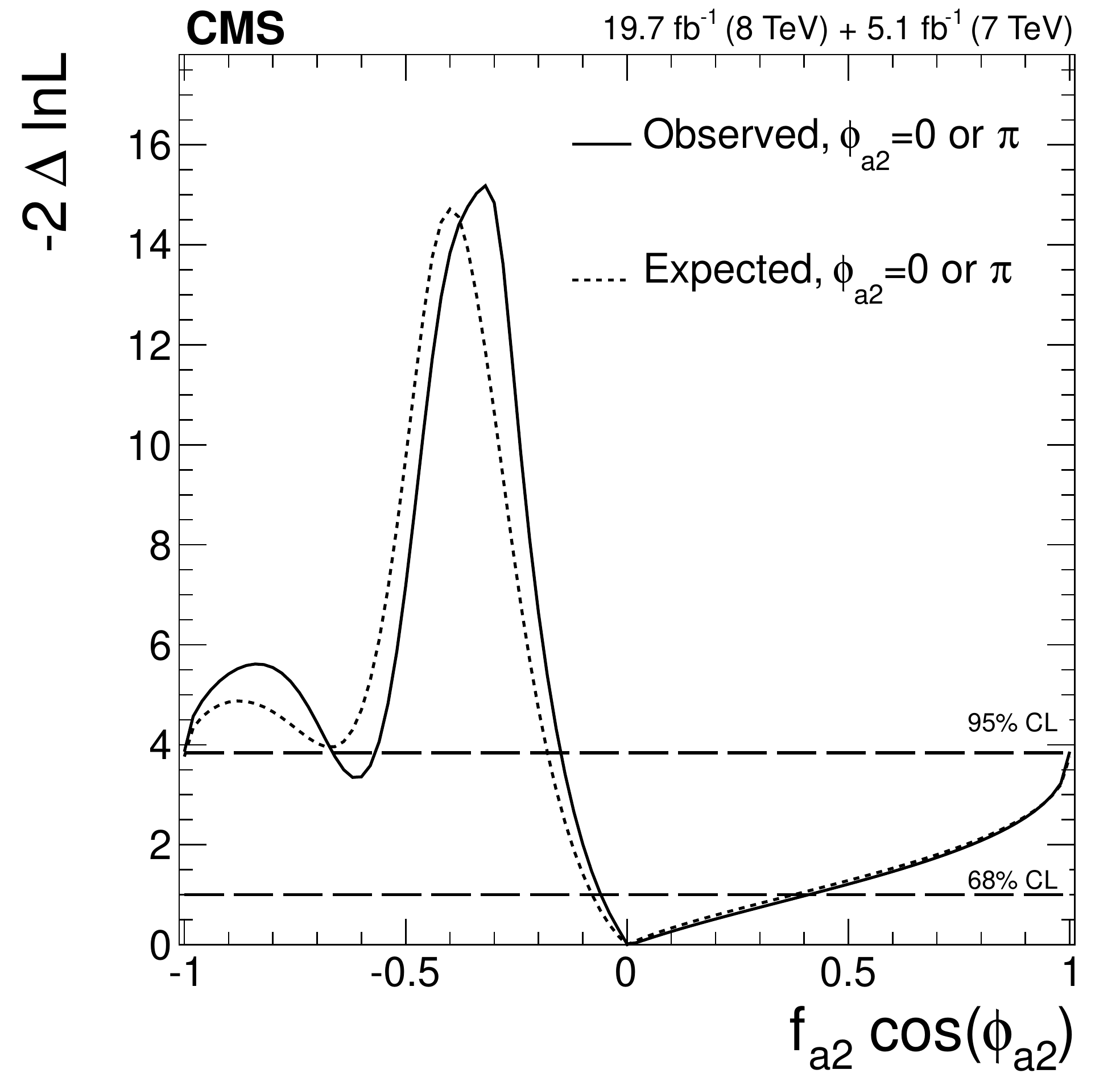}
       \includegraphics[width=0.40\textwidth]{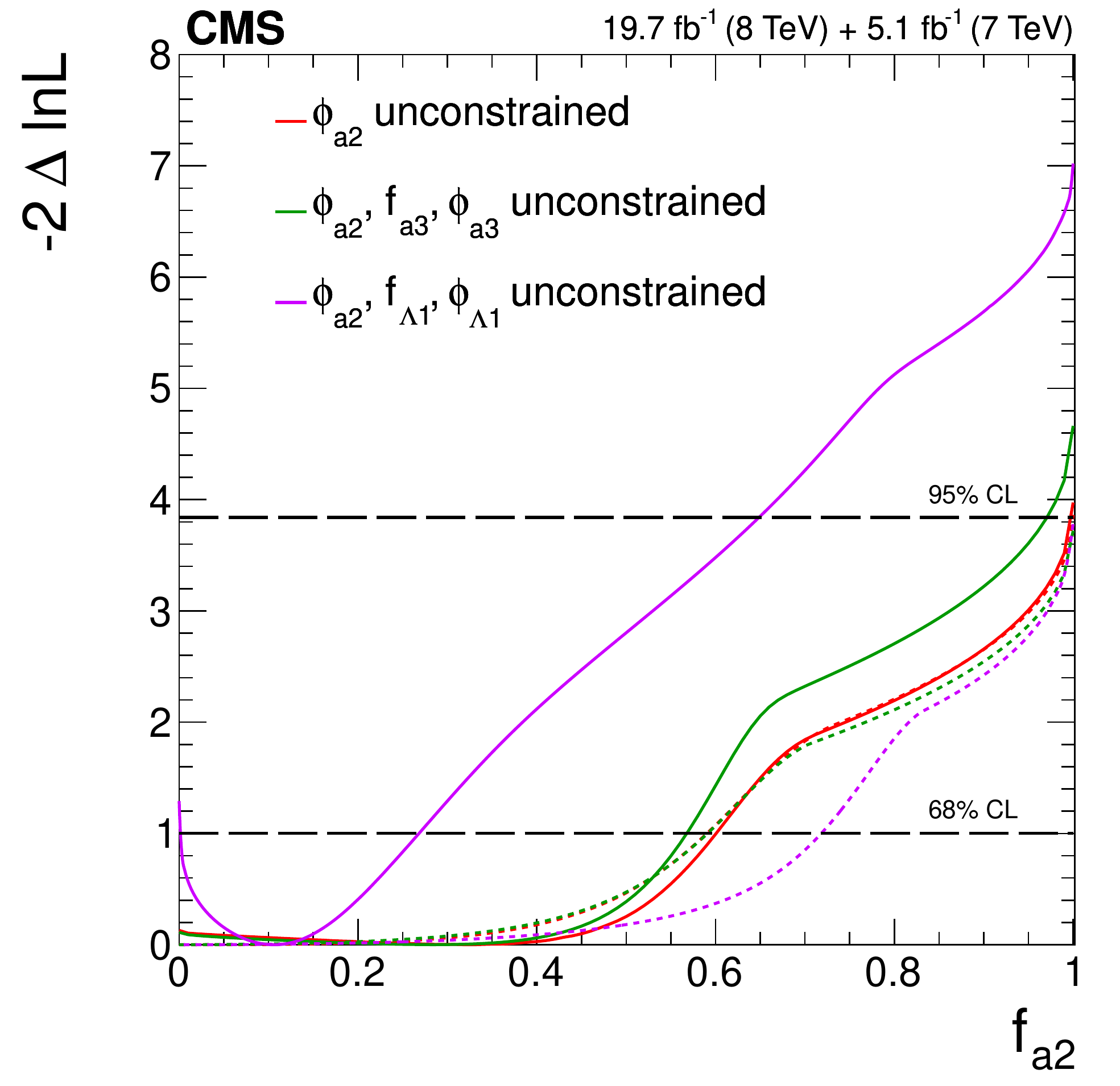} \\
       \includegraphics[width=0.40\textwidth]{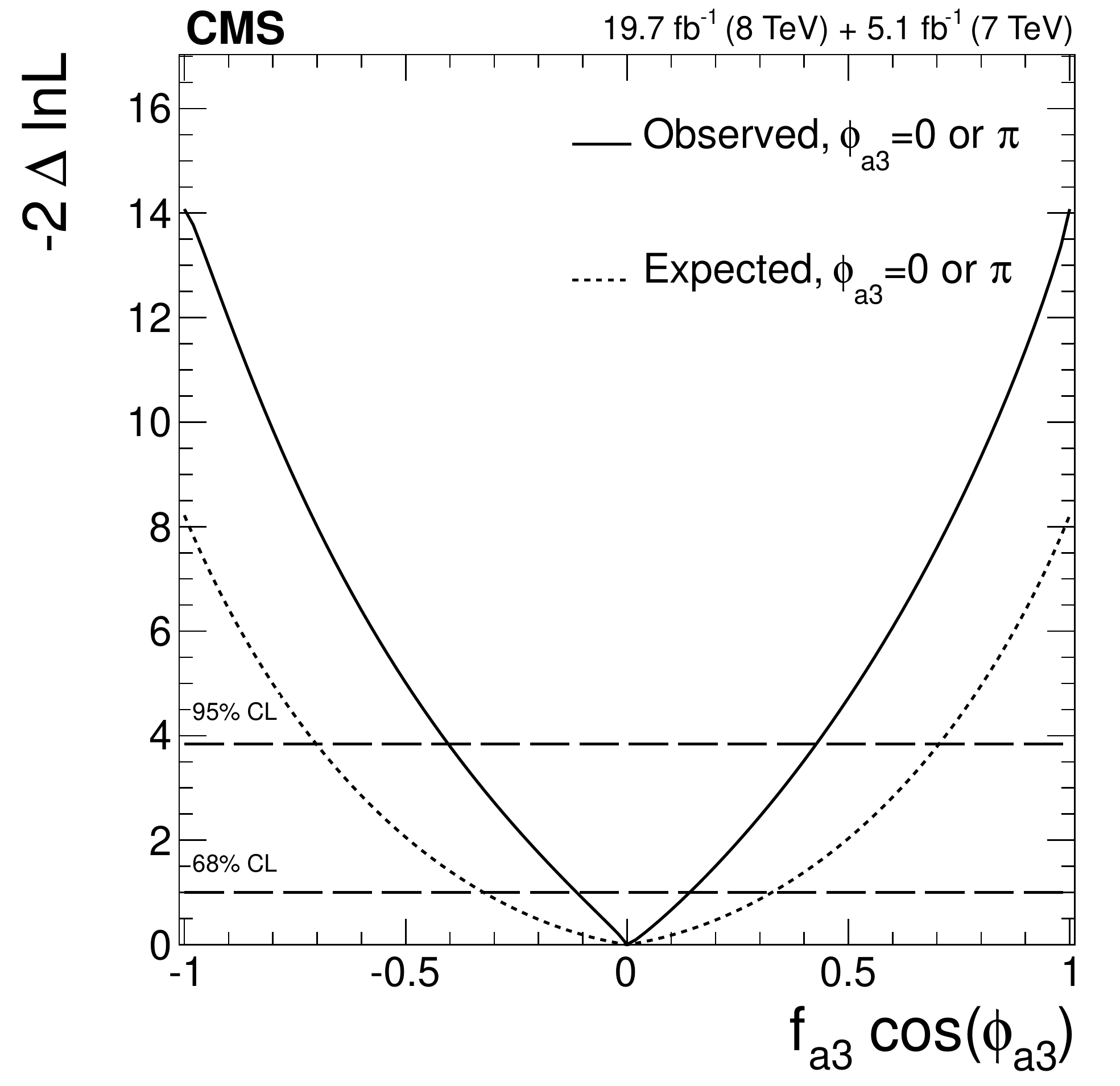}
       \includegraphics[width=0.40\textwidth]{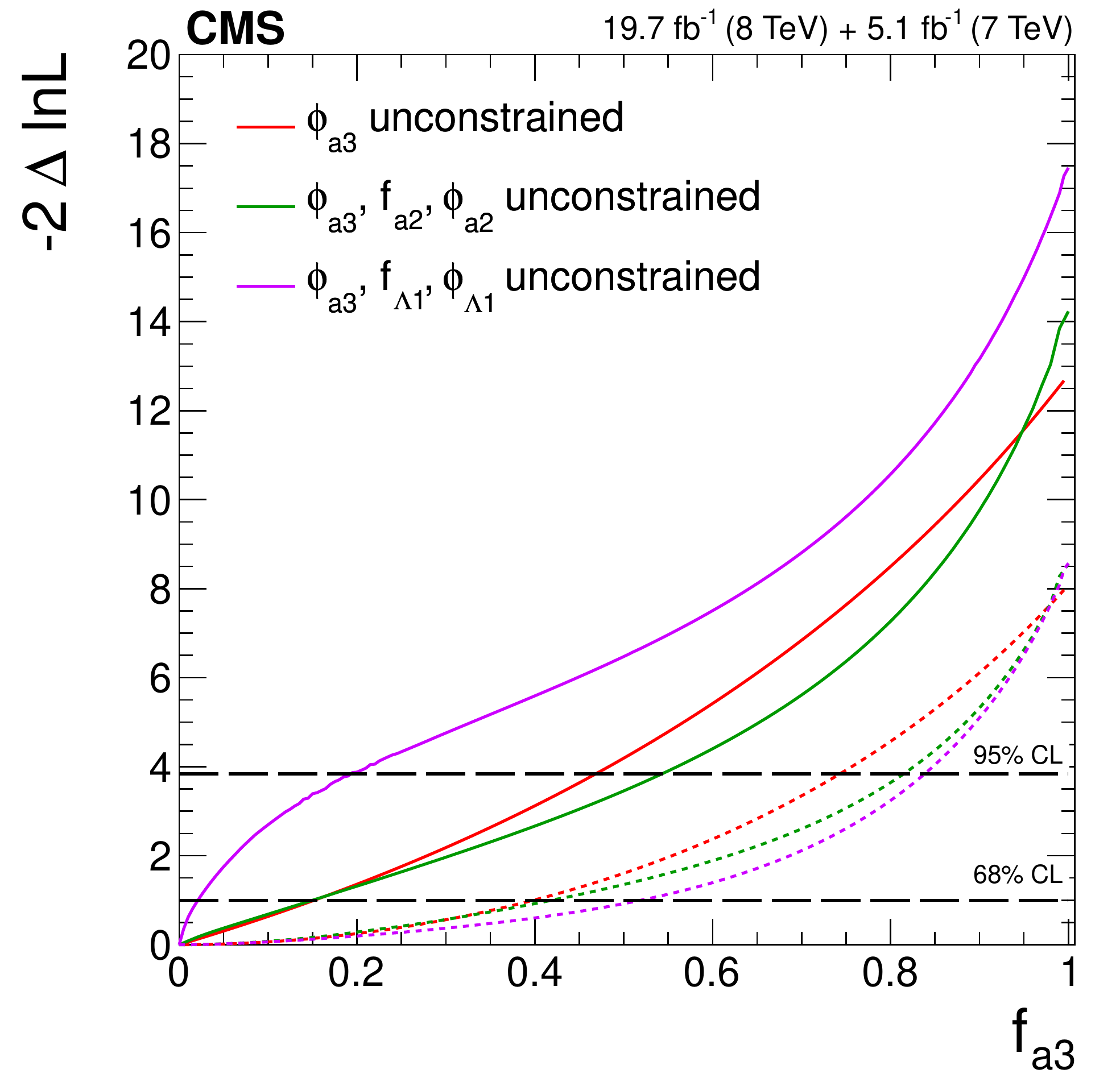} \\
	\caption{
	Expected (dashed) and observed (solid) likelihood scans using the template method for
	the effective fractions $f_{\Lambda1}$, $f_{a2}$, $f_{a3}$ (from top to bottom) describing $\PH\Z\Z$ interactions.
	Plots on the left show the results when the couplings studied are constrained to be real and all other couplings are fixed to the SM predictions.
	The $\cos\phi_{ai}$ term allows a signed quantity where $\cos\phi_{ai}=-1$ or $+1$.
	Plots on the right show the results where the phases of the anomalous couplings and
	additional $\PH\Z\Z$ couplings are left unconstrained, as indicated in the legend.
	The $f_{a3}$ result with $\phi_{a3}$ unconstrained (in the bottom-right plot) is from Ref.~\cite{Chatrchyan:2013mxa}.
	}
	\label{fig:results_ZZ_1D}
\end{figure*}

The results of the likelihood function scan for the three parameters, $f_{ai}\cos\phi_{ai}$, are shown in
Fig.~\ref{fig:results_ZZ_1D} (left), where the $\cos\phi_{ai}$ term allows for a signed quantity with
$\cos\phi_{ai}=-1$ or $+1$.
The 68\% and 95\%~\CL intervals are shown in Table~\ref{tab:summary_spin0}.
Using the transformation in
Eq.~(\ref{eq:fa_conversion}), these results can be interpreted for the coupling parameters
used in Eq.~(\ref{eq:formfact-fullampl-spin0}), as shown in Table~\ref{tab:Spin0zz_interpretation}.
Strong destructive interference of the SM and anomalous contributions
at $f_{\Lambda1}\cos(\phi_{\Lambda1})\sim+0.5$ or $f_{a2}\cos(\phi_{a2})\sim-0.5$
leads to very different kinematic distributions and exclusions with high confidence levels.
Additional features with multiple likelihood function maxima observed in the $f_{\Lambda1}$ likelihood scan
are due to the superposition of measurements in the $4\Pe/4\mu$ and $2\Pe2\mu$ channels,
which have different maxima due to the interference between the leptons.

Next, two parameters $f_{ai}$ and $\phi_{ai}$ are considered at the same time. For example, if the coupling is known
to be either positive or negative, such a scenario is considered in Table~\ref{tab:Spin0_ZZ_1D_KDb}. In this
case, constraints are set on $f_{ai}$ for a given phase value. More generally, one can allow $\phi_{ai}$ to be unconstrained,
that is, to have any value between $-\pi$ and $+\pi$ with a generally complex coupling.
Such a fit is performed for $f_{\Lambda1}$ and $f_{a2}$ using the
same configuration, but with additional $\phi_{\Lambda1}$ and $\phi_{a2}$ parameters in Eq.~(\ref{eq:fractions-general}).
The results with $\phi_{ai}$ unconstrained (any) are shown in Table~\ref{tab:Spin0_ZZ_1D_KDb} as well.
The $f_{a3}$ measurement with $\phi_{a3}$ unconstrained is performed with a different technique and is presented
in Ref.~\cite{Chatrchyan:2013mxa}, where the $\mathcal{D}_{C\!P}$ observable is removed from the fit and the result
becomes insensitive to the phase of the amplitude. This technique is adopted due to its simpler implementation and
equivalent performance.

The next step in generalizing the constraints is to consider two anomalous contributions at the same time, both with
and without the constraints that the couplings are real. Therefore, up to four parameters are considered at the same
time: $f_{ai}$, $\phi_{ai}$, $f_{aj}$, and $\phi_{aj}$. Constraints on one parameter, when other parameters are left unconstrained in
the full allowed parameter space, with $0\le f_{ai}\le 1$, are presented in Table~\ref{tab:Spin0_ZZ_1D_KDb}.
Even though the expansion with only three anomalous contributions in
Eq.~(\ref{eq:formfact-fullampl-spin0}) becomes incomplete when large values of $f_{ai}\sim 1$ are considered,
this is still a valuable test of the consistency of the data with the SM.
All of the above results, with phases fixed or unconstrained and with other anomalous couplings unconstrained
are shown in Fig.~\ref{fig:results_ZZ_1D} (right).
Some observed $f_{ai}$ constraints appear to be tighter when compared to the one-parameter fits shown in
Fig.~\ref{fig:results_ZZ_1D} (left). This happens because the values of other profiled parameters are away
from the SM expectation at the minimum of $-2\ln\mathcal{L}$, though still consistent with the SM.
The expected constraints are always weaker with additional free parameters.

\begin{table*}[htbp]
\centering
\topcaption{
Summary of the allowed 95\%~\CL intervals on the anomalous couplings in $\PH\Z\Z$ interactions
using results in Table~\ref{tab:summary_spin0}.
The coupling ratios are assumed to be real (including $\cos(\phi_{\Lambda_{1}})=0$ or $\pi$).
\label{tab:Spin0zz_interpretation}
}
\renewcommand{\arraystretch}{1.25}
\begin{scotch}{cccccc}
Parameter  & Observed & Expected    \\
\hline

$(\Lambda_{1}\sqrt{\abs{a_1}}) \cos(\phi_{\Lambda_{1}})$ &   $[-\infty,-119\GeV]\cup[104\GeV,\infty]$ & $[-\infty,50\GeV] \cup [116\GeV,\infty]$                                            \\

$ a_{2}/a_{1} $ &   $[-2.28,-1.88]\cup[-0.69,\infty]$ & $[-0.77,\infty]$                                               \\

$ a_{3}/a_{1} $ &   $[-2.05,2.19]$ & $[-3.85,3.85]$                                               \\
\end{scotch}
\end{table*}

\begin{figure*}[htbp]
\centering
	\includegraphics[width=0.40\textwidth]{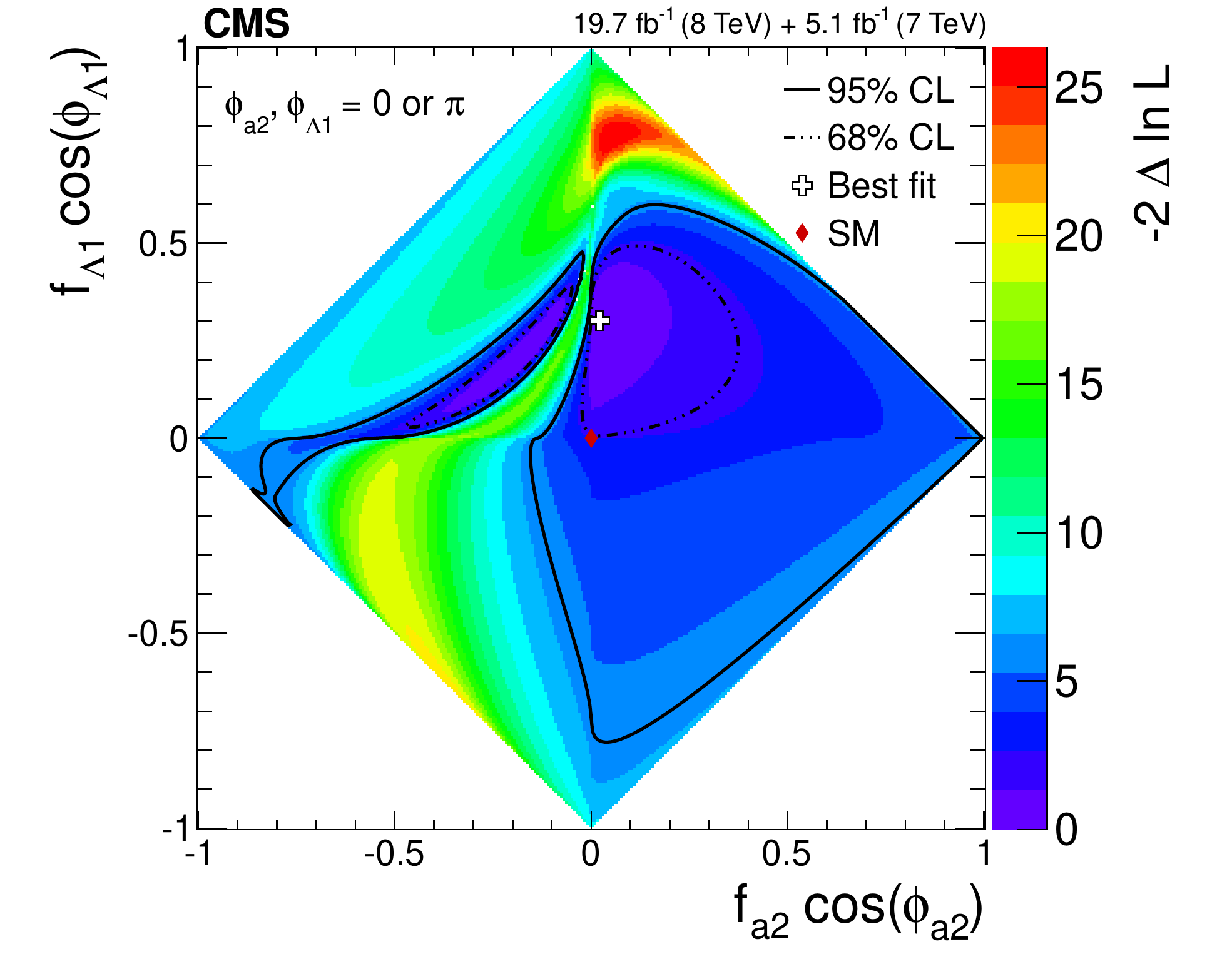}
	\includegraphics[width=0.40\textwidth]{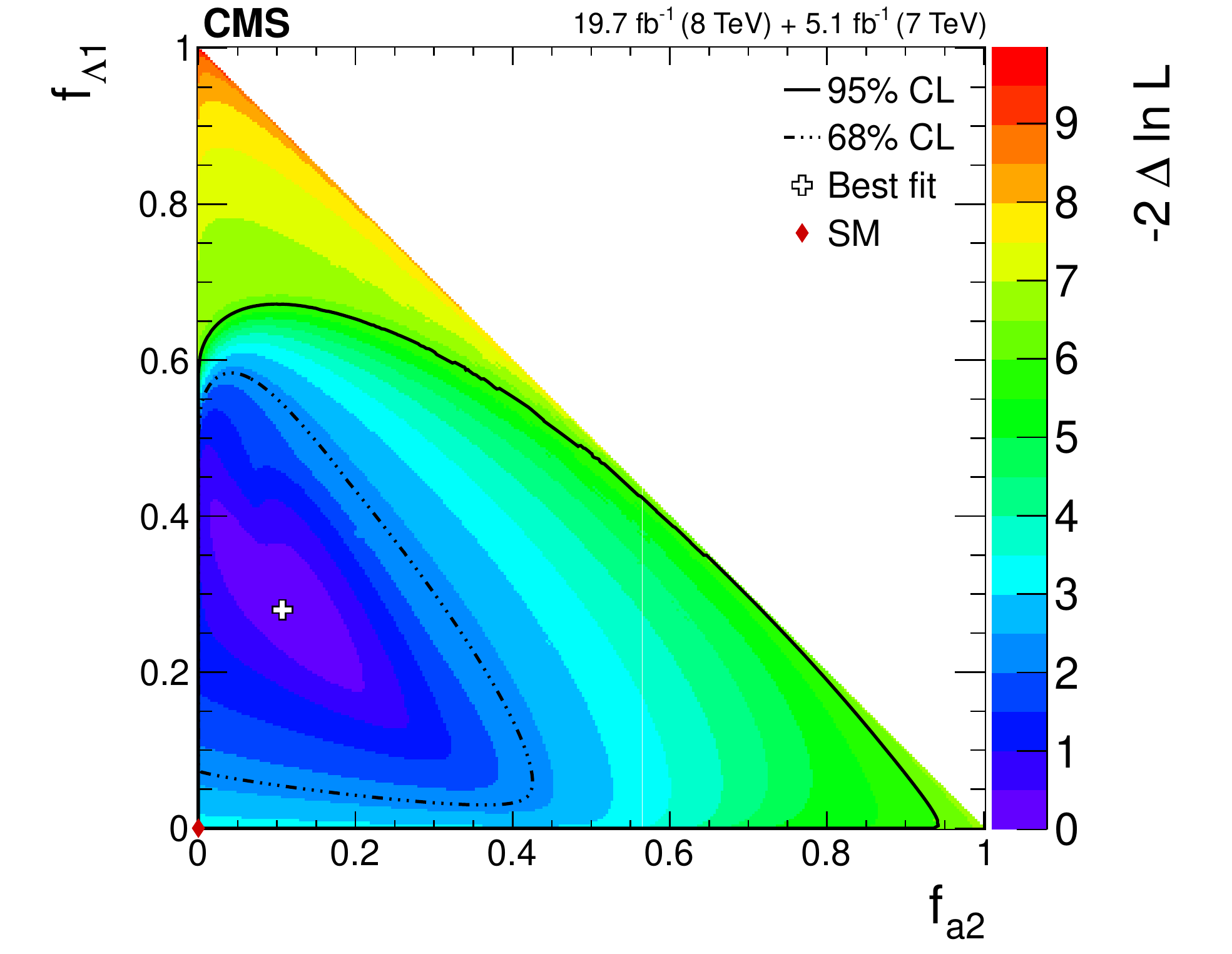} \\
	\includegraphics[width=0.40\textwidth]{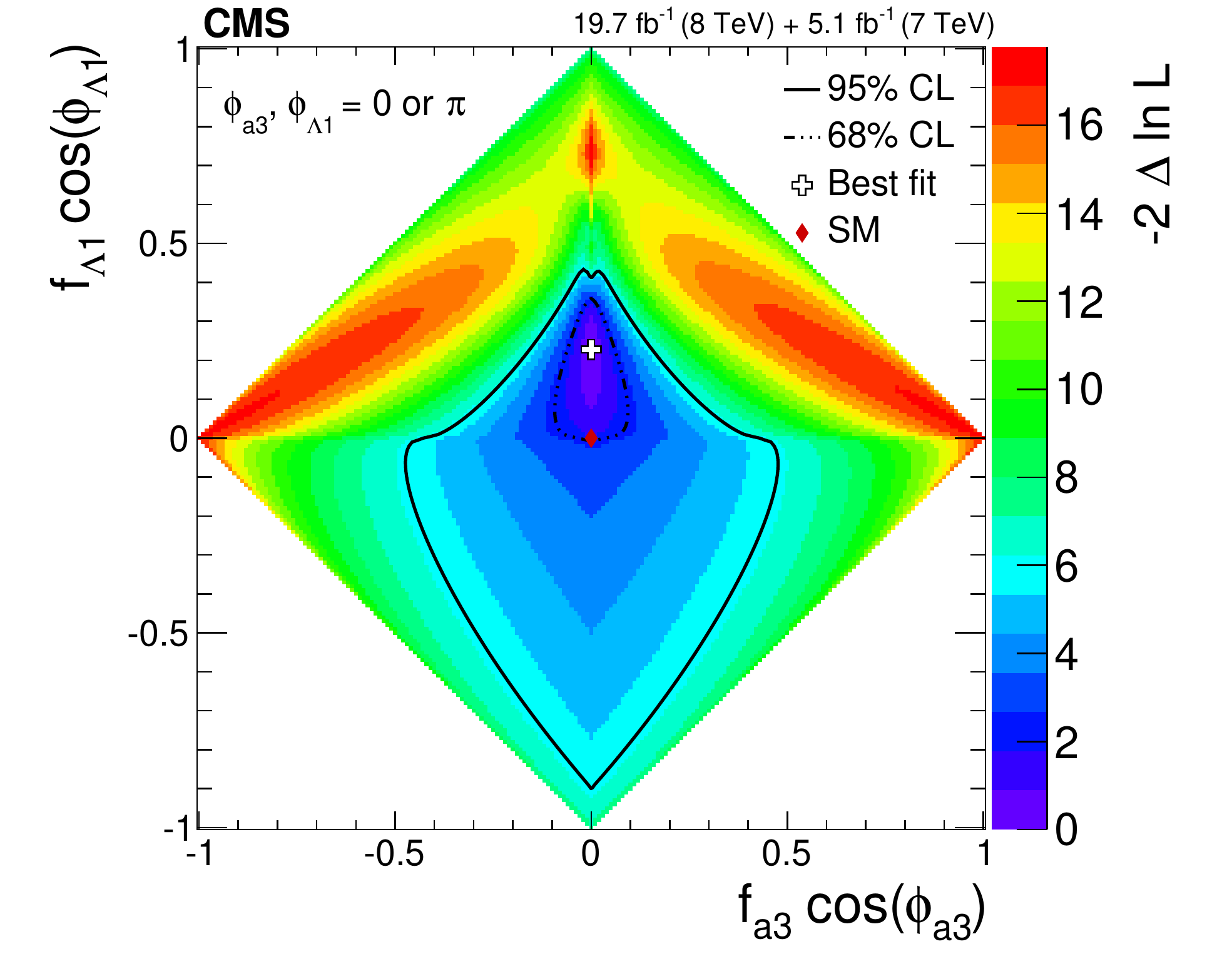}
	\includegraphics[width=0.40\textwidth]{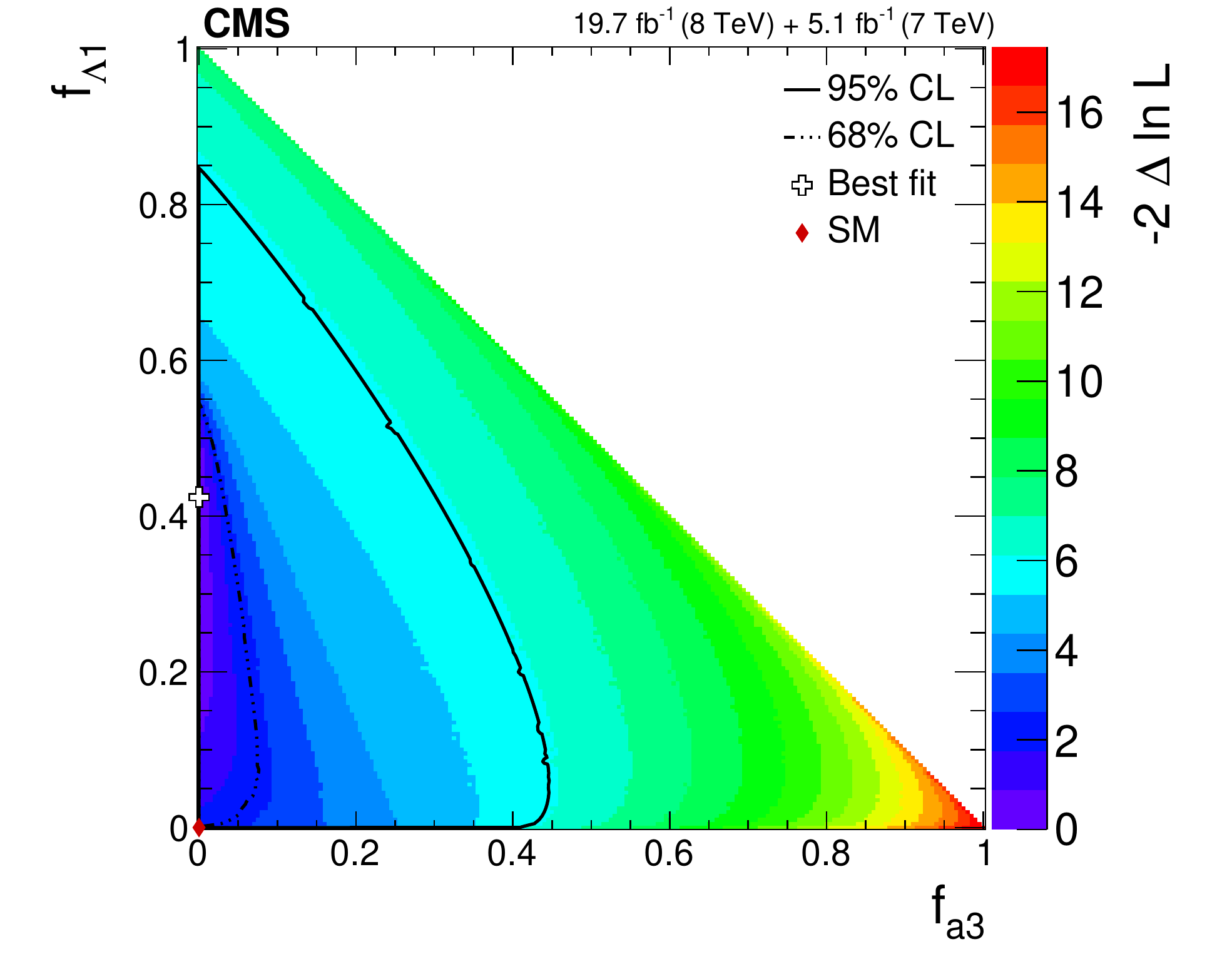}  \\
	\includegraphics[width=0.40\textwidth]{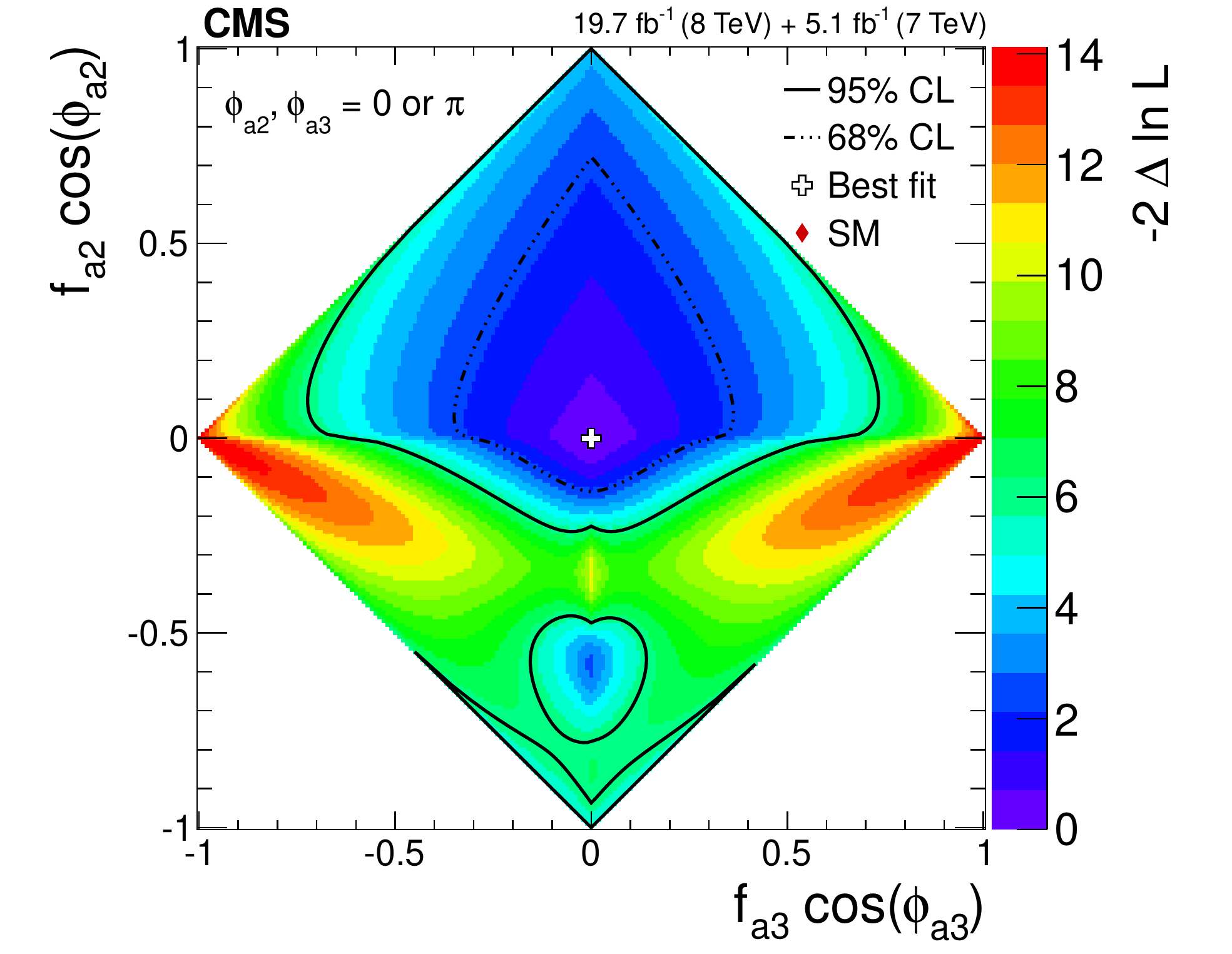}
	\includegraphics[width=0.40\textwidth]{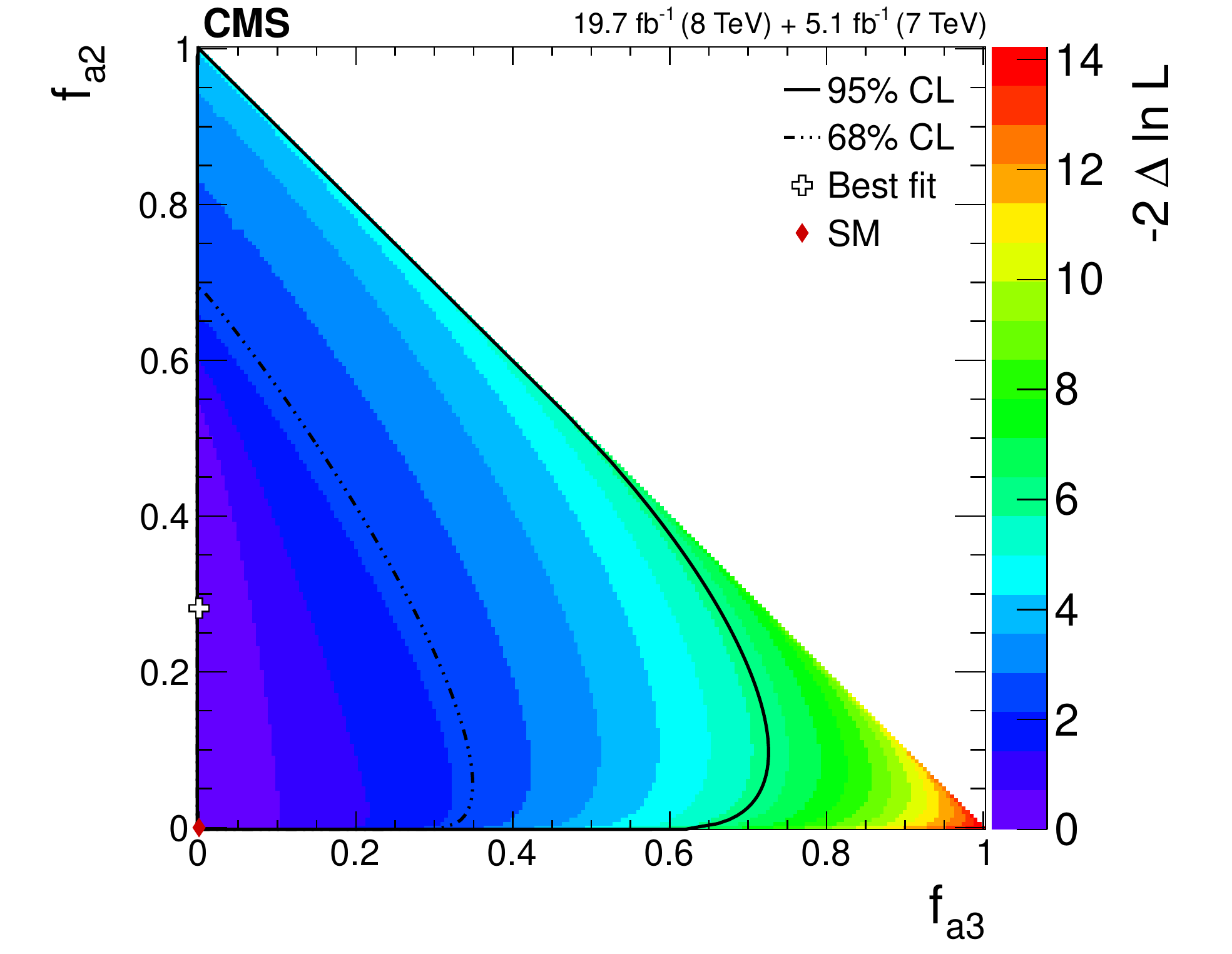}
	\caption{
	Observed likelihood scans using the template method for pairs of effective fractions $f_{\Lambda1}$ vs. $f_{a2}$,
	$f_{\Lambda1}$ vs. $f_{a3}$, and $f_{a2}$ vs. $f_{a3}$ (from top to bottom) describing $\PH\Z\Z$ interactions.
	Plots on the left show the results when the couplings studied are constrained to be real
	and all other couplings are fixed to the SM predictions.
	Plots on the right show the results when the phases of the anomalous couplings are left unconstrained.
	The SM expectations correspond to points (0,0) and the best fit values are shown with the crosses.
	The confidence level intervals are indicated by the corresponding $-2\,\Delta\ln\mathcal{L}$ contours.
	}
	\label{fig:results_ZZ_2D}
\end{figure*}

The above one-parameter measurements, with other couplings also considered to be unconstrained, are obtained
from the fit configurations used for the two-parameter measurements shown in Fig.~\ref{fig:results_ZZ_2D}.
Both options are considered, either with or without the assumption that the couplings are real. To keep
the number of observables to the maximum of three, in the template approach, the following discriminants are
used to set the constraints,
($\mathcal{D}_\text{bkg}$, $\mathcal{D}_{\Lambda1}$, $\mathcal{D}_{0h+}$),
($\mathcal{D}_\text{bkg}$, $\mathcal{D}_{\Lambda1}$, $\mathcal{D}_{0-}$),
and ($\mathcal{D}_\text{bkg}$, $\mathcal{D}_{0-}$ or $\mathcal{D}_{0h+}$),
for the measurements of
$f_{\Lambda1}$ vs. $f_{a2}$, $f_{\Lambda1}$ vs. $f_{a3}$, and $f_{a2}$ vs. $f_{a3}$, respectively.
The left set of plots in Fig.~\ref{fig:results_ZZ_2D} shows constraints on two real couplings, and
the right set of plots in Fig.~\ref{fig:results_ZZ_2D} shows constraints on two couplings that are
allowed to have any complex phase. Similarly to the one-parameter constraints,
the allowed 95\% \CL regions are formally defined using the profile likelihood function ($-2\Delta \ln\mathcal{L} = 5.99$).
The results in Table~\ref{tab:Spin0_ZZ_1D_KDb} are obtained from these two-parameter likelihood scans by profiling one parameter.

Overall, all anomalous $\PH\Z\Z$ couplings are found to be consistent with zero, which is also consistent with the expectation
from the SM where these couplings are expected to be very small, well below the current sensitivity.

\begin{table*}[htbp]
\centering
\topcaption{
Summary of the allowed 68\%~\CL (central values with uncertainties) and 95\%~\CL (ranges in square brackets)
intervals on anomalous coupling parameters in the $\PH\Z\Z$ interactions under the condition of a given phase of
the coupling (0 or $\pi$) or when the phase or other parameters are unconstrained (any value allowed).
Results are presented with the template method and expectations are quoted in parentheses
following the observed values. The results for $f_{a3}$ with $\phi_{a3}$ unconstrained are from Ref.~\cite{Chatrchyan:2013mxa}.
}
\renewcommand{\arraystretch}{1.25}
\begin{scotch}{lccc}
Measurement              &  $f_{\Lambda1}$ &  $f_{a2}$  & $f_{a3}$    \\
\hline
 $\phi_{ai}=0$
 &   $0.22^{+0.10}_{-0.16}$ $[0.00,0.37]$ &   $0.00^{+0.42}_{-0.00}$ $[0.00,1.00]$ & $0.00^{+0.14}_{-0.00}$ $[0.00,0.43]$  \\
 &   (~$0.00^{+0.16}_{-0.00}$ $[0.00,0.27]$ &   (~$0.00^{+0.35}_{-0.00}$ $[0.00,1.00]$~) & (~$0.00^{+0.33}_{-0.00}$ $[0.00,0.70]$~)  \\
 &   ~~~~~~~~~~~~~~~~$\cup [0.92,1.00]$~) &  &  \\
 $\phi_{ai}=\pi$
 &  $0.00^{+0.08}_{-0.00}$ $[0.00,0.82]$ &   $0.00^{+0.06}_{-0.00}$ $[0.00,0.15]$ &   $0.00^{+0.11}_{-0.00}$ $[0.00,0.40]$  \\
 &   &  ~~~~~~~~~~~~~~~$\cup [0.56,0.68]$ &    \\
 &   (~$0.00^{+0.87}_{-0.00}$ $[0.00,1.00]$~) &   (~$0.00^{+0.08}_{-0.00}$ $[0.00,0.18]$ & (~$0.00^{+0.32}_{-0.00}$ $[0.00,0.70]$~)  \\
 &  &   $~~~~~~~~~~~~~~~~\cup [0.62,0.73]$~) &  \\
  any $\phi_{ai}$
 &  $0.39^{+0.16}_{-0.31}$ $[0.00,0.57]$ &   $0.32^{+0.28}_{-0.32}$ $ [0.00,1.00]$ &   $0.00^{+0.17}_{-0.00}$  $[0.00,0.47]$  \\
 &   (~$0.00^{+0.85}_{-0.00}$ $[0.00,1.00]$~) &   (~$0.00^{+0.59}_{-0.00}$ $[0.00,1.00]$~) & (~$0.00^{+0.40}_{-0.00}$ $[0.00,0.74]$~)  \\
any $\phi_{ai},f_{\Lambda1},\phi_{\Lambda1}$
&   ~~~~ &   $0.11^{+0.16}_{-0.11}$ $[0.00,0.65]$ &   $0.00^{+0.02}_{-0.00}$ $[0.00,0.19]$  \\
&   ~~~~  &   (~$0.00^{+0.72}_{-0.00}$ $[0.00,1.00]$~) &(~$0.00^{+0.52}_{-0.00}$ $[0.00,0.84]$~)  \\
any $\phi_{ai},f_{a2},\phi_{a2}$
&  $0.28^{+0.21}_{-0.15}$ $[0.00,0.63]$ &    ~~~~ &   $0.00^{+0.15}_{-0.00}$ $[0.00,0.54]$  \\
&   (~$0.00^{+0.85}_{-0.00}$ $[0.00,1.00]$~) &   ~~~~ & (~$0.00^{+0.42}_{-0.00}$ $[0.00,0.81]$~)  \\
any $\phi_{ai},f_{a3},\phi_{a3}$
&  $0.42^{+0.09}_{-0.33}$ $[0.00,0.57]$ &   $0.28^{+0.29}_{-0.28}$ $[0.00,0.97]$ &    ~~~~  \\
&   (~$0.00^{+0.86}_{-0.00}$ $[0.00,1.00]$~) &   (~$0.00^{+0.59}_{-0.00}$ $[0.00,1.00]$~) & ~~~~  \\
\end{scotch}
\label{tab:Spin0_ZZ_1D_KDb}
\end{table*}

\subsection{Validation of the \texorpdfstring{$\PH\Z\Z$}{HZZ} measurements} \label{sec:Resultspinzero8d}

\begin{figure*}[htbp]
\centering
	\includegraphics[width=0.44\textwidth]{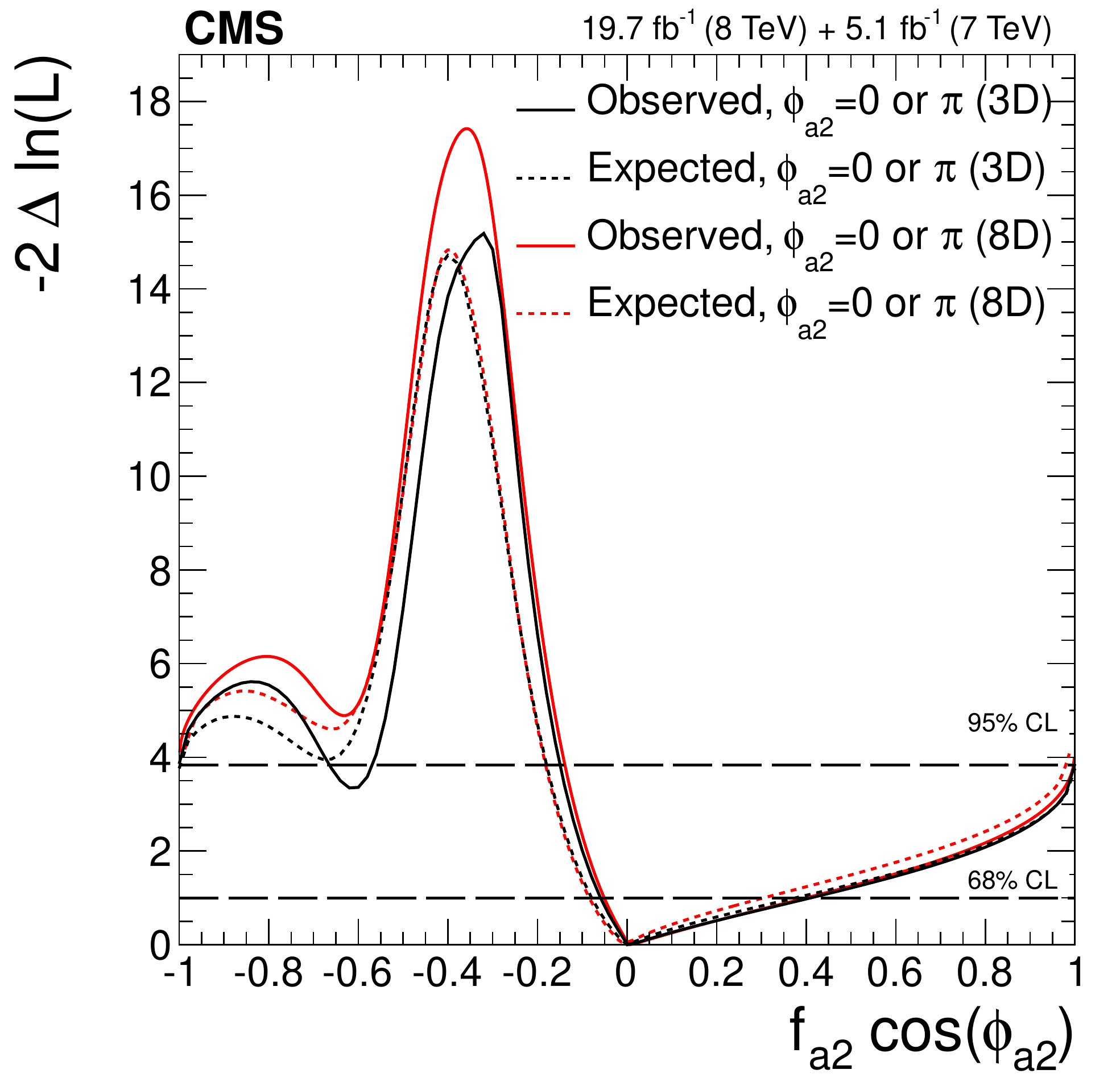}
	\includegraphics[width=0.44\textwidth]{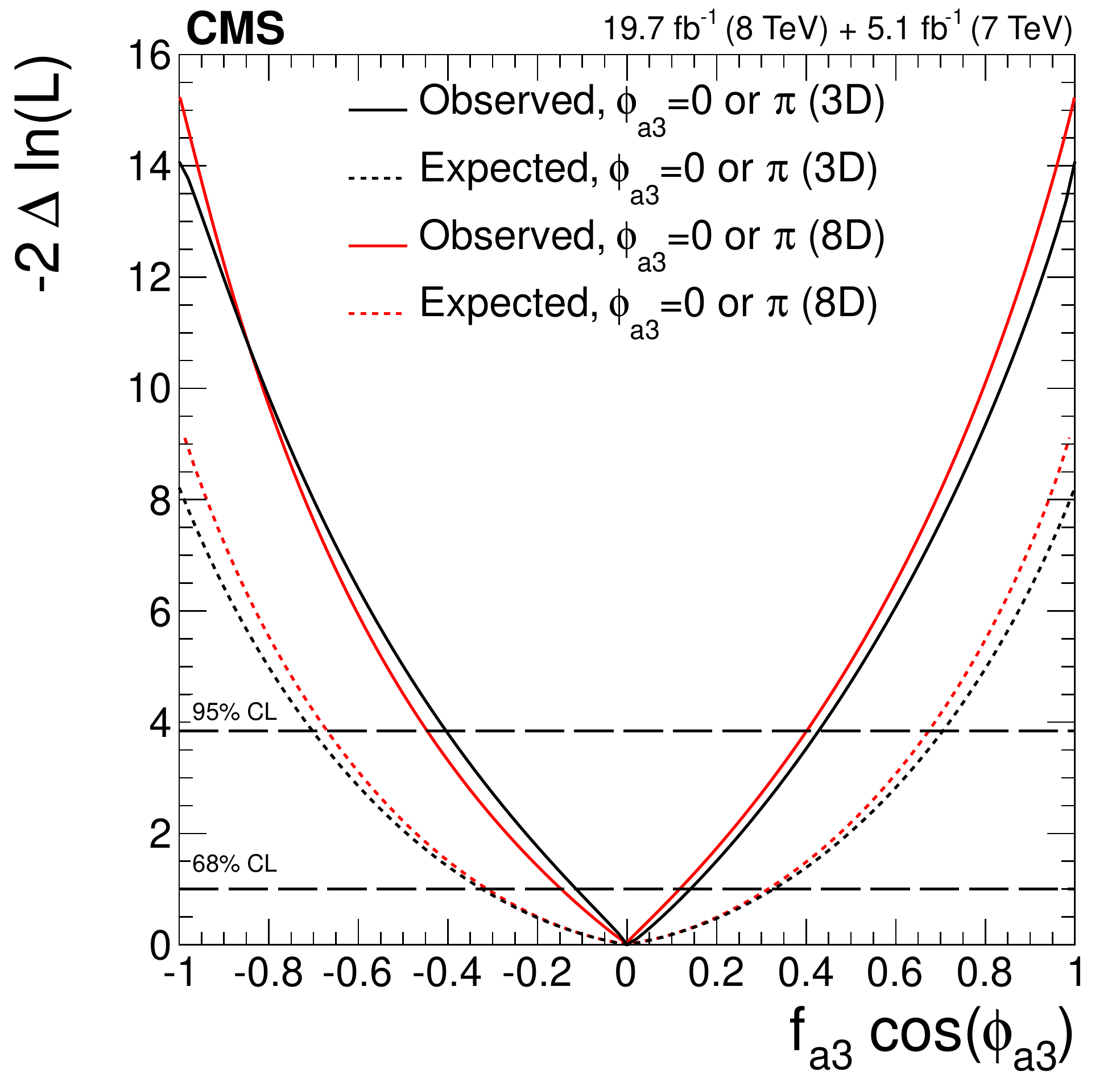}
	\includegraphics[width=0.48\textwidth]{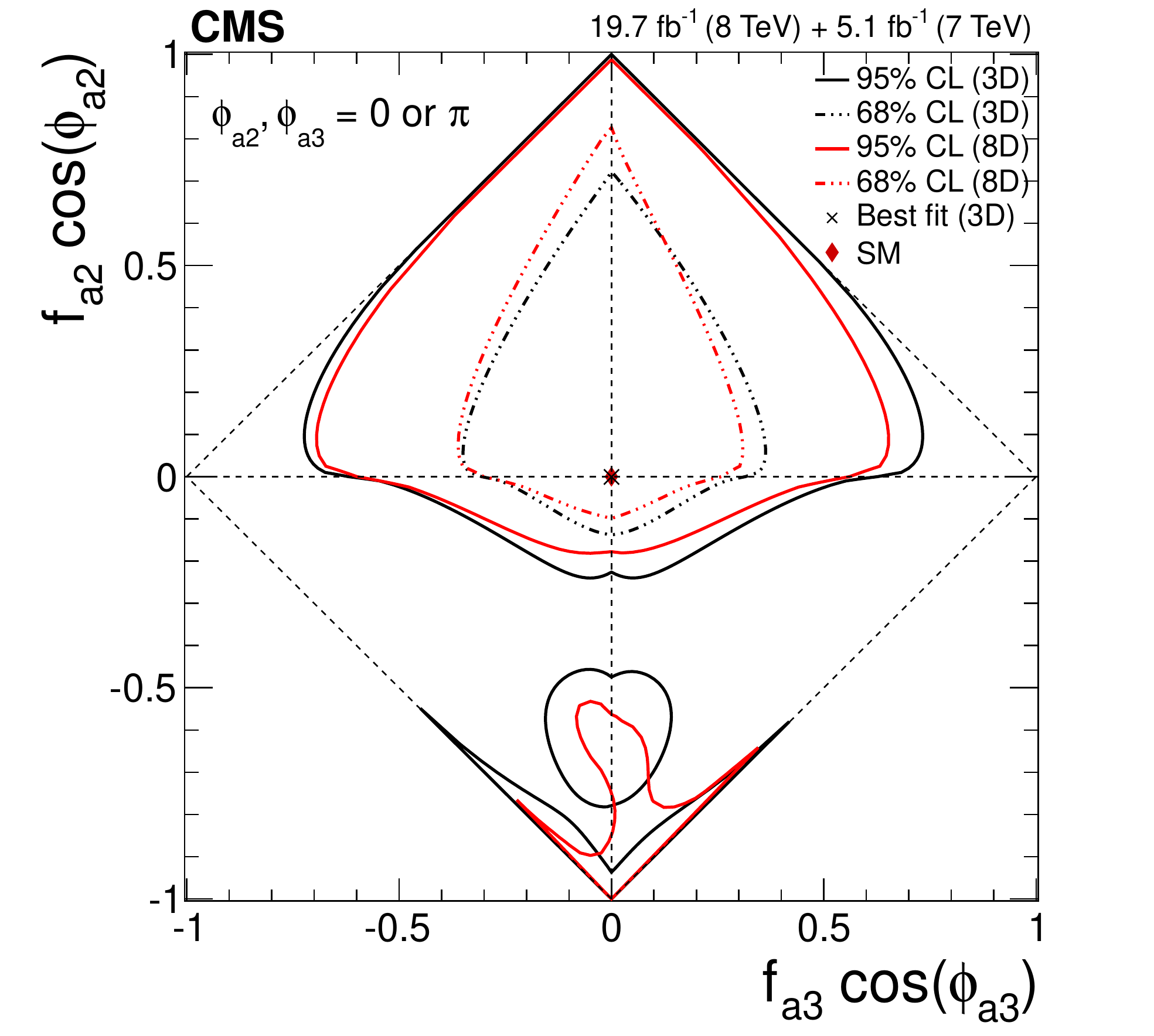}
	\caption{
	Expected (dashed) and observed (solid) likelihood scans for $f_{a2}$ (top left) and $f_{a3}$ (top right),
	and observed likelihood scan for the $f_{a2}$ vs. $f_{a3}$ fractions (bottom),
	obtained using the template method (3D, black) and the multidimensional distribution method (8D, red)
	in the study of anomalous $\PH\Z\Z$ interactions. The couplings are constrained to be real.
	}
	\label{fig:results_ZZ_1D_Compare}

\end{figure*}

It has been shown that the template method with a small set of optimal observables
and multidimensional distribution method are expected to produce equivalent results~\cite{Anderson:2013afp}.
Nonetheless, this is validated explicitly with a subset of the above $\PH\Z\Z$ measurements.
The multidimensional distribution method has been applied to the study of the
$f_{a2}$ and $f_{a3}$ parameters, as shown in Table~\ref{tab:summary_8d}.
Figure~\ref{fig:results_ZZ_1D_Compare} shows the expected
and observed likelihood scans for the effective fractions $f_{a2}$ and $f_{a3}$ under the assumption
of real couplings for both the template and multidimensional distribution methods.
The two methods have a compatible expected performance and the differences are within the systematic uncertainties
of the methods. The observed constraints are not expected to produce identical results because of the incomplete overlap
of the data, which is due to the slightly different selection requirement on $m_{4\ell}$. Also, statistical variations occur
because of the different parameterization of observables. The two methods provide consistent results.

\begin{table}[htbp]
\centering
\topcaption{
Summary of the allowed 95\%~\CL intervals on the anomalous coupling parameters in $\PH\Z\Z$ interactions
under the assumption that all the coupling ratios are real ($\phi_{ai}=0$ or $\pi$) using the multidimensional
distribution method. These results cross-check those presented in Table~\ref{tab:summary_spin0}.
}
\renewcommand{\arraystretch}{1.25}
\begin{scotch}{cccccc}
 & Parameter                                   &  \multicolumn{2}{c}{Observed} &  \multicolumn{2}{c}{Expected} \\
\hline
& $f_{a2}\cos(\phi_{a2})$                    & \multicolumn{2}{c}{$ [-0.14,1.00]$} & \multicolumn{2}{c}{$ [-0.18,0.97] $}                \\
& $f_{a3}\cos(\phi_{a3})$                    & \multicolumn{2}{c}{$ [-0.44,0.40]$} & \multicolumn{2}{c}{$ [-0.67,0.67] $}                \\
\end{scotch}
\label{tab:summary_8d}
\end{table}

\subsection{Study of \texorpdfstring{$\PH\Z\gamma$}{HZ gamma} and \texorpdfstring{$\PH\gamma\gamma$}{H gamma gamma} couplings with the \texorpdfstring{$\PH\to\V\V\to4\ell$}{H to VV to 4l} channel}  \label{sec:Resultspinzerohzg}

\begin{figure*}[htbp]
\centering
	\includegraphics[width=0.42\textwidth,angle=0]{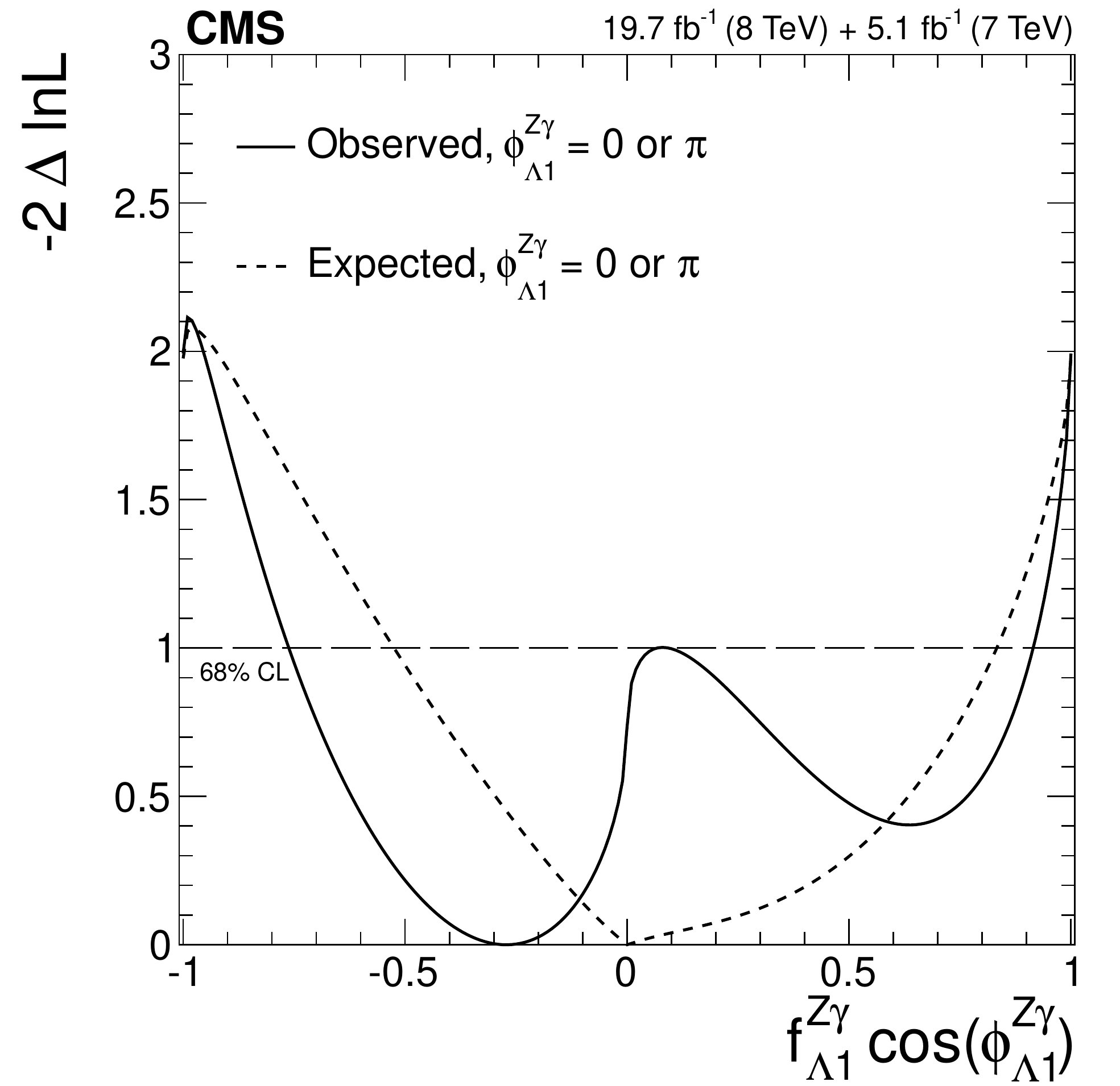} \\
	\includegraphics[width=0.42\textwidth,angle=0]{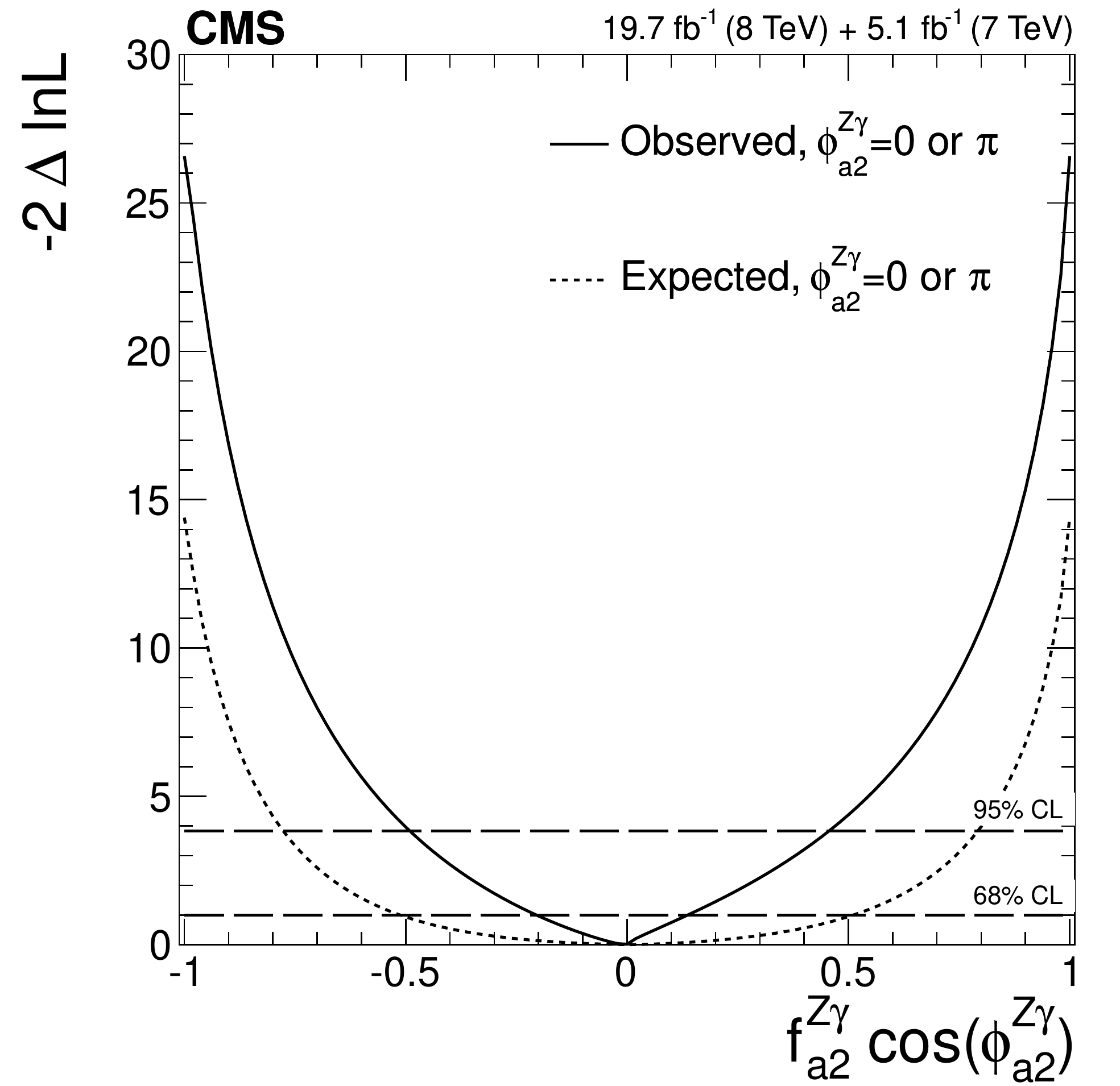}
	\includegraphics[width=0.42\textwidth,angle=0]{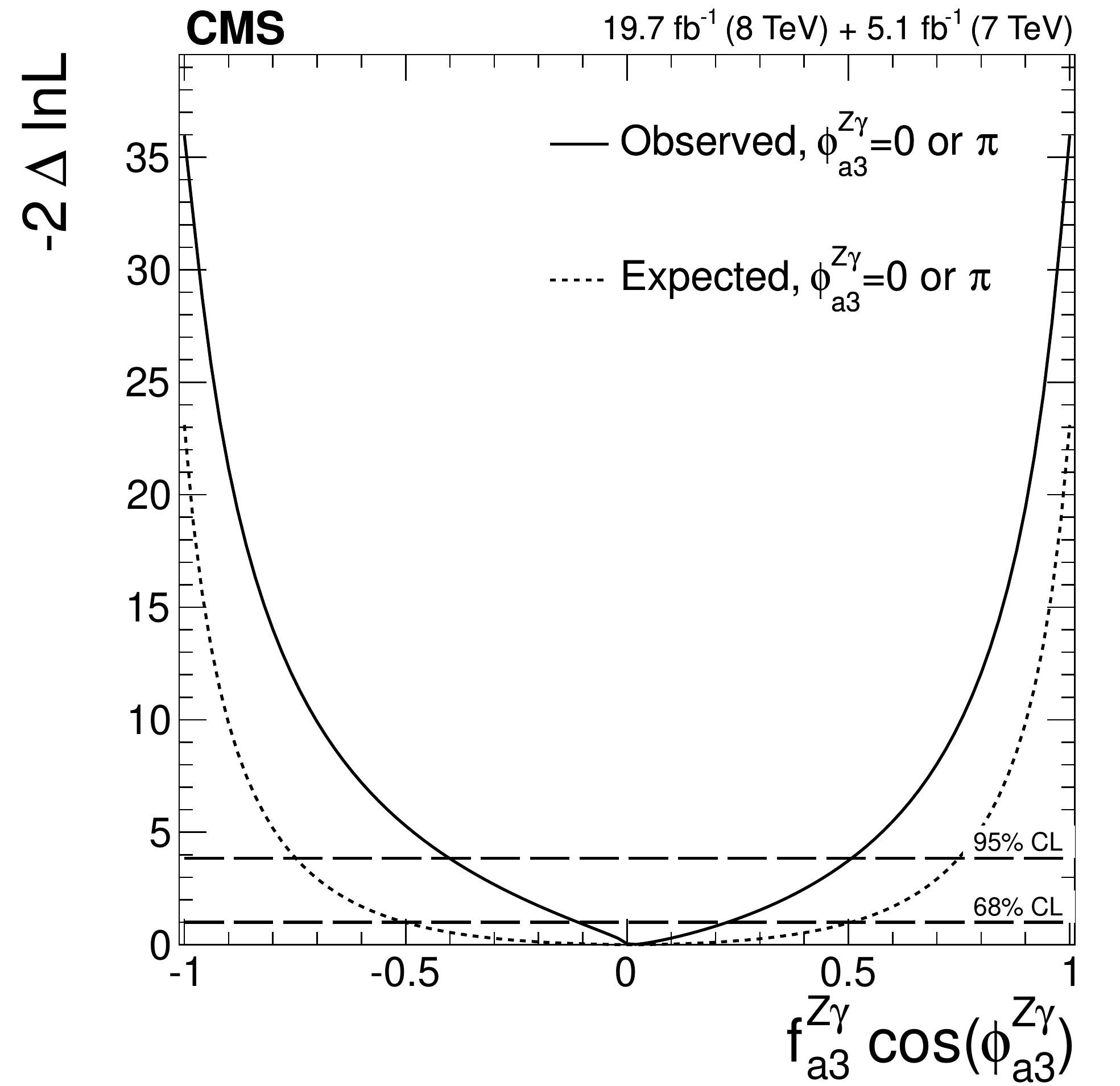} \\
	\includegraphics[width=0.42\textwidth,angle=0]{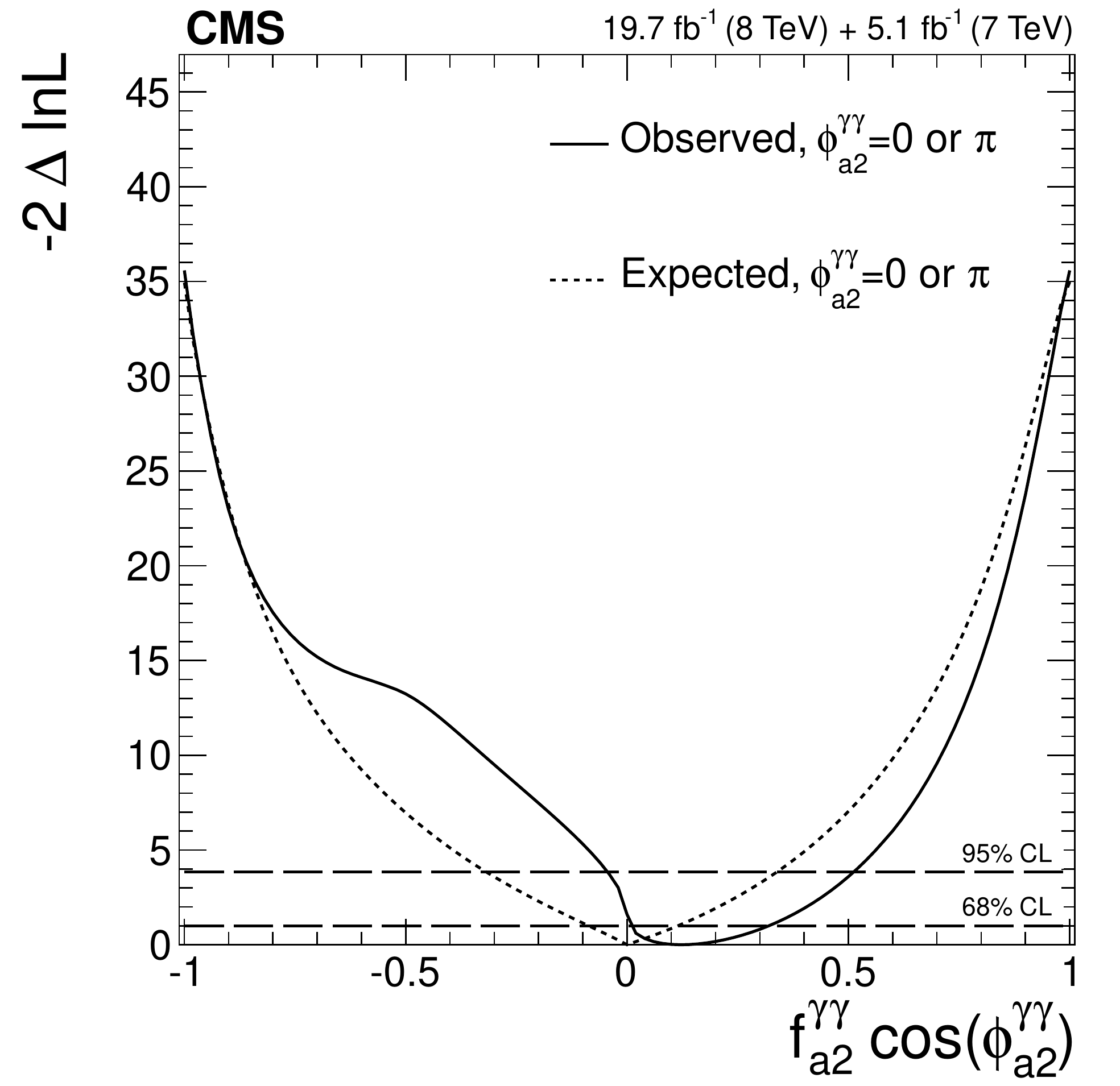}
	\includegraphics[width=0.42\textwidth,angle=0]{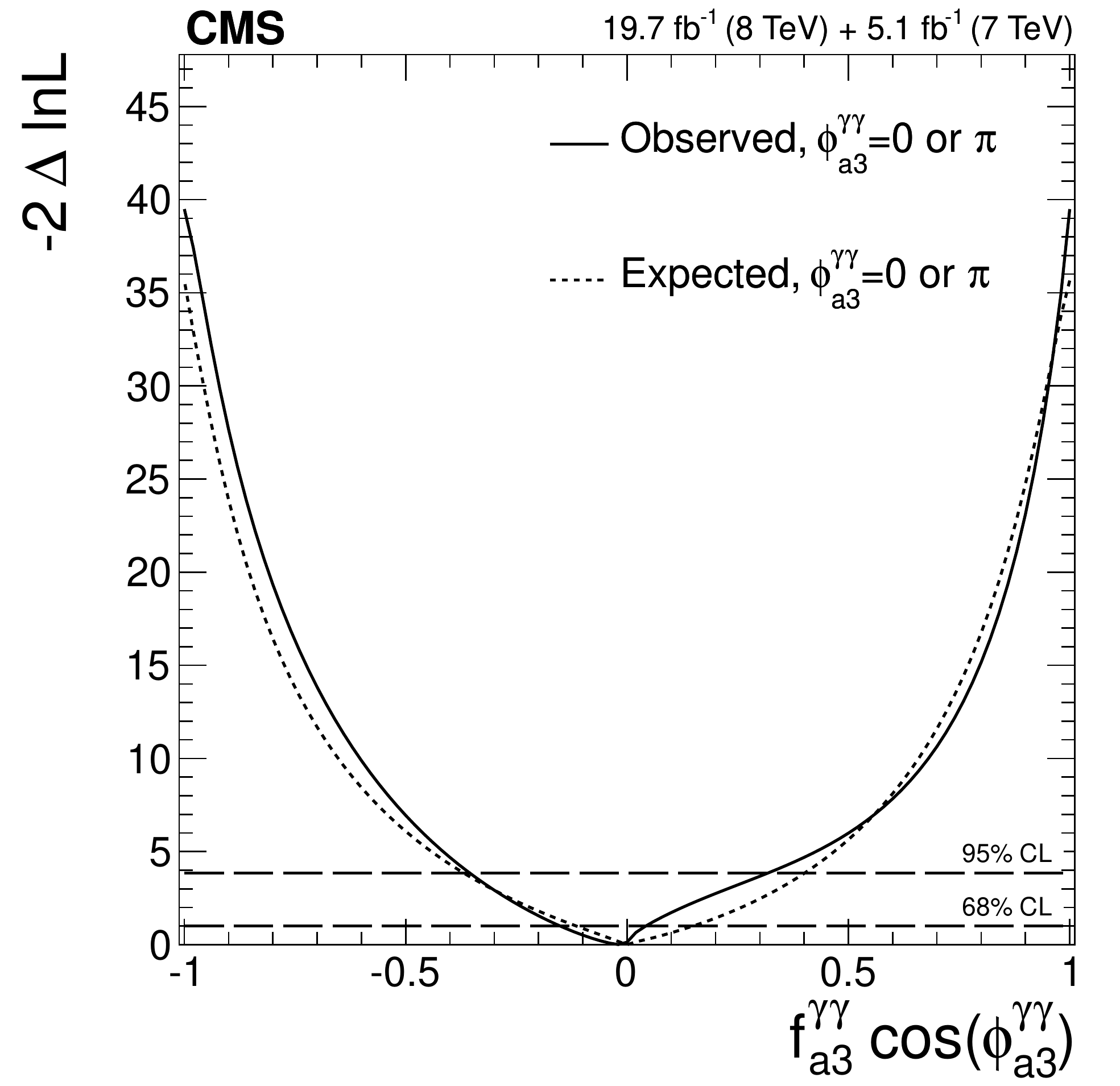}
	\caption{
	Expected (dashed) and observed (solid)  likelihood scans using the template method for the effective fractions
	$f_{\Lambda1}^{Z\gamma}$ (top), $f_{a2}^{Z\gamma}$ (middle left), $f_{a3}^{Z\gamma}$ (middle right),
	$f_{a2}^{\gamma\gamma}$ (bottom left), and  $f_{a3}^{\gamma\gamma}$ (bottom right).
	The couplings studied are constrained to be real and all other couplings are fixed to the SM predictions.
	The $\cos\phi_{ai}^{\V\V}$ term allows a signed quantity where $\cos\phi_{ai}^{\V\V}=-1$ or~$+1$.
	}
	\label{fig:results_ZA_AA_1D}
\end{figure*}

In the following, constraints on anomalous $\PH\Z\gamma$ and $\PH\gamma\gamma$ interactions
are obtained using the $\PH\to \V\V\to 4\ell$ data. Five anomalous couplings are considered,
following Eq.~(\ref{eq:formfact-fullampl-spin0}) and Table~\ref{tab:kdlist}, where the three
observables for each measurement are listed. Only real couplings, $\phi_{ai}=0$ or $\pi$,
are considered in this test. The results of the likelihood function scan for the three parameters,
$f_{ai}\cos\phi_{ai}$, are shown in Fig.~\ref{fig:results_ZA_AA_1D}, following the same
formalism presented for the $\PH\Z\Z$ couplings in Section~\ref{sec:Resultspinzerohzz}.
The 68\% and 95\%~\CL intervals are shown in Table~\ref{tab:summary_spin0}.
In the case of the $f_{\Lambda1}^{Z\gamma}$ measurement, there are two minima and only
one central value with its 68\%~\CL interval is shown in Table~\ref{tab:summary_spin0},
while both 68\%~\CL intervals are presented in Fig.~\ref{fig:spin0_summary}.

\begin{table*}[htbp]
\centering
\topcaption{
Summary of the allowed 95\%~\CL intervals on the anomalous couplings in $\PH\Z\gamma$ and $\PH\gamma\gamma$ interactions
using results obtained with the template method in Table~\ref{tab:summary_spin0}.
The coupling ratios are assumed to be real ($\cos(\phi^{\V\V}_{ai})=0$ or $\pi$).
\label{tab:Spin0zg_interpretation}
}
\renewcommand{\arraystretch}{1.25}
\begin{scotch}{cccccc}
Parameter & Observed & Expected    \\
\hline

$(\Lambda_{1}^{\Z\gamma}\sqrt{\abs{a_1}}) \cos(\phi^{\Z\gamma}_{\Lambda_{1}})$ &   $[-\infty,+\infty]$ & $[-\infty,+\infty]$ \\
$ a^{\Z\gamma}_2/a_{1} $ &   $[-0.046,0.044]$ & $[-0.089,0.092]$ \\
$ a^{\Z\gamma}_{3}/a_{1} $ &   $[-0.042,0.053]$ & $[-0.090,0.090]$ \\
$ a^{\gamma\gamma}_2/a_{1} $ &   $[-0.011,0.054]$ & $[-0.036,0.038]$ \\
$ a^{\gamma\gamma}_{3}/a_{1} $ &   $[-0.039,0.037]$ & $[-0.041,0.044]$ \\
\hline
$ (\sigma^{Z\gamma}_2/\sigma^{\Z\gamma}_\mathrm{SM})(2a_2^{\Z\gamma}/{a_1})^2\cos(\phi_{a2}^{\Z\gamma}) $ &   $[-1.7,1.6]\times10^2$ & $[-6.5,6.9]\times10^2$ \\
$ (\sigma^{Z\gamma}_{3}/\sigma^{\Z\gamma}_\mathrm{SM})(2a_3^{\Z\gamma}/{a_1})^2\cos(\phi_{a2}^{\Z\gamma})  $ &   $[-1.2, 1.9]\times10^2$ & $[-5.5, 5.5]\times10^2$ \\
$ (\sigma^{\gamma\gamma}_2/\sigma^{\gamma\gamma}_\mathrm{SM})(2a_2^{\gamma\gamma}/{a_1})^2\cos(\phi_{a2}^{\gamma\gamma}) $ &   $[-0.3, 7.3]\times10^2$ & $[-3.3,3.6]\times10^2$ \\
$ (\sigma^{\gamma\gamma}_{3}/\sigma^{\gamma\gamma}_\mathrm{SM})(2a_3^{\gamma\gamma}/{a_1})^2\cos(\phi_{a3}^{\gamma\gamma}) $ &   $[-3.8,3.3]\times10^2$ & $[-4.1,4.7]\times10^2$ \\
\end{scotch}
\end{table*}

Using the transformation in Eq.~(\ref{eq:fa_conversion}), these results can be interpreted in terms of the coupling parameters
used in Eq.~(\ref{eq:formfact-fullampl-spin0}) as shown in Table~\ref{tab:Spin0zg_interpretation}.
The ratio $(\sigma^{\V\gamma}_i/\sigma^{\V\gamma}_\mathrm{SM})(2a_i^{\V\gamma}/{a_1})^2$ approximates the ratio
$\mu = \sigma / \sigma_\mathrm{SM}$ of the measured and expected SM cross sections for a Higgs boson decay $\PH\to\V\gamma$.
The ratio $({2}/{a_1})^2$ scales this measurement with respect to the $\PH\to\Z\Z$ coupling and  is expected to be 1.0 in the SM.
As can be seen in Table~\ref{tab:Spin0zg_interpretation}, the constraints presented on these ratios
($<$170 for $\abs{a_2^{\Z\gamma}}$ or $<$730 for $\abs{a_2^{\gamma\gamma}}$ at 95\% \CL)
are about one or three orders of magnitude higher than from the analyses of the direct
$\PH\to\Z\gamma$ ($\mu < 9.5$ at 95\% \CL~\cite{Chatrchyan:2013vaa}) or
$\PH\to\gamma\gamma$ ($\mu = 1.14^{+0.26}_{-0.23}$ at 68\% \CL~\cite{Khachatryan:2014ira})
decays with on-shell photons, respectively.
Therefore, the constraints presented on $f_{a2}^{\Z\gamma}$, $f_{a3}^{\Z\gamma}$, $f_{a2}^{\gamma\gamma}$,
$f_{a3}^{\gamma\gamma}$ are not competitive compared with the direct cross-section measurements
in $\PH\to\Z\gamma$ or $\gamma\gamma$ decays.
However, eventually with sufficiently large integrated luminosity it might be possible to measure
$f_{a2}^{\V\gamma}$ and $f_{a3}^{\V\gamma}$ separately in the $\PH\to \V\V\to 4\ell$ decay,
allowing for measurements of the $CP$ properties in these couplings~\cite{Dawson:2013bba,Chen:2014gka}.
The $\PH\to\Z\gamma$ or $\gamma\gamma$ measurements with on-shell photons are sensitive only to the sum of the
 two cross-section fractions $f_{a2}^{\V\gamma}$ and $f_{a3}^{\V\gamma}$ and therefore cannot distinguish the two.
 Moreover, the $f_{\Lambda1}^{\Z\gamma}$ measurement is not possible with on-shell photons.

As in the case of the $\PH\Z\Z$ couplings, anomalous $\PH\Z\gamma$ and $\PH\gamma\gamma$
couplings are found to be consistent with zero, as expected  in the SM with the current precision.
Since the measurement of the $\PH\Z\gamma$ and $\PH\gamma\gamma$ couplings in the
$\PH\to \V\V\to 4\ell$ decay is not yet competitive with the on-shell measurements, further
investigation of several parameters simultaneously is not considered with the current data.

\subsection{Study of \texorpdfstring{$\PH\PW\PW$}{HWW} couplings with the \texorpdfstring{$\PH\to\PW\PW\to\ell\nu\ell\nu$}{H to WW to l nu l nu} channel}  \label{sec:Resultspinzerohww}

\begin{figure*}[htbp]
  \centering
      \includegraphics[width=0.42\textwidth]{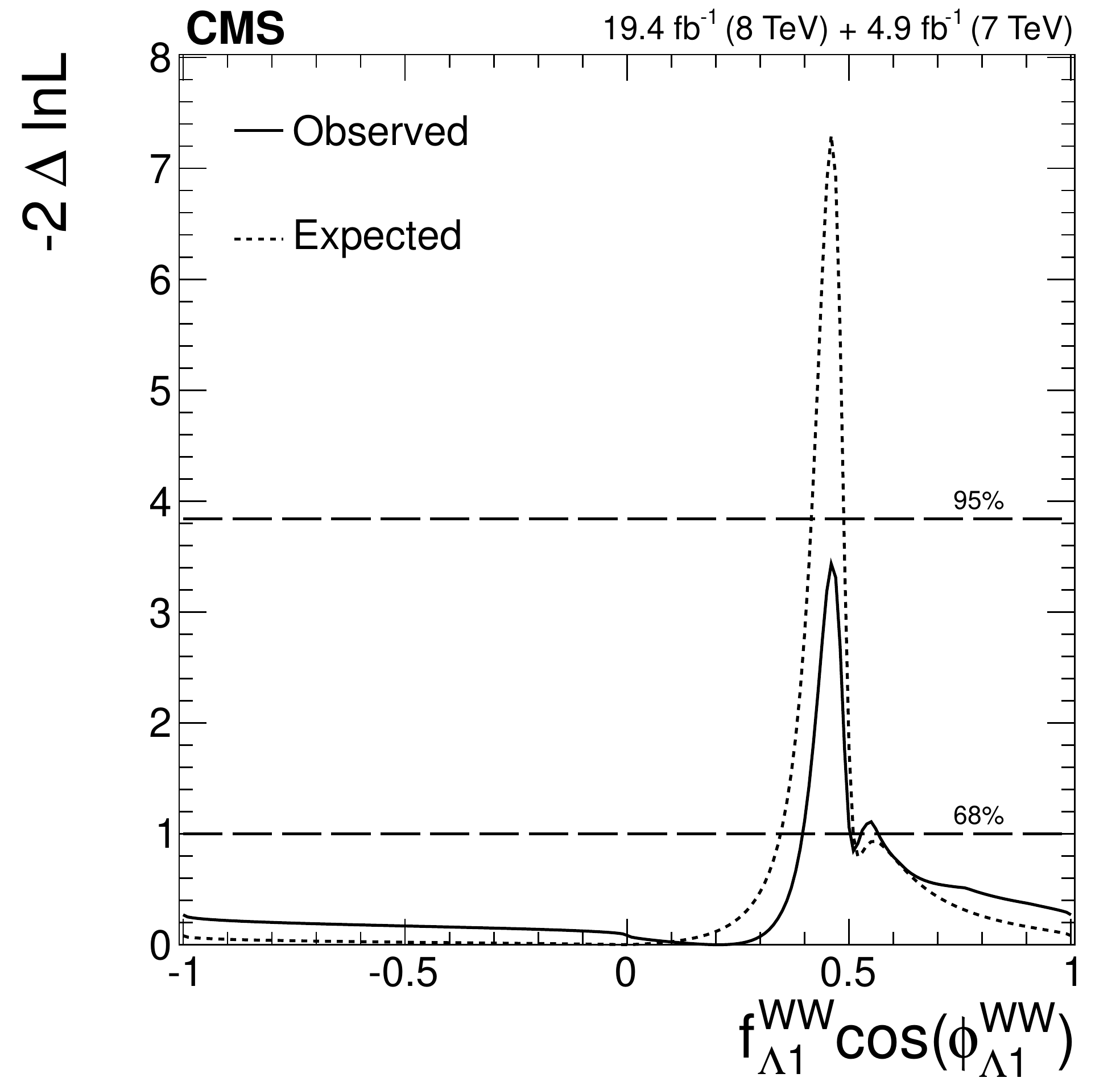}
      \includegraphics[width=0.42\textwidth]{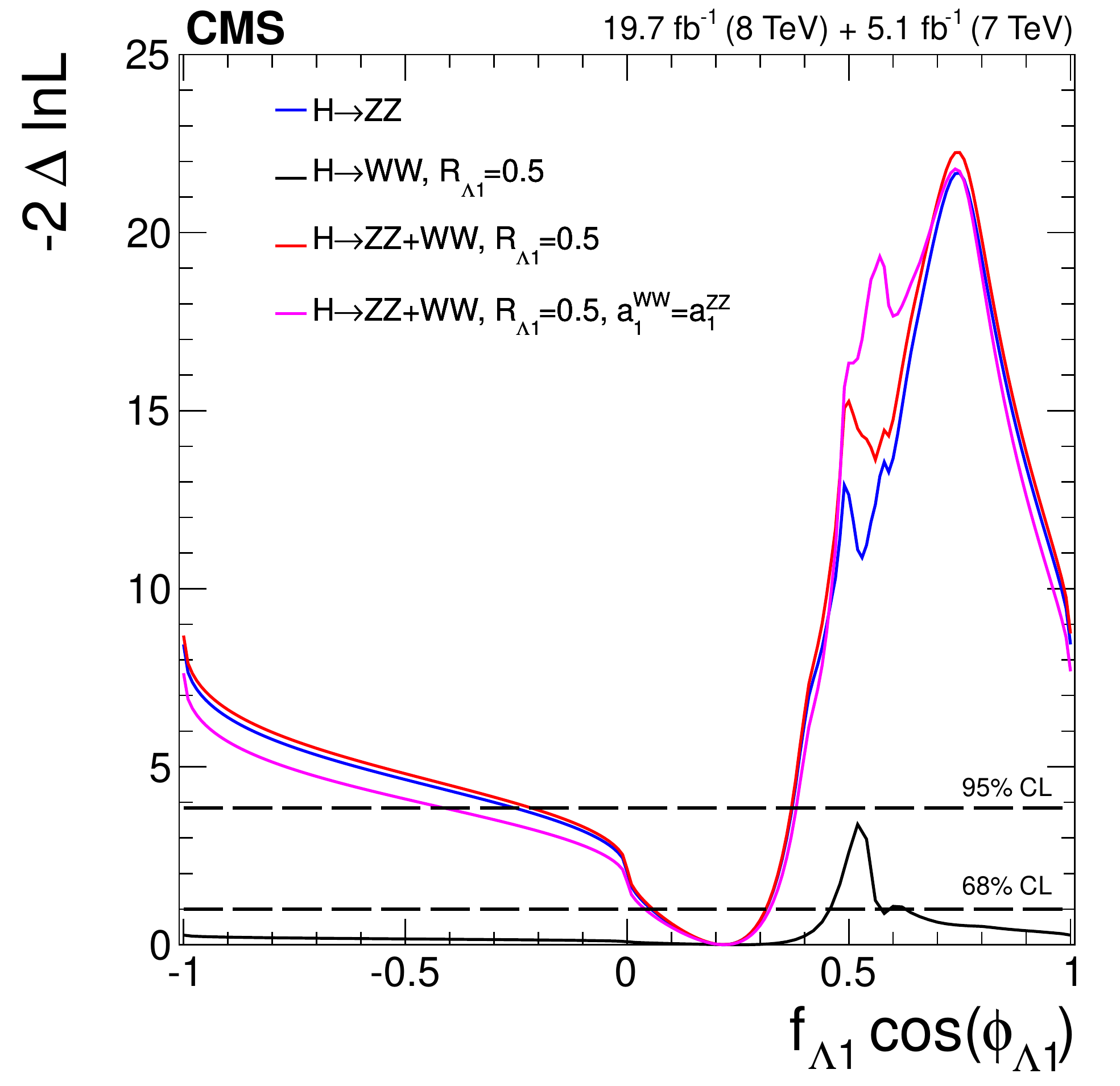} \\
      \includegraphics[width=0.42\textwidth]{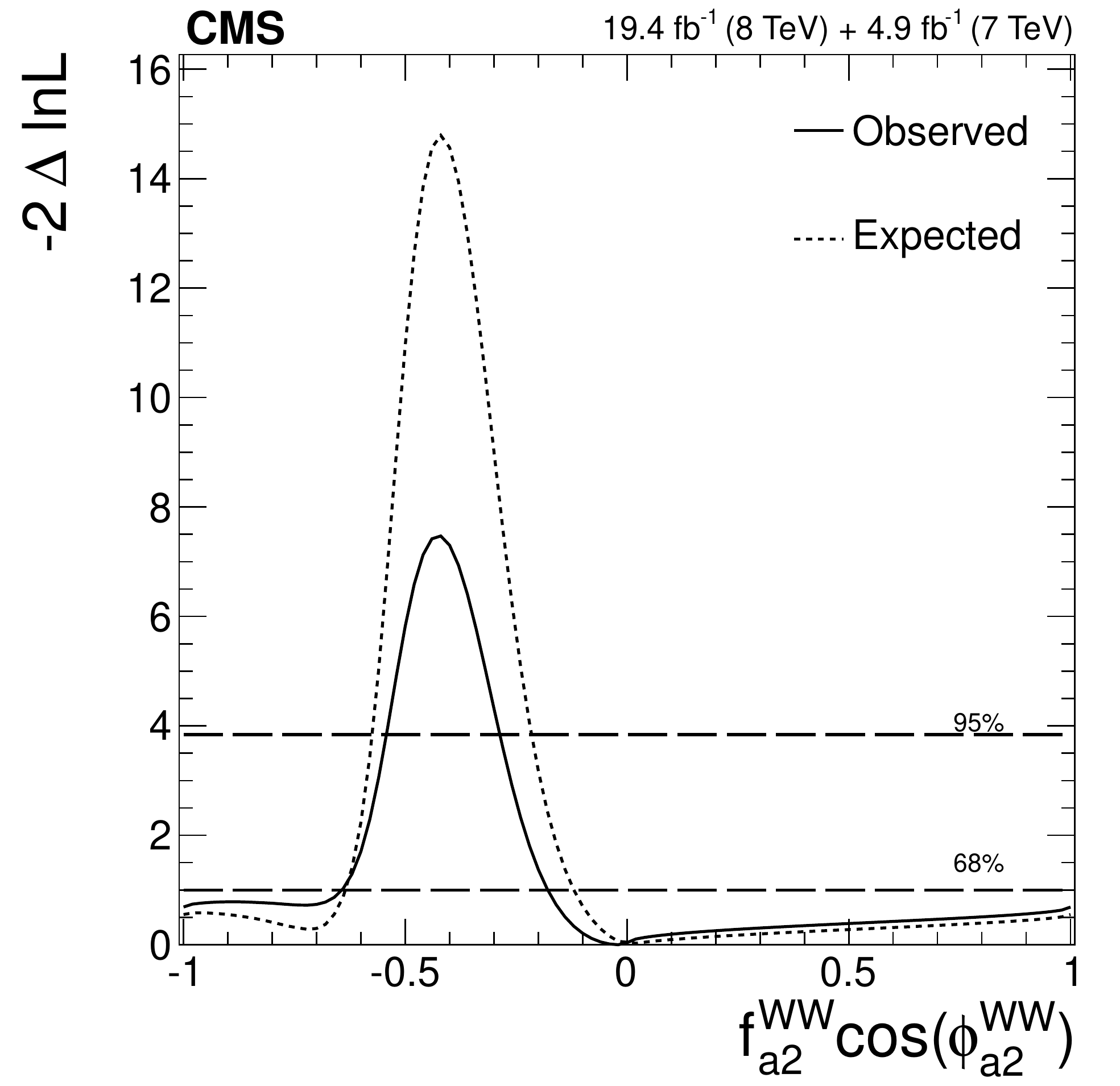}
      \includegraphics[width=0.42\textwidth]{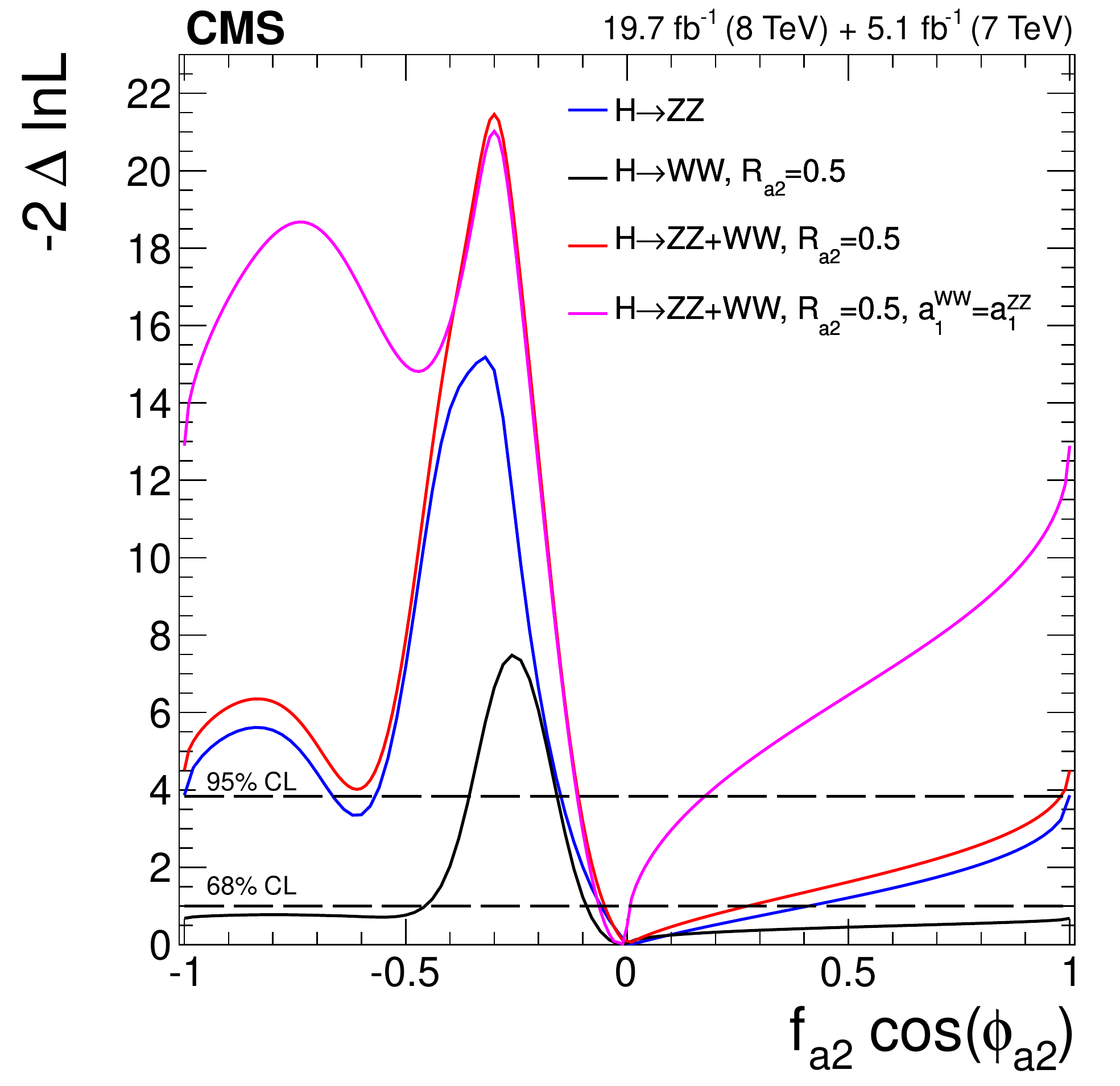} \\
      \includegraphics[width=0.42\textwidth]{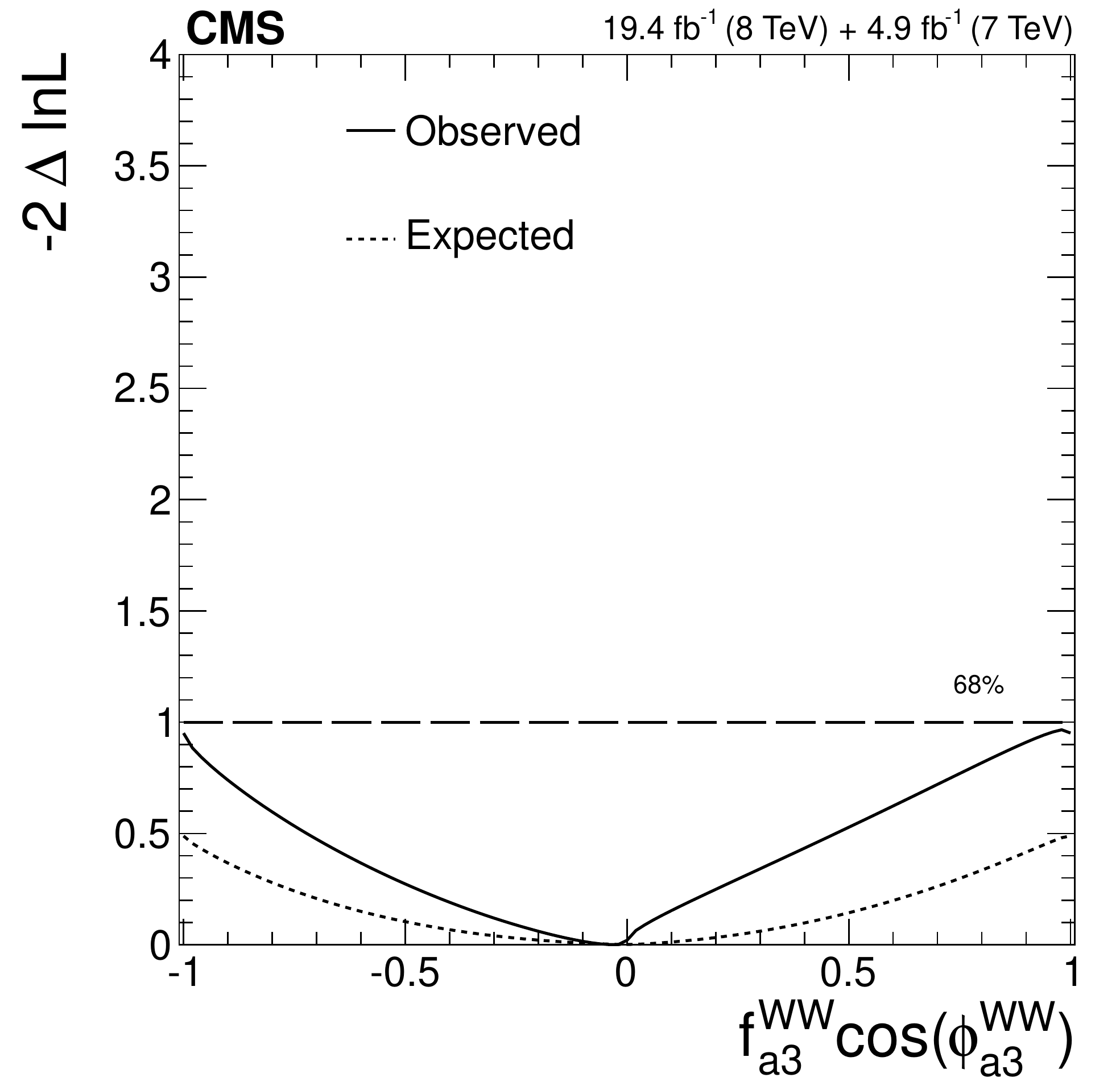}
       \includegraphics[width=0.42\textwidth]{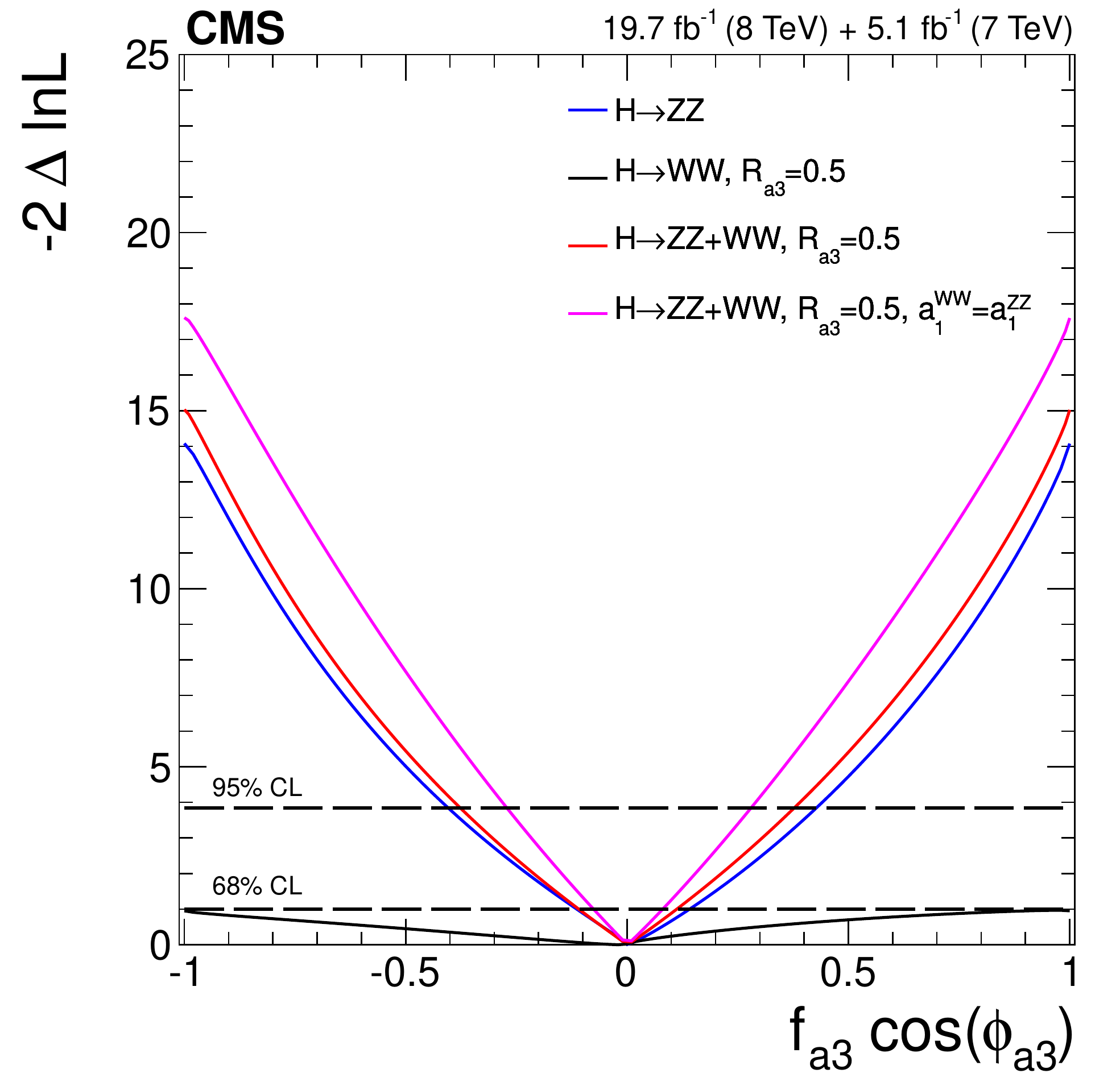}
         \caption{
         Expected (dashed) and observed (solid) likelihood scans for effective fractions
	$f_{\Lambda1}$ (top), $f_{a2}$ (middle), $f_{a3}$ (bottom).
	The couplings studied are constrained to be real and all other anomalous couplings are fixed to the SM predictions.
	The $\cos\phi_{ai}$ term allows a signed quantity where $\cos\phi_{ai}=-1$ or $+1$.
	Plots on the left show the results of the $\PH\to \PW\PW\to \ell\nu\ell\nu$ analysis expressed in terms of the $\PH\PW\PW$ couplings.
	Plots on the right show the combined  $\PH\to \PW\PW$ and $\PH\to \Z\Z$ result in terms of the $\PH\Z\Z$ couplings for $R_{ai} = 0.5$.
         Measurements are shown for each channel separately and two types of combination are present:
         using $a_1^{\PW\PW} = a_1$ (red) and without such a constraint (magenta).
	}
    \label{fig:hwwscans}

\end{figure*}

Constraints on anomalous $\PH\PW\PW$ interactions are obtained using the $\PH\to \PW\PW\to\ell\nu\ell\nu$ final state.
Three measurements are performed using the template method with the two observables, $\mt$ and $\mll$,
as summarized in Table~\ref{tab:kdlist}. Only real couplings, $\phi_{ai}^{\PW\PW}=0$ or $\pi$, are considered. The results of the likelihood function scan for the three parameters,
$f_{ai}^{\PW\PW}\cos\phi_{ai}^{\PW\PW}$, are shown in Fig.~\ref{fig:hwwscans}, following the $\PH\Z\Z$
approach presented in Section~\ref{sec:Resultspinzerohzz}.
The 68\% and 95\%~\CL intervals are shown in Table~\ref{tab:summary_spin0}.
Using the transformation in Eq.~(\ref{eq:fa_conversion}), these results could be interpreted for the coupling parameters
used in Eq.~(\ref{eq:formfact-fullampl-spin0}) as shown in Table~\ref{tab:Spin0ww_interpretation}.

Similarly to the $\PH\Z\Z$ case, strong destructive interference of the SM and anomalous contributions
at $f^{\PW\PW}_{\Lambda1}\cos(\phi^{\PW\PW}_{\Lambda1})\sim+0.5$ or $f^{\PW\PW}_{a2}\cos(\phi^{\PW\PW}_{a2})\sim-0.5$
leads to very different kinematic distributions and exclusions with high confidence levels.
Since the measurement of the $\PH\PW\PW$ anomalous couplings with the $\PH\to \PW\PW\to\ell\nu\ell\nu$ decay
is not expected to provide strong constraints with the current data, a deeper
investigation of several parameters simultaneously is not considered here.
On the other hand, the combination of the $\PH\Z\Z$ and $\PH\PW\PW$ measurements is expected to
provide an improvement in the precision of the $\PH\V\V$ couplings with certain symmetry considerations.

\begin{table*}[htbp]
\centering
\topcaption{
Summary of the allowed 95\%~\CL intervals on the anomalous couplings in $\PH\PW\PW$ interactions
using results obtained with the template method in Table~\ref{tab:summary_spin0}.
The coupling ratios are assumed to be real (including $\cos(\phi^{\PW\PW}_{\Lambda_{1}})=0$ or $\pi$).
\label{tab:Spin0ww_interpretation}
}
\renewcommand{\arraystretch}{1.25}
\begin{scotch}{cccccc}
Parameter  & Observed & Expected    \\
\hline
$(\Lambda^{\PW\PW}_{1}\sqrt{|a^{\PW\PW}_{1}|}) \cos(\phi^{\PW\PW}_{\Lambda_{1}})$ &   $[-\infty, +\infty]$ & $[-\infty, 87\GeV] \cup [93\GeV,+\infty]$      \\
$ a^{\PW\PW}_{2}/a^{\PW\PW}_{1} $ &   $[-\infty,-1.22] \cup [-0.71,+\infty]$ & $[-\infty,-1.30] \cup [-0.59,+\infty]$                                               \\
$ a^{\PW\PW}_{3}/a^{\PW\PW}_{1} $ &   $[-\infty,+\infty]$ & $[-\infty,+\infty]$                                               \\
\end{scotch}
\end{table*}

\subsection{Combination of \texorpdfstring{$\PH\Z\Z$}{HZZ} and \texorpdfstring{$\PH\PW\PW$}{HWW} results} \label{sec:Combination}  \label{sec:Spinzerocombination}

Further improvement on the $\PH\V\V$ anomalous coupling constraints can be obtained from the combination
of the $\PH \to \Z\Z \to 4\ell$ and $\PH \to \PW\PW \to \ell\nu\ell\nu$ analyses by employing symmetry
considerations between the $\PH\Z\Z$ and $\PH\PW\PW$ interactions.
Two scenarios are considered. In the first, custodial symmetry is assumed, leading to $a_1^{WW} = a_1$.
The second scenario assumes no relationship between the two couplings.
In both cases, a combined likelihood scan of $f_{ai}$ is performed for the full range of $-1\le R_{ai}\le +1$.
For a given value of $R_{ai}$ in Eq.~(\ref{eq:ratio_ww_zz}),
the $f_{ai}$ and $f_{ai}^{WW}$ values are related by Eq.~(\ref{eq:a2_conversion}) and constraints on
a single parameter can be obtained.

For the combination where custodial symmetry is assumed, the yield in the $\PH \to \PW\PW \to \ell\nu\ell\nu$
channel is related to the yield in the $\PH \to \Z\Z \to 4\ell$ channel, which leads to stronger constraints.
This yield relationship is possible if either the fraction of VBF and $\V\PH$ production is known or
the reconstruction efficiency for these and gluon fusion production mechanisms is the same. The latter is
known to be somewhat different in the  $\PH \to \PW\PW \to \ell\nu\ell\nu$ channel due to the selection being
sensitive to the associated jets, see Table~\ref{tab:hww01j_yields}. The fraction of VBF and $\V\PH$ production
has been found to be small and consistent with the SM expectation of 12\%~\cite{Khachatryan:2014jba}.
However, this constraint is performed under the assumption of the SM kinematics of associated particles.
While it is possible to obtain similar constraints on associated Higgs boson production with anomalous
couplings, that analysis is beyond the scope of this paper. Therefore, we assume that the gluon fusion
production dominates and the fraction of VBF and $\V\PH$ production is not larger than expected in the SM.
This leads to potential uncertainty on the yield relationship of about 3\%, which we neglect. Should the
fraction of VBF and $\V\PH$ production be different, the corresponding limits could be recalculated.

\begin{table*}[htb]
\centering
\topcaption{
Summary of allowed 68\%~\CL (central values with uncertainties) and 95\%~\CL (ranges in square brackets)
intervals on anomalous coupling parameters in $\PH\V\V$ interactions
in combination of $\PH\Z\Z$ and $\PH\PW\PW$ measurements
assuming the symmetry $a_i = a_i^{\PW\PW}$, including $R_{ai}=0.5$, and real coupling ratios ($\phi_{ai}^{\V\V}=0$ or $\pi$).
The last column indicates the observed (expected) confidence level of a pure anomalous coupling
corresponding to $f_{ai}^{\V\V}=1$ when compared to the SM expectation $f_{ai}^{\V\V}=0$.
The results are obtained with the template method.
}
\renewcommand{\arraystretch}{1.25}
\begin{scotch}{ccccccc}
& Parameter                                   &  \multicolumn{2}{c}{Observed} &  \multicolumn{2}{c}{Expected} & $f_{ai}^{\V\V}=1$\\
\hline
& $f_{\Lambda1}\cos(\phi_{\Lambda1})$        & \multicolumn{2}{c}{ $0.21^{+0.11}_{-0.17}$ $[-0.42,0.38]$}          & \multicolumn{2}{c}{ $0.00^{+0.15}_{-0.80}$ $[-1,0.27]\cup [0.95,1]$}& 0.56\% (13\%)                                            \\
& $f_{a2}\cos(\phi_{a2})$         & \multicolumn{2}{c}{$-0.01^{+0.02}_{-0.05}$ $[-0.11,0.17]$}     & \multicolumn{2}{c}{$0.00^{+0.08}_{-0.03}$ $[-0.07,0.51]$}            & 0.03\% (0.25\%)\\
& $f_{a3}\cos(\phi_{a3})$         & \multicolumn{2}{c}{ $0.00^{+0.08}_{-0.08}$ $[-0.27,0.28]$} & \multicolumn{2}{c}{ $0.00^{+0.23}_{-0.23}$ $[-0.53,0.53]$} & $<$0.01\% (0.08\%) \\
\end{scotch}
\label{tab:combine_spin0}
\end{table*}

\begin{table*}[htb]
\centering
\topcaption{
Summary of the allowed 95\%~\CL intervals on the anomalous couplings in $\PH\V\V$ interactions
in combination of $\PH\Z\Z$ and $\PH\PW\PW$ measurements in Table~\ref{tab:combine_spin0}
assuming the symmetry $a_i = a_i^{\PW\PW}$, including $R_{ai}=0.5$, and real coupling ratios
($\phi_{ai}^{\V\V}=0$ or $\pi$).
\label{tab:Spin0comb_interpretation}}
\renewcommand{\arraystretch}{1.25}
\begin{scotch}{cccccc}
Parameter  & Observed & Expected    \\
\hline
$(\Lambda_{1}\sqrt{|a_{1}|}) \cos(\phi_{\Lambda_{1}})$ &   $[-\infty,-100\GeV] \cup [103\GeV,\infty]$   & $[-\infty,43\GeV] \cup [116\GeV,\infty]$ \\
$ a_{2}/a_{1} $ &   $[-0.58,0.76]$ & $[-0.45,1.67]$ \\
$ a_{3}/a_{1} $ &   $[-1.54,1.57]$ & $[-2.65,2.65]$ \\
\end{scotch}
\end{table*}

\begin{figure*}[htbp]
\centering
\includegraphics[width=0.48\textwidth]{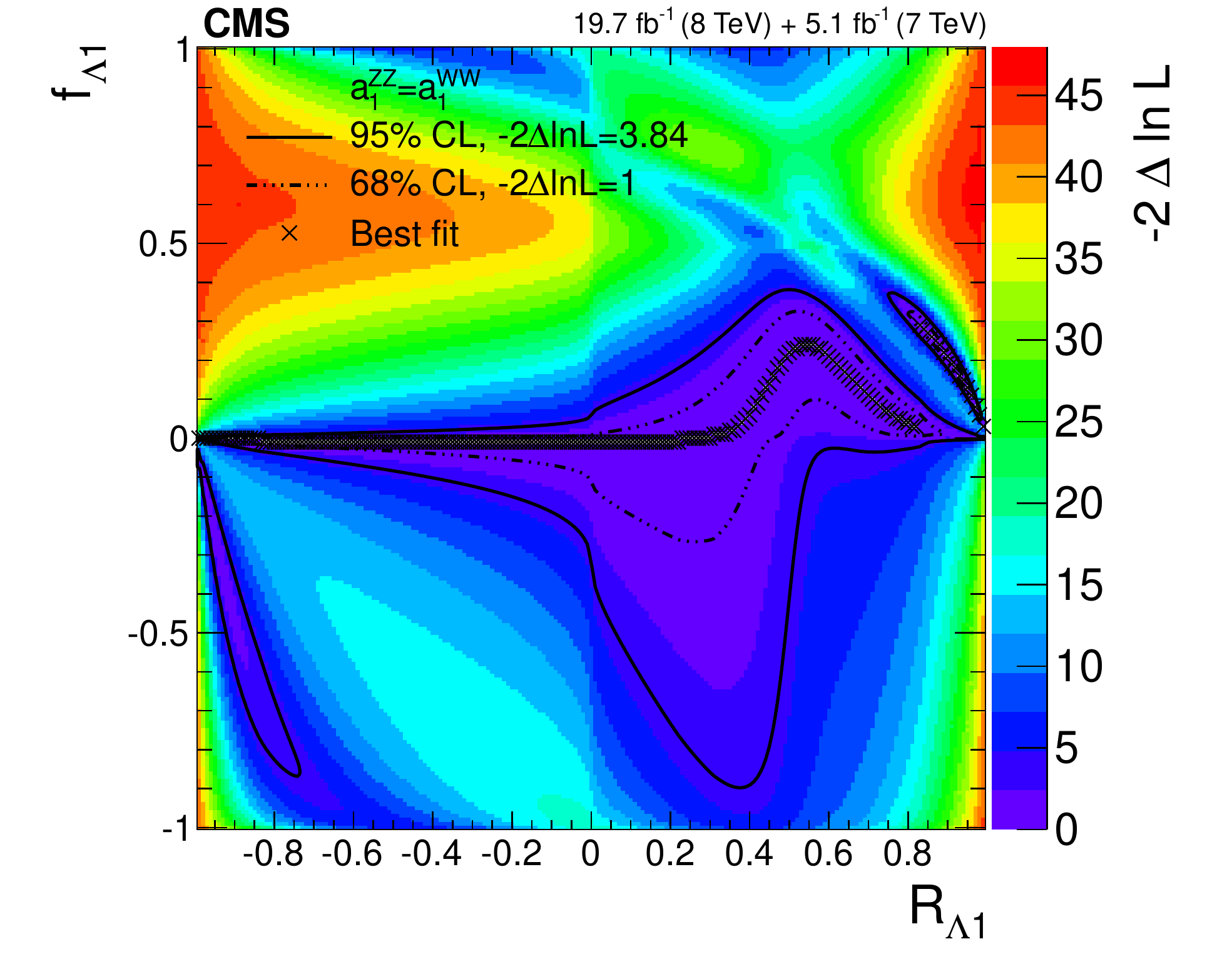}
\includegraphics[width=0.48\textwidth]{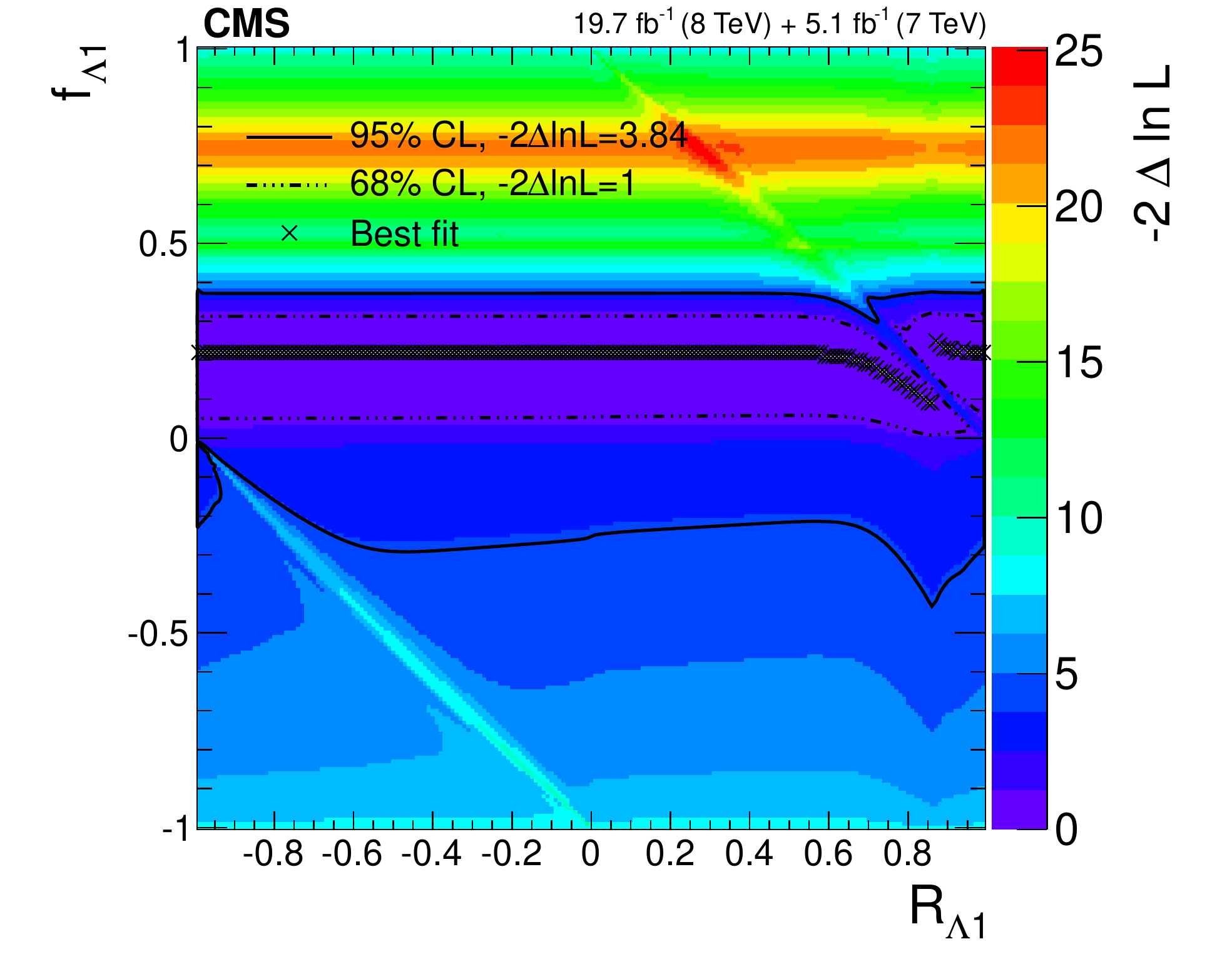}\\
\includegraphics[width=0.48\textwidth]{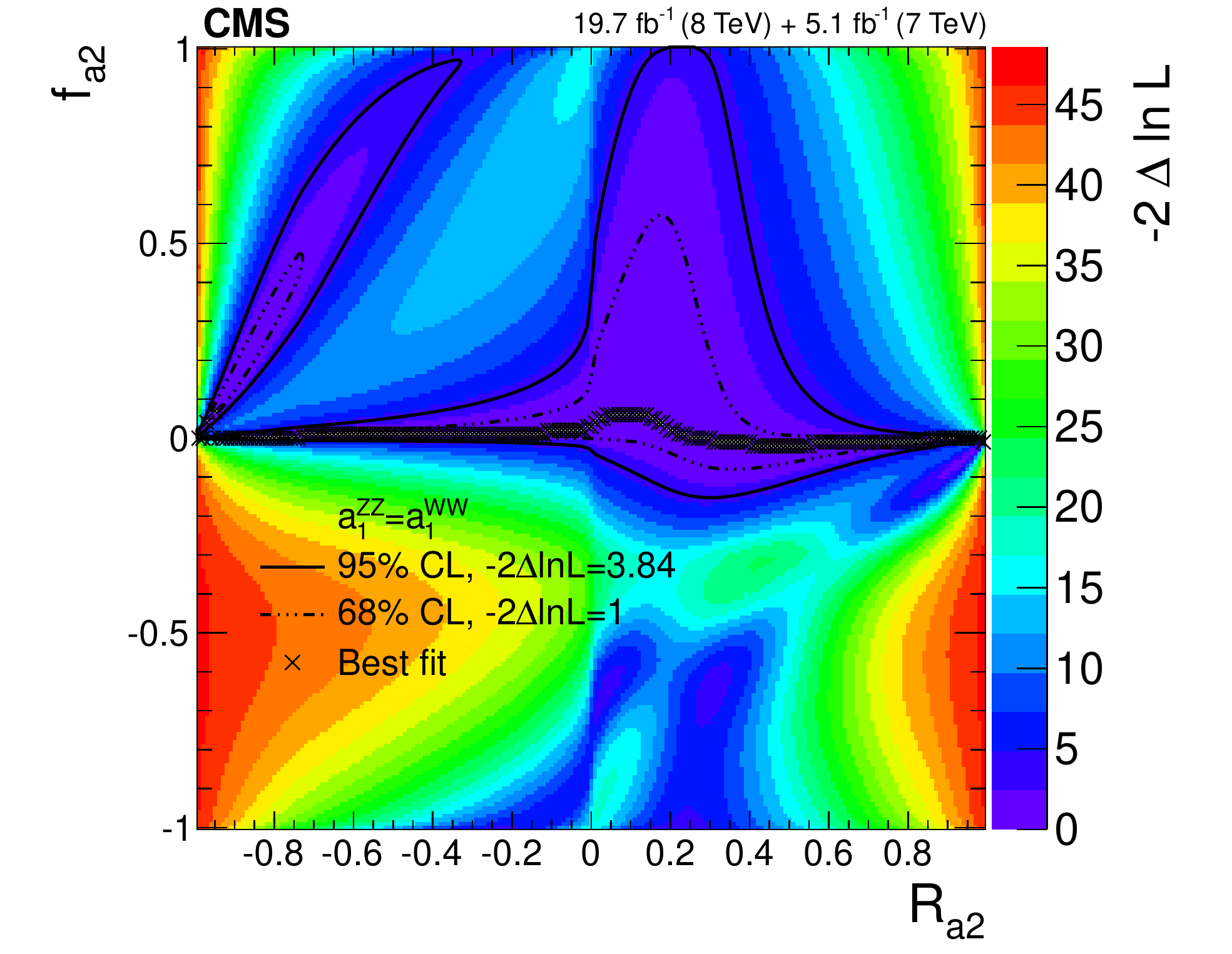}
\includegraphics[width=0.48\textwidth]{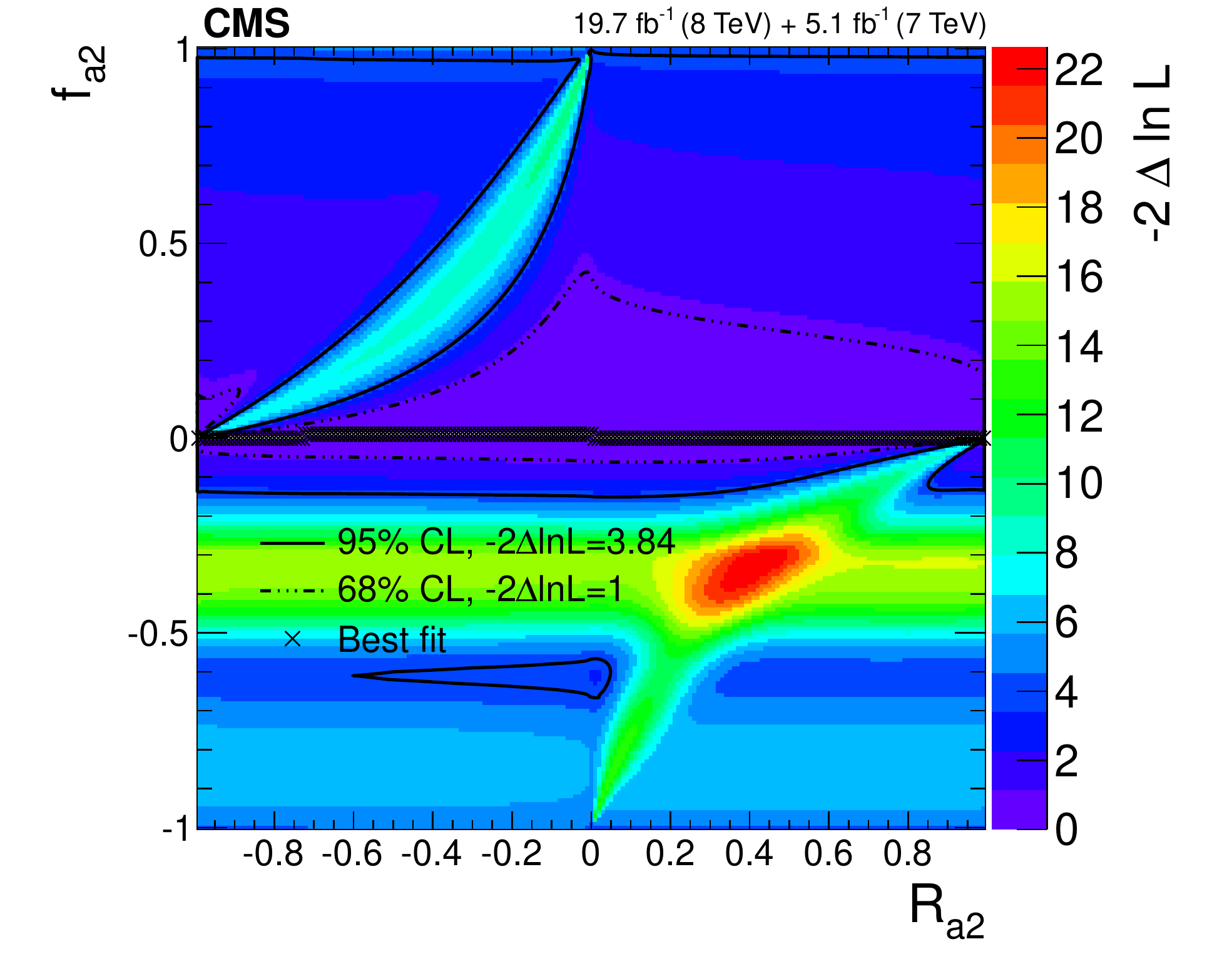}\\
\includegraphics[width=0.48\textwidth]{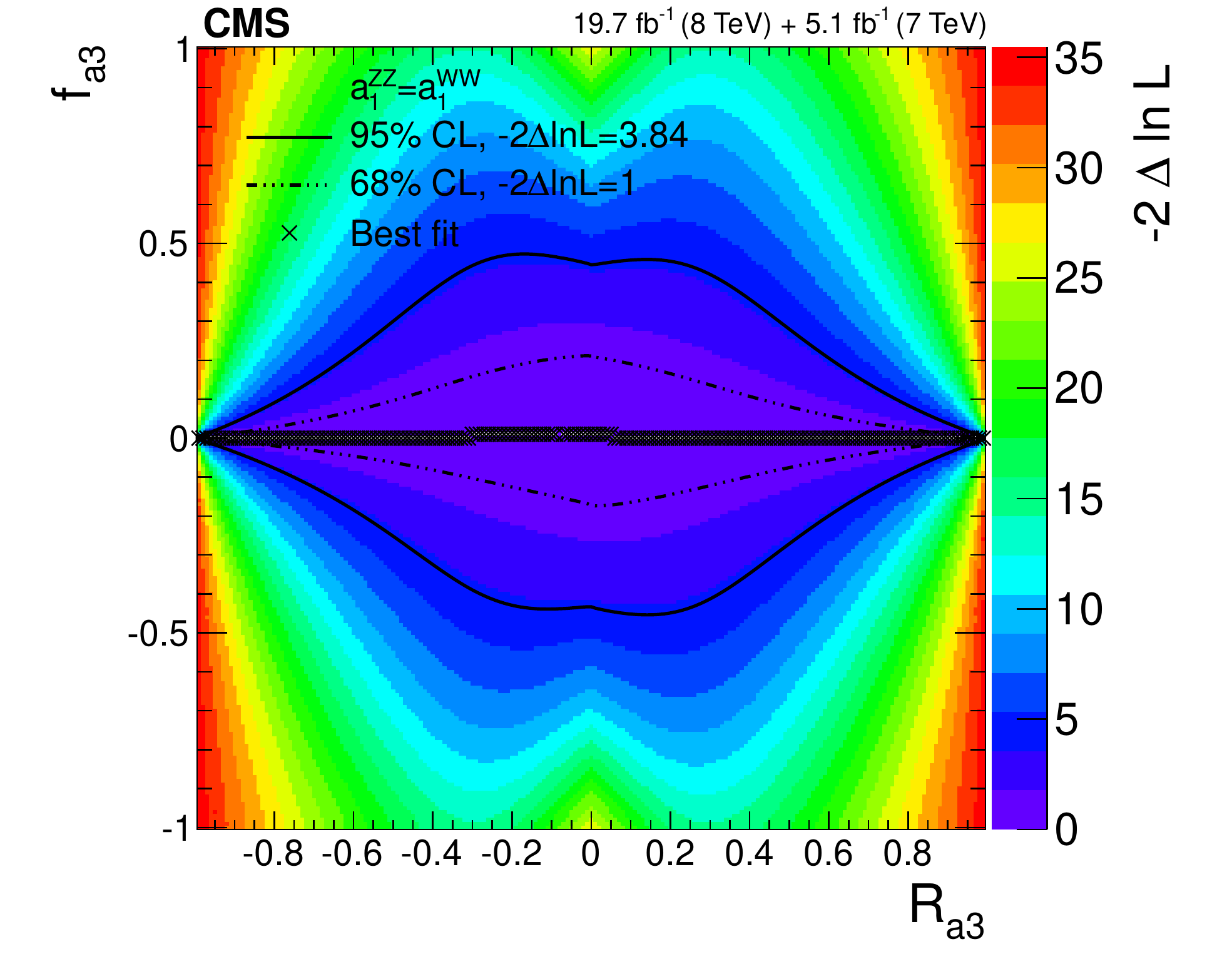}
\includegraphics[width=0.48\textwidth]{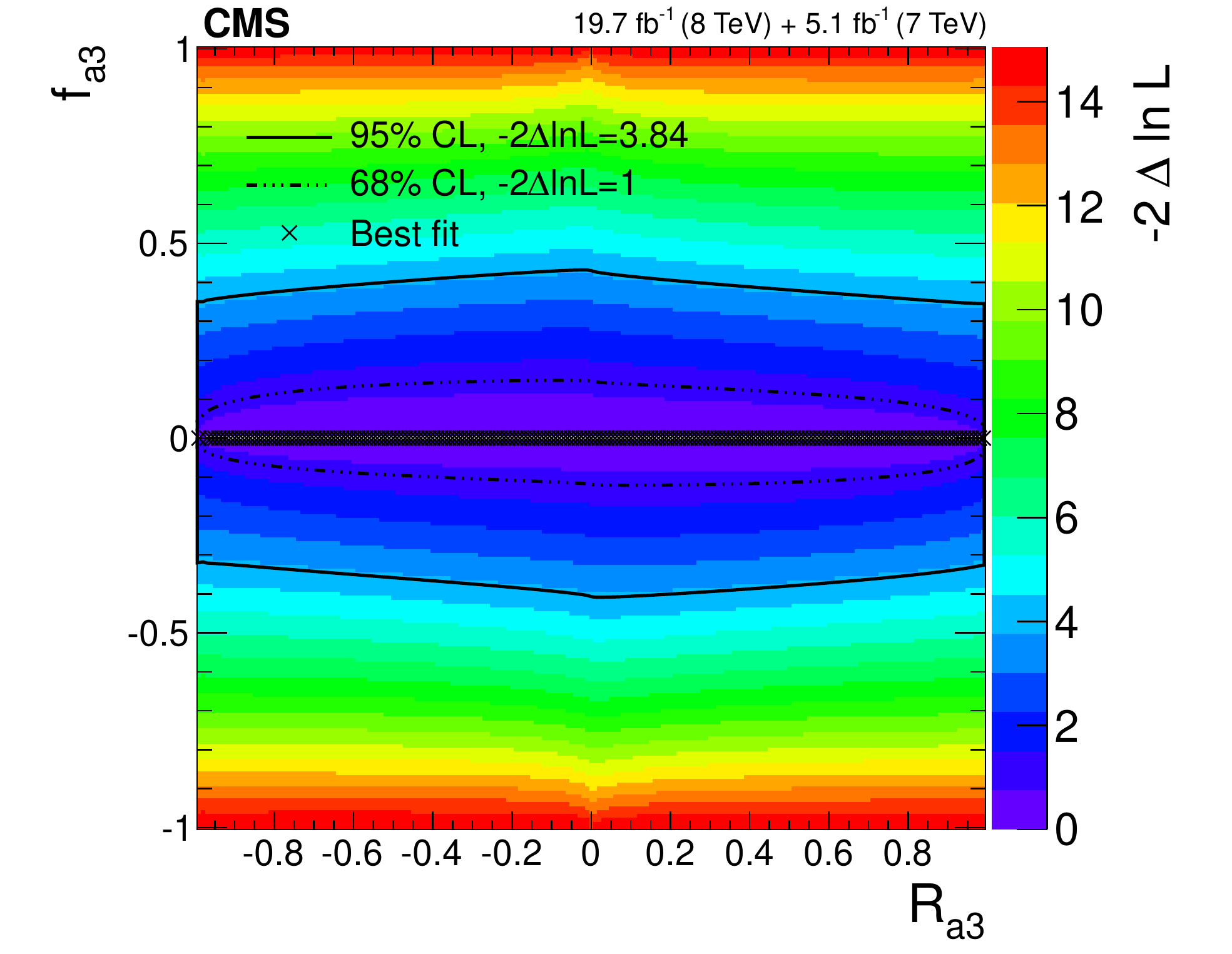}
\caption{
Observed conditional likelihood scans of $f_{\Lambda1}$ (top), $f_{a2}$ (middle), $f_{a3}$ (bottom)
for a given $R_{ai}$ value from the combined analysis of the $\PH \to \PW\PW$ and $\PH \to \Z\Z$ channels
using the template method. The results are shown with custodial symmetry $a_1=a_1^{\PW\PW}$ (left)
and without such an assumption (right).
Each cross indicates the minimum value of $-2\,\Delta\!\ln\mathcal{L}$ and the contours indicate the
one-parameter confidence intervals of $f_{ai}$ for a given value of $R_{ai}$.
}
\label{fig:fa_combine}
\end{figure*}

An example of the combination under the assumption $R_{ai}=0.5$ $(r_{ai} = 1)$ is shown in Fig.~\ref{fig:hwwscans}
and Table~\ref{tab:combine_spin0}, where the effect of using the information on the relative yield can be seen.
Both combination scenarios are shown. When the $\PH \to \Z\Z \to 4\ell$ and $\PH \to \PW\PW \to \ell\nu\ell\nu$ signal
yields are left independent, custodial symmetry is not assumed.
The increase in expected signal yield towards $f_{ai}=\pm1$ is greater in the $\PH \to \PW\PW$ channel
compared to the $\PH \to \Z\Z$ channel, leading to additional discriminating power when the yields are related.
For example, the enhancement relative to the SM is a factor of 2.4 for $f_{a2}=\pm1$.
Since the number of events observed in the $\PH \to \Z\Z \to 4\ell$ and $\PH \to \PW\PW \to \ell\nu\ell\nu$
channels is compatible with the SM, this enhancement of 2.4 is not compatible with the SM and the $f_{a2}=\pm1$
scenario is strongly excluded from the consideration of the yields alone. The combined analysis uses both yield
and kinematic information in an optimal way.
Using the transformation in Eq.~(\ref{eq:fa_conversion}), these results could be interpreted for the coupling parameters
used in Eq.~(\ref{eq:formfact-fullampl-spin0}) as shown in Table~\ref{tab:Spin0comb_interpretation}.

To present the results in a model-independent way,
conditional likelihood scans of $f_{ai}$, for a particular $R_{ai}$ value, are performed.
In this way, the confidence intervals of $f_{ai}$ are obtained for a given value of $R_{ai}$.
These results are presented in Fig.~\ref{fig:fa_combine}, and show features that arise
from the combination of the $\PH\to \Z\Z$ and $\PW\PW$ channels,
but with larger exclusion power in some areas of the parameter space.

\section{Summary} \label{sec:Conclusion}

In this paper, a comprehensive study of the spin-parity properties of the recently discovered $\PH$ boson
and of the tensor structure of its interactions with electroweak gauge bosons is presented using the
$\PH\to\Z\Z$, $\Z \gamma^*$, $\gamma^* \gamma^*\to4\ell$, $\PH\to\PW\PW\to\ell\nu\ell\nu$,
and $\PH\to\gamma\gamma$~decay modes.
The results are based on the 2011 and 2012 data from pp collisions recorded with the CMS detector at the LHC,
and correspond to an integrated luminosity of up to \usedLumiA \ at a center-of-mass energy of 7\TeV and
up to \usedLumiB at 8\TeV.

The phenomenological formulation for the interactions of a spin-zero, -one, or -two boson with
the SM particles is based on a scattering amplitude or, equivalently, an effective field theory Lagrangian,
with operators up to dimension five. The dedicated simulation and matrix element likelihood
approach for the analysis of the kinematics of $\PH$ boson production and decay in different topologies
are based on this formulation. A maximum likelihood fit of the signal and background distributions provides
constraints on the anomalous couplings of the $\PH$ boson.

The study focuses on testing for the presence of anomalous effects in $\PH\Z\Z$ and $\PH\PW\PW$ interactions
under spin-zero, -one, and -two hypotheses. The combination of the $\PH\to\Z\Z$ and $\PH\to\PW\PW$
measurements leads to tighter constraints on the $\PH$ boson spin-parity and anomalous $\PH\V\V$ interactions.
The combination with the $\PH\to\gamma\gamma$ measurements also allows tighter constraints in the spin-two case.
The $\PH\Z \gamma$ and $\PH\gamma\gamma$  interactions are probed
for the first time using the $4\ell$ final state.

The exotic-spin study covers the analysis of mixed-parity spin-one states and ten spin-two hypotheses
under the assumption of production either via gluon fusion or quark-antiquark annihilation, or without
such an assumption. The spin-one hypotheses are excluded at a greater than 99.999\% \CL
in the $\Z\Z$ and $\PW\PW$ modes, while in the $\gamma\gamma$ mode they are excluded
by the Landau-Yang theorem.
The spin-two boson with gravity-like minimal couplings is excluded at a 99.87\% \CL, and the other
spin-two hypotheses tested are excluded at a 99\% \CL or higher.

Given the exclusion of the spin-one and spin-two scenarios,
constraints are set on the contribution of eleven anomalous couplings to the
$\PH\Z\Z$, $\PH\Z\gamma$, $\PH\gamma\gamma$, and $\PH\PW\PW$ interactions
of a spin-zero $\PH$ boson, as summarized in Table~\ref{tab:summary_spin0}.
Among these is the measurement of the $f_{a3}$ parameter,
which is defined as the fractional pseudoscalar cross section in the $\PH\to\Z\Z$ channel.
The constraint is $f_{a3}<0.43$ (0.40) at a 95\% \CL for the positive (negative)
phase of the pseudoscalar coupling with respect to the dominant SM-like coupling
and $f_{a3}=1$ exclusion of a pure pseudoscalar hypothesis at a 99.98\% \CL.

All observations are consistent with the expectations for a scalar SM-like Higgs boson.
It is not presently established that the interactions of the observed state conserve $C$-parity or $CP$-parity.
However, under the assumption that both quantities are conserved, our measurements require the quantum numbers
of the new state to be $J^{PC}=0^{++}$. The positive $P$-parity follows from the $f_{a3}^{\V\V}$ measurements
in the $\PH\to\Z\Z$, $\Z \gamma^*$, $\gamma^* \gamma^*\to4\ell$, and $\PH\to\PW\PW\to\ell\nu\ell\nu$ decays
and the positive $C$-parity follows from observation of the $\PH\to\gamma\gamma$ decay.
Further measurements probing the tensor structure of the $\PH\V\V$ and $\PH f\bar f$ interactions can test the assumption of $CP$-invariance.

\begin{acknowledgments}\label{sec:Acknowledgments}
\hyphenation{Bundes-ministerium Forschungs-gemeinschaft Forschungs-zentren} We thank Markus Schulze for optimizing the \textsc{JHUGen} Monte Carlo simulation program for this analysis. We congratulate our colleagues in the CERN accelerator departments for the excellent performance of the LHC and thank the technical and administrative staffs at CERN and at other CMS institutes for their contributions to the success of the CMS effort. In addition, we gratefully acknowledge the computing centres and personnel of the Worldwide LHC Computing Grid for delivering so effectively the computing infrastructure essential to our analyses. Finally, we acknowledge the enduring support for the construction and operation of the LHC and the CMS detector provided by the following funding agencies: the Austrian Federal Ministry of Science, Research and Economy and the Austrian Science Fund; the Belgian Fonds de la Recherche Scientifique, and Fonds voor Wetenschappelijk Onderzoek; the Brazilian Funding Agencies (CNPq, CAPES, FAPERJ, and FAPESP); the Bulgarian Ministry of Education and Science; CERN; the Chinese Academy of Sciences, Ministry of Science and Technology, and National Natural Science Foundation of China; the Colombian Funding Agency (COLCIENCIAS); the Croatian Ministry of Science, Education and Sport, and the Croatian Science Foundation; the Research Promotion Foundation, Cyprus; the Ministry of Education and Research, Estonian Research Council via IUT23-4 and IUT23-6 and European Regional Development Fund, Estonia; the Academy of Finland, Finnish Ministry of Education and Culture, and Helsinki Institute of Physics; the Institut National de Physique Nucl\'eaire et de Physique des Particules~/~CNRS, and Commissariat \`a l'\'Energie Atomique et aux \'Energies Alternatives~/~CEA, France; the Bundesministerium f\"ur Bildung und Forschung, Deutsche Forschungsgemeinschaft, and Helmholtz-Gemeinschaft Deutscher Forschungszentren, Germany; the General Secretariat for Research and Technology, Greece; the National Scientific Research Foundation, and National Innovation Office, Hungary; the Department of Atomic Energy and the Department of Science and Technology, India; the Institute for Studies in Theoretical Physics and Mathematics, Iran; the Science Foundation, Ireland; the Istituto Nazionale di Fisica Nucleare, Italy; the Ministry of Science, ICT and Future Planning, and National Research Foundation (NRF), Republic of Korea; the Lithuanian Academy of Sciences; the Ministry of Education, and University of Malaya (Malaysia); the Mexican Funding Agencies (CINVESTAV, CONACYT, SEP, and UASLP-FAI); the Ministry of Business, Innovation and Employment, New Zealand; the Pakistan Atomic Energy Commission; the Ministry of Science and Higher Education and the National Science Centre, Poland; the Funda\c{c}\~ao para a Ci\^encia e a Tecnologia, Portugal; JINR, Dubna; the Ministry of Education and Science of the Russian Federation, the Federal Agency of Atomic Energy of the Russian Federation, Russian Academy of Sciences, and the Russian Foundation for Basic Research; the Ministry of Education, Science and Technological Development of Serbia; the Secretar\'{\i}a de Estado de Investigaci\'on, Desarrollo e Innovaci\'on and Programa Consolider-Ingenio 2010, Spain; the Swiss Funding Agencies (ETH Board, ETH Zurich, PSI, SNF, UniZH, Canton Zurich, and SER); the Ministry of Science and Technology, Taipei; the Thailand Center of Excellence in Physics, the Institute for the Promotion of Teaching Science and Technology of Thailand, Special Task Force for Activating Research and the National Science and Technology Development Agency of Thailand; the Scientific and Technical Research Council of Turkey, and Turkish Atomic Energy Authority; the National Academy of Sciences of Ukraine, and State Fund for Fundamental Researches, Ukraine; the Science and Technology Facilities Council, UK; the US Department of Energy, and the US National Science Foundation.

Individuals have received support from the Marie-Curie programme and the European Research Council and EPLANET (European Union); the Leventis Foundation; the A. P. Sloan Foundation; the Alexander von Humboldt Foundation; the Belgian Federal Science Policy Office; the Fonds pour la Formation \`a la Recherche dans l'Industrie et dans l'Agriculture (FRIA-Belgium); the Agentschap voor Innovatie door Wetenschap en Technologie (IWT-Belgium); the Ministry of Education, Youth and Sports (MEYS) of the Czech Republic; the Council of Science and Industrial Research, India; the HOMING PLUS programme of Foundation for Polish Science, cofinanced from European Union, Regional Development Fund; the Compagnia di San Paolo (Torino); the Consorzio per la Fisica (Trieste); MIUR project 20108T4XTM (Italy); the Thalis and Aristeia programmes cofinanced by EU-ESF and the Greek NSRF; and the National Priorities Research Program by Qatar National Research Fund.
\end{acknowledgments}

\bibliography{auto_generated}
\cleardoublepage \appendix\section{The CMS Collaboration \label{app:collab}}\begin{sloppypar}\hyphenpenalty=5000\widowpenalty=500\clubpenalty=5000\textbf{Yerevan Physics Institute,  Yerevan,  Armenia}\\*[0pt]
V.~Khachatryan, A.M.~Sirunyan, A.~Tumasyan
\vskip\cmsinstskip
\textbf{Institut f\"{u}r Hochenergiephysik der OeAW,  Wien,  Austria}\\*[0pt]
W.~Adam, T.~Bergauer, M.~Dragicevic, J.~Er\"{o}, M.~Friedl, R.~Fr\"{u}hwirth\cmsAuthorMark{1}, V.M.~Ghete, C.~Hartl, N.~H\"{o}rmann, J.~Hrubec, M.~Jeitler\cmsAuthorMark{1}, W.~Kiesenhofer, V.~Kn\"{u}nz, M.~Krammer\cmsAuthorMark{1}, I.~Kr\"{a}tschmer, D.~Liko, I.~Mikulec, D.~Rabady\cmsAuthorMark{2}, B.~Rahbaran, H.~Rohringer, R.~Sch\"{o}fbeck, J.~Strauss, W.~Treberer-Treberspurg, W.~Waltenberger, C.-E.~Wulz\cmsAuthorMark{1}
% ...
\vskip\cmsinstskip
% ...
1:~~Also at Vienna University of Technology, Vienna, Austria\\
2:~~Also at CERN, European Organization for Nuclear Research, Geneva, Switzerland\\
% ...

\end{sloppypar}
\end{document}